%% file: thesis.tex
\pdfoutput=1
\documentclass[a4paper,12pt]{report}
\usepackage[left=2cm,right=2cm,top=2.5cm,bottom=2.5cm]{geometry}
\include{thesis.preamble}

\begin{document}

\title{\LARGE {\bf The role of advection in phase-separating binary liquids}\\
 \vspace*{6mm}
}

\author{Lennon \'O N\'araigh}
\submitdate{January 2008}

\normallinespacing
\maketitle

\preface
\input{abstract/abstract}

\input{acknowledgements/acknowledgements}

\input{dedication/dedication}

\body
\input{introduction/introduction}

\input{background/background}

%
%
%
%
\input{chaotic_advection/chaotic_advection}

\input{estimating_mixedness/estimating_mixedness}

\input{analysis_thin_films/analysis_thin_films}

\input{simulations_thin_films/simulations_thin_films}

\input{singular_solutions/gilbert_eqn}

\input{conclusion/conclusion}

%
%
%
%
%

\input{bibliography/bibliography}

%
%
%
\appendix
\input{appendix_a/appendix_a}

\input{appendix_b/appendix_b}
\end{document}

%% file: thesis.preamble.tex
\usepackage{graphicx}
\usepackage{subfigure}
\usepackage{verbatim}
\usepackage{latexsym}
\usepackage{setspace}

\usepackage{amsmath}
\usepackage{amsfonts}
\usepackage{bm}
\usepackage{wasysym}
\usepackage{subfigure}
\usepackage{color}

\input{blocked.sty}
\input{uhead.sty}
\input{boxit.sty}
\input{icthesis.sty}

\graphicspath{{figs_png/}}

\newcommand{\Small}{\varepsilon}
\newcommand{\Smalll}{\delta}
\newcommand{\Diff}{D}
\newcommand{\Tr}{\gamma}
\newcommand{\Dperp}{\nabla_{\perp}}

\newcommand{\n}{n} 
\newcommand{\D}{D}
\newcommand{\Time}{T}
\newcommand{\Deform}{W}

\newcommand{\op}{\hat{Q}}

\newcommand{\ord}{N}

\newcommand{\Mag}{\bm{m}}
\newcommand{\Subm}{m}

\newcommand{\Int}{\int_{-\infty}^{\infty}}
\newcommand{\ordp}{\psi}

\newcommand{\normallinespacing}{\renewcommand{\baselinestretch}{1.5} \normalsize}

%% file: abstract/abstract.tex
\addcontentsline{toc}{chapter}{Abstract}

\begin{abstract}

Using the advective Cahn--Hilliard equation as a model, we illuminate the
role of advection in phase-separating binary liquids.  The advecting velocity
is either prescribed, or is determined by an evolution equation that accounts
for the feedback of concentration gradients into the flow.

After obtaining some general results about the existence and regularity of
solutions to the model equation, we focus on two specific cases: advection
by a chaotic flow, and coupled Navier--Stokes Cahn--Hilliard equations in
a thin geometry.

By numerically simulating chaotic flow, we show that it is possible to overwhelm
the segregation by vigorous stirring, and to create a homogeneous state.
 We analyze the mixing properties of the model: by measuring fluctuations
 of the concentration away from its mean value, we find \emph{a priori} bounds
 on the amount of homogenization achievable.

We discuss the Navier--Stokes Cahn--Hilliard equations and derive a thin-film
version of these equations.  We examine the dynamical coupling of the concentration
and velocity (backreaction).  To study long-time behaviour, we regularize
the equations with a Van der Waals potential. We obtain existence and regularity
results for the thin-film equations; the analysis also provides a nonzero
lower bound for the film's height, which prevents rupture.

We carry out numerical simulations of the thin-film equations and,
by comparing
the results with experiments on polymer blends, we show that
our model captures the qualitative features of real binary liquids.  The
outcome of the phase separation depends strongly on the backreaction, which
we demonstrate by applying a shear stress at the film's surface. When the
backreaction is small, the domain boundaries align with the direction of
the stress, while for larger backreaction strengths, the domains align in
the perpendicular direction.

Lastly, we compare and contrast the Cahn--Hilliard equation with other models
of aggregation; this leads us to investigate the orientational Holm--Putkaradze
model.  We demonstrate the emergence of singular solutions in this system,
which we interpret as the formation of magnetic particles.  Using elementary
dynamical systems arguments, we classify the interactions of these particles.

\end{abstract}

%% file: acknowledgements/acknowledgements.tex
\cleardoublepage

\addcontentsline{toc}{chapter}{Acknowledgements}

\begin{acknowledgements}
I would like to thank my supervisor Jean-Luc Thiffeault for his time, dedication,
and fruitful exchanges in respect of this research.  I am grateful to Darryl
Holm for the introduction he gave me into Cahn--Hilliard dynamics.  My other
colleagues at Imperial College have always proved ready to answer my
questions; in particular, I gratefully acknowledge the patience of Colin
Cotter and Mark Haskins, and the generosity of Grigorios Pavliotis, who enthusiastically
lent me what added up to be a significant fraction of his library.

I acknowledge the financial support of Imperial College and the
effort of John Gibbons and Jean-Luc Thiffeault in arranging this.  I am thankful
for the additional support of the Government of Ireland through the Easter
Week Scholarship scheme.  The support of EPSRC through payment of research
expenses is acknowledged.

I am grateful to my parents for their constant encouragement of my academic
career and for their exhortations never to give up, and to Goodenough College,
in whose safe confines in leafy Bloomsbury my wife and I stayed for the duration
of my research.  This leads me to thank my wife, Brooke, to whom I am indebted
for her support and encouragement during my research, and for her general
and great kindness.
\end{acknowledgements}

%% file: dedication/dedication.tex
\cleardoublepage

\begin{dedication}
  To my wife, Brooke: \emph{is duitse a scr\'iobham an tr\'achtas seo,
  a st\'oir\'in}.
\end{dedication}

%% file: introduction/introduction.tex
\chapter{Introduction}
\label{ch:introduction}

\renewcommand{\thefootnote}{\fnsymbol{footnote}}

\subsection*{Overview, review, preview}

In her column in \emph{The Guardian}, Maureen Lipman has depicted the routine
of making home-made mayonnaise:
\begin{quotation}
One day she... gave me a demonstration on how to make mayonnaise.  I had
no idea it was so technical... She whisked the mustard with one yolk for
a few minutes, then started dribbling in the oil.  As soon as any separation
appeared she whisked even faster and continued whisking and oiling for long
enough to make my wrist hurt, let alone hers.  It was riveting, like watching
an old master mixing his ochres with his burnt siennas.~\cite{Maureen_Lipman}
\end{quotation}
In this report we shall study this process in a more abstract setting, together
with the parallel problem of phase separation in thin layers.  We wish to
characterize the competing effects of homogenization and phase
separation in a stirred binary fluid.  The advective Cahn--Hilliard equation
describes precisely these mechanisms and this will be the subject of
our report.  In this chapter we shall review the literature pertaining to
both stirred and unstirred phase separation under Cahn--Hilliard (CH) dynamics
and then outline the study we have made.
 
In a seminal paper, Cahn and Hilliard~\cite{CH_orig} introduced their
eponymous equation to model the dynamics of phase separation.  They pictured
a
binary alloy in a mixed state, cooling below a critical temperature.  This
state is unstable to small perturbations so that fluctuations cause the alloy
to separate into bubbles (or domains) rich in one material or the other.
 A
small transition layer separates the bubbles.  Because the evolution of the
concentration field is an order-parameter equation, this description is
completely general.  Thus, by coupling the phase-separation model to fluid
equations, one has a broad description of binary fluid flow encompassing
polymers, immiscible binary fluids with interfacial tension, glasses, and
mayonnaise.

In industrial applications, the components of a binary mixture often need
to
be mixed in a homogeneous state~\cite{Karim2002}.  In that case, the coarsening
tendency of
multiphase fluids is undesirable.  In this report we shall examine how a
stirring flow affects this mixing process.  We are interested in stirring
not
just to limit bubble growth, as is often effected with shear flows, but to
break up the bubbles into a homogeneous mixture.  We shall also examine the
interplay between the concentration gradients and the fluid velocity, and
shall find that this coupling effect can play an important role in the control
of phase separation

We outline previous work for three increasing levels of complexity:
CH in the absence of flow, passive CH with flow, and active
CH with flow.

\emph{Cahn--Hilliard fluids without flow:} There is a substantial literature
that treats of the CH equation without flow.  For a review, see~\cite{Bray_advphys}.
 The main physical feature of this simpler model is
the existence of a single lengthscale (the bubble or domain size) such that
quantities of interest (e.g., the correlation function) are self-similar
under this scale.  This length grows in time as $t^{1/3}$, a result
seen in many numerical simulations (see, for example, Zhu et
al.~\cite{Zhu_numerics}).  Of mathematical interest are the existence of
a Lyapunov functional, and hence of a bounded solution~\cite{Elliott_Zheng},
and the failure of the maximum principle owing to fourth-order derivatives
in the equation for the concentration field~\cite{Gajewski_nonlocal,Elliott_varmob}.

\emph{Passive Cahn--Hilliard fluids:} Given an externally-imposed flow (e.g.,
a shear flow, turbulence, or a stirring motion), we have a passive tracer
equation for a CH fluid.  There have been several studies
of the (active and passive) shear-driven CH fluid~\cite{shear_Shou, shear_Berthier}.
 In these studies, it is claimed that domain elongation persists in a direction
 that asymptotically aligns with the flow direction, while domain formation
 is arrested in a direction perpendicular to this axis.  This seems to be
 confirmed in the experiments of Hashimoto~\cite{Hashimoto}, although these
 results could be due to finite-size effects~\cite{shear_Bray}, in which
 the arrest of domain growth is due to the limitation of the boundary on
 domain size.   Of interest is the existence or otherwise of an arrest scale
 and the dependence of this scale on the system parameters.

A more effective form of stirring involves chaotic flows, in which neighbouring
particle trajectories separate exponentially in time, leading to chaotic
advection~\cite{Aref1984}.  The average rate of separation is the Lyapunov
exponent, positive for such a flow~\cite{Eckmann1985,WigginsIntroDynSys}.
 One may also regard this exponent as the average rate of strain of the flow.
  By imposing a chaotic flow on the CH fluid, finite-size effects are easily
  overcome. This is because the shear direction changes in time, so that
  domain elongation in one particular direction, leading to an eventual
  steady state in which domains touch many boundaries, is no longer possible.
  Berthier et al.~\cite{chaos_Berthier} investigated the passive stirring
  of the CH fluid and observed a coarsening arrest arising from a dynamical
  balance of surface tension and advection.  The stirring by the chaotic
  flow destroys large-scale structures, and this is balanced by the domain-forming
  tendency of the CH equation.  A balancing scale is identified with the
  equilibrium domain size.

The importance of exponential stretching is amplified by the results of
Lacasta et al.~\cite{Lacasta1995}, in which a passive turbulent flow endowed
with
a tunable amplitude, correlation length, and correlation time, causes both
arrest and indefinite growth, in different limits.  We explain this heuristically
as follows: In one limit of turbulent flow (small Prandtl number), velocity
correlations are short in range compared to bubble size, fluid particles
diffuse~\cite{Taylor1921}, and bubble radii evolve as ${d R_b^2}/dt = 2\kappa_{\mathrm{eff}}
+ \text{[Cahn--Hilliard contribution]}$, where $R_b$ is the bubble radius
and $\kappa_{\mathrm{eff}}$ is the effective tracer diffusivity.  Because
$\kappa_{\mathrm{eff}} > 0$, the presence of diffusion fails to arrest bubble
growth.  In the limit of large Prandtl number however, the velocity field
has long-range correlations and exhibits exponential separation of trajectories,
so that the radius equation is ${d R_b^2}/{dt}=-2\lambda R_b^2+ \text{[Cahn--Hilliard
contribution]}$, where $\lambda$ is the average rate of strain, or Lyapunov
exponent, of the flow.  Thus, the exponential stretching due to stirring
balances the bubble growth.

\emph{Active Cahn--Hilliard fluids:} At the highest level of complexity,
one considers the reaction of the mixture on the flow by coupling the CH
equation to the Navier--Stokes equations.  This choice of coupling is not
unique, although the different models agree in the case that interests us,
namely density-matched fluids~\cite{LowenTrus, Berti2005}.
In experiments involving
turbulent binary fluids near the critical temperature, it is found that a
turbulent flow suppresses phase separation and homogenizes the fluid~\cite{Pine1984,Chan1987}.
 However, by cooling the fluid further and
maintaining the same level of forcing, phase separation is achieved.  This
result is limited to near-critical fluids, although it does suggest the
possibility of homogenization by stirring in other contexts.

The most recent work on the CH fluid (Berti et al.,~\cite{Berti2005}) focusses
on the active CH tracer.  They couple the CH concentration field to an externally-forced
velocity field and observe coarsening arrest independently of finite-size
effects.  They identify a balance of local shears and surface tension as
the mechanism for coarsening arrest.  Thus, the chaotic advection discussed
above is a sufficient condition for coarsening arrest, but not a necessary
one since shear will suffice.  Of interest again is the dependence of the
arrest scale on the mean rate of shear.

With this literature in mind, we study the case of a chaotic flow coupled
passively to the CH equation.  Our ultimate focus will be on the breakup
of bubbles by stirring, and not just coarsening arrest.  We believe this
aspect of CH flow deserves a further and more complete study, as it is a
regime of particular relevance to industry, for example, in droplet breakup
or in the mixing of polymers~\cite{Aarts2005}.

Berthier et al.~\cite{chaos_Berthier} used an alternating sine flow to study
coarsening arrest.  Because of its simplicity, the sine flow is a popular
testbed for studying chaotic mixing~\cite{lattice_PH2,Antonsen1996,Neufeld_filaments,lattice_PH1,
Thiffeault2004}.  In contrast to Berthier et al.~\cite{chaos_Berthier}, we
use a sine flow where the phase is randomized at each period,
leading to a flow that has a positive Lyapunov exponent at all stirring amplitudes
and for all initial conditions.  This avoids the coexistence of regular and
chaotic regions, which would lead to inhomeogeneous mixing characteristics
over the domain of the problem.  Because of its uniform stirring, the random-phase
sine flow is often used as a proxy for turbulent stirring at high Prandtl
number, but at much lower computational cost.  Using this flow, we study
coarsening arrest, as in previous papers~\cite{chaos_Berthier, Berti2005}.
 However, we shall also use it to study mixing due to the stirring and the
 hyperdiffusion of the CH dynamics.  Under this additional process the scaling
 hypothesis that characterizes the dynamics of bubble formation, namely that
 there is one dominant lengthscale~\cite{Bray_advphys, Berti2005}, breaks
 down.

Having found a mechanism that counteracts domain formation and promotes
mixing, we introduce a quantitative measure of homogenization, studying the
$p^{\mathrm{th}}$ power-mean fluctuation of the concentration about its average
value.  We specialize to the symmetric mixture in which equal amounts of
both binary fluid components are present.
By fluctuations about the average value, we mean spatial fluctuations around
the (constant) average spatial concentration, which we then average over
space and time.
In effect, we study the time-averaged $L^p$ norm of the concentration.
If this quantity is small, the average deviation of the system about the
homogeneous state is small, and we therefore use this quantity as a
proxy for the mixedness of the fluid.
An approach similar to this has already been taken in the theory of miscible
fluids~\cite{Thiffeault2004,Shaw2006,Thiffeault2006,Thiffeault2007}.  There,
the equation of interest is the advection-diffusion equation,
and fluctuations about the mean are measured by the variance or centred second
power-mean of the fluid concentration~\cite{Danckwerts1952,Edwards1985}.
The variance is reduced by certain stirring mechanisms.  By specifying a
source term, it is possible to state the maximum amount by which a given
flow can reduce the variance, and hence mix the fluid.  
By quantifying the variance reduction, one can classify flows according to
how effective they are at mixing.
Just as the linearity of the advection-diffusion equation suggested the variance
as a natural way of measuring fluctuations in the concentration, the nonlinearity
of the CH equation and its concomitant free energy (cubic and
quartic in the concentration, respectively) will fix our attention on the
fourth power-mean of the concentration fluctuations.  Owing to H\"older's
inequality,
a binary liquid that is well mixed in this sense will also be well mixed
in the variance sense.  The advantage we gain in considering the fourth power-mean
is the derivation of explicit bounds on our chosen measure of mixedness,
which have manifest flow- and source-dependence.

Due to the relevance of phase-separating thin films in industrial
applications~\cite{Karim2002, Smith1995}, many experiments and numerical
simulations focus on understanding how phase separation is altered if the
binary fluid forms a thin layer on a substrate.  
One potential application of phase separation in thin films
 is in self-assembly~\cite{Putkaradze2005,Xia2004, Krausch1994}.  Here
molecules (usually residing in a thin layer) respond to an energy-minimization
requirement by spontaneously forming large-scale structures.  Equations of
Cahn--Hilliard type have been proposed to explain the qualitative features
of self-assembly~\cite{Putkaradze2005,Darryl_eqn6}, and knowledge of variations
in the film height could enhance these models.
Therefore, we examine this application,
and explain the main features of these
studies, introducing a long-wavelength approximation of the coupled Navier--Stokes
Cahn--Hilliard (NSCH) equations for a thin layer of fluid with a free surface.

Several recent experiments have clarified the different regimes of domain
growth in a binary thin film.  Wang and Composto~\cite{WangH2000}
have identified early, intermediate, and late stages of evolution.    The
early stage comprises three-dimensional domain growth, while the intermediate
stage is characterized by the formation of wetting layers at the film boundaries,
the thinning of the middle layer, and significant surface roughening.  Due
to the thinning of the  middle layer, the sandwich-like structure breaks
up and matter from the wetting layer flows back into the bulk.  Thus, a late
stage is reached, consisting of bubbles coated by thin wetting layers.
 This characterization of the evolution has been seen in other experiments~\cite{ChungH2004,WangW2003,
 Klein2001},
 although clearly a variety of behaviours is possible, depending on the wetting
 properties of the mixture.  Our model captures the essential features of
 this evolution, in particular the tendency for concentration gradients to
 promote film rupture and surface roughening.
 
In a series of papers, Das \emph{et al.}~\cite{Puri2005, Puri2002, Puri2001,
Puri1997, Puri1994} numerically
investigate the behaviour of binary fluids with wetting.  In~\cite{Puri2005}
they study the wetting properties of a binary mixture in an ultra-thin film.
 The behaviour is different from that in bulk mixtures.  In bulk mixtures
\begin{figure}
\centering
\subfigure[]{
    \includegraphics[width=.25\textwidth]{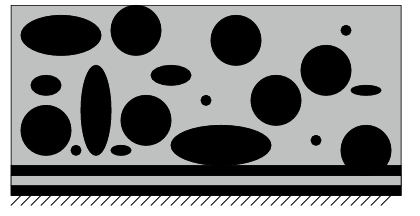}
}
\subfigure[]{
    \includegraphics[width=.25\textwidth]{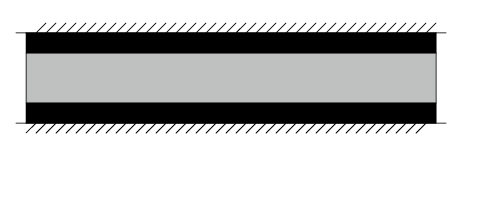}
}
\subfigure[]{
    \includegraphics[width=.25\textwidth]{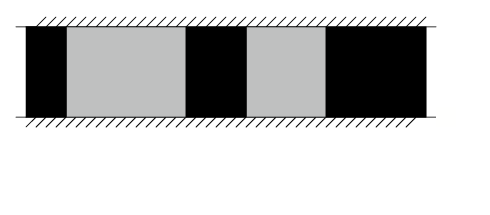}
}
\caption{Spinodal decomposition with wetting. Subfigure (a)
provides a schematic representation of the spinodal wave at the lower horizontal
wall.  This structure forms when one of the binary fluid
components wets the wall.  Layers of alternating demixed regions propagate
into the bulk until the wall effects are negligible.  This effect is suppressed
in thin films.
Subfigure (b) shows the completely
wet thin film, in which a single binary fluid component wets the horizontal
bounding walls.  The film is too thin to support a spinodal wave.
Subfigure (c)
shows the partially wet thin film, in which both binary fluid components
wet the horizontal bounding walls, and there is no vertical domain structure.
In this report we shall focus on the partially wet case.}
\label{fig:intro_wetting}
\end{figure}
where one component of the binary fluid is preferentially attracted to the
boundary, a layer rich in this component is established at the boundary,
followed by a depletion layer.  This layered structure propagates into the
bulk and is called a spinodal wave~\cite{Puri2002, Puri2001, Puri1997}. 
 In ultra-thin films, where the film
  thickness is less than a spinodal wavelength, this layered structure is
  suppressed.  Two types of  behaviour are then possible.  In the partially
  wet case, both fluid components come into contact with the film boundaries.
   The ultimate state of the system is a domain-like structure extending
   in the lateral directions.  The domains grow in time as $t^{1/3}$, indicating
   Lifshitz--Slyozov diffusion~\cite{LS}.  In the other case, complete
   wetting, only one of the fluid components is in contact with the film
   boundaries.  These different wetting regimes are portrayed schematically
   in Fig.~\ref{fig:intro_wetting}.  Our focus in this report is on the partially
   wet case.

 The papers of Das \emph{et al.} elucidate the role of wetting and film thickness
 on phase separation,
 although they do not discuss hydrodynamics or the effect
 of free-surface variations on domain formation.  We therefore
 focus on ultra-thin films with a variable free surface, and for simplicity
 we restrict our attention to the case where both fluids experience the same
 interaction with the substrate and free surface.
The model we introduce is based on the coupled NSCH equations. 
With an applied external forcing, the model highlights the effect of the
dynamical backreaction of concentration gradients
on the flow, a useful feature in applications where control of phase separation
is required~\cite{Krausch1994}.
When the binary fluid forms a thin film on a substrate, a long-wavelength
approximation
simplifies the NSCH equations, which reduce to a
pair of coupled evolution equations for the free surface and concentration.
If $h\left(\bm{x},t\right)$ is the scaled free-surface height, and $c\left(\bm{x},t\right)$
is the binary fluid concentration, then the dimensionless equations take
the form
\begin{subequations}
\begin{equation}
\frac{\partial h}{\partial t}+\Dperp\cdot\bm{J}=0,\qquad
\frac{\partial}{\partial t}\left(c h\right)+\Dperp\cdot\left(\bm{J}c\right)=\Dperp\cdot\left(h\Dperp\mu\right),
\end{equation}
where
\begin{equation}
\bm{J}=\tfrac{1}{2}h^2\Dperp\Gamma-\tfrac{1}{3}h^3\bigg\{\Dperp\left(-\frac{1}{C}\Dperp^2{h}
+\phi\right)+\frac{r}{h}\Dperp\left[h\left(\Dperp{c}\right)^2\right]\bigg\},
\end{equation}
\begin{equation}
\mu=c^3-c-\frac{C_{\mathrm{n}}^2}{h}\Dperp\cdot\left(h\Dperp{c}\right).
\end{equation}%
\label{eq:model}%
\end{subequations}%
The symbol $\Dperp=\left(\partial_x,\partial_y\right)$ denotes the gradient
operator in the lateral directions.  The positive constants $C$, $r$, and
$C_{\mathrm{n}}$ measure surface tension, coupling strength between concentration
and free-surface variations, and the scaled interfacial thickness, respectively.
Additionally, $\Gamma$
is the dimensionless, spatially-varying surface tension, and $\phi$ is the
body-force potential acting on the film.  In this report we take $\phi=-Ah^{-3}$,
$A>0$, the repulsive Van der Waals potential~\cite{Book_Parsegian2006}. 
This choice stabilizes the film and prevents rupture.  Although rupture is
in itself an important feature in thin-film equations~\cite{Oron1997, Bertozzi1996,
Bertozzi1998}, in this report we are interested in late-time phase separation
and it is therefore undesirable.

The analysis of thin-film equations was given great impetus by Bernis and
Friedman in~\cite{Friedman1990}.  They focus on the basic thin-film equation
\begin{equation}
\frac{\partial h}{\partial t}=-\frac{\partial}{\partial x}\left(h^n\frac{\partial^3{h}}{\partial{x^3}}\right),
\label{eq:basic_h}
\end{equation}
with no-flux boundary conditions on a line segment, and smooth nonnegative
initial conditions.  For $n=1$ this equation describes a thin bridge between
two masses of fluid in a Hele--Shaw cell, for $n<3$ it is used
in slip models as $h\rightarrow0$~\cite{Myers1998}, while for $n=3$ it gives
the evolution of the free surface of a thin film experiencing capillary forces~\cite{Oron1997}.
Using a decaying free-energy functional, they prove the existence of nonnegative
solutions to Eq.~\eqref{eq:basic_h} for $n\geq1$, while for $n\geq 4$, the
solution is unique, strictly positive, and is almost always bounded in the
$H^{3,2}$ norm.  This paper has inspired other work on the subject~\cite{Bertozzi1996,
Bertozzi1998,Laugesen2002}, in which the effect of a Van der Waals term on
Eq.~\eqref{eq:basic_h} is investigated.  These works provide results concerning
regularity, long-time behaviour, and film rupture, in the presence of an
attractive Van der Waals force.
More relevant to the present work is the paper by Wieland and Garcke~\cite{Wieland2006},
in which a pair of partial differential equations describes the coupled evolution
of free-surface variations, and surfactant concentration.  The authors derive
the relevant equations using the long-wavelength theory, obtain a decaying
energy functional, and prove results concerning the existence and nonnegativity
($h\geq0$) of solutions.  We shall take a similar approach in this report.

We have noted that equations of Cahn--Hilliard type have been used to model
the self-assembly of molecules.  In Ch.~\ref{ch:singular_solutions} we investigate
 the aggregation of magnetic particles using the Holm--Putkaradze model of
 aggregation.
  We compare and contrast the Holm--Putkaradze and the Cahn--Hilliard models.
  The formulation of these models through a gradient flow serves
  as a bridge between our study and more recent work in nanoscale modelling.

The modelling of nanoscale physics has become important in recent years,
both because of industrial applications~\cite{Denis2002,Veinot2002,Moller2001,Wijnhoven1998},
and because of the development of experiments that probe these small scales~\cite{WolfBook}.
In a series of papers, Holm, Putkaradze and Tronci~\cite{Darryl_eqn1,Darryl_eqn2_0,Darryl_eqn2_1,Darryl_eqn3,Darryl_eqn5,Darryl_eqn6}
have focussed on the derivation of aggregation equations that
possess emergent singular solutions.  Continuum aggregation equations have
been used
to model gravitational collapse and the subsequent emergence of stars~\cite{ChandraStars},
the localization of biological populations~\cite{KellerSegel1970,Segel1985,Topaz2006},
and the self-assembly of nanoparticles~\cite{Putkaradze2005}.  These are
complexes of atoms or molecules that form mesoscale structures with particle-like
behaviour.
The utility of the Holm--Putkaradze model lies in its emphasis on nonlocal
physics, and the emergence of singular solutions from smooth initial data.
 Because of the singular (delta-function) behaviour of the model, it is an
 appropriate way to describe the universal phenomena of 
aggregation and the subsequent formation of particle-like structures.  Indeed
in this framework, it is possible to prescribe the dynamics of the particle-like
structures after collapse.
Thus, the model
provides a description of directed self-assembly in nanoscale physics~\cite{Putkaradze2005,Xia2004},
wherein microscopic particles collapse under the potential they exert on
each other, and form mesoscopic structures that in turn behave like particles.
 Our work in this area focusses on the vector-scalar Holm--Putkaradze model
 which, although derived elsewhere~\cite{Darryl_eqn1}, has not yet been studied
 in any detail.  This model provides an appropriate description of self-assembly
 of magnetic particles.

Although the Cahn--Hilliard theory has a venerable history, studies of phase
separation with advection are more recent.  Our result concerning
homogenization by a chaotic flow is new.  The application of (standard) functional
analysis techniques to the advective Cahn--Hilliard equation, and the derivation
of lower bounds on the amount of homogenization possible
is also new.  Although
long-wavelength theory is common in applications, the thin-film Stokes Cahn--Hilliard
(SCH) equations are entirely new, as are the numerical and analytical results
concerning this model.  In particular, we have identified a novel way of
controlling phase separation by a combination of applied surface stresses
and backreaction tuning.
 The vector Holm--Putkaradze model is a recent addition to the family of
 aggregation models and, while the equations were formulated elsewhere, our
 analysis of the coupled
 vector-scalar model is new.

This report is organized as follows.  In Ch.~\ref{ch:background} we
derive the Cahn--Hilliard and Navier--Stokes Cahn--Hilliard equations from
a combination of intuition and variational techniques.  We discuss the elements
of functional analysis that we need in this report, and prove an existence
and uniqueness result for the CH equation with passive advection.
 In Ch.~\ref{ch:chaotic_advection}
we outline a numerical study of the CH equation stirred by a
passive, chaotic flow.  We characterize the mixing process by the lengthscales
 involved in the problem, by the concentration probability distribution function,
 and by a simple one-dimensional model.  In Ch.~\ref{ch:estimating_mixedness},
 we use the functional analysis results of Ch.~\ref{ch:background},
 and obtain upper and lower bounds on the level of mixedness achievable by
 any flow.  In Ch.~\ref{ch:analysis_thin_films} we derive the thin-film Stokes
 Cahn--Hilliard equations, highlighting the importance of
 the dynamical
 coupling between the concentration and fluid velocity.  We obtain existence
 and regularity results for the model equations.  Although such analysis
 gives very precise results, it is unable to tell us much about the qualitative
 features of the solutions.  Therefore, in Ch.~\ref{ch:simulations_thin_films}
 we carry out numerical simulations of the SCH equations,
 and compare the results with experimental papers.  The simple model introduced
 here explains the qualitative features observed in real phase-separating
 fluids.  
In Ch.~\ref{ch:singular_solutions} we focus on the Holm--Putkaradze model
of aggregation, which serves as a bridge between the Cahn--Hilliard model
we have studied, and more recent work in nanoscale physics.  We examine
the
equations introduced by Holm, Putkaradze, and Tronci
for the aggregation of oriented particles~\cite{Darryl_eqn1,Darryl_eqn2_0,Darryl_eqn2_1,Darryl_eqn3}.
  We investigate these equations numerically and study their evolution and
  aggregation properties.  One aspect of nonlocal problems, already mentioned
  in~\cite{Darryl_eqn6}, is the effect of competition between the lengthscales
  of nonlocality
  on the system evolution.  We highlight this effect with a linear
  stability analysis of the full density-magnetization equations. 
 Finally, in Ch.~\ref{ch:conclusions}, we present our conclusions.

 
\subsection*{Publications} 
 
Many of the results presented in this thesis have already been published.
 We summarize this work in what follows.  
Unless otherwise stated, the principal author is L. \'O N\'araigh and the
co-author is J.-L. Thiffeault.

\begin{itemize} 
\item The numerical study of phase separation and passive chaotic
advection is presented in~\cite{ONaraigh2006}.
\item The question, ``given a passive flow in a binary liquid, by what amount
can the fluid be homogenized?'' is addressed
in~\cite{ONaraigh2007_2}.  There,
the authors use functional analytic and numerical techniques to
answer this question.
\item 
Long-wavelength theory is used to derive a set of thin-film Stokes Cahn--Hilliard
(SCH) equations in~\cite{ONaraigh2007}.  The authors highlight the importance
of the backreaction when the control of phase separation is required.
\item Using a combination of numerical analysis and simple dynamical systems
 arguments, the vector Holm--Putkaradze model is investigated in~\cite{ONaraigh2007_3}.
  The first author is D.~D. Holm and the
 co-authors are L. \'O N\'araigh and C. Tronci.
\end{itemize}


\subsection*{Statement of originality}

The work presented in this thesis is my own, unless indication is given to
the contrary.

%% file: background/background.tex
\chapter{General Formulation}
\label{ch:background}

\section{Overview}
\label{sec:background:intro}

In this chapter we outline the thermodynamical formalism that gives rise
to the Cahn--Hilliard equation and demonstrate its coupling to a flow field.
 We introduce the Navier--Stokes Cahn--Hilliard equations and obtain evolution
 equations for the physical quantities associated with the dynamical variables.
  We also discuss the mathematical formalism necessary to this report.
 
\section{The Cahn--Hilliard equation}
\label{sec:background:ch}

In this report, the Cahn--Hilliard model is viewed as an order-parameter
equation, derived from thermodynamical considerations.  An order-parameter
description is independent of the detailed nature of the system being considered~\cite{KittelISSP},
and thus the equation is completely general, describing any two-component
system where the mixed state is energetically unfavorable,
and where the total amount of matter is conserved.  Let $c\left(\bm{x},t\right)$
be the order parameter, or simply put, the concentration of the binary fluid.
 We write down a free energy for the system in $n$ dimensions, as a functional
 of the concentration $c$ and temperature $T$,
\begin{equation}
F\left[c,T\right]=\varepsilon_0\int_{\Omega}{d^nx}\left[\tfrac{1}{4}c^4-\tfrac{1}{2}\left(1-\frac{T}{T_{\mathrm{c}}}\right)c^2+\tfrac{1}{2}\Tr\left|\nabla{c}\right|^2\right],
\label{eq:fe_functional}
\end{equation}
where $T_{\mathrm{c}}$ is a critical temperature, and where the gradient
energy density $\tfrac{1}{2}\Tr\left|\nabla{c}\right|^2$ penalizes the formation
of sharp interfaces.  Indeed, we shall see that the lengthscale $\sqrt{\Tr}$
represents the typical thickness of a smoothly-varying transition zone between
unmixed regions, and that the finite thickness of these zones prevents
the formation of infinite gradients in the problem.  The constant $\varepsilon_0$
appears in order to give Eq.~\eqref{eq:fe_functional} the correct units:
$\varepsilon_0=k_B T_{\mathrm{c}}/|\Omega|$ has units of energy density.
 We shall, however, omit this constant if possible.
The eager reader will
note that Eq.~\eqref{eq:fe_functional} is nothing other than the Ginzburg--Landau
free energy~\cite{KittelISSP}.  The system admits a critical temperature
$T_{\mathrm{c}}$;
this is a threshold temperature across which the system exhibits qualitatively
different behaviour.  For $T>T_{\mathrm{c}}$, the energy density
\[
f_0\left(c,T\right)=\tfrac{1}{4}c^4-\tfrac{1}{2}\left(1-\frac{T}{T_{\mathrm{c}}}\right)c^2
\]
has a single minimum at $c=0$, and thus the mixed state is energetically
favourable.  On the other hand, for $T<T_{\mathrm{c}}$, the energy density
has a double-well structure, with two stable minima at $c=\pm\sqrt{1-\left(T/T_{\mathrm{c}}\right)}$.
 These minima describe the pure phases of the mixture.
 The mixed state $c=0$ becomes a local maximum and is unstable (as
 the linear stability analysis in this section will demonstrate).  We shall
 focus
 on the subcritical case, where spontaneous demixing is possible.  Indeed,
 in the following derivation, we shall
 set $T=0$, eliminating all thermal noise from the system.  This simplification
 will be justified at the end of the derivation.
 
To minimize the free energy $F\left[c\right]\equiv F\left[c,0\right]$, the
concentration evolves according to an equation of the type
\[
\frac{\partial c}{\partial t} = \hat{O}\frac{\delta F}{\delta c},
\]
where $\hat{O}$ is an operator to be determined.  Moreover, we are interested
in conserved dynamics; that is, the total concentration $\int_\Omega{d^nx}c\left(\bm{x},t\right)$
must be constant.  If, as in many applications~\cite{CH_orig}, the system
is invariant
under changes of parity, the simplest form for the operator $\hat{O}$ is,
\[
\hat{O}\left(\cdot\right)=\Diff\nabla\cdot\left[\phi\left(c\right)\nabla\left(\cdot\right)\right],
\]
where $\Diff$ is a diffusion coefficient and $\phi\left(c\right)$ is a
nonnegative function called the mobility.  The choice of
mobility
function $\phi$ is determined by physical considerations; for many of our
applications, we shall simply set $\phi=1$, while an in-depth discussion
of the choice of functional form of $\phi$ is given in Sec.~\ref{sec:chaotic_advection:varmob}.
Under these assumptions, the Cahn--Hilliard (CH) equation is
\begin{equation}
\frac{\partial c}{\partial t}=\Diff\Delta\left(c^3-c-\Tr\Delta{c}\right).
\label{eq:ch}
\end{equation}
The order-parameter nature of this equation has resulted in its use in situations
far beyond its original purpose as a model of the phase separation
of a binary alloy~\cite{CH_orig}: it has found applications in polymer
physics~\cite{Aarts2005}, in interfacial flows~\cite{LowenTrus}, and in mathematical
biology~\cite{Murray1981}.
By identifying the chemical potential~\cite{Finn_ThermalPhysics} $\mu=c^3-c-\Tr\Delta{c}$
and the flux
\begin{equation}
\bm{J}_{\mathrm{CH}}=-D\nabla\mu,
\label{eq:flux_ch}
\end{equation}
the equation~\eqref{eq:ch} can be recast in flux-conservative form:
\[
\frac{\partial c}{\partial t}+\nabla\cdot\bm{J}_{\mathrm{CH}}=0,
\]
so that, upon applying Gauss's theorem, we have,
\[
\frac{d}{dt}\int_{\Omega}d^nxc\left(\bm{x},t\right)+\int_{\partial\Omega}dA_n\hat{\bm{n}}\cdot\bm{J}_{\mathrm{CH}}=0,
\]
where $\Omega$ is the problem domain in $n$ dimensions, $\partial\Omega$
is its boundary.  The quantity $dA_n$ is the differential element of area
and $\hat{\bm{n}}$ is the outward-pointing unit normal.
By specifying the boundary conditions (BCs)
\begin{equation}
\hat{\bm{n}}\cdot\nabla c = \hat{\bm{n}}\cdot\bm{J}_{\mathrm{CH}}=0,\qquad\text{on
}\partial\Omega,
\label{eq:bc_ch}
\end{equation}
the total concentration is conserved, and when combined with initial data,
this gives sufficient information to solve Eq.~\eqref{eq:ch}.  

We verify that the concentration evolution Eq.~\eqref{eq:ch} causes the decay
of free energy, as desired,
\begin{eqnarray*}
\frac{dF}{dt}
&=&-D\int_\Omega{d^n{x}}\left|\nabla\mu\right|^2+\int_{\partial\Omega}dA_n\left[D\mu\hat{\bm{n}}\cdot\nabla\mu+\frac{\partial{c}}{\partial{t}}\hat{\bm{n}}\cdot\nabla{c}\right].
\end{eqnarray*}
Using the BCs~\eqref{eq:bc_ch}, this is simply
\begin{equation}
\frac{dF}{dt}
=-D\int_\Omega{d^n{x}}\left|\nabla\mu\right|^2,
\label{eq:decay_fe}
\end{equation}
and the free energy is a decaying function of time.

Let us investigate the linear stability of a constant state $c\left(\bm{x},t\right)=c_0$.
 Performing a linear stability analysis of Eq.~\eqref{eq:ch} about this state,
 we obtain an equation for the perturbation $\delta c$,
\[
\frac{\partial}{\partial t}\delta c = \left(3c_0^2-1\right)\Diff\Delta\delta{c}-\Tr\Diff\Delta^2\delta{c},
\]
and by making the ansatz $\delta c\left(\bm{x},t\right) \propto e^{i\bm{k}\cdot\bm{x}}e^{\sigma
t}$, we obtain an expression for the growth rate of perturbations,
\[
\sigma = -\Diff k^2\left[\left(3c_0^2-1\right)+\Tr k^2\right].
\]
For $-1/\sqrt{3}<c_0<1/\sqrt{3}$, perturbations on scales below the critical
lengthscale
\[
\lambda_{\mathrm{crit}}=\frac{1}{2\pi}\sqrt{\frac{\gamma}{\left|3c_0^2-1\right|}}
\]
decay, while perturbations on scales above $\lambda_{\mathrm{crit}}$
grow.  If the initial data contain
a spectrum of lengthscales, this stability analysis demonstrates that small
scales
decay in amplitude, while large scales grow, suggesting the formation of
large-scale structures.  In later chapters we shall see that this is indeed
the case.  This linear stability analysis yields another piece of information:
if the box size $L$ of the problem is less than the critical scale $\lambda_{\mathrm{crit}}$,
then no large-scale structures can form, and the perturbations always decay
to zero.  This is the statement that the problem dimensions are so small
as to favour the hyperdiffusion of Eq.~\eqref{eq:ch} over phase separation,
a useful result to keep in mind when considering phase separating liquids
in ultrathin films.

The existence of equilibrium solutions to the CH equation~\eqref{eq:ch} is
well known~\cite{Furukawa_scales}, and we provide the details here
 because they will be useful in the following chapters.
 It is known that the late-time solution is one of bubbles, for which
 $c\approx\pm1$, with transition regions of thickness
 $\sqrt{\gamma}$ separating these bubbles.  It is possible to derive an equilibrium
 solution that describes these transition zones.  We write down the equilibrium
 version of the Cahn--Hilliard equation,
\[
c^3-c-\gamma\Delta c=0.
\]
 Let us set up a coordinate
 system on the bubble boundary with coordinates both perpendicular and
 parallel
 to the boundary.  If the bubble's radius of curvature is large, as indeed
 it will be at late times, then the equilibrium CH equation reduces to
\begin{equation}c^3-c-\gamma\frac{\partial^2c}{\partial g^2}=0,
\label{eq:ch_no_flow}
\end{equation}
where $g$ is the coordinate perpendicular to the bubble interface (we discuss
the effects of a finite radius of curvature in Sec.~\ref{sec:chaotic_advection:1D}).
The boundary conditions place a single isolated bubble ($c=+1$) in a sea
of negative concentration, $c=-1$.  Thus, the boundary conditions at $g=\pm\infty$
are $c=\pm1$, $c'=0$.  Multiplying Eq.~\eqref{eq:ch_no_flow} by the
integrating factor ${\partial c}/{\partial g}$ and using the boundary conditions,
we obtain the equation
\begin{equation}
f_0\left(c\right)=\frac{\gamma}{2}\left(\frac{\partial c}{\partial g}\right)^2,
\label{eq:balance_energy}
\end{equation}
with solution 
\begin{equation}c\left(g\right)=\tanh\left(g/\sqrt{2\gamma}\right),
\label{eq:tanh_sln}
\end{equation}
consistent with the given boundary conditions.  Note that for $\gamma\rightarrow0$,
Eq.~\eqref{eq:tanh_sln} becomes a step function, with delta-function
first derivative.  Such infinite gradients are undesirable in our work, and
we shall therefore always take $\Tr>0$.

In $n$ dimensions, the surface tension is the free energy associated with
the bubble interface, per unit interfacial area.  That is,
\[
\Gamma=\Small_0\int_{-\infty}^{\infty}dg\left[f_0\left(c\right)+\frac{\gamma}{2}\left(\frac{\partial{c}}{\partial{g}}\right)^2\right].
\]
Upon substitution of the solution~\eqref{eq:tanh_sln}, the surface tension
reduces to $\Gamma=\Small_0\sqrt{{8\gamma}/{9}}$.  Thus, $\left[\Gamma\right]=\left[\text{Energy}\right]L^{-n+1}$,
and $\Gamma$ has the correct dimensions of surface tension.

Finally, since $g$ is the coordinate perpendicular to the interface, the
gradient $\nabla c$ has the form $\bm{\hat{g}}{\partial c}/{\partial g}$,
where $\bm{\hat{g}}$ is a unit vector perpendicular to the interface.  Thus,
it follows from Eq.~\eqref{eq:balance_energy} that
\[
f_0\left(c\right)=\frac{\gamma}{2}\left|\nabla c\right|^2,
\]
and the kinetic and potential energy densities are in equipartition near
the transition region.  Moreover, far from the transition region, $c=\pm1$,
and thus local equipartition also holds in the bulk.  
Integration over the domain of the problem
therefore yields equipartition of kinetic and potential energy.

We examine briefly the evolution equation at finite temperatures.  At finite
temperatures, we are forced to take into account thermal fluctuations in
the concentration field.  Equation~\eqref{eq:ch} is modified by the addition
of a noise term,
\begin{subequations}
\begin{equation}
\frac{\partial c}{\partial t}=\Diff\Delta\left(c^3-c-\Tr\Delta{c}\right)+\xi\left(\bm{x},t\right),
\end{equation}
where $\xi\left(\bm{x},t\right)$ is a white-noise term satisfying the relation
\begin{equation}
\langle\xi\left(\bm{x},t\right)\xi\left(\bm{x}',t'\right)\rangle=-2{\Diff}k_BT\Delta\delta\left(\bm{x}-\bm{x}'\right)\delta\left(t-t'\right).
\end{equation}%
\label{eq:chc}%
\end{subequations}%
Equation~\eqref{eq:chc} is the Cahn--Hilliard--Cooke equation~\cite{Elder1988}.
 Bray~\cite{Bray_RNG} shows that the scaling laws that characterize phase
 separation are the same for the Cahn--Hilliard and Cahn--Hilliard--Cooke
 equations, and that temperature is thus an irrelevant variable, at least
 when discussing the asymptotic (i.e. as $t\rightarrow\infty$) features of
 phase
 separation.  We are therefore
 justified in ignoring it in this report.

\section{The Navier--Stokes Cahn--Hilliard equations}
\label{sec:background:NSCH}

Using a combination of variational principles and physical intuition, we
obtain a multicomponent flow model that couples fluid velocities to a concentration
field.  This approach is based loosely on the work of Lowengrub and Truskinowsky~\cite{LowenTrus}.

Consider a two-component
binary liquid, with components labelled by the index $i=1,2$.  In a small
volume $V$, the $i^{\mathrm{th}}$ component occupies a volume
$V_i$, with $V_1+V_2=V$.  In the same way, if the volume $V$ has total mass
$M$, with the $i^{\mathrm{th}}$ fluid component giving a contribution $M_i$,
then $M_1+M_2=M$.  We introduce the real mass density $\rho_i=M_i/V_i$, in
contrast to the apparent mass density $\tilde{\rho}_i=M_i/V$.  Then,
$\tilde{\rho}_i=M_i/V=\rho_i V_i/V=\rho_i\chi_i$, where $\chi_i$
is the $i^{\mathrm{th}}$ volume fraction.  That is, given the characteristic
function $X_i\left(\bm{x}\right)$, with $X_i\left(\bm{x}\right)=1$
if $\bm{x}$ is in the $i^{\mathrm{th}}$ phase and zero otherwise, we have
\[
\chi_i\left(V\right)=\frac{1}{|V|}\int_V X_i\left(\bm{x}\right)d^nx,\qquad i=1,2,
\]
where $\chi_2=1-\chi_1$.
With this knowledge, we introduce the binary liquid density,
\[
\rho=\frac{M}{V}=\frac{M_1}{V}+\frac{M_2}{V}=\rho_1\chi_1+\rho_2\chi_2.
\]
To obtain a more convenient expression for $\rho$, we
make use of the following expression for the volume fraction $\chi_i$,
\[
\chi_i=\frac{V_i}{V}=\frac{V_i\rho_i}{V\rho}\frac{\rho}{\rho_i}=\frac{M_i}{M}\frac{\rho}{\rho_i}:=c_i\frac{\rho}{\rho_i},
\]
where we have defined the mass fraction $c_i$, 
\[
\chi_i\rho_i=\frac{M_i\rho}{M}=c_i\rho,\qquad{i=1,2}.
\]
Thus, we obtain two equations for $\chi_1$,
\[
\rho_1\chi_1=c_1\rho,\qquad \rho=\chi_1\left(\rho_1-\rho_2\right)+\rho_2.
\]
Eliminating $\chi_1$ between the equations, we obtain an expression for the
density $\rho$ which depends on the mass fraction $c_1\equiv c$,
\begin{equation}
\frac{1}{\rho}=\frac{c}{\rho_1}+\frac{1-c}{\rho_2}.
\label{eq:simple_mix}
\end{equation}
This is the so-called simple-mixture equation.  Note that the density so
defined is equivalent to the identity $\rho=\tilde{\rho}_1+\tilde{\rho}_2$.

Let $\bm{v}_i$ be the velocity at which the $i^{\mathrm{th}}$ component travels
in the mixture.  We require that the mass of each binary fluid component
be conserved,
\[
\frac{\partial\tilde{\rho}_i}{\partial t}+\nabla\cdot\left(\bm{v}_i\tilde{\rho}_i\right)=0.
\]
By identifying the mass-weighted velocity field
\begin{equation}
\bm{v}=\frac{\tilde{\rho}_1\bm{v}_1+\tilde{\rho}_2\bm{v}_2}{\tilde{\rho}_1+\tilde{\rho}_2}.
\end{equation}
we obtain a conservation law for the mixture variables $\left(\bm{v},\rho=\tilde{\rho}_1+\tilde{\rho}_2\right)$,
\begin{equation}
\frac{\partial\rho}{\partial t}+\nabla\cdot\left(\bm{v}\rho\right)=0.
\label{eq:mass_consv_ch2}
\end{equation}
In this report we focus on density-matched fluids, for which $\rho_1=\rho_2=\mathrm{constant}$.
 Then the mixture density is simply
\[
\frac{1}{\rho}=\frac{1}{\rho_1}:=\frac{1}{\rho_0},
\]
and thus the mass-conservation law~\eqref{eq:mass_consv_ch2} reduces to the incompressibility
condition
\begin{equation}
\nabla\cdot\bm{v}=0.
\label{eq:incompressibility}
\end{equation}
Moreover, we postulate an evolution equation for the concentration (mass
fraction) $c\left(\bm{x},t\right)$, independently of the analysis in Sec.~\ref{sec:background:ch},
\begin{eqnarray}
\rho\left(\frac{\partial c}{\partial t}+\bm{v}\cdot\nabla{c}\right)=\tilde{\Diff}\Delta\mu_D
\end{eqnarray}
where $\mu_\Diff$ is a chemical potential to be determined, and $\tilde{\Diff}$
is a constant with the dimensions $\left[\mathrm{Length}\right]^{2-n}\left[\mathrm{Mass}\right]\left[\mathrm{T}\right]^{-1}$.

To find the dynamics of the concentration and velocity fields, we momentarily
set $\tilde{\Diff}$ to zero, and assume that the concentration $c\left(\bm{x},t\right)$
is in fact conserved; that is, it satisfies
\[
\frac{\partial c}{\partial t}+\bm{v}\cdot\nabla c=0.
\]
In the Lagrangian picture of fluid mechanics, this expression has the form
\[
\left(\frac{\partial c}{\partial \tau}\right)_{\bm{a}}=0,
\]
where the partial derivative $\left(\partial/\partial \tau\right)_{\bm{a}}$
indicates time differentiation with respect to a fixed Lagrangian label $\bm{a}$,
and where 
%
%
%
%
%
the fluid parcel with label $\bm{a}$ has the velocity
\[
\left(\frac{\partial\bm{x}}{\partial \tau}\right)_{\bm{a}}=\bm{v}\left(\bm{x},\tau\right).
\]
%
%
%
%
%
%
%
%
%
We write down a Lagrangian for the fluid velocity and phase-separating concentration,
\[
L=\tfrac{1}{2}\int_\Omega d^na\left(\frac{\partial\bm{x}}{\partial\tau}\right)^2
-\frac{\varepsilon_0}{\rho_0}\int_\Omega d^na\left[f_0\left(c\right)+\tfrac{1}{2}\gamma|\nabla{c}|^2\right]
+\int_\Omega d^na \left(1-\frac{\rho}{\rho_0}\right)\frac{p}{\rho_0},
\]
where $d^na=dm=\rho d^n x$, and $\bm{a}$ is a Lagrangian label.
The density $\rho$ is then given as the Jacobian
\[
\rho=\frac{\partial\left(a_1,...,a_n\right)}{\partial\left(x_1,...,x_n\right)},
\]
and the term $\int_\Omega d^n a \left[1-\left(\rho/\rho_0\right)\right]\left(p/\rho_0\right)$
is the Lagrangian multiplier that enforces the constant-density condition.
The Lagrangian action is then
\[
S=\int_{t_1}^{t_2}{d\tau}L.
\]
Following standard practice~\cite{LandauCourseTP1}, stationarity of this
action with respect to variations in the Lagrangian trajectories $\bm{x}\left(\bm{a},\tau\right)$
yields the equations of motion.  The variations along trajectories take place
at fixed Lagrangian label, and the variation in the density has the form
\[
\delta\rho=-\rho\nabla\cdot\delta\bm{x},
\]
which follows from the constancy of $\rho d^n{x}$ along particle trajectories,
$\left[\partial_{\tau}\left(\rho d^n{x}\right)\right]_{\bm{a}}=0$.
Moreover, the related variations in concentration have the form,
\[
\delta c=\nabla c\cdot\delta\bm{x},\qquad
\partial_i \delta c = \left(\partial_i\partial_j{c}\right)\delta x_j +\partial_j{c}\left(\partial_i\delta{x_j}\right),\qquad
\delta\partial_i c = \left(\partial_i\partial_j c\right)\delta x_j,
\]
and hence
\[
\delta\left(\partial_i c\right)=\partial_i\left(\delta c\right)-\partial_j{c}\left(\partial_i\delta
x_j\right).
\]
Throughout the report, unless indication is given to the contrary, we use
the summation convention and sum over repeated indices.

Using these relations, we vary the action $S$ over particle trajectories
$\bm{x}\left(\bm{a},\tau\right)$, at fixed particle label.  We choose variations
$\delta\bm{x}$ that vanish at the end times $t_1$ and $t_2$, and at the container
boundary $\partial\Omega$.  Thus,
\begin{multline*}
\delta S=-\int_{t_1}^{t_2}dt\int_\Omega d^na\left[\frac{\partial^2\bm{x}}{\partial\tau^2}+\frac{1}{\rho}\nabla\left(\frac{p\rho^2}{\rho_0^2}\right)\right]\cdot\delta\bm{x}+\int_{t_1}^{t_2}dt\int_\Omega{d^na}\frac{\delta{p}}{\rho_0}\left(1-\frac{\rho}{\rho_0}\right),\\
-\frac{\varepsilon_0}{\rho_0}\int_{t_1}^{t_2}dt\int_\Omega{d^na}\frac{\gamma}{\rho}\partial_i\left(\rho\partial_ic\partial_jc\right)\delta{x}_j
-\frac{\varepsilon_0}{\rho_0}\int_{t_1}^{t_2}dt\int_\Omega{d^na}\left[f_0'\left(c\right)-\frac{\gamma}{\rho}\nabla\cdot\left(\rho\nabla{c}\right)\right]\delta{c}.
\end{multline*}
Using $\delta S/\delta p =0$, we obtain the constraint $\rho=\rho_0$.  Then
using the result $\delta S/\delta\bm{x}=0$, we obtain the equation of motion
\[
\frac{\partial^2{\bm{x}}}{\partial\tau^2}=-\frac{1}{\rho_0}\nabla{p}-\frac{\varepsilon_0\gamma}{\rho_0}\nabla\cdot\left(\nabla{c}\nabla{c}\right)
\]
Finally, the stationarity result $\delta S/\delta c=0$ gives the constraint
\[
f_0'\left(c\right)-\gamma\Delta c=0.
\]

We rewrite the equations in Eulerian form,
\begin{eqnarray*}
\frac{\partial\bm{v}}{\partial{t}}+\bm{v}\cdot\nabla\bm{v}&=&-\frac{1}{\rho_0}\nabla{p}-\frac{\varepsilon_0\gamma}{\rho_0}\nabla\cdot\left(\nabla{c}\nabla{c}\right),\\
0&=&\mu\equiv f_0'\left(c\right)-\gamma\Delta c.
\end{eqnarray*}
By making the identification $\mu_D\equiv\mu$ and restoring $\tilde{\Diff}\neq0$,
we obtain the equations
\begin{subequations}
\begin{equation}
\frac{\partial\bm{v}}{\partial{t}}+\bm{v}\cdot\nabla\bm{v}=-\frac{1}{\rho_0}\nabla{p}-\frac{\varepsilon_0\gamma}{\rho_0}\nabla\cdot\left(\nabla{c}\nabla{c}\right)
\label{eq:euler_ch_v}
\end{equation}
\begin{equation}
\frac{\partial{c}}{\partial t}+\bm{v}\cdot\nabla{c}=\frac{\tilde{\Diff}}{\rho_0}\Delta\left(f_0'\left(c\right)-\gamma\Delta{c}\right),
\label{eq:euler_ch_c}
\end{equation}
\begin{equation}
\nabla\cdot\bm{v}=0.
\label{eq:euler_ch_mass}
\end{equation}%
\end{subequations}%
We observe that the Lagrange multiplier that enforces the constant density
plays the role of the pressure in these equations.  The Lagrange multiplier
is not arbitrary: it is fixed by the equation
\[
p=-\rho_0\Delta^{-1}\left[\partial_i\left(v_j\partial_j{v}_i\right)+\frac{\varepsilon_0\gamma}{\rho_0}\partial_i\partial_j\left(\partial_i{c}\partial_j{c}\right)\right].
\]
Finally, we model the effect of viscosity by adding a diffusion term to 
Eq.~\eqref{eq:euler_ch_v}, which gives rise to the following constant-density
Navier--Stokes Cahn--Hilliard (NSCH) equations,
\begin{subequations}
\begin{equation}
\frac{\partial\bm{v}}{\partial{t}}+\bm{v}\cdot\nabla\bm{v}=-\frac{1}{\rho_0}\nabla{p}-\frac{\varepsilon_0\gamma}{\rho_0}\nabla\cdot\left(\nabla{c}\nabla{c}\right)+\nu\Delta\bm{v},
\label{eq:NSCH_v_nontensorial}
\end{equation}
\begin{equation}
\frac{\partial{c}}{\partial t}+\bm{v}\cdot\nabla{c}=\frac{\tilde{\Diff}}{\rho_0}\Delta\left(f_0'\left(c\right)-\Tr\Delta{c}\right),
\label{eq:NSCH_c_nontensorial}
\end{equation}
\begin{equation}
\nabla\cdot\bm{v}=0,
\label{eq:NSCH_incomp_nontensorial}
\end{equation}%
\label{eq:NSCH1}%
\end{subequations}%
where $\nu$ is the kinematic viscosity of the mixture.  By specifying
the viscosity term as we have done, we confine ourselves to Newtonian fluids.
 These equations possess an energy $E\left(t\right)$,
\[
E=\frac{\rho_0}{2}\int_\Omega{d^nx}\bm{v}^2+\varepsilon_0\int_\Omega{d^nx}\left[f_0\left(c\right)+\tfrac{1}{2}\Tr\left|\nabla{c}\right|^2\right],
\]
which is dissipated according to the relation
\[
\frac{dE}{dt}=-\nu\int_\Omega{d^nx}\left|\nabla\bm{v}\right|^2-\frac{\varepsilon_0\tilde{D}}{\rho_0}\int_\Omega{d^nx}\left|\nabla\mu\right|^2\leq0.
\]

We comment on the symmetries of Eq.~\eqref{eq:NSCH1}.  By identifying
a tensor
\begin{equation}
T_{ij}=-\frac{p}{\rho_0}\delta_{ij}+\nu\left(\frac{\partial{v}_i}{\partial{x}_j}+\frac{\partial{v}_j}{\partial{x}_i}\right)-\frac{\varepsilon_0\gamma}{\rho_0}\frac{\partial{c}}{\partial{x}_i}\frac{\partial{c}}{\partial{x}_j},
\label{eq:stress_tensor}
\end{equation}
Eqs.~\eqref{eq:NSCH1} can be rewritten as
\begin{subequations}
\begin{equation}
\frac{\partial\bm{v}}{\partial{t}}+\bm{v}\cdot\nabla\bm{v}=\nabla\cdot\bm{T},
\label{eq:NSCH_v}
\end{equation}
\begin{equation}
\frac{\partial{c}}{\partial t}+\bm{v}\cdot\nabla{c}=\frac{\tilde{\Diff}}{\rho_0}\Delta\left(f_0'\left(c\right)-\Tr\Delta{c}\right),
\label{eq:NSCH_c}
\end{equation}
\begin{equation}
\nabla\cdot\bm{v}=0.
\end{equation}%
\label{eq:NSCH2}%
\end{subequations}%
Choosing $f_0'\left(c\right)=c^3-c$ as in Sec.~\ref{sec:background:ch},
this
set of equations is manifestly invariant under the change $c\rightarrow-c$,
a symmetry present in the basic CH equation.  The form of the
tensor in Eq.~\eqref{eq:stress_tensor} is essential for preserving this symmetry.
 One way to picture
this symmetry is to imagine that the concentration $c\left(\bm{x},t\right)$
describes a binary mixture of a red and a green fluid, with $c=+1(-1)$ indicating
a region of pure red (green).  The invariance of the equations~\eqref{eq:NSCH2}
is the statement that a person with dichromatism can be trusted to perform
experiments on the system.  Furthermore, owing to the form of the tensor
in
Eq.~\eqref{eq:stress_tensor}, the equations~\eqref{eq:NSCH2} are rotationally
invariant.  Indeed, Eqs.~\eqref{eq:NSCH2} provide a simple model in which
these symmetries hold, and this reasoning provides another route to obtaining
model equations for advection and phase separation.

We note that the passive version of Eq.~\eqref{eq:NSCH2} is obtained by setting
the backreaction term in the tensor $T_{ij}$ to zero.  In this limit, it
is possible to imagine the incompressible velocity field $\bm{v}\left(\bm{x},t\right)$
as being prescribed, and we consider the much simpler equation
\[
\frac{\partial{c}}{\partial t}+\bm{v}\cdot\nabla{c}=\Diff\Delta\left(f_0'\left(c\right)-\Tr\Delta{c}\right),\qquad\nabla\cdot\bm{v}=0,
\]
where $\Diff=\tilde{D}/\rho_0$ is a diffusion coefficient.  We shall be concerned
with this \emph{passive advection equation} in Chapters~\ref{ch:chaotic_advection}
and~\ref{ch:estimating_mixedness}.

\section{Inequalities: Relations between function spaces}
\label{sec:background:inequalities}

In this section we outline some of the basic tools necessary for the mathematical
analysis of partial differential equations (PDEs).  We shall study a periodic
domain $\Omega$ in $\mathbb{R}^n$, although these results require only that
$\Omega$ be compact with Lipschitz boundary $\partial\Omega$.  We consider
classes of functions that form a vector space with respect to a given norm.
 By deriving inequalities for the norms, we shall obtain
 relations between the vector spaces of functions.

We study functions of type
\[
f:\Omega\rightarrow\mathbb{R}.
\]
The $L^p$ norm of the function $f$ is the number
\begin{equation}
\|f\|_p=\left(\int_\Omega{d^nx}|f|^p\right)^{1/p},\qquad p\geq1,
\label{eq:Lp}
\end{equation}
provided the integral exists.  If the integral does exist, we write $f\in
L^p\left(\Omega\right)$.  The set $L^p\left(\Omega\right)$ is closed under
addition and scalar multiplication of functions and therefore forms a vector
space.  This closure result will be useful in Ch.~\ref{ch:analysis_thin_films}.
 We also have the Sobolev $\left(q,p\right)$ norm of the function $f$,
\begin{equation}
\|f\|_{q,p}=\left(\int_\Omega{d^nx}\sum_{\alpha_1+...+\alpha_n\leq q}\left|D^{\alpha}f\right|^p\right)^{1/p},\qquad
q\geq0,\phantom{a}p\geq 1
\label{eq:sobolev}
\end{equation}
provided the integrals exist.  Here $D^\alpha$ refers to the derivative
\[
{D}^\alpha=\frac{\partial^{\alpha_1+...+\alpha_n}}{\partial x_1^{\alpha_1}...\partial
x_n^{\alpha_n}}.
\]
If the integrals in Eq.~\eqref{eq:sobolev} exist we write $f\in H^{q,p}\left(\Omega\right)$.
 Again, it can be shown that $H^{q,p}\left(\Omega\right)$ forms a vector
 space, called the Sobolev $\left(q,p\right)$ space.  Finally, for functions
 of the type
\[
f:\Omega\times\left[0,T\right]\rightarrow\mathbb{R},
\]
we introduce the norm
\[
\|f\|_{L^r\left(0,T;H^{q,p}\left(\Omega\right)\right)}=\left(\int_0^T{dt}\|f\|_{q,p}^r\right)^{1/r},\qquad
r\geq1
\]
provided the integral exists.  If the integral exists we write $f\in L^r\left(0,T;H^{q,p}\left(\Omega\right)\right)$,
and identify another function space, $L^r\left(0,T;H^{q,p}\left(\Omega\right)\right)$.

To make progress in deriving relations between these norms, we introduce
the following inequalities for real numbers.
\begin{itemize}
\item
\textbf{Triangle inequality.}  If $a$ and $b$ are real numbers, then
\[
|a+b|\leq |a|+|b|.
\]
\item
\textbf{Young's first inequality.}  If $a$ and $b$ are real numbers, and
if $\kappa$ is any positive number, then
\[
|ab|\leq \kappa a^2 + \frac{b^2}{4\kappa}.
\]
\textit{Proof:}  Since the square of any real number is nonnegative,
\[
0\leq \left(\sqrt{\kappa}|a|-\frac{1}{2\sqrt{\kappa}}|b|\right)^2=\kappa
a^2+\frac{b^2}{4\kappa}-|ab|,
\]
and the result follows.\newline
\item
\textbf{Young's second inequality.}  If $a$ and $b$ are positve real numbers,
and if $p,q>1$ satisfy ${1}/{p}+{1}/{q}=1$, then
\[
ab\leq \frac{a^p}{p}+\frac{b^q}{q}.
\]
\textit{Proof:}  Using the string of identities
\[
ab=\exp\left(\log\left(ab\right)\right)=\exp\left(\log\left(a\right)+\log\left(b\right)\right)=\exp\left(\frac{\log\left(a^p\right)}{p}+\frac{\log\left(b^q\right)}{q}\right),
\]
and the convexity of the exponential function, $e^{tx+\left(1-t\right)y}\leq
te^x+\left(1-t\right)e^y$, for $t\in\left(0,1\right)$, we obtain, by identifying
$t=1/p$, $1/q=1-t$,
\[
ab\leq\frac{a^p}{p}+\frac{b^q}{q},
\]
as required.
\end{itemize}
Using this knowledge, we prove some results for function spaces.
\begin{itemize}
\item
\textbf{H\"older's inequality.}  Let $f\in L^p\left(\Omega\right)$ and $g\in
L^q\left(\Omega\right)$, with $p,q > 1$ and $1/p+1/q=1$.  Then
\[
\|fg\|_1\leq \|f\|_p\|g\|_q.
\]
\textit{Proof:} Without loss of generality, we shall prove the inequality
for the case when $\|f\|_p=\|g\|_q=1$.  Then it suffices to show that $\|fg\|_1\leq1$.
 By Young's second inequality,
\[
|fg|\leq\frac{|f|^p}{p}+\frac{|g|^q}{q}.
\]
Integration over the domain $\Omega$ gives the required result.
\item
\textbf{Cauchy--Schwarz inequality.}  When $p=q=2$ in H\"older's inequality,
we have the result
\[
\|fg\|_1\leq \|f\|_2\|g\|_2.
\]
\item
\textbf{H\"older's inequality for} $p=\infty$, $q=1$\textbf{.}  If the function
$f$ is in the class $L^{\infty}\left(\Omega\right)$, that is, if
\[
\|f\|_{\infty}=\text{ess-sup}_{x\in\Omega}|f\left(x\right)|<\infty,
\]
then H\"older's inequality holds with $p=\infty$ and $q=1$,
\[
\|fg\|_1\leq\|f\|_\infty\|g\|_1.
\]
\item\textbf{Monotonicity of norms.}  By setting $g\left(x\right)=1$ in H\"older's
inequality, and $f=|\phi|^r$, where $r\geq1$, we obtain the result
\[
\|\phi\|_r\leq\left|\Omega\right|^{\left(1/r\right)-\left(1/s\right)}\|\phi\|_s,\qquad
r\leq s.
\]
A function that is bounded in the $L^{s}\left(\Omega\right)$ sense is therefore
bounded in the $L^{r}\left(\Omega\right)$ sense.  Such a result is called
a continuous embedding~\cite{Kantorovich}.  We write,
\[
L^s\left(\Omega\right)\subset L^r\left(\Omega\right)\qquad r\leq s.
\]
\item
\textbf{Minkowski's inequality.}  Let $f,g\in L^p\left(\Omega\right)$.  Then,
\[
\|f+g\|_p\leq \|f\|_p + \|g\|_p.
\]
\textit{Proof:} Using the triangle inequality, we have,
\[
\int_\Omega{d^nx}|f+g|^p\leq \int_\Omega{d^nx}|f+g|^{p-1}|f|+\int_\Omega{d^nx}|f+g|^{p-1}|g|.
\]
Applying H\"older's inequality to this relation, we have,
\begin{multline*}
\int_\Omega{d^nx}|f+g|^p\\
\leq \left(\int_\Omega{d^nx}|f+g|^p\right)^{1-\left(1/p\right)}\left[\left(\int_\Omega{d^nx}|f|^p\right)^{1/p}+\left(\int_\Omega{d^nx}|g|^p\right)^{1/p}\right],
\end{multline*}
and the result follows.
\item
\textbf{Poincar\'e's inequality.}  Let the function $f$
be a smooth, square-integrable function on the periodic domain
$\Omega=\left[0,L\right]^n$.  Then we have the relation
\[
\|f-\bar{f}\|_2\leq \left(\frac{L}{2\pi}\right)^2\|\nabla{f}\|_2^2,
\]
where $\overline{f}=|\Omega|^{-1}\int_\Omega{d^nx}f\left(x\right)$
is the mean value of the function $f$.\newline
\textit{Proof:} Let $\bm{m}$ be a variable running over all values in $\mathbb{N}_0^n=\left(\mathbb{N}\cup\{0\}\right)^n$.
 By Parseval's identity,
\begin{eqnarray*}
\|\nabla f\|_2^2&=&\left(\frac{2\pi}{L}\right)^2\sum_{\bm{m}\in\mathbb{N}_0^n,\bm{m}\neq0}\bm{m}^2\left(a_{\bm{m}}^2+b_{\bm{m}}^2\right),\\
&\geq&\left(\frac{2\pi}{L}\right)^2\sum_{\bm{m}\in\mathbb{N}_0^n,\bm{m}\neq0}\left(a_{\bm{m}}^2+b_{\bm{m}}^2\right),\\
&=&\left(\frac{2\pi}{L}\right)^2\left[\|f\|_2^2-\overline{f^{\phantom{2}}}^2\right].
\end{eqnarray*}
\item
\textbf{Gagliardo--Nirenberg--Sobolev inequality.}  Let the function $f$
be a mean-zero, smooth, square-integrable function on the periodic domain
$\Omega=\left[0,L\right]^n$.  Then for $1\leq q,r\leq\infty$ and $j$ and
$m$ integers satisfying $0\leq j<m$, there is a constant $0<c_1<\infty$ such
that
\[
\|D^jf\|_p\leq c_1 \|D^m f\|_r^a\|f\|_q^{1-a},
\]
where 
\[
\frac{1}{p}=\frac{j}{n}+a\left(\frac{1}{r}-\frac{m}{n}\right)+\frac{1-a}{q},
\]
and for $a$ in the interval
\[
\frac{j}{m}\leq a<1,
\]
with the restriction that if $m-j-n/r$ is a nonnegative integer, then $a$
must equal $j/m$.\newline
\textit{Proof:} The proof is given in~\cite{Adams_SobolevSpace}.
\item
\textbf{Gagliardo--Nirenberg--Sobolev inequality (celebrated special case).}
  Let the function $f$ be a smooth, square-integrable function on the periodic
  domain $\Omega=\left[0,L\right]^n$.  For $j=0$, $m=1$, $a=\tfrac{1}{2}$,
  $r<n$, and
\[
p=q=\frac{nr}{n-r}\equiv r^*>r,
\]
the Gagliardo--Nirenberg--Sobolev inequality reduces to
\[
\|f-\overline{f}\|_{r^*}\leq c_1\|\nabla f\|_r,
\]
where $r^*$ is called the Sobolev conjugate of $r$.  Rewriting this 
\[
\|f\|_{L^{r^*}\left(\Omega\right)}\leq c_2\left(\|\nabla f\|_{L^{r}\left(\Omega\right)}+|\overline{f}|\right),
\]
we have the embedding
\[
H^{1,r}\left(\Omega\right)\subset L^{r^*}\left(\Omega\right).
\]
Here $c_1$ and $c_2$ are constants that are independent of the function $f$.
\end{itemize}
Using the Minkowski inequality, and the result
\[
\|f\|_{L^p\left(\Omega\right)}=0\text{ iff }f=0\text{
almost everywhere},
\]
it follows that $\|\cdot\|_{L^p\left(\Omega\right)}$ and $\|\cdot\|_{H^{q,p}\left(\Omega\right)}$
are norms on the vector spaces $L^p\left(\Omega\right)$ and $H^{q,p}\left(\Omega\right)$
respectively.  These normed vector spaces are, moreover, complete, and therefore
form Banach spaces~\cite{Sell_dynamics}.

As a conclusion to this theory section, we provide a list nonstandard inequalities
that will be vital in Ch.~\ref{ch:analysis_thin_films} in analyzing the thin-film
Stokes Cahn--Hilliard equations  in one spatial dimension.
\begin{itemize}
\item Let $\Omega$ be a periodic domain in $\mathbb{R}$ and let $f:\Omega\rightarrow\mathbb{R}$
belong to the class $H^{1,1}\left(\Omega\right)$.  Then the following inequality
holds,
\begin{equation}
\sup_\Omega |f|\leq \frac{1}{L}\|f\|_1 + \|f_x\|_1.
\label{eq:result1}
\end{equation}
\emph{Proof:} Using the Fundamental Theorem of Calculus, we have
\[
f\left(x\right)=f\left(a\right)+\int_a^x{ds}\frac{\partial f}{\partial
s}.
\]
Let $|f\left(x\right)|$ be the maximum of the function $|f|$ over the
set $\Omega$.  That is, the maximum value of $|f|$ is attained at $x$.  Then
\[
\sup_\Omega|f|\leq |f\left(a\right)|+\int_a^{x}{ds}\left|\frac{\partial f}{\partial
s}\right|\leq  |f\left(a\right)|+\int_\Omega{ds}\left|\frac{\partial f}{\partial
s}\right|.
\]
Since this is true for any $a\in\Omega$, by integrating over $a$, we obtain
the inequality
\begin{eqnarray*}
\sup_\Omega|f|&\leq&\frac{1}{L}\|f\|_1+\|f_x\|_1,\\
&\leq&\frac{1}{\sqrt{L}}\|f\|_2+\sqrt{L}\|f_x\|_2,
\end{eqnarray*}%
as required.
\item Let $f:\Omega\rightarrow\mathbb{R}$ belong
to the class $H^{2,1}\left(\Omega\right)$.  Then the following inequality
holds,
\begin{equation}
\|f_{x}\|_2^2\leq L\|f_{xx}\|_1^2 + \frac{4}{L}\|f\|_1\|f_{xx}\|_1.
\label{eq:result2}
\end{equation}
\emph{Proof:}  We have the identity $\int_\Omega f_x^2{dx}=-\int_\Omega ff_{xx}$,
true
for any function $f$ with periodic boundary conditions (in general, this
identity is true when $ff_x|_{\partial\Omega}=0$).  With H\"older's
inequality, this yields
\[
\|f_x\|_2^2\leq \|f\|_\infty\|f_{xx}\|_1.
\]
Using the relation~\eqref{eq:result1}, this becomes
\begin{eqnarray*}
\|f_x\|_2^2&\leq&\left[\frac{1}{L}\|f\|_1+\|f_x\|_1\right]\|f_{xx}\|_1,\\
&\leq&\left[\frac{1}{L}\|f\|_1+\sqrt{L}\|f_x\|_2\right]\|f_{xx}\|_1,
\end{eqnarray*}%
which is a quadratic inequality in $\|f_x\|_2$, with solution
\[
\|f_x\|_2\leq\tfrac{1}{2}\left[\sqrt{L}\|f_{xx}\|_1+\sqrt{L\|f_{xx}\|_1^2+4L^{-1}\|f\|_1\|f_{xx}\|_1}\right].
\]
By sacrificing the sharpness of the bound, we obtain a simpler one,
\[
\|f_x\|_2^2\leq {L}\|f_{xx}\|_1^2 + \frac{4}{L}\| f\|_1\|f_{xx}\|_1,
\]
as required.
\end{itemize}

We shall use some of this formalism in the next section in proving the existence
and uniqueness of solutions for a given PDE.

\section{Galerkin approximations}
\label{sec:background:Galerkin}

In this section we study the typical fourth-order partial differential
equation
\begin{equation}
\frac{\partial u}{\partial t}+\Delta^2u=f\left(x,t\right),
\label{eq:pde_linear}
\end{equation}
on the periodic domain $\left[0,L\right]^n$ in $\mathbb{R}^n$, although the
discussion is equally valid for any smooth bounded domain together with appropriate
boundary conditions.  The function
$f\left(x,t\right)$ belongs to the class $L^2\left(0,T;L^2\left(\Omega\right)\right)$,
for any $T>0$.  Although the equation that forms the subject of this report
is nonlinear, the intuition we gain in analyzing the linear equation~\eqref{eq:pde_linear}
will enable us to study the advective Cahn--Hilliard equation~\eqref{eq:ch_adv},
and to prove the uniqueness result of Ch.~\ref{ch:analysis_thin_films}.

We specify the initial condition
\[
u\left(x,0\right)=u_0\left(x\right)\in H^{2,2}\left(\Omega\right),
\]
and introduce the $L^2$ pairing

\begin{equation}
\left(u,v\right)=\int_\Omega{d^nx}u\left(x\right)v\left(x\right).
\label{eq:L2_paring}
\end{equation}

The strategy we take in analyzing Eq.~\eqref{eq:pde_linear} is to define
weak solutions, construct a sequence of approximate weak solutions, show
that this sequence converges to a limit that we identify as the solution,
and obtain uniqueness results for this construction.
\begin{itemize}
\item
A function $u\left(x,t\right)$ satisfies Eq.~\eqref{eq:pde_linear} weakly
on $\Omega\times\left(0,T\right]$ if the following integral identity holds,
\begin{equation}
\frac{d}{dt}\left(u,v\right)+\left(u,Q^*v\right)=\left(f,v\right),
\label{eq:weak_sln_Q}
\end{equation}
$v\left(x\right)$ is a smooth periodic test function, and $Q^*=Q=\Delta^2$.
\item
We construct approximate solutions of Eq.~\eqref{eq:pde_linear} as Galerkin
sums,
\begin{equation}
u_\ord\left(x,t\right)=\sum_{i=0}^\ord\phi_i\left(x\right)\eta_i\left(t\right),
\label{eq:galerkin}
\end{equation}
where the $\phi_i$'s are eigenfunctions of the Laplace operator on $\Omega$,
with eigenvalues $-\lambda_i^2$, and where $\phi_0$ is the constant eigenfunction.
 The set $\{\phi_0,...,\phi_\ord,...\}$ is a complete orthonormal basis with
 respect to the $L^2$ pairing~\eqref{eq:L2_paring}.
If the equation~\eqref{eq:pde_linear} has initial data
\[
u\left(x,0\right)=u_0\left(x\right)=\sum_{i=0}^{\infty}\phi_i\left(x\right)\eta_i^0,
\]
then the approximate solution $u_\ord\left(x,t\right)$ is seeded with the
initial data
\[
u_\ord\left(x,0\right)=u_\ord^0\left(x\right)=\sum_{i=0}^\ord\phi_i\left(x\right)\eta_i^0,
\]
and $u_\ord^0\left(x\right)$ converges strongly to $u_0\left(x\right)$ in
the $L^2$ norm.
\item To obtain the $\eta_i$'s, we demand that the Galerkin approximation~\eqref{eq:galerkin}
satisfy the following weak form of the PDE,
\begin{equation}
\frac{d}{dt}\left(u_\ord,\phi_i\right)+\left(u_\ord,Q^*\phi_i\right)=\left(f,\phi_i\right),
\label{eq:weak_galerkin}
\end{equation}
for $\phi_i\in\{\phi_0,...,\phi_\ord\}$.  This is simply
\[
\frac{d\eta_i}{dt}=f_i-\sum_{j=0}^\ord\eta_j\left(Q\phi_j,\phi_i\right),
\]
where the $f_i$'s are given by the relation $f=\sum_{i=0}^{\infty} f_i\left(t\right)\phi_i\left(x\right)$.
 In matrix form, this is
\begin{equation}
\frac{d\bm{\eta}}{dt}=\bm{f}-A\bm{\eta},
\label{eq:lin_alg}
\end{equation}
where $A_{ij}=\left(Q\phi_i,\phi_j\right)$.  In view of the  boundedness
of the operator $Q$, the right-hand side of this expression is globally Lipschitz
in the vector $\bm{\eta}$, and thus local existence theory~\cite{DoeringGibbon}
guarantees that Eq.~\eqref{eq:lin_alg} has a unique solution that is Lipschitz
continuous
in time, in a small interval $\left[0,\sigma\right)$.  Thus, we have prescribed
$u_\ord\left(x,t\right)$ in a small time-interval $\left[0,\sigma\right)$.
\item To continue this solution to another time interval $\left[\sigma,\sigma_1\right)$,
and thus to the whole interval $\left[0,T\right)$, we must find a bound on
$\|u_\ord\|_2$ that depends only on the initial data.   To do this, let us
first of all replace $\phi_i$ with $u_\ord$ in Eq.~\eqref{eq:weak_galerkin},
and obtain
\begin{equation}
\tfrac{1}{2}\frac{d}{dt}\|u_\ord\|_2^2+\|\Delta u_\ord\|_2^2\leq \frac{1}{4\kappa}\|f\|_2^2+\kappa\|u_\ord\|_2^2,
\label{eq:uniform_bd1}
\end{equation}
where $\kappa$ is an arbitrary positive constant that comes from Young's
first inequality.
Hence,
\[
\left(1-2\kappa T\right)\left(\sup_{t\in\left[0,T\right]}\|u_\ord\|_2^2\left(t\right)\right)
\leq \sup_{\ord\in\left[0,\infty\right)}\|u_\ord\|_2^2\left(0\right)+\frac{1}{2\kappa}\int_0^T\|f\|_2^2\left(s\right)ds.
\]
We choose $\kappa$ such that $2\kappa T<1$.  Then,
\begin{eqnarray}
\sup_{t\in\left[0,T\right]}\|u_\ord\|_2^2\left(t\right)
&\leq&\frac{1}{\left(1-2\kappa T\right)}\left[\sup_{\ord\in\left[0,\infty\right)}\|u_\ord\|_2^2\left(0\right)+\frac{1}{2\kappa}\int_0^T\|f\|_2^2\left(s\right)ds\right],\nonumber\\
&\leq&k_1<\infty,\qquad 0<t\leq\sigma\leq T,
\label{eq:uniform_bd2}
\end{eqnarray}
where the number $k_1$ depends only on the initial conditions, and on the
time
$T$.

Additional bounds are obtained by replacing
$\phi_i$ with $\Delta^2 u_\ord$ in Eq.~\eqref{eq:weak_galerkin}.  Then we
obtain the inequality
\begin{equation}
\tfrac{1}{2}\frac{d}{dt}\|\Delta u_\ord\|_2^2+\|\Delta^2 u_\ord\|_2^2\leq
\frac{1}{4\kappa}\|f\|_2^2+\kappa\|\Delta^2 u_\ord\|_2^2,
\label{eq:uniform_bd3}
\end{equation}
where $\kappa$ is an arbitrary positive constant that comes from Young's
first inequality.  Choosing $\kappa=1$ gives
\begin{eqnarray}
\sup_{t\in\left[0,\sigma\right]}\|\Delta u_\ord\|_2^2\left(t\right)&\leq&\sup_{\ord\in\left[0,\infty\right)}\|\Delta{u}_\ord\|_2^2\left(0\right)+\tfrac{1}{2}\int_0^T\|f\|_2^2\left(s\right)ds,\nonumber\\
&\leq&k_2<\infty,\qquad 0<\sigma\leq T,
\label{eq:uniform_bd4}
\end{eqnarray}
where the number $k_2$ depends only on the initial conditions, and on the
time $T$.  Choosing $\kappa=\tfrac{1}{2}$ gives
\begin{equation}
\int_0^t\|\Delta^2u_\ord\|_2^2\left(s\right)ds\leq\sup_{\ord\in\left[0,\infty\right)}\|\Delta{u}_\ord\|_2^2\left(0\right)+\int_0^t\|f\|_2^2\left(s\right)ds,\qquad
0<t\leq\sigma\leq T.
\label{eq:uniform_bd5}
\end{equation}
\item The uniform bounds~\eqref{eq:uniform_bd2},~\eqref{eq:uniform_bd4},
and~\eqref{eq:uniform_bd5}
enable us to continue the approximate solution $u_\ord\left(x,t\right)$
to the whole time interval $\left(0,T\right]$.  Moreover, a uniformly bounded
(in $L^2\left(\Omega\right)$) sequence of functions $\{u_\ord\}_{\ord=0}^{\infty}$
contains a subsequence (still denoted by $\{u_\ord\}$) which converges weakly
to a limit $u$, in the following 
sense\footnote{A bounded sequence on a
Hilbert space always has a weakly convergent subsequence.  Moreover, for weak convergence
on a Hilbert space, $x_n\rightharpoonup x$, we have $\|x\|\leq \liminf_{n\rightarrow\infty}\|x_n\|$.},
\[
\left(u_\ord,v\right)\rightarrow\left(u,v\right),\qquad
\text{ as }\ord\rightarrow\infty,\qquad v\in L^2\left(\Omega\right).
\]
Since the bound~\eqref{eq:uniform_bd2} is independent
of $\ord$ and $\sigma$, it holds in the limit $u_\ord \rightharpoonup u$,
\[
\|u\|_2^2\left(t\right)\leq k_1<\infty,\qquad 0<t\leq T.
\]
Thus, by Poincar\'e's inequality, we have a hierarchy of weak convergence
results,
\begin{multline*}
\left(u_\ord,v\right)\rightarrow\left(u,v\right),\qquad
\left(\nabla u_\ord,v\right)\rightarrow\left(\nabla u,v\right),\qquad
\left(\Delta u_\ord,v\right)\rightarrow\left(\Delta u,v\right),\\
v\in L^2\left(\Omega\right),
\end{multline*}
\begin{multline*}
\int_0^T{dt}\,\left(\nabla\Delta u_\ord,\varphi\right)\rightarrow\int_0^T{dt}\,\left(\nabla
u,\varphi\right),\\
\int_0^T{dt}\,\left(\Delta^2 u_\ord,\varphi\right)\rightarrow\int_0^T{dt}\,\left(\Delta
u,\varphi\right),\\
\text{as }\ord\rightarrow\infty,\qquad \varphi \in L^2\left(0,T;L^2\left(\Omega\right)\right).
\end{multline*}
Since the bounds we have obtained are independent of $\ord$, we have the
conditions $\nabla\Delta{u},\Delta^2{u}\in
L^2\left(0,T;L^2\left(\Omega\right)\right)$, in addition to the fact that
$u,\nabla u,\Delta u\in L^{\infty}\left(0,T;L^2\left(\Omega\right)\right)$.
\item
By multiplying the Galerkin weak solution
\begin{equation}
\frac{d}{dt}\left(u_\ord,\phi_i\right)+\left(u_\ord,Q^*\phi_i\right)=\left(f,\phi_i\right),
\end{equation}
by a function $\tau\left(t\right)$ that vanishes at $t=0,T$ and integrating
the result over time, using the limiting relation $u_\ord\rightharpoonup u$,
we obtain the relation
\begin{equation}
-\int_0^{T}{dt}\left(u,\phi_i\right)\frac{d\tau}{dt}+\int_0^T{dt}\left(u,Q^*\phi_i\right)\tau\left(t\right)=\int_0^T{dt}\left(f,\phi_i\right)\tau\left(t\right),
\end{equation}
By superposition, we can replace $\phi_i$ by a smooth function $v\left(x\right)$.
 Then, taking $\tau\left(t\right)$ to be a distribution on $\left(0,T\right)$,
 we obtain the result
\begin{equation}
\frac{d}{dt}\left(u,v\right)+\left(u,Q^*v\right)=\left(f,v\right),
\end{equation}
which is Eq.~\eqref{eq:weak_sln_Q}
\item Inspection of Eq.~\eqref{eq:weak_sln_Q} shows that $\left(u,v\right)$
is in the class $H^{1,2}\left(\left[0,T\right]\right)$, for any smooth $v\left(x\right)$,
and it is therefore continuous.  It follows that $u\left(x,0\right)=u_0\left(x\right)$,
as required.
\item Extra regularity is obtained by replacing the function $v$ in Eq.~\eqref{eq:weak_galerkin}
with $\partial u_\ord/\partial t$.  Then,
%
%
%
%
%
\begin{eqnarray*}
\int_0^T{dt}\,\bigg\|\frac{\partial u_\ord}{\partial t}\bigg\|_2^2&\leq&\sup_{N\in\left[0,\infty\right)}\|\Delta
u_\ord\|_2^2\left(0\right)+\int_0^T{dt}\,\|f\|_2^2,
\end{eqnarray*}
%
%
%
%
%
%
%
%
\end{itemize}

To characterize the solution further, we make use of the theory of semigroups~\cite{TaylorPDEbook}.
 Formally, a solution to the equation
\[
\frac{\partial u}{\partial t}+\Delta^2 u = f\left(x,t\right),\qquad u\left(x,0\right)=u_0\left(x\right)
\]
can be written as
\begin{equation}
u\left(\cdot,t\right)=e^{-t\Delta^2}u_0+\int_0^t{ds}e^{-\left(t-s\right)\Delta^2}f\left(\cdot,s\right).
\label{eq:semigroup_sln}
\end{equation}
The object $P_t:=e^{-t\Delta^2}$, $t\geq0$ is a map from the Banach space
$L^2\left(\Omega\right)$ to itself, and possesses the semigroup property
\[
P_tP_s=P_{t+s},\qquad t,s\geq0.
\]
If $u_0\in H^{2,2}\left(\Omega\right)$ and if $f\left(\cdot,\cdot\right)$
is $C^\infty$ in its 
arguments\footnote{A weaker condition on the function
$f\left(\cdot,\cdot\right)$ will suffice: if $f\left(\cdot,\cdot\right)$
is $L^2$ in space and H\"older continuous in time, then Eq.~\eqref{eq:semigroup_sln}
is still the unique, strong solution of the equation~\eqref{eq:pde_linear}.
 Relaxing the assumption on $f$ even further gives rise to a less smooth
 solution.
}, 
then it can be shown~\cite{TaylorPDEbook,HenryPDEbook} that
the solution~\eqref{eq:semigroup_sln}
is a unique, strong solution of the equation~\eqref{eq:pde_linear}, and coincides
with the solution constructed by the Galerkin approximation.  
Indeed, the semigroup equation~\eqref{eq:semigroup_sln} holds for $f=\left(x,t,\nabla{u},\Delta{u},\nabla\Delta{u}\right)$,
where $f\left(\cdot,\cdot,\cdot,\cdot,\cdot\right)$ is a $C^{\infty}$ function
of its arguments.  In this manner, we have obtained the unique, strong solution
to the equation~\eqref{eq:pde_linear}, which satisfies the prescribed initial
conditions and belongs to the class
\[
u\in L^{\infty}\left(0,T;H^{2,2}\left(\Omega\right)\right)\cap L^2\left(0,T;H^{4,2}\left(\Omega\right)\right)\cap
H^{1,2}\left(0,T;L^2\left(\Omega\right)\right).
\]
We shall use this result, and this method of constructing solutions, throughout
this report.
%
%
%
%
%
%
%
%
%
%

\section{A strong existence theory for the advective Cahn--Hilliard equation}
\label{sec:background:existence_ch}

In this section we shall prove that solutions to the equation
\[
\frac{\partial c}{\partial t}+\bm{v}\cdot\nabla{c}=\Diff\Delta\left(c^3-c-\Tr\Delta{c}\right),
\]
exist and are unique, once a sufficiently regular starting condition and
velocity field $\bm{v}\left(\bm{x},t\right)$
are prescribed.   We work on a periodic domain $\Omega=\left[0,L\right]^n$
in $\mathbb{R}^n$, although this restriction can be lifted.  We shall use
the machinery developed in Secs.~\ref{sec:background:inequalities} and~\ref{sec:background:Galerkin}.
We present this analysis for two reasons.  Firstly,  it is important
to develop understanding of the basic advective Cahn--Hilliard equation before
carrying out further numerical and analytical studies.  Secondly, since the
results of this section are a combination of old analysis of a new equation,
this section serves as a bridge between the well-known material of this chapter,
and the new results that follow.
To our knowledge, no analysis of the advective Cahn--Hilliard equation~\eqref{eq:ch_analysis}
exists in the literature, although the proof presented here is similar to
Elliott's and Zheng's~\cite{Elliott_Zheng} for the nonadvective case.  
The introduction of advection necessitates a new ingredient in the proof,
which we provide in~\eqref{eq:bound_on_bad_bit_existence}.
%
%
%
%
We shall provide analysis of the long-time behaviour of
Eq.~\eqref{eq:ch_analysis} in Ch.~\ref{ch:estimating_mixedness}.
Here prove the following statement,
\begin{quote}
\textit{Given the advective Cahn--Hilliard equation
\begin{equation}
\frac{\partial c}{\partial t}+\bm{v}\cdot\nabla{c}=\Diff\Delta\left(c^3-c-\Tr\Delta{c}\right),\qquad
\nabla\cdot\bm{v}=0,
\label{eq:ch_analysis}
\end{equation}
defined on the periodic domain $\Omega=\left[0,L\right]^n$, where $\Diff$
and $\Tr$ are positive constants, where $\bm{v}\left(\bm{x},t\right)$ is
bounded in space and time in the following sense,
\[
\bm{v}\in H^{1,\infty}\left(\Omega\right),
\]
and where the initial data are defined by the identity
\[
c\left(x,0\right)=c_0\left(x\right)\in H^{2,2}\left(\Omega\right),
\]
there exists a unique strong solution to Eq.~\eqref{eq:ch_analysis} that
lives in the class
\[
c\in L^{\infty}\left(0,T;H^{2,2}\left(\Omega\right)\right)\cap L^2\left(0,T;H^{4,2}\left(\Omega\right)\right)\cap
H^{1,2}\left(0,T;L^2\left(\Omega\right)\right),
\]
for any $T>0$.
}
\end{quote}

We begin by constructing a Galerkin approximation to the weak problem
\begin{equation}
\frac{d}{dt}\left(c,u\right)+\left(c,\Tr\Diff\Delta^2u-\bm{v}\cdot\nabla{u}\right)=\Diff\left(c^3-c,\Delta{u}\right),
\label{eq:ch_weak}
\end{equation}
where $u\left(x\right)$ is a smooth function on $\Omega$.
We form the Galerkin approximation
\[
c_\ord=\sum_{i=1}^{\ord}\eta_i\left(t\right)\phi_i\left(x\right),
\]
with initial data
\[
c_0\left(x,0\right)=\sum_{i=0}^\infty \eta_i^0\phi_i\left(x\right),\qquad
c_\ord\left(x,0\right)=c_\ord^0\left(x\right)=\sum_{i=0}^\ord\eta_i^0\phi_i\left(x\right),
\]
where the $\phi_i$'s are the eigenfunctions of $\Delta$ on the domain $\Omega$,
with corresponding eigenvalues $-\lambda_i^2$; $\phi_0$ is the constant eigenfunction.
 The Galerkin approximation $c_\ord\left(x,t\right)$ is  then required to
 satisfy the equation
\[
\frac{d}{dt}\left(c_\ord,\phi_i\right)+\left(c_\ord,\Tr\Diff\Delta^2\phi_i-\bm{v}\cdot\nabla{\phi_i}\right)=\Diff\left(c_\ord^3-c_\ord,\Delta{\phi_i}\right),
\] 
or in vector form,
\begin{equation}
\frac{d\eta_i}{dt}=\sum_{j=0}^\ord A_{ij}\eta_j-\Diff\lambda_i^2\sum_{j,k,l=0}^{\ord}\eta_j\eta_k\eta_l
B_{ijkl},
\label{eq:vector_ch}
\end{equation}
where $A_{ij}$ is the matrix
\begin{eqnarray*}
A_{ij}&=&\left(\phi_i,-\Diff\Delta\phi_j+\Tr\Diff\Delta^2\phi_j-\bm{v}\cdot\nabla\phi_j\right),\\
&=&\left(\phi_i,\Diff\lambda_j^2\phi_j+\Tr\Diff\lambda_j^4\phi_j-\sqrt{-1}\bm{v}\cdot\bm{k}_j\phi_j\right),\\
&=&\delta_{ij}\left(\Diff\lambda_j^2+\Tr\Diff\lambda_j^4-\sqrt{-1}\bm{v}\cdot\bm{k}_j\right),\qquad\text{
no sum over }j,
\end{eqnarray*}
and thus $\|A\|_2$ exists (of course, any linear operator on a finite-dimensional
vector space has finite norm).  On the other hand, we have the totally symmetric
tensor
\[
B_{ijkl}=\int_\Omega{d^nx}\phi_i\phi_j\phi_k\phi_l,
\]
which is bounded in sense that $\sup_{ijkl}|B_{ijkl}|<\infty$.  It follows
that the function $\sum_{jkl}B_{ijkl}\eta_i\eta_j\eta_l$ is Lipschitz in the
vector $\left(\eta_0,...,\eta_\ord\right)$, and local existence theory then
guarantees that a solution to Eq.~\eqref{eq:vector_ch} exists in a small
time-interval $\left(0,\sigma\right)$.

Next, we obtain a uniform bound on $c_\ord\left(x,t\right)$.  Testing the equation~\eqref{eq:ch_weak}
with $v=c_\ord$, we obtain,
\begin{eqnarray*}
\tfrac{1}{2}\frac{d}{dt}\|c_\ord\|_2^2+\Tr\Diff\|\Delta c_\ord\|_2^2 &=&
\Diff\left(c_\ord^3-c_\ord,\Delta{c}_\ord\right),\\
&=&\left(\nabla c_\ord,\nabla c_\ord\right)-\left(3c_\ord^2,|\nabla{c}_\ord|^2\right),\\
&\leq &\Diff\|\nabla{c}_\ord\|_2^2,\\
&\leq &\Diff\|\Delta{c}_\ord\|_2\|c_\ord\|_2,\\
&\leq &\Tr\Diff\|\Delta{c}_\ord\|_2^2 + \frac{D}{4\gamma}\|c_\ord\|_2^2,
\end{eqnarray*}%
where the last line makes use of Young's first inequality with $\kappa=\gamma$.
Thus,
\begin{eqnarray}
\|c_\ord\|_2^2\left(T\right)&\leq&\left(\sup_{\ord\in\left[0,\infty\right)}\|c_\ord\|_2^2\left(0\right)\right)e^{\Diff{T}/2\Tr},\nonumber\\
\Tr\Diff\int_0^T{dt}\|\Delta{c}_\ord\|_2^2&\leq& \left(\sup_{\ord\in\left[0,\infty\right)}\|c_\ord\|_2^2\left(0\right)\right)\left(e^{\Diff{T}/2\Tr}-1\right),
\label{eq:useless_bounds}%
\end{eqnarray}
where $0<T<\sigma$.

Now we consider the Galerkin approximation
\[
\frac{d}{dt}\left(c_\ord,\phi_i\right)+\left(c_\ord,\Tr\Diff\Delta^2\phi_i-\bm{v}\cdot\nabla{\phi_i}\right)=\Diff\left(c_\ord^3-c_\ord,\Delta{\phi_i}\right),
\] 
and the chemical potential $\mu_\ord=c_\ord^3-c_\ord-\Tr\Delta c_\ord$, and
replace the test function $\phi_i$ by $\mu_\ord$.  Thus,
\[
\left(\frac{\partial c_\ord}{\partial t},\mu_\ord\right)+\left(\bm{v}\cdot\nabla{c}_\ord,\mu_\ord\right)=-\Diff\left(\nabla\mu_\ord,\nabla\mu_\ord\right),
\]
or
\[
\frac{dF_\ord}{dt}+\left(\bm{v}\cdot\nabla{c}_\ord,\mu_\ord\right)=-\Diff\left(\nabla\mu_\ord,\nabla\mu_\ord\right),
\]
where $F_\ord=\int_\Omega{d^nx}\left[\tfrac{1}{4}\left(c_\ord^2-1\right)^2+\tfrac{1}{2}\Tr\left|\nabla{c}_\ord\right|^2\right]$
is the approximate free energy.  We examine the term $\left(\bm{v}\cdot\nabla{c}_\ord,\mu_\ord\right)$.
 The parts of $\mu_\ord$ that are powers of $c_\ord$ drop out of this expression,
 owing to the antisymmetry of the operator $\bm{v}\cdot\nabla$.  We are left
 with
\begin{eqnarray*}
\left(\bm{v}\cdot\nabla c_\ord,\Delta c_\ord\right)&=&\int_\Omega\left(\partial_i\partial_ic_\ord\right)\left(v_j\partial_jc_\ord\right)d^{\n}x,\\
&=&-\int_\Omega\left(\partial_ic_\ord\right)\left[\partial_i\left(v_j\partial_jc_\ord\right)\right]d^{\n}x,\\
&=&-\int_\Omega\left(\partial_ic_\ord\right)\left(\partial_iv_j\right)\left(\partial_jc_\ord\right)d^{\n}x-
\int_\Omega\left(\partial_ic_\ord\right)\left(\bm{v}\cdot\nabla\right)\left(\partial_ic_\ord\right)d^{\n}x,\\
&=&-\int_\Omega\bm{w}\Deform\bm{w}^Td^\n x,\qquad\bm{w}=\nabla c_\ord,\qquad
{\Deform}_{ij}=\tfrac{1}{2}\left(\partial_iv_j+\partial_jv_i\right).
\end{eqnarray*}
The quadratic form $\bm{w}\Deform\bm{w}^T$ satisfies 
$\left|\bm{w}\Deform\bm{w}^T\right|\leq{\n}\max_{ij}|\Deform_{ij}|\|\bm{w}\|_2^2$,
which gives rise to the inequality
\begin{equation}
\left|\int_\Omega\Delta c_\ord\bm{v}\cdot\nabla c_\ord d^\n x\right|\leq{\n}\left(\sup_{\Omega,i,j}\left|\Deform_{ij}\right|\right)\int_{\Omega}\left|\nabla
c_\ord\right|^2d^\n x.
\label{eq:bound_on_bad_bit_existence}
\end{equation}
Hence, 
\[
\frac{dF_\ord}{dt}+\|\nabla \mu_\ord\|_2^2\leq n \Deform_{\infty}\|\nabla
c_\ord\|_2^2,
\]
where $W_{\infty}=\sup_{t\in\left[0,\infty\right),\Omega,i,j}|W_{ij}|$.
Integration over time yields
\[
\int_0^T{dt}\|\nabla\mu_\ord\|_2^2\leq \sup_{\ord\in\left[0,\infty\right)}F_\ord\left(0\right)+n\Deform_{\infty}\int_0^T{dt}\|\nabla
c_\ord\|_2^2,
\]
and since $c_\ord\in L^{\infty}\left(0;T;L^2\left(\Omega\right)\right)$ and
$\Delta c_\ord\in L^2\left(0,T;L^2\left(\Omega\right)\right)$, it follows
that 
\[
\nabla \mu_\ord,\nabla\Delta c_\ord \in L^2\left(0,T;L^2\left(\Omega\right)\right),
\]
independently of the order of the Galerkin approximation $\ord$.

The same integration yields
\[
\sup_{t\in\left[0,T\right]} F_\ord\left(t\right) + \int_0^T{dt}\|\nabla\mu_\ord\|_2^2\leq
\sup_{\ord\in\left[0,\infty\right)}
F_\ord\left(0\right)+ n \Deform_{\infty}\int_0^T{dt}\|\nabla{c}_\ord\|_2^2
\]
Since $\int_0^T{dt}\|\nabla c_\ord\|_2^2 < \infty$, independently of $\ord$,
we have the bound
\[
\tfrac{1}{2}\Tr\sup_{\left[0,T\right]}\|\nabla c_\ord\|_2^2 < \infty,
\]
independently of $\ord$.
Hence, from the theory developed in Sec.~\ref{sec:background:Galerkin}, it
follows that
\begin{itemize}
\item
$\left(c_\ord,\nabla{c}_\ord,\Delta c_\ord,\mu_\ord,\nabla\mu_\ord\right)\rightharpoonup\left(c,\nabla{c},\Delta{c},\mu,\nabla\mu\right)$,
weakly in $L^2\left(0,T;L^2\left(\Omega\right)\right)$;
\item $c \in L^{\infty}\left(0,T; H^{1,2}\left(\Omega\right)\right)$.
\end{itemize}
We shall now prove that $\Delta^2 c_\ord\in L^2\left(0,T;L^2\left(\Omega\right)\right)$,
and hence that $\Delta\left(c_\ord^3-c_\ord\right)\equiv \Delta\mu_{0,\ord}\in
L^2\left(0,T;L^2\left(\Omega\right)\right)$.
 We shall work in dimension $n=1$, although the generalization to dimensions
 $n=2$ and $n=3$ is straightforward. 
We multiply the advective Cahn--Hilliard equation~\eqref{eq:ch_analysis}
by $\Delta^2 c_\ord$ and integrate over space, to obtain
\begin{eqnarray*}
\frac{d}{dt}\|\Delta c_\ord\|_2^2&+&\Diff\Tr\|\Delta^2 c_\ord\|_2^2\\
&=&\Diff\left(\Delta\mu_{0,\ord},\Delta^2{c}_\ord\right)-\left(\bm{v}\cdot\nabla{c}_\ord,\Delta^2{c}_\ord\right),\qquad
\mu_{0,N}=c_{0,N}^3-c_{0,N}\\
&\leq&\Diff\left(\Delta\mu_{0,\ord},\Delta^2{c}_\ord\right)+\|\bm{v}\|_p\|\nabla{c}_\ord\|_q\|\Delta^2{c}_\ord\|_2,\qquad\frac{1}{p}+\frac{1}{q}=\tfrac{1}{2}\\
&\leq&\Diff\|\Delta\mu_{0,\ord}\|_2\|\Delta^2 c_\ord\|_2+\|\bm{v}\|_p\|\nabla{c}_\ord\|_q^2\|\Delta^2{c}_\ord\|_2,\\
&=&D\|\left[\left(3c_\ord^2-1\right)\Delta{c}_\ord+6c_\ord\left|\nabla{c}_\ord\right|^2\right]\|_2\|\Delta^2{c}_\ord\|_2+\|\bm{v}\|_p\|\nabla{c}_\ord\|_q\|\Delta^2{c}_\ord\|_2.
\end{eqnarray*} 
Let us consider the terms on the right-hand side.  The first term is
\begin{eqnarray*}
D\|\left(3c_\ord^2-1\right)\Delta{c}_\ord\|_2\|\Delta^2{c}_\ord\|_2&\leq&\|3c_\ord^2-1\|_\infty\|\Delta{c}_\ord\|_2\|\Delta^2{c}_\ord\|_2,\\
&\leq& \left(\sup_{t\in\left[0,T\right]}\|3c_\ord^2-1\|_\infty^2\right)\left(\frac{1}{4\kappa_1}\|\Delta{c}_\ord\|_2^2+\kappa_1\|\Delta^2{c}_\ord\|_2^2\right),
\end{eqnarray*} 
where $0<\kappa_1<\infty$ is an arbitrary positive constant.  The second
term is
\begin{eqnarray*}
6D\|c_\ord \left|\nabla c_\ord\right|^2\|_2\|\|\Delta^2{c}_\ord\|_2&\leq&6D\|{c}_\ord\|_\infty\|\nabla{c}_\ord\|_4\|\|\Delta^2{c}_\ord\|_2,\\
&\leq&6DL^{3/4}\|{c}_\ord\|_\infty \|\Delta c_\ord\|_2\|\Delta^2{c}_\ord\|_2,\qquad{n=1},\\
&\leq & 6DL^{3/4}\|{c}_\ord\|_\infty \left(\|c_\ord\|_2\|\Delta^2 c_\ord\|_2\right)^{1/2}\|\Delta^2{c}_\ord\|_2,\\
&\leq& 6DL\|c_\ord\|_\infty^{3/2}\|\Delta^2c_\ord\|_2^{3/2},\\
&\leq& \frac{\kappa_3}{\kappa_2^3}\|c_\ord\|_\infty^6
+ \kappa_2\|\Delta^2 c_\ord\|_2^2,
\end{eqnarray*}
where we have used Young's second inequality and the Gagliardo--Nirenberg--Sobolev
inequality in one spatial dimension.  In particular, we have used the following
Gagliardo--Nirenberg--Sobolev estimates,
\[
\|\nabla c_\ord\|_4\leq L^{\frac{3}{4}}\|\Delta c_\ord\|_2,\qquad n=1,
\]
and 
\[
\|c_\ord\|_\infty\leq \frac{1}{\sqrt{L}}\|c_\ord\|_2+\sqrt{L}\|\nabla c_\ord\|_2<\infty,\qquad
n=1;
\]
the procedure for obtaining these results is sketched in Sec.~\ref{sec:background:inequalities}.
 The same approach, with appropriate Gagliardo--Nirenberg--Sobolev estimates,
 gives the necessary bounds in higher dimensions.
%
%
%
%
%
%
%
Finally, the third term is
\begin{eqnarray*}
\|\bm{v}\|_p\|\nabla{c}_\ord\|_q\|\Delta^2{c}_\ord\|_2&\leq&\frac{1}{4\kappa_4}\|\bm{v}\|_p^2\|\nabla{c}_\ord\|_q^2+\kappa_4\|\Delta^2{c}_\ord\|_2^2.
\end{eqnarray*} 
Choosing $\kappa_1 \left(\sup_{t\in\left[0,T\right]}\|3c_\ord^2-1\|_\infty^2\right)+\kappa_2+\kappa_4=\tfrac{1}{2}\Tr\Diff$,
and
integrating over $\left[0,T\right]$ gives the inequality
\begin{multline*}
\|\Delta c_\ord\|_2^2\left(T\right)+\tfrac{1}{2}\Tr\Diff\int_0^T{dt}\|\Delta^2{c}_{\ord}\|_2^2\\
\leq \|\Delta c_\ord\|_2^2\left(0\right)
+\frac{1}{4\kappa_1}\left(\sup_{t\in\left[0,T\right]}\|3c_\ord^2-1\|_\infty^2\right)\int_0^T{dt}\|\Delta{c}_\ord\|_2^2\\
+ \frac{\kappa_3}{\kappa_2^3}\left(\sup_{t\in\left[0,T\right]}\|c_\ord\|_\infty^6\right)
+\frac{1}{4\kappa_4}\int_0^T{dt}\|\bm{v}\|_p^2\|\nabla{c}_\ord\|_q^2.
\end{multline*}
A sufficient condition for the boundedness of $\|\Delta c_\ord\|_2$, and $\int_0^T{dt}\|\Delta^2
c_\ord\|_2$ is obtained when $p=\infty$ and $q=2$.  We require that
\[
\int_0^T{dt}\|\bm{v}\|_{\infty}^2<\infty.
\]
Then we have the bound
\begin{multline*}
\|\Delta c_\ord\|_2^2\left(T\right)+\tfrac{1}{2}\Tr\Diff\int_0^T{dt}\|\Delta^2{c}_\ord\|_2^2\\
\leq \sup_{\ord\in\left[0,\infty\right)}\|\Delta c_\ord\|_2^2\left(0\right)
+\frac{1}{4\kappa_1}\left(\sup_{t\in\left[0,T\right]}\|3c_\ord^2-1\|_\infty^2\right)\int_0^T{dt}\|\Delta{c}_\ord\|_2^2\\
+ \frac{\kappa_3}{\kappa_2^3}\left(\sup_{t\in\left[0,T\right]}\|c_\ord\|_\infty^6\right)
+\frac{1}{4\kappa_4}\left(\sup_{t\in\left[0,T\right]}\|\nabla{c}_\ord\|_2^2\right)\int_0^T{dt}\|\bm{v}\|_\infty^2.
\end{multline*}
Owing to the $\ord$-independence of these results we obtain further information
about the regularity of the weak solution $c=\lim_{\ord\rightarrow\infty}
c_\ord$,
\begin{itemize}
\item $\Delta{c}\in L^{\infty}\left(0,T;L^2\left(\Omega\right)\right)$, 
\item $\Delta^2{c}\in L^2\left(0,T;L^{2}\left(\Omega\right)\right)$.
\end{itemize}
It follows that $\Delta \mu_0 \in L^2\left(0,T;L^2\left(\Omega\right)\right)$,
which paves the way for our key result.  This result also holds in dimensions
$n=2$ and $n=3$.

We obtain our final result by rewriting the advective Cahn--Hilliard equation
as
\[
\frac{\partial c}{\partial t}+\Tr\Diff\Delta^2 c = \Diff\Delta\underbrace{\left(c^3-c\right)}_{=\mu_0}
- \bm{v}\cdot\nabla{c}\equiv{f}.
\]
We have shown that the function $f$ resides in $L^2\left(0,T;L^2\left(\Omega\right)\right)$.
 Using the result of Sec.~\ref{sec:background:Galerkin}, it follows that
 the solution
 $c\left(\bm{x},t\right)$ is in fact strong, unique, and belongs to the class
\[
c\in L^{\infty}\left(0,T;H^{2,2}\left(\Omega\right)\right)\cap L^2\left(0,T;H^{4,2}\left(\Omega\right)\right)\cap
H^{1,2}\left(0,T;L^2\left(\Omega\right)\right).
\]

Although the results in this section hold for any finite time-interval $\left[0,T\right]$,
they tell us nothing about the late-time behaviour of the solution to Eq.~\eqref{eq:ch_analysis}.
 Indeed, the exponential-in-time bounds~\eqref{eq:useless_bounds} are distinctly
 unhelpful in
 answering this
 question.  In Ch.~\ref{ch:estimating_mixedness} we shall prove the existence
 of long-time averages of the free energy $F$.  Let us, however, take this
 existence on trust for now,
and perform numerical
simulations on Eq.~\eqref{eq:ch_analysis} for the case where $\bm{v}\left(\bm{x},t\right)$
is a chaotic flow.

%% file: chaotic_advection/chaotic_advection.tex
\chapter{Chaotic stirring of a Cahn--Hilliard fluid}
\label{ch:chaotic_advection}

\section{Overview}

In this chapter we study the advective Cahn--Hilliard (CH) equation
\begin{equation}
\frac{\partial c}{\partial t} +\bm{v}\cdot\nabla c =\Diff\Delta\left(c^3-c-\gamma\Delta{c}\right),\qquad\nabla\cdot\bm{v}=0,
\label{eq:ch_adv}
\end{equation}
for the case where $\bm{v}\left(\bm{x},t\right)$ is a prescribed chaotic
flow.   We focus our attention on the symmetric mixture, in which equal
amounts of both binary fluid components are present.   This
special case involves a wide range of phenomena~\cite{Zhu_numerics,Hashimoto,Berti2005,Krekhov2004,Tong1989}
and we are therefore justified in considering it here.  However, the results
in this chapter do not depend on this assumption.  By numerically simulating
a chaotic flow by the random-phase sine map, we characterize the effects
of chaotic flow on the late-time concentration morphology.  The key tool
in this study is the use of scaling laws derived
from dimensional analysis.

\section{Scaling laws for the advective Cahn--Hilliard equation}
\label{sec:chaotic_advection:model}
In this section we discuss the notion of dynamical equilibrium for the CH
equation without flow in terms of the structure function.  We then examine
the lengthscales that arise in the presence of flow.

The CH equation without flow exhibits dynamical equilibrium in the following
sense.  Starting from a concentration field fluctuating around the unstable
equilibrium $c\left(\bm{x},t\right)=0$, the late-time concentration field
has properties that are time independent when lengths are measured in units
of typical bubble size $R_{\mathrm{b}}$.  Such properties include the structure
function \cite{Toral_scaling} which we now introduce.
A measure of the correlation of concentration between neighbouring points,
given in Fourier space, is the following:
\begin{equation}S\left(\bm{k},t\right) = \frac{1}{|\Omega|}\int_\Omega d^n
x\int_\Omega d^n x' e^{-i\bm{k}\cdot\bm{x}}\left[c\left(\bm{x}+\bm{x'}\right)c\left(\bm{x}\right)-\overline{c^{\phantom{2}}}^2\right],
\end{equation}
where $\overline{\left(\cdot\right)}$ denotes the spatial average.  We normalize
this function and compute its spherical average to obtain the structure function
\begin{equation}
s\left(k,t\right)=\frac{1}{\left(2\pi\right)^n}\frac{\tilde{S}\left(k,t\right)}{\overline{c^2}-\overline{c^{\phantom{2}}}^2}
\label{eq:structure_fn}
\end{equation}
The spherical average $\tilde{\phi}\left(k\right)$ any function $\phi\left(\bm{k}\right)$
is defined as
\begin{equation}
        \tilde{\phi}\left(k\right) = \frac{1}{\omega_n}\int d\omega_n \phi\left(\bm{k}\right),
\end{equation}
where $d\omega_n$ is the element of solid angle in $n$ dimensions and $\omega_n$
is its integral.  Thus $\tilde{c}_k$ is the spherical average of the Fourier
coefficient $c_{\bm{k}}$.  For a symmetric binary fluid $\overline{c^{\phantom{2}}}
= 0$,
the structure function is simply the spherically-averaged power spectrum:
\begin{equation}
        s\left(k,t\right)=\frac{1}{\left(2\pi\right)^n}\frac{\left|\tilde{c}_k\right|^2}{\overline{c^2}},\qquad\text{for
        }\overline{c^{\phantom{2}}}=0.
\end{equation}   
The dominant lengthscale is identified with the reciprocal of the most important
$k$-value, defined as the first moment of the distribution $s\left(k,t\right)$,
\begin{equation}
k_1=\frac{\int_0^\infty k s\left(k,t\right)dk}{\int_0^\infty
s\left(k,t\right)dk}.
\end{equation}
Since the system we study is isotropic except on scales comparable
to the size of the problem domain, this lengthscale is a measure of scale
size in each spatial direction.  That is, we have lost no information in
performing the spherical average in Eq.~\eqref{eq:structure_fn}.  This
lengthscale is in turn identified with the bubble size: $R_{\mathrm{b}}\propto1/k_1$.
  In two-dimensional simulations ($n=2$) it is found \cite{Zhu_numerics,
  Toral_scaling} that while $k_1$ and $s\left(k,t\right)$ depend on time,
  $k_1^n s\left(k/k_1,t\right)$ is a time-independent function with a single
  sharp maximum, confirming that the system is in dynamical
equilibrium and that a dominant scale exists.  Indeed, the growth law $R_{\mathrm{b}}\sim
t^{1/3}$ is obtained, in agreement with the evaporation-condensation picture
of phase separation proposed by Lifshitz and Slyozov (LS exponent)~\cite{LS}.

We present some simple scaling arguments to reproduce the $t^{1/3}$ scaling
law and to illuminate the effect of flow.  For $\bm{v}=0$, the equilibrium
solution of Eq.~\eqref{eq:ch_adv} is $c \approx \pm1$ in domains, with small
transition regions of width $\sqrt{\gamma}$ in between.  Across these transition
regions, it can be shown \cite{LowenTrus} that
\[\mu = -{\Gamma\kappa}/2,\]
where 
\begin{equation}
        \Gamma = \sqrt{{8\gamma}/9}
        \label{eq:integral_st}        
\end{equation}
is the surface tension and $\kappa$ is the radius of curvature.
 Thus, if $R_{\mathrm{b}}$ is a typical bubble size, that is, a length over
 which $c$ is constant, the chemical potential associated with the bubble
 is
\begin{equation}\delta\mu\sim\Gamma / R_{\mathrm{b}}.\end{equation}
Balancing terms in the Cahn--Hilliard equation $\partial_t{c}+\bm{v}\cdot\nabla{c}=\Diff\Delta\mu$,
it follows that the time $t$ required for a bubble to grow to a size $R_{\mathrm{b}}$
is $1/t\sim {\Gamma D}/{R_{\mathrm{b}}^3}$, implying the LS growth law $R_{\mathrm{b}}\sim
\left(\Gamma D t\right)^{1/3}$.

Now the CH free energy $F[c]$ is the system energy owing to the presence
of bubbles:
\[
F[c]=\int_\Omega d^nx \left[\tfrac{1}{4}\left(c^2-1\right)^2
    +\tfrac{1}{2}{\gamma}\left|\nabla
    c\right|^2\right].
\]
The surface tension $\Gamma$ is the free energy per unit area.  A bubble
carries free energy $\Gamma R_{\mathrm{b}}^{n-1}$, where prefactors due to
angular integration are omitted.  The total free energy $F\left[c\right]$
in the motionless case is then $\Gamma R_{\mathrm{b}}^{n-1} \times N_{\mathrm{b}}$,
where $N_{\mathrm{b}}$ is the total number of bubbles.  Because the system
is isotropic and because there is a well-defined bubble size, we estimate
$N_{\mathrm{b}}$ by $|\Omega|/{R_{\mathrm{b}}^n}$.  The free energy then
has the
scale dependence $F/|\Omega|\sim\Gamma/R_{\mathrm{b}}$, and using the growth
law for the length $R_{\mathrm{b}}$ we obtain the asymptotic free energy
relation
\begin{equation}
  \frac{F}{|\Omega|}\sim \left(\frac{\Gamma^2}{D}\right)^{1/3}t^{-1/3}.
  \label{eq:fe_decay_mixing}
\end{equation}
By introducing the tracer variance
\begin{equation}
\sigma^2=\overline{c^2}-\overline{c^{\phantom{2}}}^2
\label{eq:sigsq}
\end{equation}
and restricting to a symmetric mixture $\overline{c^{\phantom{2}}}= 0$, we
identify $\sigma^2\simeq \Omega_{\mathrm{b}}/|\Omega|$, where $\Omega_{\mathrm{b}}
= N_{\mathrm{b}} R_{\mathrm{b}}^n$ is the volume occupied by bubbles.  Since
$F=\Gamma R_{\mathrm{b}}^{n-1} N_{\mathrm{b}}$, it follows that
\begin{equation}
        R_{\mathrm{b}}\sim\sigma^2 / F.
        \label{eq:Rdsig}
\end{equation}
We shall verify these results in Section~\ref{sec:chaotic_advection:numerics}.
Equation~\eqref{eq:Rdsig} will provide a useful measure of bubble size when
a flow is imposed, since no assumption is made about the number of bubbles
per unit volume.  In a dynamical scaling regime, the lengthscales calculated
from these energy considerations must agree with that computed from the first
moment of the structure function, since there is only one lengthscale in
such a regime \cite{Furukawa_scales}.

Stirring a CH fluid introduces new lengthscales.  The flow will alter the
sharp power spectrum found above and so it may not be possible to extract
definite scales from experiments or numerical simulations.  We might expect
further ambiguity of scales during a regime change (crossover), for example,
as the bubbles are broken up and diffusion takes over.  Nevertheless, for
a given flow, it is possible to construct lengthscales from
the system parameters.  We shall impose a flow $\bm{v}\left(\bm{x},t\right)$
that is chaotic in the sense that nearby particle trajectories separate
exponentially in time at a mean rate given by the Lyapunov exponent.

If the advection term and the surface tension term have the same order of
magnitude in a chaotic flow, by balancing the terms $\bm{v}\cdot\nabla c$
and $D\nabla^2\mu$ we obtain a scale
\begin{equation}
  R_{\mathrm{st}} = \left(\Gamma D / \lambda\right)^{1/3},
  \label{eq:Rdsim}
\end{equation}
where $\lambda$ is the Lyapunov exponent of the flow.  Because the term
$\bm{v}\cdot\nabla c$ gives an exponential amplification of gradients, the
balance of segregation and hyperdiffusion in the term $\mu = f_0'\left(c\right)-\gamma\nabla^2c$
may be broken and hyperdiffusion may overcome segregration.  This will happen
on a scale~$R_{\mathrm{diff}}$ given by the balance of the terms $\bm{v}\cdot\nabla
c$ and $\gamma\nabla^4c$,
\begin{equation}
        R_{\mathrm{diff}}= \left({\gamma D} / \lambda \right)^{1/4}.
        \label{eq:R_diff}        
\end{equation}
In this regime, we expect mixing owing to the presence of the advection and
the (hyper) diffusion.  Using Eqs.~\eqref{eq:integral_st},~\eqref{eq:Rdsim}
and~\eqref{eq:R_diff}, the crossover between the bubbly and the diffusive
regimes takes place when $\lambda\simeq D/\gamma$.  In the following sections
we shall examine these scales and look for this crossover between bubbles
and filaments, the latter being characteristic of mixing by advection-diffusion.

\section{A model stirring flow}
\label{sec:chaotic_advection:model_stirring}
In order to run simulations at high resolution, we make use of the sine
flow model~\eqref{eq:sineflow} in two dimensions.  
We are interested in the following velocity field, defined on a
square domain $\left[0,L\right]^2$ with periodic boundary conditions:
\begin{equation}
\begin{split}
  v_x\left(x,y,t\right) &= V_0\sin\left(\frac{2\pi y}{L}+\phi_j\right),\qquad
  v_y=0,\qquad j\tau\leq t<\left(j+\tfrac{1}{2}\right)\tau,\\
  v_y\left(x,y,t\right) &= V_0\sin\left(\frac{2\pi x}{L}+\psi_j\right),\qquad
  v_x=0,\qquad \left(j+\tfrac{1}{2}\right)\tau \leq t<\left(j+1\right)\tau,
\end{split}
\label{eq:sineflow_mixing}
\end{equation}
where $\phi_j$ and $\psi_j$ are phases that are randomized once during each
flow period $\tau$ and the integer $j$ labels the period.  We solve
Eq.~\eqref{eq:ch_adv} with and without the flow~\eqref{eq:sineflow} to investigate
the scaling laws presented in Section~\ref{sec:model}.

Equation~\eqref{eq:ch_adv} is integrated by an operator splitting technique,
in which a specialized advective step followed by a phase-separation step
are performed at each iteration.  The phase-separation step is the semi-implicit
spectral algorithm proposed by Zhu et al.~\cite{Zhu_numerics} for the case
without flow.  The semi-implicitness of the algorithm enables us to use a
reasonably large timestep, while the spectral nature of the phase-separation
step enables us to work at high resolution, namely $512^2$ gridpoints.  By
imposing the sine flow~\eqref{eq:sineflow},
we may use Pierrehumbert's lattice method for advection~\cite{lattice_PH1,
lattice_PH2}.  With the problem defined on a discrete grid, this method is
exact for the advection step.  The sine flow has the added advantage that
its correlation length is of order of the box size $L$, which is the largest
 scale in the problem.  Thus, using the prediction of Lacasta et al.~\cite{Lacasta1995}
 explained in Ch.~\ref{ch:introduction},
 we expect this velocity field to produce coarsening arrest even at very
 small stirring amplitudes.
The parameters $\Diff$ and $\Tr$ in Eq.~\eqref{eq:ch_adv} are
dimensional, and we therefore nondimensionalize the equation before undertaking
numerical simulations.  
Let $\Time$
be a timescale associated with the velocity $\bm{v}\left(\bm{x},t\right)$,
let $V_0$ be the magnitude of $\bm{v}\left(\bm{x},t\right)$, and finally,
let $L$ be the box size.  It is then possible to write
down Eq.~\eqref{eq:ch} using a nondimensional time $t'=t/\Time$ and a nondimensional
spatial variable $\bm{x}'=\bm{x}/L$,
\begin{equation*}
\frac{\partial c}{\partial t'}+\alpha\tilde{\bm{v}}\cdot\nabla' c = {\Diff}'\Delta'\left(c^3-c-\gamma'\Delta'
c\right),\\
\end{equation*}
where $\tilde{\bm{v}}$ is a dimensionless shape function, $\alpha = \Time
V_0/L$,
$\Diff' = \Diff\Time/L^2$, and where $\gamma'=\gamma/L^2$  The quantity $\Diff'
= \Diff\Time/L^2$ is identified
with the ratio $\Time/\Time_D$, being the ratio of the velocity timescale
to the diffusion timescale.  For ease of notation, we shall henceforth take
$\bm{v}$ to mean $\alpha\tilde{\bm{v}}$, and drop the primes from the nondimensionalized
equation.
Now the lattice method with its splitting of the advection
and diffusion
steps, is effective only when the spread of matter due to diffusion is much
slower than the spread due to advection, that is, $\Time/\Time_D\ll1$.
 We therefore set $\Diff=10^{-5}$.  We shall also set $\tau=1$ and $L=2\pi$.
 Following standard practice~\cite{Berti2005, chaos_Berthier}, we choose
 $\gamma\sim\Delta x^2$, the gridsize.

\section{The Lyapunov exponent for the sine map}
\label{sec:lyapunov}
Loosely speaking, the Lyapunov exponent for a flow is the flow's mean rate
of strain.  Here we define the Lyapuov exponent exactly and compute it for
the special case of the two-dimensional sine flow.  We provide an asymptotic
expressions for the exponent, in the limit of small and large stirring amplitudes.

The integral of the period-one sine flow~\eqref{eq:sineflow_mixing} gives
a map from the 2-torus to itself:
\begin{equation}
\begin{split}
  x_{n+1} &= x_n+a\alpha\sin\left(y_n+\phi_n\right),\\
  y_{n+1} &= y_n+\alpha\sin\left(x_{n+1}+\psi_n\right).
\end{split}
\label{eq:sinemap}
\end{equation}
By studying this map we may compute the Lyapunov exponent of the flow~\eqref{eq:sineflow_mixing}.
 Here $\alpha$ and $a$ are stirring parameters and $\phi_n$ and $\psi_n$
 are phases that are randomized once per period.
The chaoticity of this map is quantified by the function $\gamma_q$:
\begin{equation}
\gamma_q\left(a,\alpha\right) = 
\lim_{n\to\infty}\frac{1}{n q} \log \Big\langle\left(\frac{w^T{J_n^{\mathrm{cum}}}^T
J_n^{\mathrm{cum}} w}{w^T w}\right)^{q/2}\Big\rangle,\qquad q\in\left[0,\infty\right),
\end{equation}
where $w = \left(x_0,y_0\right)$ is a two-vector of initial conditions, $\alpha$
is the stirring parameter and $J_n^{\mathrm{cum}}$ is the matrix
\begin{equation}
J_n^{\mathrm{cum}}(w,\alpha)=\frac{\left(\partial
x_{n+1},\partial y_{n+1}\right)}{\left(\partial x_n,\partial y_n \right)}\frac{\left(\partial
x_{n},\partial y_{n}\right)}{\left(\partial x_{n-1},\partial y_{n-1} \right)}...\frac{\left(\partial
x_{1},\partial y_{1}\right)}{\left(\partial x_0,\partial y_0 \right)},
\end{equation}
with 
$\left(\partial x_{1},\partial y_{1}\right)/\left(\partial x_0,\partial
y_0 \right)=\bm{1}_{2\times 2}$.  
In what follows, for brevity, let
\[
J_n = \frac{\left(\partial x_{n+1},\partial y_{n+1}\right)}{\left(\partial
x_n,\partial y_n \right)},\qquad
L_n=\frac{w^T{J_n^{\mathrm{cum}}}^T J_n^{\mathrm{cum}}w}{w^T w}.
\]  
Finally, let the angle brackets $\langle..\rangle$ denote an ensemble average
over all possible phase angles.

We study the function $\gamma_q$ for particular values of $q$.  For example,
by setting $q = 0$ we arrive at an expression of type $0/0$, and applying
L'H\^opital's rule we replace $\frac{1}{q}\log\langle L^{q/2}\rangle$ with
\begin{equation}
\frac{d}{dq}\log\langle L^{q/2}\rangle=\frac{\langle L^{q/2}\log L^{1/2}\rangle}{\langle
L^{q/2}\rangle}.
\end{equation}
Thus,
\begin{equation}
\gamma_0\left(a,\alpha\right) \equiv\lambda = \lim_{n\to\infty}\frac{1}{2n}\langle\log
L_n\rangle;\qquad\gamma_2\left(a,\alpha\right) = \lim_{n\to\infty}\frac{1}{2n}\log\langle
L_n\rangle.
\end{equation}
The quantity $\lambda$ is the Lyapunov exponent of the map.  In general,
$\gamma_q$
is a monotone increasing function of $q$ (see for example, Finn and Ott~\cite{Finn1988}),
so that $\gamma_2$ is an upper bound for $\lambda$.  This nonequality also
indicates the importance of the order in which the averaging is performed.
\begin{figure}[htb]
\centering
\noindent
\subfigure[]{
  \includegraphics[width=.3\textwidth]{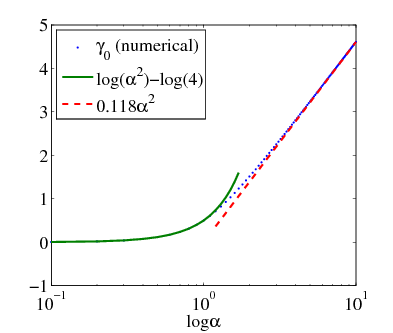}
}
\subfigure[]{
    \includegraphics[width=.3\textwidth]{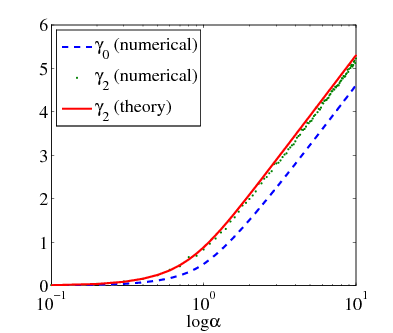}
}
\subfigure[]{
   \includegraphics[width=.3\textwidth]{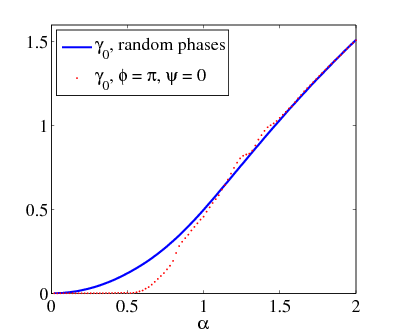}
}
\caption{%
(a) Lyapunov exponent $\gamma_0$ and its small- and large-amplitude forms;
(b) Comparison between the Lyapunov exponent and $\gamma_2$;
(c) Comparison between the Lyapunov exponent for the sine map with and without
the randomization of phases.
}
\label{fig:leps}
\end{figure}

It is straightforward to compute $\gamma_2$: we obtain
the exact result
\begin{equation}
\gamma_2 = \frac{1}{2}\log\left[1+\frac{\beta^4}{8}+\frac{\beta^2}{8}\sqrt{16+\beta^4}\right],
\label{eq:gamma2}
\end{equation}
where $\beta = \alpha\sqrt{a}$.  The evaluation of $\gamma_0$ is not so straighforward
and so we have recourse to asymptotics.  Thus, we replace the matrix $J_n$
by its large-amplitude approximation
\[
\left( 
\begin{array}{cc}
0 & 0 \\
0 & a\alpha^2A_n B_n \end{array} 
\right),
\]
where $A_n=\cos(y_n+\phi_n)$ and $B_n=\cos(x_{n+1} + \psi_n)$.  The quantity
$L_n$ is easily calculated: it is
\[\frac{y_0^2}{x_0^2+y_0^2}\prod_{i=1}^n \left(\alpha^2 a\right)^2 A_n^2
B_n^2.
\]  
The order in which the computation is carried out is important: we take the
log first and then ensemble average over all phase angles.  Thus,
\[\frac{1}{2 n}\log L_n = \log\alpha^2 + \log a + \frac{1}{2 n}\sum_{i=1}^n\log
A_n^2 B_n^2 + \frac{1}{2 n}\log\frac{y_0^2}{x_0^2+y_0^2}.
\]
It remains to compute the phase average $\langle\sum_{i=1}^n\log
A_n^2 B_n^2\rangle  = \langle\log A_1^2 B_1^2\rangle+...+\langle\log A_n^2
B_n^2\rangle$.  We start with the last term and integrate over
$\phi_n$ and $\psi_n$ first, since this is the only term depending on these
variables.  Hence,
\[
\langle\log A_n^2 B_n^2\rangle = \frac{1}{4\pi^2}\int_0^{2\pi}d\phi_n\int_0^{2\pi}d\psi_n
\log\left[\cos^2\left(y_n + \phi_n\right)\cos^2\left(x_{n+1}+\psi_n\right)\right],
\]
or
\begin{multline*}%
\langle\log A_n^2 B_n^2\rangle \\
= \frac{1}{4\pi^2}\left[2\pi\int_0^{2\pi}d\phi_n\log\cos^2\left(y_n+\phi_n\right)+\int_0^{2\pi}d\phi_n\int_0^{2\pi}d\psi_n\log\cos^2\left(x_{n+1}\left(\phi_n\right)+\psi_n\right)
\right].
\end{multline*}
We use the integral\footnote{For a derivation of this integral relation,
see Appendix A.}
\[\int_0^{2\pi}du\log\cos^2 u = -4\pi\log2\]
to obtain the average $\langle\log A_n^2 B_n^2\rangle=-4\log 2$, so that
$\langle\frac{1}{2n}\sum_{i=1}^n\log A_i^2 B_i^2 \rangle = -2\log 2$, and
hence the result
\begin{equation}
\lambda\sim\log\alpha^2+\log a-\log 4,\qquad\alpha\gg1
\end{equation}
The calculation in the limit $\alpha\to 0$ is not as simple, and
so we resort to numerical methods.

We compute $\gamma_2$ numerically and compare the result 
with the theoretical prediction Eq.~\eqref{eq:gamma2}.  Hereafter $a =
1$.  We find that $\gamma_2$ fluctuates noticeably compared with $\gamma_0$.
 It is systematically below the theoretical value.  However, by
 increasing the number of realizations in the ensemble average, this difference
 disappears.  These results are presented in Fig.~\ref{fig:leps}.  The reasonable
 agreement of the theory curve with the numerical one confirms the usefulness
 of the numerical calculation.  Of significance is the fact that $\gamma_0$
 is nonnegative even for very small $\alpha$.  Thus, the sine map is chaotic
 for all parameter values.

The large-$\alpha$ result for $\gamma_0$ is checked against a numerical simulation
and a perfect agreement is obtained.  In Section~\ref{sec:chaotic_advection:numerics}
we shall
need a formula for the strain rate for moderate values of the stirring amplitude
$\alpha$ and so we fit a curve to $\gamma_0\left(\alpha\right)$
for the parameter range $\alpha\in [0,1]$.  To make the system (1) invariant
under $\alpha\to -\alpha$, the fitted curve should be a polynomial in $\alpha^2$.
 The equation
\begin{equation}\lambda\left(a,\alpha\right) = 0.118a\alpha^2\end{equation}
is found to be optimal.  Going to higher order in $\alpha^2$ does not
improve the accuracy of the fit.
 
As a final computation we contrast the sine map with random phases with one
in which the phases are constant and set to $\phi=\pi$, $\psi=0$.  This contrast
is shown in Fig.~\ref{fig:leps}(c).  For large $\alpha$, these maps have
the same behaviour, growing as $\log\alpha^2$, and this is explained by the
analysis of this section.  For small $\alpha$ there is an approximate parameter
range $\alpha\in\left[0,0.5\right]$ for which the constant-phase map does
not exhibit
chaos.  The asymptotic form $\lambda=\lim_{n\to\infty}\left(1/{2n}\right)\langle\log
L_n\rangle$ is arrived at after 40 iterations for the random-phase map and
after 10,000 iterations for the constant-phase map.  Thus, in contrast to
the constant-phase map, the random-phase map is efficient in separating trajectories
at all stirring amplitudes and over all but very short timescales.

\section{Numerical simulations}
\label{sec:chaotic_advection:numerics}

We integrate Eq.~\eqref{eq:ch_adv} in the manner explained in Sec.~\ref{sec:chaotic_advection:model_stirring}.
 The initial
conditions are chosen to represent the sudden cooling of the binary
fluid: above a certain temperature $T_{\mathrm{c}}$ the homogeneous state
$c=0$ is stable, while below this temperature the mixture free energy changes
character to become $\tfrac{1}{4}\left(c^2-1\right)^2$, which makes the homogeneous
state unstable.  Thus, at temperatures $T>T_{\mathrm{c}}$, $c\left(\bm{x}\right)=0+
\text{[fluctuations]}$, and the sudden cooling of the system below $T_{\mathrm{c}}$
leaves the system in this state.   In order for the CH equation without thermal
noise~\eqref{eq:ch_adv} to describe the evolution of these initial conditions,
the cooling we have imposed must take the system to $T=0$.  Thus, we are
working in the so-called deep-quench limit.  With the initial conditions
$c\left(\bm{x},0\right)=0+\text{[fluctuations]}$, both components of the
fluid are present in equal amounts and the spatial average of the concentration
field is zero.

Although the case without flow has been investigated before \cite{Zhu_numerics,
Toral_scaling}, we reproduce it for two reasons.  First, it serves as a validation
of our algorithm.  Second, we shall confirm the free energy
laws~\eqref{eq:fe_decay_mixing} and~\eqref{eq:Rdsig} which to our knowledge have
not been examined before.  Finding an appropriate measure of bubble growth
will be important for understanding the behaviour of the bubbles when we introduce
stirring.

\begin{figure}[htb]
\centering
\noindent
\subfigure[]{
    \includegraphics[width=.35\textwidth]{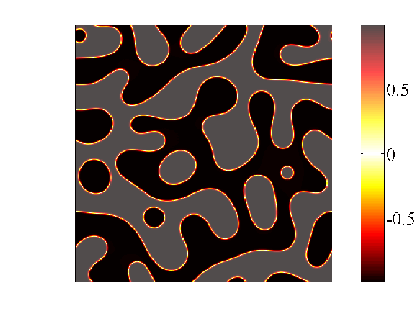}
}
\subfigure[]{
    \includegraphics[width=.35\textwidth]{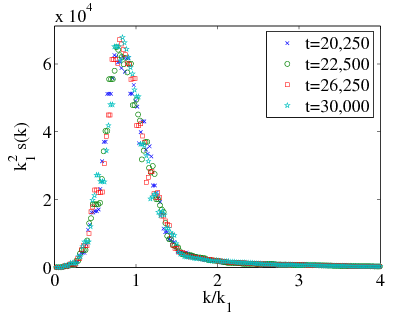}
}
\subfigure[]{
    \includegraphics[width=.35\textwidth]{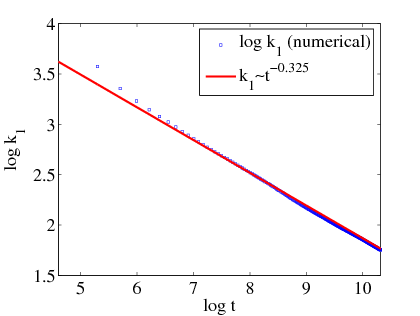}
}
\subfigure[]{
    \includegraphics[width=.35\textwidth]{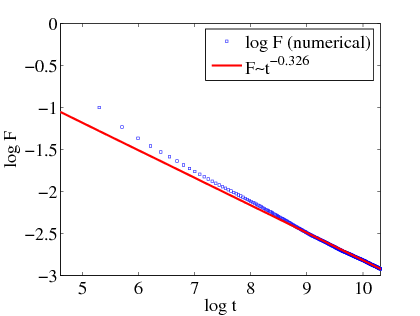}
}
\caption{%
(a) Concentration, at $t=30,000$;
(b) Structure function $k_1^2 s\left(k/k_1,t\right)$ in the scaling regime
$t\geq20,000$; (c) Time dependence of $k_1$, with power law exponent close
to the LS exponent $1/3$;
(d) Free energy, $F$, exhibiting the time dependence $F\sim t^{-0.326}$,
close to the LS exponent.
}
\label{fig:CHnoflow}
\end{figure}

The results of these simulations are presented in Fig.~\ref{fig:CHnoflow}.
The scaling function $k_1^2s\left(k/k_1,t\right)$ is approximately time independent
%
%
for $t>20,000$ as evidenced by Fig.~\ref{fig:CHnoflow}(b), implying that
the scaling exponents for $k_1$ and $F$ are to be extracted from the late-time
data with $t>20,000$.  For $t>30,000$ finite-size effects spoil the scaling
laws.  Thus, while the scaling exponents are extracted from a small window,
the fit is good and this suggests that the power laws obtained give the true
time-dependence of $k_1$ and $F$ at late times.  In this way we recover the
dynamical scaling regime in which $k_1^2 s\left(k/k_1,t\right)$ is time independent,
in addition to the decay law $k_1\sim t^{-1/3}$ \cite{Zhu_numerics}.  Running
the simulation repeatedly for an ensemble of random initial conditions improves
the accuracy of the exponent.  Furthermore, we obtain the energy laws $F\sim
c_1 t^{-1/3}$, and $1-\sigma^2\sim c_2t^{-1/3}$.  Thus,
\begin{equation}
\frac{\sigma^2}{F}=\frac{1-c_2 t^{-1/3}}{c_1t^{-1/3}}\sim\frac{1}{c_1}t^{1/3},\qquad\text{for
}t\gg1,
\end{equation}
so that $\sigma^2/F$, $1/F$ and $1/k_1$ all grow as $t^{1/3}$ and hence provide
identical measures of scale growth, in agreement with the classical assumption
that there is a unique lengthscale in the problem \cite{Bray_advphys, Zhu_numerics}.

\begin{figure}[htb]
\centering
\noindent
\subfigure[]  {\includegraphics[width=.3\textwidth]{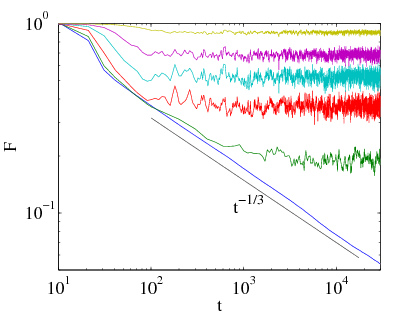}}
\subfigure[]  {\includegraphics[width=.3\textwidth]{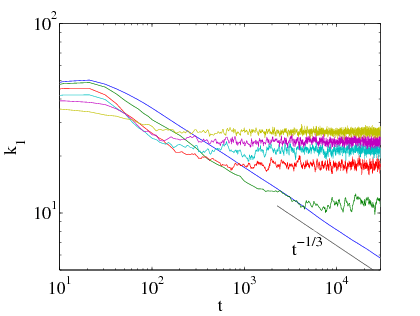}}
\subfigure[]  {\includegraphics[width=.3\textwidth]{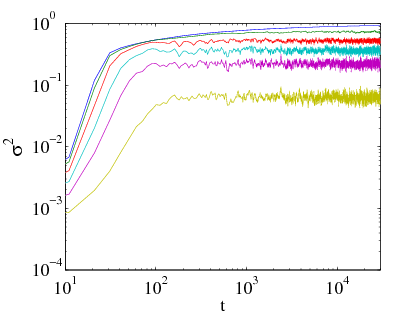}}
\caption{(a) $F$ vs $t$ for (from bottom) $\alpha=0,0.1,0.3,0.5,0.7,1.0$;
(b) $k_1$ vs $t$ for the same values of $\alpha$; (c) Tracer variance $\sigma^2$
for (from bottom) $\alpha = 1.0,0.7,0.5,0.3,0.1,0$.}
\label{fig:fluctuations}
\end{figure}

\begin{figure}[htb]
\centering
\noindent
\subfigure[]{  \includegraphics[width=.3\textwidth]{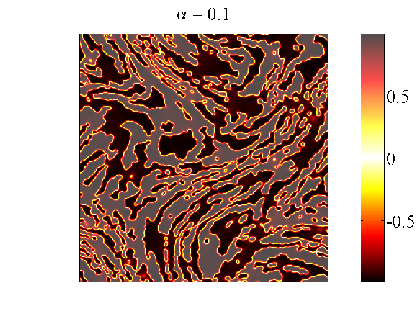}}
\subfigure[]{  \includegraphics[width=.3\textwidth]{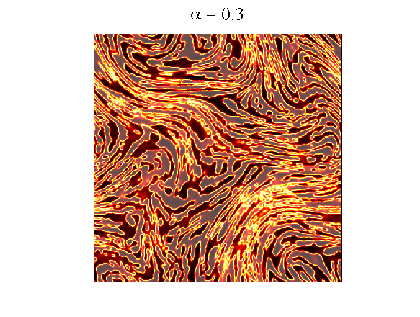}}
\subfigure[]{  \includegraphics[width=.3\textwidth]{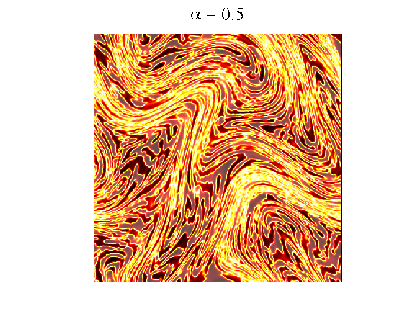}}
\subfigure[]{  \includegraphics[width=.3\textwidth]{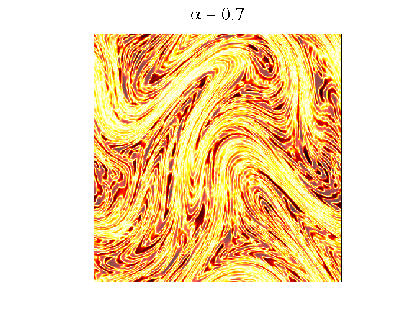}}
\subfigure[]{  \includegraphics[width=.3\textwidth]{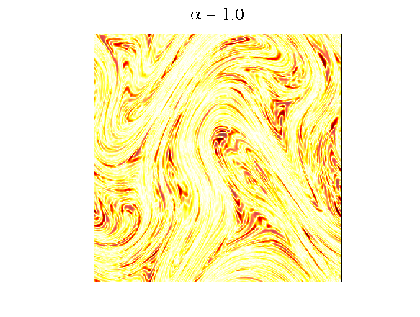}}
\caption{The steady-state concentration field after $t=30,000$ timesteps,
for various stirring amplitudes $\alpha$.}
\label{fig:CHflow}
\end{figure}

We investigate the stirred case in a similar manner and obtain the results
below, after $t=30,000$ timesteps.  In order to ensure that we are in a
steady state, we study the inverse lengths $F$ and $k_1$
and the tracer variance $\sigma^2$ (see Eq.~\eqref{eq:sigsq}) which measures
the homogeneity of the concentration ($\sigma^2=0$ for a homogeneous
mixture)~\cite{lattice_PH2,lattice_PH1}.  These quantities are time  dependent
and have the property that after some transience, they fluctuate around a
mean value, without secular trend.  This can be seen in Fig.~\ref{fig:fluctuations}.
%
%
%
%
%
In Fig.~\ref{fig:CHflow}, for $\alpha = 0.1$, the concentration field at
late time looks similar to that for $\alpha = 0$.  This is because for small
$\alpha$, the effect of advection is  diffusive, with small movements of
particles
%
%
%
%
%
%
giving rise to bubbles with jagged boundaries, but no breakup.  The PDFs
of the concentration for $\alpha\in [0,0.01,...,0.2]$ have the same structure.
 Nevertheless, it is clear from Fig.~\ref{fig:fluctuations}(b) that coarsening
 arrest takes place for $\alpha = [0.01,...,0.2]$, in contrast to $\alpha=0$.
  As $\alpha$ increases, the bubbles become less and less evident.  For large
  $\alpha$, we see regions containing filaments of similar concentration.
   After a period of transience, the free energy $F$, the mean wavenumber
   $k_1$ and the variance $\sigma^2$ fluctuate around mean values, confirming
   the existence of the steady state.  The bubble size is extracted from
   the
   quantity $\langle{\sigma^2/F}\rangle$, where $\langle\cdot\rangle$ denotes
   a time average over the fluctuations.  We investigate how the mean lengths
   scale with the Lyapunov exponent $\lambda$.

\begin{figure}[htb]
\centering
\subfigure[]{
    \includegraphics[width=.32\textwidth]{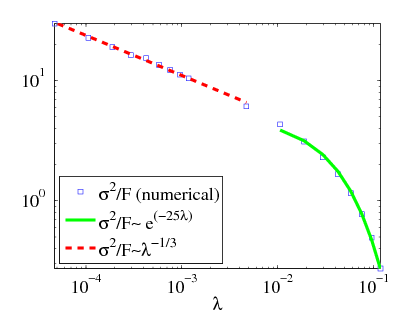}
}
\subfigure[]{
    \includegraphics[width=.32\textwidth]{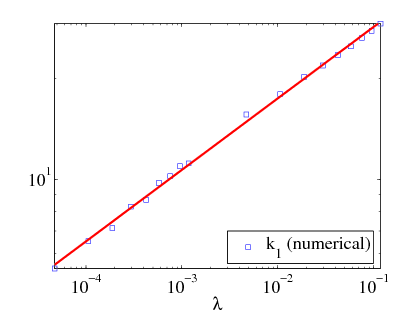}
}
\subfigure[]{
    \includegraphics[width=.32\textwidth]{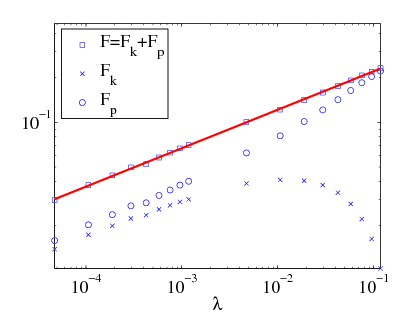}
}
\subfigure[]{
    \includegraphics[width=.32\textwidth]{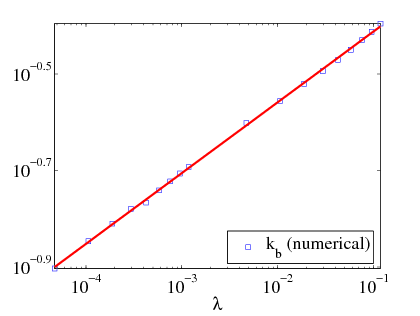}
}
\caption{(a) 
The plot of $\log\langle{\sigma^2/F}\rangle$ against $\log\lambda$
suggests the scaling law $\langle{\sigma^2/F}\rangle\sim\lambda^{-1/3}$ for
small stirring amplitudes while for large amplitudes this proxy for the bubble
radius decays exponentially with increasing Lyapunov exponent.
(b) Graph of $\log\langle{k_1}\rangle$ against $\log\lambda$, with approximate
power law behaviour $\langle{k_1}\rangle\sim \lambda^{0.21}$.  The exponent
is not the same over the entire range, however.
(c) The graph of $\log\langle{F}\rangle$ exhibits a much cleaner power law
behaviour $\langle{F}\rangle\sim\lambda^{0.26}$.
(d)  The Batchelor wavenumber $k_b=\big[{\overline{\left|\nabla c\right|^2}}/{\overline{
c^2}}\big]^{1/2}$ also possesses a clear power law behaviour $\langle{k_b}\rangle\sim\lambda^{0.15}$.
}
\label{fig:CHscale}
\end{figure}

In Fig.~\ref{fig:CHscale} we see that for small $\lambda$ the proxy $\langle{\sigma^2/F}\rangle$
for bubble radius scales with the
%
%
%
%
%
%
LS exponent as $\lambda^{-1/3}$, suggesting an equilibrium bubble size on
this scale.  For larger $\lambda$, the hyperdiffusion breaks up these bubbles
and mixes the fluid.  This process leads to a faster-than-algebraic decay
of bubble size, visible in the figure.  Other measures of the bubble size,
such as the mean wavenumber $\langle{k_1}\rangle$ or the quantity $1-\langle{\sigma^2}\rangle$
do not possess a clear scaling law.  For example, in Fig.~\ref{fig:CHscale}(b)
we plot $\log\langle{k_1}\rangle$ against $\log\lambda$.  The exponent changes
over the three decades of data, indicating that the wavenumber $\langle{k_1}\rangle$
does not have a clear scaling law with the Lyapunov exponent $\lambda$.
%
%
%
%
%
%
%
%
%
\begin{figure}[htb]
\centering
\subfigure[]
{
    \includegraphics[width=.3\textwidth]{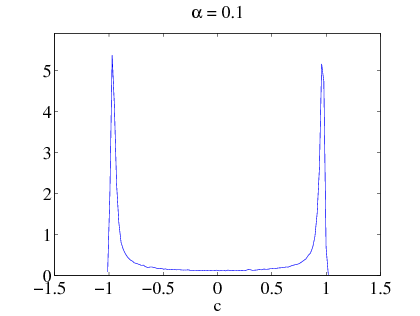}}
\subfigure[]
{
    \includegraphics[width=.3\textwidth]{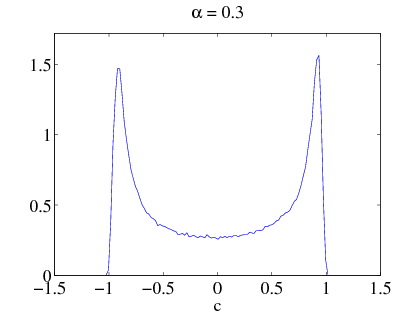}}
\subfigure[]
{
    \includegraphics[width=.3\textwidth]{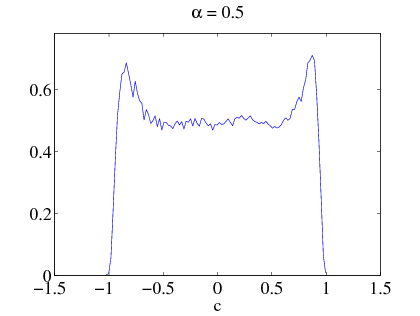}}
\subfigure[]
{
    \includegraphics[width=.3\textwidth]{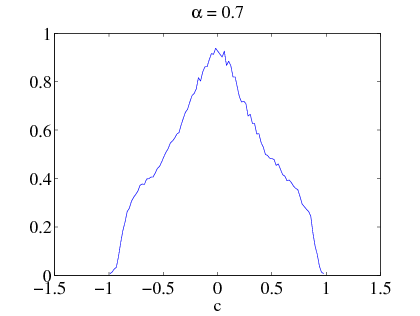}}
\subfigure[]
{
    \includegraphics[width=.3\textwidth]{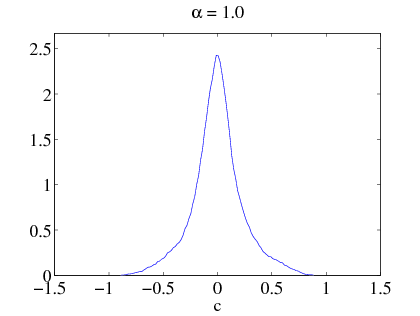}}
\subfigure[]
{
    \includegraphics[width=.3\textwidth]{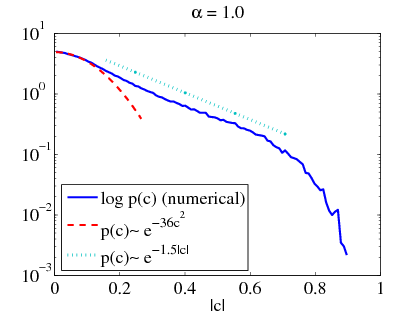}}
\caption{%
Normalized PDF of concentration in the steady state. 
(a), (b) Segregation-dominated flow ($\alpha = 0.1, 0.3$); %
(c), (d) Crossover to quasi-diffusive regime ($\alpha = 0.5, 0.7$); %
(e) Quasi-diffusive regime ($\alpha = 1.0$); %
(f) Semilog plot of PDF for $\alpha=1.0$.  Note Gaussian core and exponential
 decay down to the cutoff at $\left|c\right|\apprle1$.}
\label{fig:pdfs}
\end{figure}    
However, the dependence of the time-averaged free energy on the Lyapunov
exponent has the power law behaviour $\langle{F}\rangle\sim\lambda^{0.26}$
over
three decades, suggesting a genuine power law relationship.  The time average
of the Batchelor
wavenumber  $k_b=[{\overline{\left|\nabla c\right|^2}}/{\overline{c^2}}]^{1/2}$~\cite{Batchelor1959}
also possesses a clean power law over three decades.

In Sec.~\ref{sec:background:ch} we showed that  in a situation of dynamical
equilibrium characterized by a single lengthscale, the potential part of
the free energy $F_{\mathrm{p}}=\tfrac{1}{4}\int d^2 x\left(c^2-1\right)^2$
should be in equipartition with the kinetic part $F_{\mathrm{k}}=\tfrac{1}{2}\int
d^2x\gamma\left|\nabla c\right|^2$.  In Fig.~\ref{fig:CHscale} we see that
this is not the case when the bubbles start to break up as the stirring is
increased.  This is an indication of a crossover between the bubbly regime
and the well-mixed one in which the variance is reduced by stirring.

In Fig.~\ref{fig:CHscale} the breakdown in the power law relationship $\langle{\sigma^2/F}\rangle\sim\lambda^{-1/3}$
for large stirring amplitudes suggests a crossover between bubbly and diffusive
regimes.  Another way of seeing the transition between these regimes is to
study the stationary probability distribution function (PDF) of the field
$c\left(\bm{x},t\right)$.  We do this in Fig.~\ref{fig:pdfs}.  For small
values of $\alpha$ we note that the PDF has sharp peaks at $\pm1$, indicating
the effectiveness of phase separation at these stirring amplitudes.  For
$\alpha = 1$ however, the PDF has a Gaussian core, indicating that genuine
mixing by advection-hyperdiffusion is taking place.  For intermediate values
of $\alpha$ the PDF is a combination of these two different distributions
of concentration.

%
%
\section{A one-dimensional equilibrium problem}
\label{sec:chaotic_advection:1D}
Having observed the crossover between the bubbly and the well-mixed regimes,
we study a one-dimensional model to shed further light on the process
by which this occurs.

We examine the archetypal hyperbolic flow $\bm{v}=\left(-\lambda x,
\lambda y\right)$, with strain rate $\lambda$.  This flow tends to
homogenize the concentration in the $y$-direction through stretching. At
late times the problem then becomes one dimensional:
\begin{equation}
  \frac{\partial c}{\partial t}-\lambda x\frac{\partial c}{\partial x}
  = D\frac{\partial^2}{\partial x^2}\left(c^3-c\right)-\gamma
  D\frac{\partial^4 c}{\partial x^4}.
\label{eq:C_H1D}
\end{equation}
Scaling lengths by $\sqrt{\gamma}$ and restricting to the steady case,
Eq.~\eqref{eq:C_H1D} becomes
\begin{equation}
  -\lambda\left(\frac{\gamma}{D}\right) x\frac{d c}{d x}
  = \frac{d^2}{d x^2}\left(c^3-c\right)-\frac{d^4 c}{d x^4}.
  \label{eq:CH1Dsteady}
\end{equation}
There are two sets of boundary conditions (BCs) that are of interest,
\begin{itemize}
\item Bubble BCs:  $c\left(\pm\infty\right)=-1+\delta$,
 $c_x\left(\pm\infty\right)=0$, and $\delta>0$.
\item Front BCs: $c\left(\pm\infty\right)=\pm 1$, and $c'\left(\pm\infty\right)=0$.
\end{itemize}
The front solution has the interpretation of being the transition layer between
filaments, the width of the layer giving the width of the filament.
Before carrying out numerical work, we shall examine two asymptotic regimes
where analytical progress is possible:  the case without stirring, and that
without phase separation.
\subsection*{No stirring}
\begin{figure}[htb]
\centering
\subfigure[]{
      \includegraphics[width=.32\textwidth]{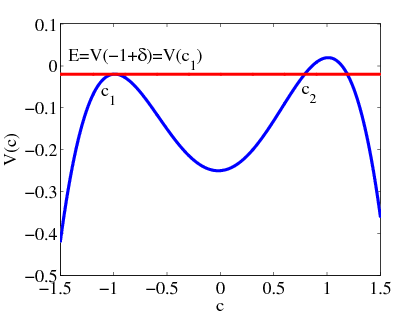}
}
\subfigure[]{
      \includegraphics[width=.32\textwidth]{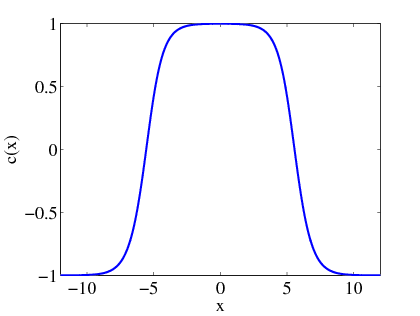}
}
\subfigure[]{
      \includegraphics[width=.32\textwidth]{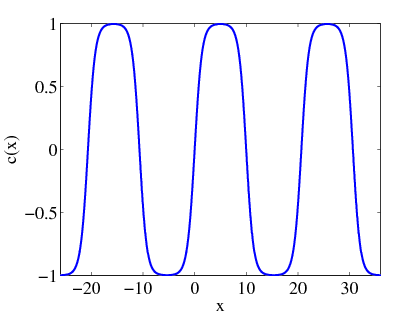}
}
\subfigure[]{
      \includegraphics[width=.32\textwidth]{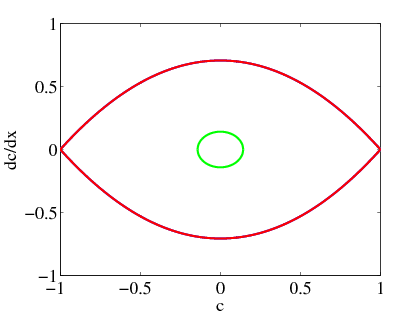}
}
\caption{%
(a) The potential function $V\left(c\right)=-\tfrac{1}{4}\left(c^2-1\right)^2+Bc$.
 Here we set $B=\left(-1+\delta\right)^3-\left(-1+\delta\right)$.
  Notice the well extremities at $c_1=-1+\delta$ and $c_2$;
(b) Integration of Eq.~\eqref{eq:homoclinic} for $\delta=0.001$ giving
the bubble profile (homoclinic orbit);
(c) Periodic solution for $E\apprle V\left(c_1\right)$, $c_1<c<c_2$, which
is close to the homoclinic orbit.
(d) Phase space diagram of solutions: the homoclinic and large-amplitude
periodic solutions nearly coincide.  A small-amplitude periodic solution
is included for reference.  In the homoclinic case, the zeros of $c_x$ lie
at the turning points $c_1$ and $c_2$.
}
\label{fig:mechanics}
\end{figure}

In the absence of advection ($\lambda=0$) the model equation reduces
to
\begin{equation}
\frac{d^2}{dx^2}\left(c^3-c-\frac{d^2c}{dx^2}\right)=0,
\end{equation}
which integrates to
\begin{equation}
c^3-c-\frac{d^2c}{dx^2}=Ax+B,
\end{equation}
where $A$ and $B$ are constants to be determined by the boundary conditions.
 We set bubble BCs $c\left(\pm\infty\right)=-1+\delta$, and
 $c_x\left(\pm\infty\right)=0$.
Using the bubble BCs and the
requirement that the second derivative of $c$ should not diverge at infinity,
$A$ is set to zero.  Thus we have the Newton equation
\begin{equation}
\frac{d^2c}{dx^2}=-\frac{\partial}{\partial c}\left[-\tfrac{1}{4}\left(c^2-1\right)^2+Bc\right].
\label{eq:newton}
\end{equation}
By multiplying both sides by ${dc}/{dx}$ we obtain the energy relation
\begin{equation}
\tfrac{1}{2}\left(\frac{dc}{dx}\right)^2-\tfrac{1}{4}\left(c^2-1\right)^2+Bc=E,
\end{equation}
where we identify the potential energy $V\left(c\right)=-\tfrac{1}{4}\left(c^2-1\right)^2+Bc$
and the total energy $E$.  The bubble boundary conditions then fix the values
of $E$ and $B$.  The bubble or homoclinic solution is obtained when two roots
of the equation $E=V\left(c\right)$ coincide with one extremum of $V$, given
by
\[
c^3-c=B.
\]
We pick out the homoclinic orbit by letting this extremum be at $c=-1+\delta$,
giving $B_{\delta} = \left(-1+\delta\right)^3-\left(1-\delta\right)$.  In
addition, we specify the energy to be
\[
E=V\left(-1+\delta\right).
\]
Then in an interval $\left(c_1=\left(-1+\delta\right),c_2\right)$, indicated
in Fig.~\ref{fig:mechanics}~(a),
the energy $E$ exceeds $V\left(c\right)$, and $E-V>0$, with equality at the
interval boundary.  Thus, the homoclinic solution is given by
\[
x-x_{\mathrm{r}}=\pm\frac{1}{\sqrt{2}}\int_{c_\mathrm{r}}^c\frac{dc}{\sqrt{V\left(-1+\delta\right)-B_\delta{c}+\tfrac{1}{4}\left(c^2-1\right)^2}}.
\]
By taking the reference level $x_{\mathrm{r}}=0$, we obtain the equation
\begin{equation}
x=\pm\frac{1}{\sqrt{2}}\int_{c\left(0\right)}^{c\left(x\right)}\frac{dc}{\sqrt{V\left(-1+\delta\right)-B_\delta{c}+\tfrac{1}{4}\left(c^2-1\right)^2}}.
\label{eq:homoclinic}
\end{equation}
Equation~\ref{eq:homoclinic} specifies $c\left(x\right)$, which can then
be obtained numerically, as in Fig.~\ref{fig:mechanics}~(b).
In the language of dynamical systems, the solution is a homoclinic orbit:
the system returns to its initial state at $x=-\infty$ after infinite time.
 The phase portrait of the solution is presented in Fig.~\ref{fig:mechanics}(d).

Finally we note that Eq.~\eqref{eq:newton} possesses a host of different
solutions, depending on the boundary conditions imposed.  Let us take $B$
as before but leave the energy unspecified.  We study the Newton equation
$c''\left(x\right)=-{\partial V}/{\partial c}$, where $V\left(c\right)$ is
the potential  $V\left(c\right)=-\tfrac{1}{4}\left(c^2-1\right)^2+B_\delta
c$.  Periodic
solutions exist when the system is placed in the potential well of Fig.~\ref{fig:mechanics}~(a),
and when $E<V\left(-1+\delta\right)$.  To obtain a solution with the flat
peaks characteristic of a bubble, the system must access the anharmonic part
of the potential, so we take $E\apprle V\left(-1+\delta\right)$.
 Thus the phase portrait of this periodic bubble solution almost coincides
 with the homoclinic orbit.  On the other hand, for energies close to the
 well minimum $V\left(c_0\right)\equiv V_0$ the periodic solution is a small
 sinusoidal undulation.  These results are best summarized by introducing
 the period of oscillation, given by~\cite{LandauCourseTP1}
\[
\tfrac{1}{2}T=\frac{1}{\sqrt{2}}\int_{c_1}^{c_2}\frac{dc}{\sqrt{E-V\left(c\right)}},
\]
where $c_1$ and $c_2$ are those roots of the equation $V\left(c\right)-E=0$
indicated in Fig.~\ref{fig:mechanics}~(a).
\begin{itemize}
\item For $E\apprge V_0$ the oscillations are small and the period is ${2\pi}\sqrt{V''\left(c_0\right)}$.
\item For $E\apprle V\left(-1+\delta\right) $ the solution is still periodic
with energy-dependent period.  This is the one-dimensional gas of Fig.~\ref{fig:mechanics}(c).
\item For $E\rightarrow V\left(-1+\delta\right)$, $T\rightarrow+\infty$,
giving the homoclinic solution.
\item For $\delta\rightarrow0$ and $E\rightarrow V\left(-1+\delta\right)$,
$T\rightarrow-i\infty$, which corresponds to the heteroclinic or front solution.
\end{itemize}
\subsection*{A quasi two-dimensional problem}
Following the mechanical intuition developed in the previous section,
we outline briefly what happens in the quasi one-dimensional setting when
the full two-dimensional problem has azimuthal symmetry.  In this case, the
stationary CH equation is
\[
\nabla_r^2\left[c^3-c-\nabla_r^2 c\right]=0,
\]
where $\nabla_r^2$ is the radial part of the Laplacian,
\begin{equation}
\nabla_r^2=\frac{1}{r}\frac{\partial}{\partial r}r\frac{\partial}{\partial
r}.
\end{equation}
Using the same analysis as before, we obtain the equation $c^3-c-\nabla_r^2c=B$,
which is a Newton equation with damping:
\begin{equation}
\frac{d^2 c}{dr^2}+\frac{1}{r}\frac{dc}{dr}=-\frac{\partial V}{\partial c},
\end{equation}
where $V\left(c\right)=-\tfrac{1}{4}\left(c^2-1\right)^2+Bc$ is the potential.
 The large-amplitude and homoclinic solutions presented previously are modified
 by the presence of damping: at small $r$ there is significant dissipation
 proportional to $1/r$ while the far-field problem is one of small oscillations
\begin{figure}[htb]
\centering
\subfigure[]{
      \includegraphics[width=.35\textwidth]{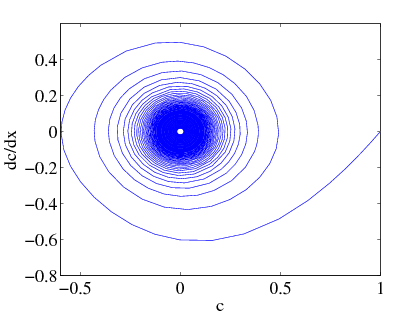}
}
\subfigure[]{
      \includegraphics[width=.31\textwidth]{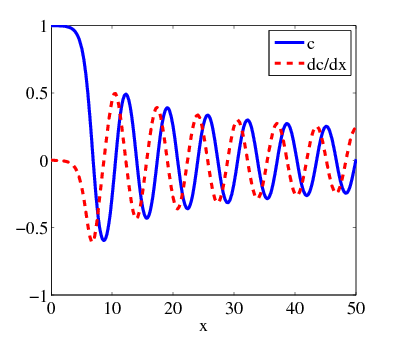}
}
\subfigure[]{
      \includegraphics[width=.35\textwidth]{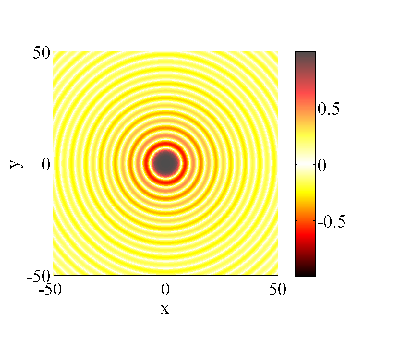}
}
\caption{%
(a) Phase portrait for the formerly homoclinic case.  The solution spirals
towards the attractor $\left(c_0,0\right)$;
(b) Profile of the solution and its derivative;
(c) The full two-dimensional solution with azimuthal symmetry.
}
\label{fig:dissipation}
\end{figure}
with damping.  These oscillations have period $2\pi\sqrt{V''\left(c_0\right)}$
($c_0$ is the well minimum), and decay to the attractor $\left(c_0,0\right)$.
 These results are shown succinctly in the phase portrait of Fig.~\ref{fig:dissipation}(a).

\subsection*{The no-segregation limit}
\begin{figure}[htb]
\centering
\subfigure[]{
      \includegraphics[width=.32\textwidth]{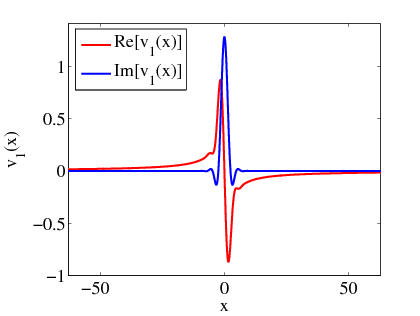}
}
\subfigure[]{
      \includegraphics[width=.3\textwidth]{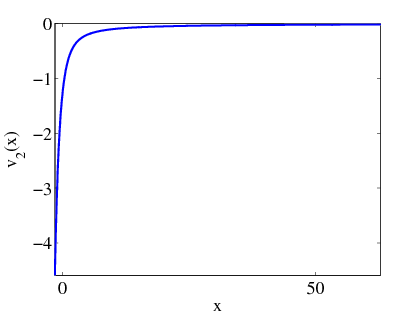}
}
\caption{%
(a) Generalized Airy function $v_1\left(x\right)$, with real and imaginary
parts shown; (b) The function $v_2\left(x\right)$, which is pure real.
}
\label{fig:airy}
\end{figure}

By measuring lengths in terms of the diffusion lengthscale $\left(\gamma
D/\lambda\right)^{1/4}$, Eq.~\eqref{eq:CH1Dsteady} becomes
\begin{equation}
-x\frac{dc}{dx}=\sqrt{{\gamma D}/{\lambda}}\frac{d^2}{dx^2}\left(c^3-c\right)-\frac{d^4c}{dx^4},
\end{equation}
which in the limit of large $\lambda$ reduces to
\begin{equation}
xv=\frac{d^3v}{dx^3},
\label{eq:gen_Airy}
\end{equation}
where $c\left(x\right)=c\left(0\right)+\int_0^x dx' v\left(x'\right)$.  Equation~\eqref{eq:gen_Airy} is a generalized Airy equation with solution~\cite{Polyanin_odes}
\begin{equation}
v\left(x\right)=\sum_{\nu=0}^{\mathrm{3}} C_{\nu}v_{\nu}\left(x\right),
\end{equation}
where the functions $v_{\nu}\left(x\right)$ are given by
\begin{equation}
v_{\nu}\left(x\right)=\epsilon_{\nu}\int_0^{\infty}\exp\left(\epsilon_{\nu} xt-t^4/4\right)dt;\qquad\nu=0,1,2,3.
\label{eq:gen_Airy_fn}
\end{equation}
The phases $\epsilon_{\nu}$ are the fourth roots of unity: $\epsilon_{\nu}=\exp\left(\pi\nu
i/2\right)$ and the constants $C_{\nu}$ must satisfy $\sum_{\nu}C_{\nu}=0$.
 A plot of $v_1$ and $v_2$ is shown in Fig.~\ref{fig:airy}.
By studying the phases $\epsilon_{\nu}$, we see that $v_{1}$ and $v_{2}$
are related to $v_3$ and $v_{0}$ respectively by a parity transformation.
 
We set $C_{1}=-C_3=1$ and $C_0=C_2=0$ to obtain the front solution of the
hyperdiffusion equation:
\begin{equation}
c_{\mathrm{front}}\left(x\right)=\int_0^{x}\left[v_1\left(x'\right)-v_3\left(x'\right)\right]dx'.
\end{equation}
Simplifying and restoring the dimensional units, this is
\begin{equation}
c_{\mathrm{front}}\left(x\right)=\int_{-\infty}^{\infty}\frac{1}{t}\sin\left(tx\right)\exp\left[-\frac{t^4}{4\lambda/\gamma
D}\right]dt.
\label{eq:front_sln}
\end{equation}
By inspection of the phases $\epsilon_{\nu}$, it is clear that the hyperdiffusion
equation has no bubble solutions on the whole real line.
\begin{figure}[htb]
\centering
    \includegraphics[width=.35\textwidth]{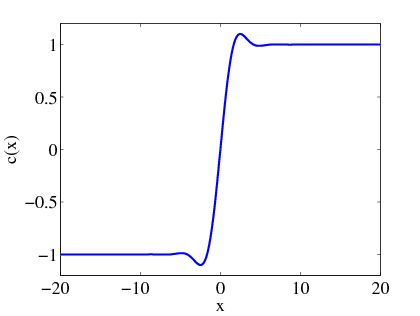}
\caption{(a) Front solution~\eqref{eq:front_sln} obtained through quadrature.}
\label{fig:quadrature}
\end{figure}
In Fig.~\ref{fig:quadrature} we present front solution. 

Finally we note that the generalized Airy functions $v_{\nu}\left(x\right)$
give the solution of the nonstationary advection-hyperdiffusion equation
\[
\frac{\partial c}{\partial t}-x\frac{\partial c}{\partial x}=-\frac{\partial^4
c}{\partial x^4},
\]
where we have again expressed lengths in terms of the diffusion scale $\left(\gamma
D / \lambda\right)^{1/4}$.  Time is expressed in terms of the Lyapunov time
$1/\lambda$.  By linearity, we may separate the variables, writing $c\left(x,t\right)=e^{-t}\tilde{c}\left(x\right)$.
 The Airy functions $v_{\nu}\left(x\right)=\epsilon_{\nu}\int_0^{\infty}e^{\phi\left(t;x\right)}
 dt$ satisfy
\begin{equation}
\frac{\partial^4 v_{\nu}}{\partial x^4}-x\frac{\partial v_{\nu}}{\partial
x} = -\epsilon_{\nu}\int_0^{\infty} t\frac{\partial \phi\left(t;x\right)}{\partial
t}e^{\phi\left(t;x\right)} dt =\epsilon_{\nu}\int_0^{\infty} e^{\phi\left(t;x\right)}dt
= v_{\nu}\left(x\right),
\end{equation}
which is the equation for $\tilde{c}\left(x\right)$.  Since $v_0$ and $v_2$
do not have the correct far-field behaviour, we are restricted to the dimensional
solutions
\begin{equation}
c\left(x,t\right)=e^{-\lambda t}v_{1,3}\left[\frac{x}{\left(\gamma D / \lambda\right)^{1/4}}\right].
\end{equation}
The functions $v_{1,3}\left(x\right)$ have the asymptotic form $v_{1,3}\left(x\right)\sim
-1/x+6i/x^4$, for $\left|x\right|\gg1$~\cite{WongAsymptotics}, and therefore
have the correct far-field behaviour.
%
%
%
%
Thus, in the absence of segregation, a bubble profile will decay in time
owing to the twin processes of advection and hyperdiffusion.

\subsection*{Numerical simulations of the boundary value problem}
Armed with this knowledge, we attempt a numerical solution of the boundary-value
problem (BVP)
\begin{equation}
  -\lambda\left(\frac{\gamma}{D}\right) x\frac{d c}{d x}
  = \frac{d^2}{d x^2}\left(c^3-c\right)-\frac{d^4 c}{d x^4},
\label{eq:front}
\end{equation}
where lengths are measured in terms of the transition layer width $\sqrt{\gamma}$.
 The analytic study of the hyperdiffusive limit suggested that seeking a
 bubble solution may not be sensible, and indeed it is not: solving the time-dependent
version of Eq.~\eqref{eq:front} with bubble initial conditions, the concentration
profile always decays to a constant.  We therefore focus on the frontal boundary
conditions.

\begin{figure}[htb]
\centering
\subfigure[]
{
    \includegraphics[width=.32\textwidth]{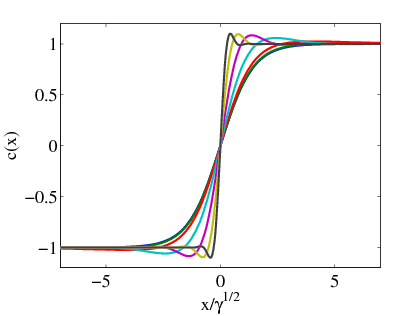}
}
\subfigure[]
{
    \includegraphics[width=.3\textwidth]{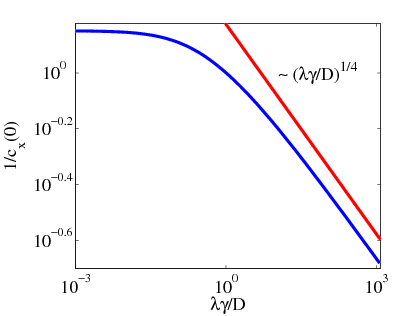}
}
\caption{(a) Front profile for $\lambda\gamma/D = 0.001,0.01,0.1,1,10,100,1000$.
 The front steepens as $\lambda\gamma/D$ increases; (b) Front width as a
 function of the parameter $\lambda\gamma/D$: for large parameter values,
 the front width scales hyperdiffusively as $\left(\lambda\gamma/D\right)^{1/4}$.}
\label{fig:1D}
\end{figure}

We solve the BVP~\eqref{eq:front} with frontal boundary conditions in Fig.~\ref{fig:1D}.
 The front steepens with increasing $\lambda\gamma/D$.  Physically, this
 corresponds to a strengthening of local shear in the velocity, which forces
 the filaments of fluid to become narrower.  The numerical solution
 agrees exactly with the analytical solutions in the limit of large and small
 $\lambda$.  At large $\lambda$, the front width has the hyperdiffusive scaling,
 indicating a regime dominated by hyperdiffusion, in agreement with the full
 two-dimensional numerical simulations in Sec.~\ref{sec:chaotic_advection:numerics}.
We note that the concentration profile takes values
$\left|c\right|>1$ in some places, a consequence of the fact that Eq.~\eqref{eq:CH1Dsteady}
has no maximum principle, in contrast (say) to the advection-diffusion
equation.  Again, this result is consistent with the numerical results of
Section~\ref{sec:chaotic_advection:numerics},
where we find $\mathrm{max}_{\bm{x}\in \Omega}\left|c\left(\bm{x},t\right)\right|>1$.
This
superabundance of one binary fluid element is unreasonable on physical
grounds, pointing to a shortcoming in the advective CH model.   However,
we shall see in the next section that the qualitative features of the constant-mobility
model and those of the more realistic variable-mobility model are in agreement,
suggesting that the simpler model gives an adequate description of advective
phase-separation dynamics.
\section{The variable-mobility case}
\label{sec:chaotic_advection:varmob}
To gain further insight into the dependence of the bubble size on
the Lifshitz-Slyozov and Lyapunov exponents, we study the variable-mobility
Cahn--Hilliard equation \cite{Langer}
\begin{equation}\frac{\partial c}{\partial t} +\bm{v}\cdot\nabla c= \nabla\cdot\left[D\left(c\right)\nabla\mu\right],\end{equation}
where as before $\mu = c^3-c-\gamma\nabla^2c$.  By introducing
this additional feature, we recover a system that, at least in the case
without flow, has a solution $c\left(\bm{x},t\right)$ confined to the
range $[-1,1]$ \cite{Elliott_varmob}.  

We specify the functional form of the mobility as 
\begin{equation}
D\left(c\right) = D_0\left(1-\eta c^2\right),
\end{equation}
fixing $\eta = 1$.  This modification corresponds to interface-driven coarsening
because inside bubbles ($c=\pm 1$) the mobility is zero.  For other values
of $\eta$ we have a mixture of bulk- and interface-driven coarsening.  This
equation has been investigated by Bray and Emmott \cite{Bray_LSW} for $\bm{v}=0$,
in an evaporation-condensation picture.  They show that the typical bubble
size $R_{\mathrm{b}}\left(t\right)$ grows as $t^{1/4}$, a result limited
to dimensions greater than two.  Numerical simulations \cite{Zhu_numerics}
suggest that this growth law also holds in two dimensions.  We couple the
modified segregation dynamics to the sine flow.

For $\bm{v}=0$, we obtain the growth law for the scale $R_{\mathrm{b}}\equiv
1/k_1$, namely $R_{\mathrm{b}}\sim t^{0.264}$.  The growth exponent is close
to the value $1/4$ predicted by the LS theory.  In addition, we obtain the
free energy decay laws $F\sim t^{-0.267}$ again close to the LS value.  Thus,
as in the constant-mobility case, a description of the problem lengthscales
can be given in terms of the bubble energy or in terms of $R_{\mathrm{b}}$,
with identical results.

Because the variable-mobility calculation is computationally slower, we shall
perform the numerical experiments with flow at lower resolution.  We do not
anticipate that this will change the results, because in integrations of
the constant-mobility equations at resolution $256^2$ the scaling exponents
were unaffected.  Switching on the chaotic flow, we observe the steady state,
characterized by the fluctuation of the free energy, the wavenumber $k_1$,
\begin{figure}[htb]
\centering
\subfigure[]
{
    \includegraphics[width=.3\textwidth]{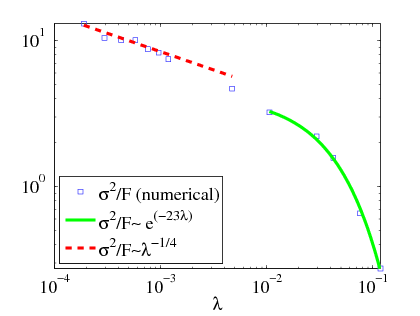}
}
\hspace{0.01cm}
\subfigure[]
{
    \includegraphics[width=.3\textwidth]{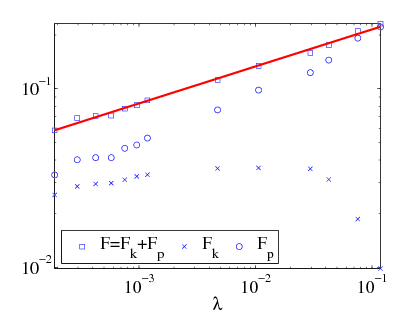}
}
\caption{(a) 
The plot of $\log\langle{\sigma^2/F}\rangle$ against $\log\lambda$ suggests
the scaling law $\langle{\sigma^2/F}\rangle\sim\lambda^{-1/4}$ for small
stirring
amplitudes, while for large amplitudes this proxy for the bubble radius decays
exponentially with increasing Lyapunov exponent.
(b) The graph of $\log\langle{F}\rangle$ exhibits power law behaviour $\langle{F}\rangle\sim\lambda^{0.21}$.
 The predominance of kinetic free energy over potential free energy at large
 $\lambda$ is visible.
}
\label{fig:varmob}
\end{figure}
and the variance $\sigma^2$ around mean values.  As in the constant-mobility
case, the graph of the time-averaged free energy and time-averaged Batchelor
scale possess clear scaling laws, while the graph of the mean wavenumber
$\langle{k_1}\rangle$ does not.  For small $\lambda$, the proxy for bubble
radius $\langle{\sigma^2/F}\rangle$ scales approximately as $\lambda^{-1/4}$,
suggesting a decay of bubble radius according to the LS exponent.  Thus,
the analogous result of Section~\ref{sec:chaotic_advection:numerics} is not
fortuitous.
For large $\lambda$
this quantity decays exponentially to zero at a rate (with respect to $\lambda$)
close to that of the constant-mobility system.  In each case, the fit of
the data is not as good as the constant-mobility case, the computational
error being larger here because we have run the simulations at resolution
$256^2$.  We present these results in Fig.~\ref{fig:varmob}.

The results are an exact replica of those for the constant-mobility case.
The arrest of bubble growth for small $\lambda$ is therefore a genuine phenomenon
whose behaviour depends on the LS exponent of the corresponding unstirred
dynamics.  The mixing at large $\lambda$ is identical to that for which the
mobility is constant.  Thus, while the variable-mobility model is more
realistic than its constant-mobility counterpart, their properties with regard
to chaotic stirring and eventual mixing are the same.

\section{Summary}
We have shown how, in the presence of an external chaotic stirring, the
coarsening dynamics of the Cahn--Hilliard equation are arrested and bubbles
form on a particular lengthscale.  The system reaches a steady state,
characterized by the fluctuation of the free energy $F$, the variance
$\sigma^2$, and the wavenumber $k_1$ around mean values.  A measure of the
typical bubble size is given by the time average of $\sigma^2/F$.  For
sufficiently large stirring intensities, the bubbles diminish in prominence,
to be replaced by filament structures.  These are due to the combination
of the advection and hyperdiffusion terms in the CH demixing mechanism. 
In this regime, the concentration tends to homogenize, so that the stirring
stabilizes the previously unstable mixed state.

We have explained with a one-dimensional model how this transition arises
and how the filament width in the homogeneous regime depends on the Lyapunov
exponent of the chaotic flow.  This latter investigation exhibits the local
superabundance of one of the binary fluid components,
which we reject as unreasonable.  In the case of no flow this difficulty
was overcome by the
addition of a variable mobility~\cite{Elliott_varmob}.  However, the
quasi-diffusive regime identified in our simulations is robust in the sense
that it is present in both the constant- and the more realistic variable-mobility
cases.  We therefore expect to see this remixing phenomenon in real
binary fluids.

%% file: estimating_mixedness/estimating_mixedness.tex
\chapter{Estimating mixedness in a stirred Cahn--Hilliard fluid}
\label{ch:estimating_mixedness}

\section{Overview}
\label{sec:estimating_mixedness:overview}
In this chapter we study the advective Cahn--Hilliard equation in the passive
setting, in the presence of sources and sinks of matter.
We focus on the symmetric mixture, which contains equal amounts of each binary
fluid component.  We have mentioned the often undesirable coarsening tendency
of the binary fluid, and the efforts made to suppress it.  
We therefore introduce a quantitative measure of coarsening suppression
by studying the $p^{\mathrm{th}}$ power-mean fluctuation of the concentration
about its average value.  
By fluctuations about the average value, we mean spatial fluctuations around
the average spatial concentration (which is zero for a symmetric mixture);
we then average these fluctuations over space and time.
In effect, we study the time-averaged $L^p$ norm of the concentration.
If this quantity is small, the average deviation of the system about the
well-mixed state $c=0$ is small, and we therefore use this quantity as a
proxy for the level of mixedness of the fluid.

In many applications, the Cahn--Hilliard equation is driven by some external
effect, for example, chemically patterned substrates~\cite{Lu2000}, electrically
charged media~\cite{Lu2005}, or the subject of this report, stirring.
In this chapter, we focus on phase separation in the presence of advection
and a source term in the Cahn--Hilliard equation.
This source term arises in a number of ways.  In~\cite{Krekhov2004} it is
maintained by thermal diffusion, through the Ludwig--Soret
effect~\cite{Craig2004}, in which concentration gradients are induced by
imposed temperature gradients. One can also produce concentration gradients
by simply injecting matter into the system.  
%
%
%
%
%
The approach we take here is motivated by studies of the advection-diffusion
equation~\cite{Thiffeault2004, Shaw2006,Thiffeault2006,Thiffeault2007}. 
There, the problem is a linear one, and describes miscible liquids, while
fluctuations about the mean are measured by the variance or centred second
power-mean of the fluid concentration~\cite{Danckwerts1952,Edwards1985}.
The variance is reduced by certain stirring mechanisms.  By specifying a
source term, it is possible to state the maximum amount by which a given
flow can reduce the variance, and hence mix the fluid.  
By quantifying the variance reduction, we can classify flows according to
how effective they are at mixing.
%
%


\section{Measures of mixedness}
\label{sec:estimating_mixedness:measures}
In this section, we introduce the advective Cahn--Hilliard equation
with sources and recall some the properties of this equation.  We outline
the tools and notation we shall use to analyze concentration fluctuations.
The results will be valid in any spatial dimension.

We study the advective Cahn--Hilliard equation in the presence of flow, and
for prescribed sources and sinks,
\begin{equation}
\frac{\partial c}{\partial t}+\bm{v}\cdot\nabla c = \Diff\Delta\left(c^3-c-\Tr\Delta
c\right)+s\left(\bm{x}\right).\\
\label{eq:ch_source}
\end{equation}
Here $\Diff$ is Cahn--Hilliard diffusion and $\sqrt{\Tr}$ is the typical
thickness of transition zones between phase-separated regions of the binary
fluid.
 
In this chapter we work with a nondimensionalization of Eq.~\eqref{eq:ch_source}
that leaves three control parameters in the problem.  Therefore, we can unambiguously
study limits where control parameters take large or small values.  Let ${\Time}_0$
be a timescale associated with the velocity $\bm{v}\left(\bm{x},t\right)$,
and let $V_0$ be the magnitude of $\bm{v}\left(\bm{x},t\right)$.  Let $S_0$
be the magnitude of the source variations and finally, let
$L$ be a lengthscale in the problem; for example, if the problem
is solved in a hypercube with periodic boundary conditions,
we take the lengthscale $L$ to be the cube length.  It is then possible to
write down Eq.~\eqref{eq:ch_source} using a nondimensional time $t'=t/{\Time}_0$ and
a nondimensional spatial variable $\bm{x}'=\bm{x}/L$,
\begin{equation*}
\frac{\partial c}{\partial t'}+V_0'\tilde{\bm{v}}\cdot\nabla' c = \Diff'\Delta'\left(c^3-c-\Tr'\Delta'
c\right)+S_0'\tilde{s}\left(\bm{x}'\right),\\
\end{equation*}
$\tilde{\bm{v}}$ and $\tilde{s}$ are dimensionless shape functions, $V_0'
= {\Time}_0 V_0/L$, $\Diff' = \Diff{\Time}_0/L^2$, $\Tr'=\Tr/L^2$, and where $S_0'=S_0{\Time}_0$.
 As before, the quantity $\Diff' = \Diff{\Time}_0/L^2$ is identified with the
 ratio ${\Time}_0/{\Time}_D$, being the ratio of the velocity timescale to the
 diffusion timescale.  Following standard practice, we shall now
 work with the dimensionless version of the equation, and omit the prime
 notation.  For ease of notation, we shall henceforth take $\bm{v}$ to mean
 $V_0'\tilde{\bm{v}}$.

We recall the following properties of~\eqref{eq:ch_source} from Ch.~\ref{ch:background}.
 These will be useful in what follows.
\begin{itemize}
\item If the source $s\left(\bm{x}\right)$ is chosen to have spatial mean
zero, then the total mass is conserved,
\begin{multline*}
\frac{d}{dt}\int_{\Omega}c\left(\bm{x},t\right)d^n x = \int_{\Omega}D\Delta\left(c^3-c-\gamma\Delta
c\right)d^n x+\int_{\Omega}s\left(\bm{x}\right)d^n x\\
=0+\left[\text{boundary
terms}\right],
\end{multline*}
where $\Omega$ is the problem domain in $n$ dimensions and $|\Omega|$ is
its volume.  The boundary terms in this equation vanish
on choosing no-flux boundary conditions $\hat{\bm{n}}\cdot\nabla{c}=\hat{\bm{n}}\cdot\nabla{\mu}=0$
on
$\partial\Omega$, or
periodic boundary conditions.  Here $\hat{\bm{n}}\cdot\nabla$ is the outward normal derivative.
 For simplicity we consider the periodic case, although this is not necessary.
\item There is a free-energy functional
\begin{equation}
F\left[c\right]=\int_\Omega\left[\tfrac{1}{4}\left(c^2-1\right)^2+\tfrac{1}{2}\gamma\left|\nabla
c\right|^2\right]d^nx,\qquad\mu=\frac{\delta F}{\delta c}=c^3-c-\gamma\Delta
c,
\label{eq:fe}
\end{equation}
where $\mu$ is the chemical potential of the system.
For a smooth concentration field $c\left(\bm{x},t\right)$, the free energy
satisfies the evolution equation
\[
\dot{F}\equiv\frac{dF}{dt}=-D\int_\Omega\left|\nabla\mu\right|^2d^nx+\int_\Omega\mu\left(-\bm{v}\cdot\nabla
c+s\right)d^{n}x,
\]
and in the absence of sources and stirring decays in time.
\end{itemize}

To study the spatial fluctuations in concentration, we consider the power
means of the quantity $c\left(\bm{x},t\right)-|\Omega|^{-1}\int_\Omega{c\left(\bm{x},t\right)}d^nx$,
\begin{equation}
{M}_p\left(t\right)=\bigg\{\int_\Omega\left|c\left(\bm{x},t\right)-\frac{1}{|\Omega|}\int_{\Omega}c\left(\bm{x},t\right)d^nx\right|^p\bigg\}^{\frac{1}{p}}.
\label{eq:power_mean}
\end{equation}  
For a symmetric mixture in which $\int_{\Omega}c\left(\bm{x},t\right)d^{\n}x=0$,
this is simply
\[
{M}_p\left(t\right)=\bigg\{\int_\Omega\left|c\left(\bm{x},t\right)\right|^pd^nx\bigg\}^{\frac{1}{p}}=\|c\|_p,
\]
where we have introduced the $L^p$ norm of the concentration, $\|c\|_p$.
 The
quantity $M_p$ is a measure of the magnitude of spatial fluctuations in the
concentration about the mean, at a given time.  Since
we are interested in the ultimate state of the system, we study the long-time
average of concentration fluctuations.  We therefore focus on the power-mean
fluctuations
\[
m_p=\langle M_p^p\rangle^{\frac{1}{p}},
\]
where $\langle\cdot\rangle$ is the long-time average
\[
\langle\cdot\rangle=\lim_{t\rightarrow\infty}\frac{1}{t}\int_0^t\left(\cdot\right){ds},
\]
provided the limit exists.  We shall repeatedly use the following results
for the monotonicity of norms,
\begin{eqnarray}
\|f\|_p&\leq&\left|\Omega\right|^{\frac{1}{p}-\frac{1}{q}}\|f\|_q,\qquad\qquad
1\leq p\leq q,f\in L^q\left(\Omega\right),\nonumber\\
\frac{1}{t}\int_0^t \left|g\left(s\right)\right|ds&\leq&
\left[\frac{1}{t}\int_0^t\left|g\left(s\right)\right|^q ds\right]^{\frac{1}{q}},
\qquad q\geq 1,g\in L^q\left(\left[0,t\right]\right),
\label{eq:monotonicity_norms}
\end{eqnarray}
discussed in Ch.~\ref{ch:background}.

The Cahn--Hilliard equation and its free energy functional contain high powers
of the concentration $c$ ($c^3$ and $c^4$ respectively), and we can therefore
estimate $m_p$ for specific $p$-values.  In particular, in the following
sections, we shall prove the following result in $\n$ dimensions:

\begin{quote}
\textit{Given a smooth solution to the advective Cahn--Hilliard equation,
the long-time average of the free energy exists, and therefore $m_p$ exists,
for $p\in\left[1,4\right]$. 
}
\end{quote}

 The only constraints we impose on the external forces
are that the velocity, its derivatives, and the source term be always and
everywhere bounded.
 We take our
result one step further by explicitly evaluating upper and lower bounds for
$m_4$, and this gives a way of quantifying concentration fluctuations in
the stirred binary fluid.

\section{Existence of long-time averages}
\label{sec:existence}

\noindent In this section, we prove a result concerning the
existence of the long-time average of the free energy, and of the power means
$m_p$, for $p\in\left[1,4\right]$.

\begin{quote}
\textit{Given the velocity field $\bm{v}\left(\bm{x},t\right)\in H^{1,\infty}\left(\Omega\right)$,
the source
$s\left(\bm{x}\right)\in L^2\left(\Omega\right)$, and smooth initial data
for the advective Cahn--Hilliard~\eqref{eq:ch_source}, the long-time average of
the free energy exists, and thus $m_p$ exists, for $p\in\left[1,4\right]$.
}
\end{quote}

The proof relies on the free-energy evolution equation.
 Using this law, we find uniform bounds on the finite-time means $\langle
 F\rangle_t$ and $\langle M_4^4\rangle_t$, where
\[
\langle\cdot\rangle_t=\frac{1}{t}\int_0^t\left(\cdot\right)ds,\qquad\langle\cdot\rangle=\lim_{t\rightarrow\infty}\langle\cdot\rangle_t.
\]
Using the monotonicity of norms, the uniform boundedness of $\langle M_p^p\rangle_t$,
follows, for $p\in\left[1,4\right]$.  The proof proceeds in multiple steps,
which we outline below.

\subsection*{Step 1: Analysis of the free-energy evolution equation}

\noindent Given the conditions on the external forces outlined above, and
the results in Ch.~\ref{ch:background}, there is a unique smooth solution
$c\left(\bm{x},t\right)$ to the advective Cahn--Hilliard equation~\eqref{eq:ch_source},
at least for finite times.  Thus, we turn to the question of the long-time
behaviour of solutions.  
We exploit the smoothness property of the concentration field $c\left(\bm{x},t\right)$
in formulating an evolution equation for the free energy.  The free energy
of the Cahn--Hilliard fluid is
\[
F\left[c\right]=\int_\Omega\left[\tfrac{1}{4}\left(c^2-1\right)^2+\tfrac{1}{2}\gamma\left|\nabla{c}\right|^2\right]d^{\n}x.
\]
Given the smooth, finite-time solution $c\left(\bm{x},t\right)$, we differentiate
the functional $F\left[c\right]$ with respect to time and obtain the relation
\[
\frac{dF}{dt}=-D\int_\Omega\left|\nabla\mu\right|^2d^nx+\int_\Omega\mu\left(-\bm{v}\cdot\nabla
c+s\right)d^{n}x,
\]
using the no-flux or periodic boundary conditions.  By averaging
this equation over finite times, we obtain the identity
\begin{equation}
\langle{\dot{F}}\rangle_t+\D\bigg\langle\int_\Omega\left|\nabla\mu\right|^2d^\n
x\bigg\rangle_t=\bigg\langle\int_\Omega\mu
sd^\n x\bigg\rangle_t-\bigg\langle\int_\Omega\mu\bm{v}\cdot\nabla cd^\n x\bigg\rangle_t,
\label{eq:fe_dissipation}
\end{equation}
We single out the quantity $\langle\dot{F}\rangle$ for study.
 Owing to the nonnegativity of $F\left(t\right)$, we have the inequality
 $\langle\dot{F}\rangle\geq0$.  Therefore, we need only consider
 two possible
 cases: $\langle\dot{F}\rangle=0$, and $\langle\dot{F}\rangle>0$.  We shall
 show that $\langle\dot{F}\rangle\neq0$ is not possible, and in doing so,
 we shall produce a uniform ($t$-independent) upper bound on $\langle{F}\rangle_t$.

 Let us assume for contradiction that $\langle\dot{F}\rangle>0$.  Then,
   given any $\varepsilon$ in the range $0<\varepsilon<\langle\dot{F}\rangle$,
   there is
   a time $T_\varepsilon$ such that $\langle\dot{F}\rangle-\varepsilon<\langle{\dot{F}}\rangle_t<\langle\dot{F}\rangle+\varepsilon$,
   for all times $t>T_\varepsilon$.  Thus, for times $t>T_\varepsilon$, the
   time average $\langle{\dot{F}}\rangle_t$ is strictly positive.
Henceforth, the inequality $t>T_\varepsilon$ is assumed.
 Using $\nabla\cdot\bm{v}=0$, Eq.~\eqref{eq:fe_dissipation} becomes
\begin{equation}
\langle{\dot{F}}\rangle_t+\D\bigg\langle\int_\Omega\left|\nabla\mu\right|^2d^\n
x\bigg\rangle_t=\bigg\langle\int_\Omega\mu sd^\n x\bigg\rangle_t+\bigg\langle\gamma\int_\Omega\Delta
c\bm{v}\cdot\nabla cd^\n x\bigg\rangle_t.
\label{eq:dissipation}
\end{equation}
As in Sec.~\ref{sec:background:existence_ch}, we use the following string
of relations,
\begin{eqnarray*}
\int_\Omega\Delta c\bm{v}\cdot\nabla cd^\n x&=&\int_\Omega\left(\partial_i\partial_ic\right)\left(v_j\partial_jc\right)d^{\n}x,\\
&=&-\int_\Omega\left(\partial_ic\right)\left[\partial_i\left(v_j\partial_jc\right)\right]d^{\n}x,\\
&=&-\int_\Omega\left(\partial_ic\right)\left(\partial_iv_j\right)\left(\partial_jc\right)d^{\n}x-
\int_\Omega\left(\partial_ic\right)\left(\bm{v}\cdot\nabla\right)\left(\partial_ic\right)d^{\n}x,\\
&=&-\int_\Omega\bm{w}\Deform\bm{w}^Td^\n x,\qquad\bm{w}=\nabla c,\qquad
{\Deform}_{ij}=\tfrac{1}{2}\left(\partial_iv_j+\partial_jv_i\right),
\end{eqnarray*}
where we have used the antisymmetry of the operator $\bm{v}\cdot\nabla$ to
deduce that
\[
\int_\Omega \phi\bm{v}\nabla\cdot\phi d^nx=0,
\] 
for any function $\phi\left(\bm{x},t\right)$.
The quadratic form $\bm{w}\Deform\bm{w}^T$ satisfies 
\[\left|\bm{w}\Deform\bm{w}^T\right|\leq{\n}\max_{ij}|\Deform_{ij}|\|\bm{w}\|_2^2,
\]
which gives rise to the inequality
\begin{equation}
\left|\int_\Omega\Delta c\bm{v}\cdot\nabla cd^\n x\right|\leq{\n}\left(\sup_{\Omega,i,j}\left|\Deform_{ij}\right|\right)\int\left|\nabla
c\right|^2d^\n x.
\label{eq:bound_on_bad_bit}
\end{equation}
%
%
%
%
%
%
The matrix ${\Deform}$ is the stretching tensor~\cite{OttinoBook}, the appearance
of which highlights
the importance of shear and stretching in the development of the morphology
of the concentration field~\cite{Berti2005}.

For each time $t'\in\left[0,t\right]$, we split the chemical potential $\mu$
into a part with
mean zero, and a mean component: $\mu = \overline{\mu}\left(t'\right)+\mu'\left(\bm{x},t'\right)$,
where $\int_\Omega\mu'\left(\bm{x},t'\right)d^n x=0$.  Then, for any function
$\phi\left(\bm{x},t\right)$ with spatial mean zero, we have the relation
$\int_\Omega\phi\mu d^{\n}x=\int_\Omega\phi\mu'd^{\n}x$.  Using this projection
relation, Eq.~\eqref{eq:dissipation} becomes
\[
\langle{\dot{F}}\rangle_t+\D\bigg\langle\int_\Omega\left|\nabla\mu'\right|^2d^\n
x\bigg\rangle_t=
\bigg\langle\int_\Omega\mu' sd^\n x\bigg\rangle_t+\bigg\langle\gamma\int_\Omega\Delta
c\bm{v}\cdot\nabla cd^\n x\bigg\rangle_t.
\]
Owing to the positivity of $\langle{\dot{F}}\rangle_t$, we have the inequality
\begin{equation}
\D\bigg\langle\int_\Omega\left|\nabla\mu'\right|^2d^{\n}x\bigg\rangle_t\leq
%
%
\bigg\langle\int_\Omega\mu' sd^\n x\bigg\rangle_t+\bigg\langle\gamma\int_\Omega\Delta
c\bm{v}\cdot\nabla cd^\n x\bigg\rangle_t.
\label{eq:dissipation1}
\end{equation}
Finally, we employ the Poincar\'e inequality for mean-zero functions
on a periodic domain $\Omega=\left[0,L\right]^2$:
%
%
\begin{equation}
\|\mu'\|_2^2\leq \left(\frac{L}{2\pi}\right)^2\|\nabla \mu'\|_2^2.
\label{eq:Poincare}
\end{equation}
Combining Eqs.~\eqref{eq:bound_on_bad_bit},~\eqref{eq:dissipation1}, and~\eqref{eq:Poincare}
gives the following inequality:
\begin{equation}
\D\left(\frac{2\pi}{L}\right)^2\langle\|\mu'\|_2^2\rangle_t\leq
\langle\|\mu'\|_2^2\rangle_t^{\frac{1}{2}}\|s\|_2+
\n \Deform_{\infty}\bigg\langle\int_\Omega\gamma\left|\nabla c\right|^2
d^\n x\bigg\rangle_t,
\label{eq:inequality_for_mu}
\end{equation}
where $\Deform_{\infty}=\text{sup}_{t\in\left[0,\infty\right),\Omega,i,j}\left|\Deform_{ij}\right|$.
 There are no angle brackets around the source term because $s\left(\bm{x}\right)$
 is independent of time.
\subsection*{Step 2: Obtaining a bound on $\langle\|\mu'\|_2^2\rangle_t$}
\noindent Using the string of relations
\[
\int_\Omega\gamma\left|\nabla c\right|^2d^{\n}x=\int_\Omega\left[\mu c+c^2-c^4\right]d^{\n}x
%
%
=\int_\Omega\left[\mu' c+c^2-c^4\right]d^{\n}x
%
%
\leq\|\mu'\|_2\|c\|_2+\|c\|_2^2,
\]
we obtain the inequality 
\begin{equation}
\int\gamma\left|\nabla c\right|^2d^\n x\leq\left|\Omega\right|^{\frac{1}{4}}\|\mu'\|_2\|c\|_4+\left|\Omega\right|^{\frac{1}{2}}\|c\|_4^2.
\label{eq:inequality_grad_c}
\end{equation}
Combining Eqs.~\eqref{eq:inequality_for_mu} and~\eqref{eq:inequality_grad_c}
gives rise to the inequality
\[
\D\left(\frac{2\pi}{L}\right)^2\langle\|\mu'\|_2^2\rangle_t\leq
\langle\|\mu'\|_2^2\rangle_t^{\frac{1}{2}}\left[\|s\|_2+\n\left|\Omega\right|^{\frac{1}{4}}\Deform_\infty\langle\|c\|_4^4\rangle_t^{\frac{1}{4}}\right]+
2\left|\Omega\right|^{\frac{1}{2}}\Deform_\infty\langle\|c\|_4^4\rangle_t^{\frac{1}{2}},
\]
a quadratic inequality in $\langle\|\mu'\|_2\rangle_t^{\frac{1}{2}}$.  Hence,
\begin{multline*}
\langle\|\mu'\|_2^2\rangle_t^{\frac{1}{2}}\leq\frac{1}{2D}\left(\frac{L}{2\pi}\right)^2
\left(\|s\|_2+\n\left|\Omega\right|^{\frac{1}{4}}\Deform_{\infty}\langle\|c\|_4^4\rangle_t^{\frac{1}{4}}\right)\\
+
\frac{1}{2D}\left(\frac{L}{2\pi}\right)^2
\sqrt{\left(\|s\|_2+\n\left|\Omega\right|^{\frac{1}{4}}\Deform_{\infty}\langle\|c\|_4^4\rangle_t^{\frac{1}{4}}\right)^2+{8D\left|\Omega\right|^{\frac{1}{2}}}\left(\frac{2\pi}{L}\right)^2\Deform_\infty\langle\|c\|_4^4\rangle_t^{\frac{1}{2}}.
}
\end{multline*}
A less sharp bound is given by
%
%
%
%
%
%
%
%
%
%
%
\begin{equation}
\langle\|\mu'\|_2^2\rangle_t\leq
\frac{1}{D^2}\left(\frac{L}{2\pi}\right)^4 
\left(\|s\|_2+\n\left|\Omega\right|^{\frac{1}{4}}\Deform_\infty\langle\|c\|_4^4\rangle_t^{\frac{1}{4}}\right)^2+\frac{8\left|\Omega\right|^{\frac{1}{2}}}{D}\left(\frac{L}{2\pi}\right)^2\Deform_\infty\langle\|c\|_4^4\rangle_t^{\frac{1}{2}},
\label{eq:bound_on_mu}
\end{equation}
which is an upper bound for $\langle\|\mu'\|_2^2\rangle_t$, in terms of the
forcing parameters, and $\langle\|c\|_4^4\rangle_t$.
\subsection*{Step 3: An upper bound on $m_4$}
\noindent We have the free energy
\begin{multline*}
F\left[c\right]=\int_\Omega\left[\tfrac{1}{4}\left(c^2-1\right)^2+\tfrac{1}{2}\gamma\left|\nabla
c\right|^2\right]d^\n x = 
\int_\Omega\left[\tfrac{1}{2}c\mu-\tfrac{1}{4}c^4\right]d^\n x+\tfrac{1}{4}\left|\Omega\right|\\
%
=\int_\Omega\left[\tfrac{1}{2}c\mu'\left(\bm{x},t'\right)-\tfrac{1}{4}c^4\right]d^{\n}x+\tfrac{1}{4}\left|\Omega\right|\geq0.
\end{multline*}
Hence,
\[
\int_\Omega c^4d^\n x\leq2\int_\Omega c\mu' d^\n x + \left|\Omega\right|
\leq2\|c\|_2\|\mu'\|_2+\left|\Omega\right|.
\]
Time averaging both sides and using the monotonicity of norms~\eqref{eq:monotonicity_norms},
we obtain the result
\[
\langle\|c\|_4^4\rangle_t\leq\left|\Omega\right|+2\left|\Omega\right|^{\frac{1}{4}}\langle\|c\|_4^4\rangle_t^{\frac{1}{4}}\langle\|\mu'\|_2^2\rangle_t^{\frac{1}{2}}.
\]
Using the bound for $\langle\|\mu'\|_2^2\rangle_t$ in~\eqref{eq:bound_on_mu},
this becomes
\begin{small}
\begin{multline*}
\langle\|c\|_4^4\rangle_t\leq\left|\Omega\right|
\\
+\frac{2\left|\Omega\right|^{\frac{1}{4}}}{D}\left(\frac{ L}{2\pi}\right)^2\langle\|c\|_4^4\rangle_t^{\frac{1}{4}}
\bigg[\left(\|s\|_2+\n\left|\Omega\right|^{\frac{1}{4}}\Deform_\infty
\langle\|c\|_4^4\rangle_t^{\frac{1}{4}}\right)^2+{4\n D\left|\Omega\right|^{\frac{1}{2}}}\left(\frac{2\pi}{L}\right)^2\Deform_\infty\langle\|c\|_4^4\rangle_t^{\frac{1}{2}}\bigg]^{\frac{1}{2}}.
\end{multline*}
\end{small}
We therefore have a $t$-independent equation for the upper bound on $\langle\|c\|_4^4\rangle_t$,
\begin{equation}
\langle\|c\|_4^4\rangle_t\leq m_{4}^{\mathrm{max}}\left[\bm{v},s,D\right],
\label{eq:first_in_chain}
\end{equation}
where $m_4^{\mathrm{max}}$ solves the polynomial
\begin{small}
\begin{multline}
\left(m_{4}^{\mathrm{max}}\right)^4=\left|\Omega\right|
\\
+\frac{2\left|\Omega\right|^{\frac{1}{4}}}{D}\left(\frac{L}{2\pi}\right)^2m_{4}^{\mathrm{max}}
\bigg[\left(\|s\|_2+\n\left|\Omega\right|^{\frac{1}{4}}\Deform_\infty{m_{4}}^{\mathrm{max}}\right)^2+{4\n
D\left|\Omega\right|^{\frac{1}{2}}}\left(\frac{2\pi}{L}\right)^2\Deform_\infty\left({m_{4}}^{\mathrm{max}}\right)^2\bigg]^{\frac{1}{2}},
\label{eq:eqn_for_x}
\end{multline}
\end{small}
The highest power of $m_{4}^{\mathrm{max}}$ on the left-hand side
is $\left(m_{4}^{\mathrm{max}}\right)^4$, while the highest
power of $m_{4}^{\mathrm{max}}$ on the right-hand side is $\left(m_{4}^{\mathrm{max}}\right)^{\frac{3}{2}}$.
 Thus, this equation always has a positive solution, and moreover this positive
 solution is unique.
 
We obtain the following chain of uniform ($t$-independent) bounds.  Each
bound follows
from the previous bounds in the chain, and the first bound follows from Eq.~\eqref{eq:first_in_chain}.
\begin{itemize}
\item $\langle\|c\|_4^4\rangle_t$ is uniformly bounded,
\item $\langle\|c\|_2^2\rangle_t$ is uniformly bounded,
\item $\langle\|\mu'\|_2^2\rangle_t$ is uniformly bounded,
\item $\langle\|\nabla{c}\|_2^2\rangle_t$ is uniformly bounded,
\item $\langle F\rangle_t$ is uniformly bounded,
\begin{equation}\label{eq:bounds}\end{equation}
\end{itemize}
\noindent for all $t>T_\varepsilon$.  
Owing to the uniformity of these bounds, they hold in the limit $t\rightarrow\infty$.
The result $\langle F\rangle<\infty$ implies the existence of a uniform bound
for $F\left(t\right)$, almost everywhere.  Given the differentiability
of $F\left(t\right)$, this implies that $F\left(t\right)$ is everywhere uniformly
bounded, and thus, $\langle\dot{F}\rangle=0$, which is a contradiction. 
Therefore, the only possibility for $\langle\dot{F}\rangle$ is that it be
zero.
 It is straightforward to verify that by taking $\langle\dot{F}\rangle=0$,
 and making
 slight alterations in Steps 1--3, the bounds in Eq.~\eqref{eq:bounds} still
 hold.

Let us examine the significance of our result. We have shown that for sufficiently
regular
flows and source terms (specifically, $\bm{v}\left(\bm{x},t\right)\in H^{1,\infty}\left(\Omega\right)$,
for all $t\in\left[0,\infty\right)$,  and $s\left(\bm{x}\right)\in L^2\left(\Omega\right)$),
there is an \emph{a priori} bound on the free energy
$\langle F\left[c\right]\rangle$.  We have shown that the system always reaches
a steady state, in the sense that $\langle\dot{F}\rangle=0$.
 We have also found an upper bound for the $m_4$ measure of concentration
 fluctuations,
 as the unique positive root of the polynomial equation Eq.~\eqref{eq:eqn_for_x}.
This bound depends only on the source amplitude, the diffusion constant,
and the maximum stretching $\Deform_\infty$.  Using the monotonicity of norms,
this number serves also as an upper bound on $m_p$ for $p\in\left[1,4\right]$.
Let us comment briefly on the volume term in the equation $\left(m_{4}^{\mathrm{max}}\right)^4=\left|\Omega\right|+...$.
 Since this upper bound includes many situations, it must take into account
 the case where both the velocity and the source vanish.  Then $c\sim\pm1$
 as $t\rightarrow\infty$, and by definition, $m_4\sim|\Omega|^{\frac{1}{4}}$,
 which is in agreement with Eq.~\eqref{eq:eqn_for_x}.  These results enhance
 the existence theory obtained in Sec.~\ref{sec:background:existence_ch},
 and give information about the long-time behaviour of solutions.
 
 As mentioned in Ch.~\ref{ch:introduction}, it is desirable
 in many applications to suppress concentration fluctuations, since this
 leads
 to a homogeneous mixture.  In this report, we use advection
 as a suppression mechanism, and we would therefore
 like to know the maximum suppression achievable for a given flow.  This
 inspires us to seek lower bounds on $m_p$, in addition to the upper bounds
 found in this section.

\section{Lower bounds on the concentration fluctuations}
\label{sec:estimating_mixedness:lower_bds}

\noindent In this section we discuss the significance of the lower bound
on the $m_p$ measure of concentration fluctuations.  Due to the powers of
the
concentration that appear in the Cahn--Hilliard equation, it is possible
to obtain an explicit lower bound for $m_4$, which we then use to discuss
stirring as a mechanism to suppress concentration fluctuations.  Using H\"older's
inequality,
a flow that suppresses concentration fluctuations in the $m_4$ sense
will also suppress them in the $m_p$ sense, for $p\in\left[1,4\right]$, and
thus our discussion of the suppression of concentration fluctuations by stirring
is valid in this broader sense.

As discussed in Sec.~\ref{sec:model}, a suitable
measure of concentration fluctuations for a symmetric mixture is
\[
m_p=\langle\|c\|_p^p\rangle^{\frac{1}{p}},
\]
where $c\left(\bm{x},t\right)$ is the concentration of the binary mixture
and
$\|c\|_p$ is its $L^p$ norm.  For $p=2$, this gives the usual variance, used
in the theory of miscible fluid mixing~\cite{Shaw2006, Thiffeault2004,Thiffeault2006,Thiffeault2007}.
In that case, the choice $p=2$ is a natural one suggested by the linearity
of the advection-diffusion equation.  In the following analysis of the advective
Cahn--Hilliard
equation, it is possible to find an explicit lower bound for $m_4$ and we
therefore use this quantity to study the suppression of concentration fluctuations
due to the imposed velocity field.  Given this formula, we can compare the
suppression achieved by a given flow with the ideal level of suppression,
and decide on the best strategy to homogenize the binary fluid.

To estimate $m_4$, we take Eq.~\eqref{eq:ch}, multiply it
by an arbitrary, spatially-varying test function $\phi\left(\bm{x}\right)$,
and then integrate over space and time, which yields
\begin{equation}
\Big\langle\int_\Omega c\left[\op\phi+Dc^2\Delta\phi\right]d^nx\Big\rangle
=-\int_{\Omega}s{\phi}d^{\n}x,
\label{eq:constraint_exact}
\end{equation}
where $\op$ is the linear operator $\bm{v}\cdot\nabla - D\Delta-\gamma D\Delta^2$.
 Using the constraint equation~\eqref{eq:constraint_exact},
the monotonicity of norms, and the Cauchy--Schwarz inequality,
we obtain the following string of inequalities,
\[
\left|\int_\Omega s\phi d^\n x\right|
%
%
\leq\langle\|c\|_2\|\op\phi+\D{c^2}\Delta\phi\|_2\rangle
\leq\langle\|c\|_2^2\rangle^{\frac{1}{2}}\langle\|\op\phi+\D c^2\Delta\phi\|_2^2\rangle^{\frac{1}{2}},
\]
which gives the relation
\[
\langle\|c\|_2^2\rangle^{\frac{1}{2}}\geq\frac{\left|\int_\Omega s{\phi}d^{\n}x\right|}
{\langle\|\op\phi+\D{c^2}\Delta\phi\|_2^2\rangle^{\frac{1}{2}}}.
\]
We study the denominator
\begin{eqnarray*}
\langle\|\op\phi+\D{c^2}\Delta\phi\|_2^2\rangle^{\frac{1}{2}}&\leq&\langle\|\op\phi\|_2^2\rangle^{\frac{1}{2}}+\D\langle\|c^2\Delta\phi\|_2^2\rangle^{\frac{1}{2}},\\
&\leq&\Big\langle\int_\Omega( \op\phi)^2d^\n x\Big\rangle^{\frac{1}{2}}+\D\|\Delta\phi\|_\infty\langle\|c\|_4^4\rangle^{\frac{1}{2}},
\end{eqnarray*}
where this bound follows from the triangle and H\"older inequalities.
 Thus we have the result
\[
\langle\|c\|_2^2\rangle^{\frac{1}{2}}\geq\frac{\left|\int_\Omega s\phi d^{\n}x\right|}{\langle\int_\Omega(\op\phi)^2d^\n
x\rangle^{\frac{1}{2}}+\D\|\Delta\phi\|_\infty\langle\|c\|_4^4\rangle^{\frac{1}{2}}},
\]
or
\[
\langle\|c\|_2^2\rangle^{\frac{1}{2}}\left[\Big\langle\int_\Omega(\op\phi)^2d^{\n}x\Big\rangle^{\frac{1}{2}}
+\D\|\Delta\phi\|_\infty\langle\|c\|_4^4\rangle^{\frac{1}{2}}\right]\geq\left|\int_\Omega
s\phi d^\n x\right|.
\]
Using the monotonicity of norms~\eqref{eq:monotonicity_norms}, we recast
this inequality as one involving only a single power-mean,
\[
\left|\Omega\right|^{\frac{1}{4}}\langle\|c\|_4^4\rangle^{\frac{1}{4}}\left[\Big\langle\int_\Omega(\op\phi)^2d^\n
x\Big\rangle^{\frac{1}{2}}+\D\|\Delta\phi\|_\infty\langle\|c\|_4^4\rangle^{\frac{1}{2}}\right]\geq\left|\int_\Omega
s\phi d^\n x\right|.
\]
Therefore, we have the following inequality for  $m_4=\langle\|c\|_4^4\rangle^{\tfrac{1}{4}}$,
\[
m_4\left[q_0\left(\bm{v},D,\gamma\right)+\D\|\Delta\phi\|_\infty m_4^2\right]\geq\left|\Omega\right|^{-\frac{1}{4}}\left|\int_\Omega
s\phi d^\n x\right|,
\]
where
\[
q_0\left(\bm{v},D,\gamma\right)=\Big\langle\int_\Omega\left[\bm{v}\cdot\nabla\phi-\D\Delta\phi-\D\gamma\Delta^2\phi\right]^2d^\n
x\Big\rangle^{\frac{1}{2}}.
\]
Thus, we obtain a lower bound for the $m_4$ measure of concentration
fluctuations,
\begin{equation}
m_4\geq m_4^{\mathrm{min}},\qquad \D\|\Delta\phi\|_\infty\left(m_4^{\mathrm{min}}\right)^3+q_0\left(\bm{v},D,\gamma\right)m_4^{\mathrm{min}}-\left|\Omega\right|^{-\frac{1}{4}}\left|\int_{\Omega}
s\phi d^\n x\right|=0.
\label{eq:funky_variance}
\end{equation}
The cubic equation satisfied by $m_4^{\mathrm{min}}$ has a unique positive
root. 

To probe the asymptotic forms of~\eqref{eq:funky_variance}, we
rewrite the forcing terms $\bm{v}\left(\bm{x},t\right)$ and $s\left(\bm{x}\right)$
as an amplitude, multiplied by a dimensionless shape function.  Thus,
\begin{eqnarray*}
\bm{v}=V_0\tilde{\bm{v}}&,&\qquad V_0 = |\Omega|^{-\frac{1}{2}}\langle\|\bm{v}\|_2^2\rangle^{\frac{1}{2}},\\
s = S_0\tilde{s}&,&\qquad S_0 = |\Omega|^{-\frac{1}{2}}\|s\|_2.
\end{eqnarray*}
Then for a fixed value of $S_0$ and $D$, and $V_0\gg1$ (large stirring),
we have
$q_0\sim V_0\langle\int\left(\tilde{\bm{v}}\cdot\nabla\phi\right)^2d^{\n}x\rangle^{\frac{1}{2}}$,
and the lower bound $m_4^{\mathrm{min}}$ takes the form
\[
m_4^{\mathrm{min}}\sim\frac{S_0}{V_0}\frac{|\Omega|^{-\frac{1}{4}}\left|\int_\Omega \tilde{s}\phi
d^{\n}x\right|}{\langle\int_\Omega\left(\tilde{\bm{v}}\cdot\nabla\phi\right)^2d^\n
x\rangle^{\frac{1}{2}}|\Omega|^{\frac{1}{4}}},\qquad V_0\gg1.
\]
On the other hand, for fixed $S_0$ and $V_0$, and $D\gg1$ (large diffusion),
the lower bound takes the form
\[
m_4^{\mathrm{min}}\sim
\frac{S_0}{D}\frac{|\Omega|^{-\frac{1}{4}}\left|\int_\Omega
\tilde{s}\phi d^\n x\right|}{\left[\int_\Omega\left(\Delta\phi+\gamma\Delta^2\phi\right)^2d^\n
x\right]^{\frac{1}{2}}|\Omega|^{\frac{1}{4}}},\qquad D\gg1.
\]

It is possible to obtain similar asymptotic expressions for $m_2$, by minimizing
a constrained functional of the concentration.  Apart from a volume factor,
the asymptotic form of $m_2$ agrees exactly with the asymptotic form of
$m_4$ just obtained.  The functional to be minimized is
\[
\Phi\left[c\right]=\tfrac{1}{2}\Big\langle\int_\Omega c^2d^{\n}x\Big\rangle-\lambda\Big\langle\int_\Omega\left(
c\op\phi+\D c^3\Delta\phi+s\phi\right)d^{\n}x\Big\rangle,
\]
where $\phi\left(\bm{x}\right)$ is a test function and $\lambda$ is the Lagrange
multiplier for the constraint.  This approach has been
taken in~\cite{Thiffeault2006} for the advection-diffusion equation.  Setting
$\delta \Phi/\delta c=0$ gives
\begin{equation}
c=\frac{1-\sqrt{1-12\lambda^2\D\Delta\phi \op\phi}}{6\lambda \D\Delta\phi}.
\label{eq:c_sqrt}
\end{equation}
Evaluation of $\delta^2 \Phi/\delta c\delta c'$ shows that Eq.~\eqref{eq:c_sqrt}
produces a minimum of $\Phi\left[c\right]$.  Given the expression $\op=V_0\tilde{\bm{v}}\cdot\nabla-D\Delta-\gamma
D\Delta^2$, the minimum Eq.~\eqref{eq:c_sqrt} is $\lambda V_0\tilde{\bm{v}}\cdot\nabla\phi$
at large $V_0$.  Substitution of this expression into the constraint
\[
\Big\langle\int_\Omega\left[c\op\phi+\D c^3\Delta\phi+s\phi\right]d^{\n}x\Big\rangle=0
\]
gives $\lambda=-\left(S_0/V_0^2\right)\left[\langle \int_\Omega \tilde{s}\phi
d^{\n}x\rangle/ \langle\int_\Omega\left(\tilde{\bm{v}}\cdot\nabla\phi\right)^2d^{\n}x\rangle\right]$,
and hence
\[
m_2^{\mathrm{min}}
\sim
\frac{S_0}{V_0}
\frac{\left|\int_\Omega \tilde{s}\phi d^\n x\right|}
{\langle\int_\Omega\left(\tilde{\bm{v}}\cdot\nabla{c}\right)^2d^{\n}x\rangle^{\frac{1}{2}}},
\qquad V_0\gg1.
\]
For fixed $V_0$ and $S_0$ and large $D$, a similar calculation gives
\[
m_2^{\mathrm{min}}\sim
\frac{S_0}{D}\frac{\left|\int_\Omega
\tilde{s}\phi d^\n x\right|}{\left[\int_\Omega\left(\Delta\phi+\gamma\Delta^2\phi\right)^2d^\n
x\right]^{\frac{1}{2}}},\qquad D\gg1.
\]

These expressions show that, apart from a volume factor, the lower bounds
on the $m_2$ and $m_4$
measures of concentration fluctuations are identical in the physically interesting
limits of high stirring strength, or high diffusion.  In particular, the
expression
\[
m_2^{\mathrm{min}},\phantom{a}|\Omega|^{\frac{1}{4}}m_4^{\mathrm{min}}\sim\frac{S_0}{V_0}\frac{\left|\int_\Omega
\tilde{s}\phi d^{\n}x\right|}{\langle\int_\Omega\left(\tilde{\bm{v}}\cdot\nabla\phi\right)^2d^\n
x\rangle^{\frac{1}{2}}},\qquad \text{for large }V_0,
\]
indicates that if a flow can be found that saturates the lower bound $m_{2,4}^{\mathrm{min}}$,
the suppression of concentration fluctuations can be enhanced by a factor
of $V_0^{-1}$ at large stirring amplitudes.  Such a flow would then be an
efficient way of mixing the binary fluid.  


\section{Scaling laws for $m_4$}
\label{sec:scaling}
In this section we investigate the dependence of the $m_4$ measure
of concentration fluctuations on the parameters of the problem, namely the
stirring velocity $\bm{v}$, the source $s$, and the diffusion constant $D$.
 For simplicity, we shall restrict our interest to a certain class of flows,
 which enables us to compute long-time averages explicitly.

\begin{figure}
\centering
\subfigure[]{
    \includegraphics[width=.32\textwidth]{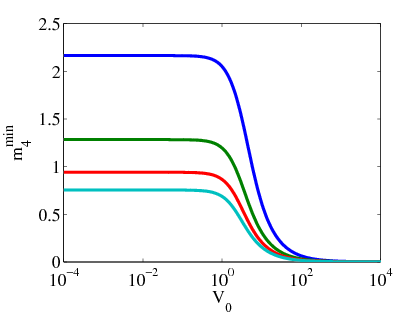}
}
\subfigure[]{
    \includegraphics[width=.32\textwidth]{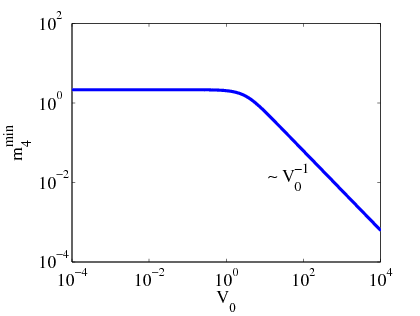}
}
\caption{(a) The lower bound for $m_4^{\mathrm{min}}$ as a function of $V_0$
for a monochromatic source; (b) the dependence of $m_4^{\mathrm{min}}$ on
the velocity amplitude $V_0$.  In (a), the scale of the source variation
decreases in integer multiples from $k_{\mathrm{s}}=2\pi/L$ in the uppermost
curve, to $k_{\mathrm{s}}=8\pi/L$ in the lowermost curve, while in (b) the
source scale is set to $2\pi/L$.  In both figures, we have set $D=S_0=1$.}
\label{fig:monochromatic}
\end{figure}

The lower bound for the $m_4$ measure of concentration fluctuations is the
unique positive root of the polynomial
\begin{equation}
\D\|\Delta\phi\|_\infty\left(m_4^{\mathrm{min}}\right)^3+q_0\left(\bm{v},D,\gamma\right)m_4^{\mathrm{min}}-\left|\Omega\right|^{-\frac{1}{4}}\left|\int_{\Omega}
s\phi d^\n x\right|=0,
\label{eq:lower_bd_monochromatic}
\end{equation}
where $\phi\left(\bm{x}\right)$ is a test function and
\[
q_0\left(\bm{v},D,\gamma\right)=\Big\langle\int_\Omega\left[\bm{v}\cdot\nabla\phi-\D\Delta\phi-\D\gamma\Delta^2\phi\right]^2d^\n
x\Big\rangle^{\frac{1}{2}}.
\]
We specialize to velocity fields whose time average has the following properties,
\begin{equation}
\langle v_i\left(\bm{x},\cdot\right)\rangle=0,
\qquad\langle v_i\left(\bm{x},\cdot\right)v_j\left(\bm{x},\cdot\right)\rangle=\frac{V_0^2}{\n}\delta_{ij}.
\label{eq:statistical_hit}
\end{equation}
The flow $\bm{v}\left(\bm{x},t\right)$ is defined on the $n$-torus $\left[0,L\right]^n$.
 A statistically homogeneous and isotropic turbulent velocity field automatically
 satisfies the relations~\eqref{eq:statistical_hit}, although it is not necessary
 for $\bm{v}$ to be of this type.  The source we consider is monochromatic
 (that is, it contains contains a single spatial scale) and varies in a single
 direction,
\begin{equation}
s=\sqrt{2}S_0\sin\left(k_{\mathrm{s}}x\right).
\label{eq:monochromatic}
\end{equation}
Our choice of velocity field makes the evaluation of
$q_0\left(\bm{v},D,\gamma\right)$ particularly easy; it is 
\[
q_0\left(\bm{v},D,\gamma\right)=\left[\frac{V_0^2}{\n}\int_{\Omega}\left|\nabla\phi\right|^2d^{\n}x+D^2\int_{\Omega}\left(\Delta\phi+\gamma\Delta^2\phi\right)^2d^{\n}x\right]^{\frac{1}{2}}.
\]
In studies of the advection-diffusion equation~\cite{Thiffeault2006}, it
is possible to find an explicit test function $\phi$ that sharpens the lower
bound on $m_2$.  The procedure for doing this depends on the linearity of
the equation.
Here, this is not possible, and for simplicity we set $\phi=s$.  This choice
of $\phi$ certainly gives a lower bound for the $m_4$ measure of concentration
fluctuations, with the added advantage of enabling explicit computations.
 Having specified the coefficients of the polynomial in Eq.~\eqref{eq:lower_bd_monochromatic}
 completely, we extract the positive root of this equation, and find the
 lower bound $m_4^{\mathrm{max}}$, as a function of $V_0$.  The results of
 this procedure are shown in Fig.~\ref{fig:monochromatic}.

\begin{figure}[htb]
\centering
    \includegraphics[width=.35\textwidth]{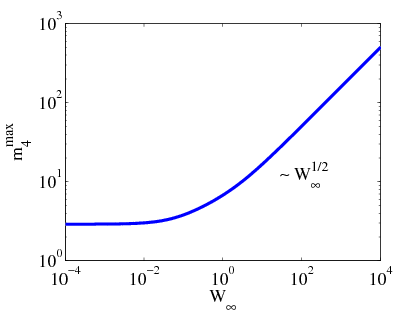}
\caption{The dependence of the upper bound $m_4^{\mathrm{max}}$ on the maximum
stretching $W_{\infty}$.  The source affects the upper bound only through
its square mean $S_0=|\Omega|^{-\frac{1}{2}}\|s\|_2$.}
\label{fig:upper_bd}
\end{figure}

Let us examine briefly the scaling of the upper bound $m_4^{\mathrm{max}}$
with the problem parameters.  The upper bound satisfies the polynomial equation
\begin{small}
\begin{multline}
\left(m_{4}^{\mathrm{max}}\right)^4=\left|\Omega\right|
\\
+\frac{2\left|\Omega\right|^{\frac{1}{4}}}{D}\left(\frac{L}{2\pi}\right)^2m_{4}^{\mathrm{max}}
\bigg[\left(S_0|\Omega|^{\frac{1}{2}}+\n\left|\Omega\right|^{\frac{1}{4}}\Deform_\infty{m_{4}}^{\mathrm{max}}\right)^2+{4\n
D\left|\Omega\right|^{\frac{1}{2}}}\left(\frac{2\pi}{L}\right)^2\Deform_\infty\left({m_{4}}^{\mathrm{max}}\right)^2\bigg]^{\frac{1}{2}},
\label{eq:upper_bd_monochromatic}
\end{multline} 
\end{small}
which depends only on the diffusion $D$, the source amplitude $S_0$, and
the maximum stretching $\Deform_\infty$.  For large $\Deform_\infty$, the
flow-dependence of the upper bound is $m_4^{\mathrm{max}}\sim\Deform_\infty^{\frac{1}{2}}$.
 This dependence is verified by obtaining the positive root of Eq.~\eqref{eq:upper_bd_monochromatic},
 which is as a function of $W_\infty$.  The results are shown in Fig.~\ref{fig:upper_bd}.

\subsection*{Mixedness at small scales}
In the same manner, we find a lower bound for the quantity $\langle\|\nabla
c\|_4^4\rangle^{\frac{1}{4}}$.
 A similar gradient norm has arisen in the problem of quantifying mixedness
 at small scales in the theory of miscible fluids~\cite{Thiffeault2006}.
  Using the constraint equation~\eqref{eq:constraint_exact} again, we find
\[
\langle\|\nabla c\|_2^2\rangle^{\frac{1}{2}}\geq\frac{\left|\int_\Omega
s\phi d^\n x\right|}{\langle\|\op_{\nabla}\phi+{3c^2}\Diff\nabla\phi\|_2^2\rangle^{\frac{1}{2}}},
\]
where $\op_{\nabla}=\bm{v}-\Diff\nabla-\Diff\Tr\Delta\nabla$.  We study the
denominator
\begin{eqnarray}
\langle\|\op_{\nabla}\phi+3\Diff{c^2}\nabla\phi\|_2^2\rangle^{\frac{1}{2}}&\leq&\langle\|\op_{\nabla}\phi\|_2^2\rangle^{\frac{1}{2}}+3\Diff\langle\|c^2\nabla\phi\|_2^2\rangle^{\frac{1}{2}},\nonumber\\
&\leq&\langle\|\op_{\nabla}\phi\|_2^2\rangle^{\frac{1}{2}}+3\Diff\|\nabla\phi\|_{\infty}\langle\|c^2\|_2^2\rangle^{\frac{1}{2}},\\
&=&\Big\langle\int_\Omega(\op_{\nabla}\phi)^2d^\n x\Big\rangle^{\frac{1}{2}}+3\Diff\|\nabla\phi\|_\infty\langle\|c\|_4^4\rangle^{\frac{1}{2}}.\nonumber
\label{eq:nabla_c}
\end{eqnarray}
Using the Gagliardo--Nirenberg--Sobolev inequality on the periodic domain
$\Omega=\left[0,L\right]^n$, we obtain 
\[
\|c\|_4\leq C_G\|\nabla c\|_{\frac{4n}{n+4}}\leq C_G|\Omega|^{\frac{n+4}{4n}-\frac{1}{2}}\|\nabla{c}\|_2,\qquad
n\leq 4.
\]
Thus, the denominator is bounded above by
\[
\Big\langle\int_\Omega(\op_{\nabla}\phi)^2d^\n x\Big\rangle^{\frac{1}{2}}+3\Diff\|\nabla\phi\|_\infty
k_\Omega\langle\|\nabla c\|_4^4\rangle^{\frac{1}{2}},\qquad k_\Omega =C_G^2|\Omega|^{\frac{n+4}{2n}-1}.
\]
We therefore have the inequality
\[
\left|\Omega\right|^{\frac{1}{4}}\langle\|\nabla c\|_4^4\rangle^{\frac{1}{4}}\left[\Big\langle\int_\Omega(\op_\nabla\phi)^2{d^{\n}x}\Big\rangle^{\frac{1}{2}}+3\Diff\|\nabla\phi\|_\infty k_\Omega\langle\|\nabla{c}\|_4^4\rangle^{\frac{1}{2}}\right]\geq\left|\int_\Omega
s\phi d^\n x\right|,
\]
and hence
\begin{equation}
\langle\|\nabla{c}\|_4^4\rangle^{\frac{1}{4}}\geq{m},\qquad 3D\|\nabla\phi\|_{\infty}k_{\Omega}m^3+q_{0,\nabla}\left(\bm{v},\Diff,\Tr\right)m-\left|\Omega\right|^{-\frac{1}{4}}\left|\int_{\Omega}
s\phi d^\n x\right|=0,
\end{equation}
where
\[
q_{0,\nabla}\left(\bm{v},\Diff,\Tr\right)=\bigg\langle\int_\Omega\left[\bm{v}-\Diff\nabla\phi-\Diff\Tr\Delta\nabla\phi\right]^2d^\n
x\bigg\rangle^{\frac{1}{2}}.
\]
For  a velocity field with the time-average relation $\langle v_i\left(\bm{x},\cdot\right)v_j\left(\bm{x},\cdot\right)\rangle=V_0^2{\n}^{-1}\delta_{ij}$,
this is
\[
q_{0,\nabla}\left(\bm{v},\Diff,\Tr\right)=\bigg\{\frac{V_0^2}{\n}\left|\Omega\right|+\Diff^2\int_\Omega\left(\nabla\phi-\gamma\Delta\nabla\phi\right)^2d^\n
x\bigg\}^{\frac{1}{2}}.
\]

This concludes our investigation into lower bounds on the level of mixedness
achievable in a stirred, phase-separating fluid.
We note briefly that for any nonzero source, the lower bound $m_4^{\mathrm{min}}$
is nonzero, meaning that no matter how hard one stirs, there will always
be some inhomogeneity in the fluid, and this is in fact true for any flow.
However, the quantity $m_4^{\mathrm{min}}$ tells us how much homogeneity we
can achieve and is therefore a yardstick for stirring protocols.  We use
this yardstick to test model flows in the next section.

\section{Numerical simulations}
\label{sec:estimating_mixedness:numerics}

\noindent In this section we solve Eq.~\eqref{eq:ch} numerically for two
flows, and verify the bounds obtained in Secs.~\ref{sec:existence}--\ref{sec:scaling}.
 We use the sinusoidal source term in Eq.~\eqref{eq:monochromatic} with periodic
 boundary conditions, and the source scale $k_{\mathrm{s}}$ therefore takes
 the form $\left(2\pi/L\right)j$, where $L$ is the box size and $j$ is an
 integer.  We specialize to two dimensions and study two standard flows that
 are used in the analysis of mixing: the random-phase sine flow of Sec.~\ref{sec:chaotic_advection:model_stirring},
 and the constant flow~\cite{Thiffeault2007}.

\subsection*{Random-phase sine flow}

\noindent The random-phase sine flow is the time-dependent two-dimensional
flow
\begin{equation}
\begin{split}
  v_x\left(x,y,t\right) &= \sqrt{2}V_0\sin\left(k_{\mathrm{v}}y+\phi_j\right),\qquad
  v_y=0,\qquad j\tau\leq t<\left(j+\tfrac{1}{2}\right)\tau,\\
  v_y\left(x,y,t\right) &= \sqrt{2}V_0\sin\left(k_{\mathrm{v}}x+\psi_j\right),\qquad
  v_x=0,\qquad \left(j+\tfrac{1}{2}\right)\tau \leq t<\left(j+1\right)\tau,
\end{split}
\label{eq:sineflow}
\end{equation}
where $\phi_j$ and $\psi_j$ are phases that are randomized once during each
flow period $\tau$, and where the integer $j$ labels the period.  The flow
is defined on the two-dimensional torus $\left[0,L\right]^2$.  The time average
of this velocity field has the properties listed in Eq.~\eqref{eq:statistical_hit}.
We solve Eq.~\eqref{eq:ch} with the flow in Eq.~\eqref{eq:sineflow} as in
Sec.~\ref{sec:chaotic_advection:model_stirring}.
Recall that we used the lattice method for advection~\cite{lattice_PH2,lattice_PH1},
which was effective only when the spread of matter due to diffusion is much
slower than the spread due to advection, that is, ${\tau}/\Time_D\ll1$.
 Thus, as before, we therefore set $D=10^{-5}$, while we set $\tau=L=1$.
 A numerical experiment with $V_0=0$ shows that $S_0=5\times10^{-4}$ gives
 rise to a morphology that is qualitatively different from the sourceless
 morphology, and we therefore work with this source amplitude.  Finally,
 following standard practice~\cite{chaos_Berthier,Berti2005}, we choose
 $\gamma\sim\Delta x^2$, the gridsize.

 Using this nondimensionalization, and the identity $\Deform_\infty=\sqrt{2}V_0k_{\mathrm{v}}$
 for the sine flow, we recall the large-stirring forms of $m_4^{\mathrm{max,min}}$.
For $V_0\gg\D$, the lower bound has the form
\begin{equation}
m_4^{\mathrm{min}}\sim\frac{S_0}{V_0}\frac{\int_\Omega\sin\left(k_{\mathrm{s}}x\right)\phi
d^2x}{\left[\int_\Omega\left|\nabla\phi\right|^2d^2x\right]^{\frac{1}{2}}},
\label{eq:lower_bd_asymp}
\end{equation}
with the power-law relationship $m_4^{\mathrm{min}}\sim V_0^{-1}$,
while for $V_0\gg1$ the upper bound has the form
\begin{equation}
m_4^{\mathrm{max}}\sim\left(\frac{2^{\frac{1}{2}}k_{\mathrm{v}}}{\pi^2D}\right)^{\frac{1}{2}}V_0^{\frac{1}{2}}.
\label{eq:upper_bd_asymp}
\end{equation}
These scaling results are identical to those for the advection-diffusion
problem~\cite{Thiffeault2004}.

Before studying the case with flow, we integrate Eq.~\eqref{eq:ch}
without flow, to verify the effect of the source.
\begin{figure}[htb]
\centering
\subfigure[]{
 \includegraphics[width=.2\textwidth]{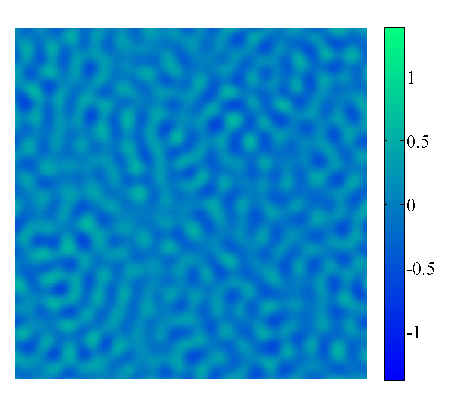}
}
\subfigure[]{
 \includegraphics[width=.2\textwidth]{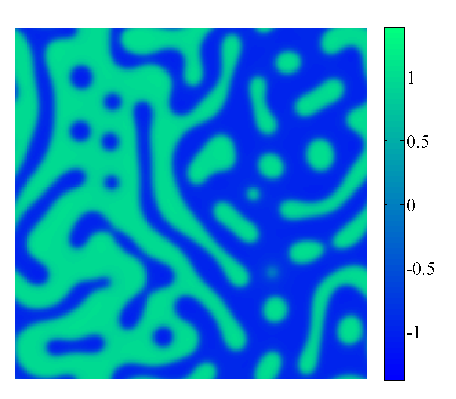}
}
\subfigure[]{
 \includegraphics[width=.2\textwidth]{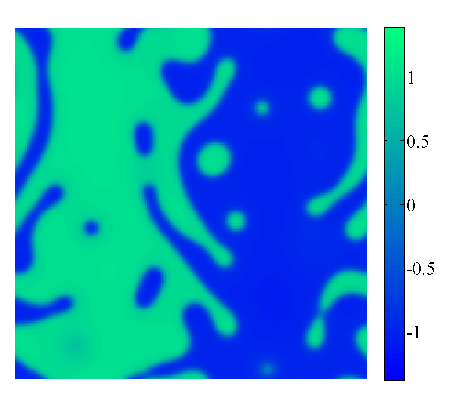}
}
\subfigure[]{
 \includegraphics[width=.2\textwidth]{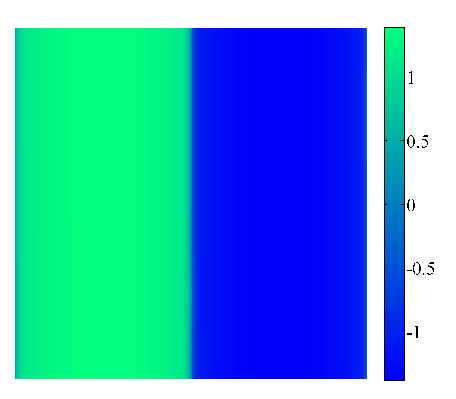}
}
\subfigure[]{
    \includegraphics[width=.35\textwidth]{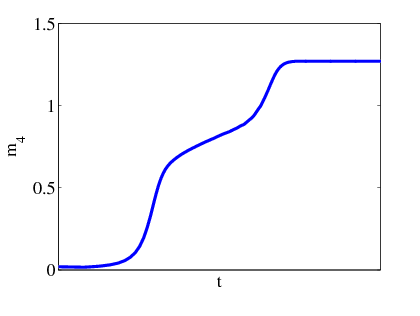}
}
\caption{The concentration field for $S_0 = 5\times10^{-4}$ and
(a) $t=100$; (b) $t=1000$; (c) $t=2000$; (d) $t=8000$.  A steady state is
reached in (d), evidenced by the time dependence of $m_4$ in
(e), where $m_4$ is constant for $t\gg1$.}
\label{fig:no_flow}
\end{figure}
For a sufficiently large source amplitude, the concentration phase separates
and forms domains rich in either binary fluid component.  These domains are
aligned with variations in the source.  A steady state
is reached and $m_4$ attains a constant value, as seen in Fig.~\ref{fig:no_flow}.
 On the other hand, for small source amplitudes, we have verified that that
 the domains do not align with the source, and their growth
 does not saturate.  We do not consider this case here, since we are interested
 in sources that qualitatively alter the phase separation.  These different
 regimes are discussed in~\cite{Krekhov2004}.

\begin{figure}[htb]
\centering
\subfigure[]{
    \includegraphics[width=.25\textwidth]{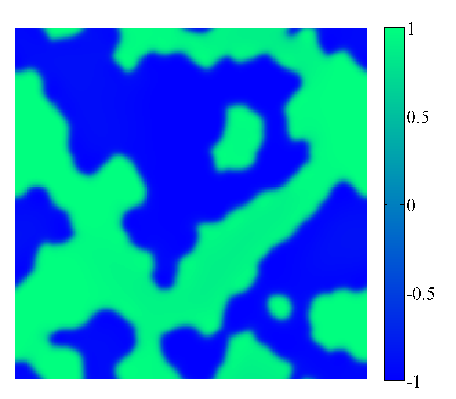}
}
\subfigure[]{
    \includegraphics[width=.25\textwidth]{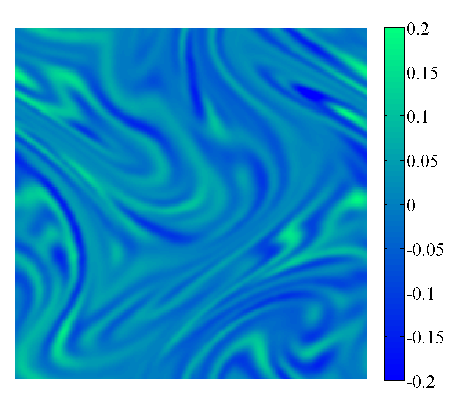}
}
\subfigure[]{
    \includegraphics[width=.25\textwidth]{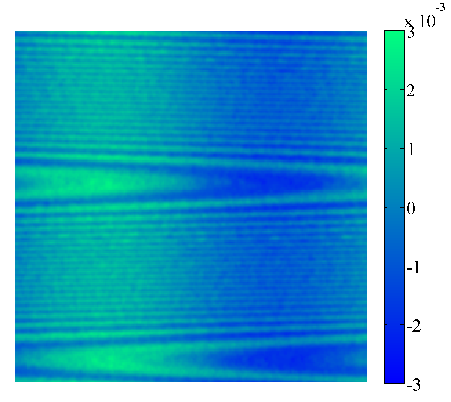}
}
\caption{A snapshot of the steady-state concentration field for (a)
$V_0=0.001$; (b) $V_0=0.1$; (c) $V_0=10$.  The source parameters are $S_0=5\times
10^{-4}$ and $k_{\mathrm{s}}=2\pi$.  In (a) domain growth
is arrested, in (b) the domains are destroyed and the binary fluid mixes,
while in (c) $m_4$ measure of concentration fluctuations is minimized, and
the source structure is visible.}
\label{fig:C_flow}
\end{figure}

We consider the case with flow by varying $V_0$, and find results that
are similar to those found in Sec.~\ref{sec:chaotic_advection:numerics},
for the same stirring
mechanism without sources.  For all values of $V_0$, the concentration
reaches a steady state, in which $\|c\|_4$, effectively the pre-averaged
form of $m_4$, fluctuates around a constant value.
For small values of $V_0$, the domain growth
is arrested due to a balance between the advection and phase-separation
terms in the equation, while for moderate values of $V_0$, the domains
are broken up and a mixed state is obtained.  At large values of $V_0$,
the $m_4$ measure of concentration fluctuations saturates: further
increases in $V_0$ do not produce further decreases in $m_4$.
 At these large values of $V_0$, the source structure is visible in snapshots
 of the concentration, as evidenced by Fig.~\ref{fig:C_flow}.

\begin{figure}
\centering
\subfigure[]{
    \includegraphics[width=.35\textwidth]{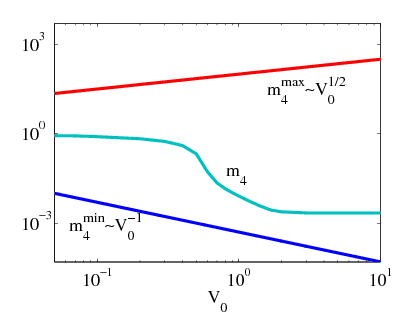}
}
\subfigure[]{
    \includegraphics[width=.35\textwidth]{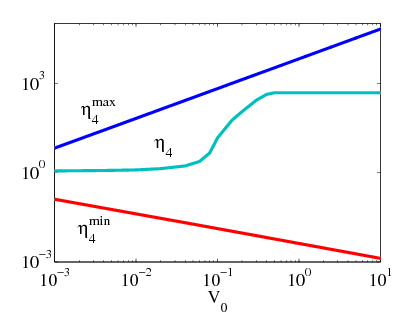}
}
\caption{(a) The $m_4$ measure of concentration fluctuations or mixing for
the
sine flow, as a function of the stirring parameter $V_0$.  The values of
$D$ and $S_0$ are given in the text.  The upper and lower bounds are shown
for comparison; (b) The mixing efficacy $\eta_4$ for the sine flow, with
the upper and lower bounds shown for comparison.}
\label{fig:min_max_sineflow}
\end{figure}

We investigate the dependence of concentration fluctuations
on the stirring strength $V_0$, and show the results in Fig.~\ref{fig:min_max_sineflow}.
 The theoretical upper and lower bounds on $m_4$ depend on $V_0$ and are
 obtained as as roots of Eqs.~\eqref{eq:lower_bd_monochromatic} and~\eqref{eq:upper_bd_monochromatic}.
  In the limit of large $V_0$, these bounds have
  power-law behaviour, with $m_4^{\mathrm{max}}\sim V_0^{\frac{1}{2}}$
  and $m_4^{\mathrm{min}}\sim V_0^{-1}$, as demonstrated by
  Eqs.~\eqref{eq:lower_bd_asymp} and~\eqref{eq:upper_bd_asymp}.  The numerical
  values of $m_4$ are indeed bounded by these limiting values,
  although the $V_0$-dependence is not a power law.  Instead, the function
  $m_4\left(V_0\right)$
is a nonincreasing function, with a sharp drop occurring in a small range
of $V_0$-values.  Thus, the fluid becomes more homogeneous with increasing
$V_0$.  We discuss the  effect of stirring on the inhomogeneity of the
fluid by introducing the notion of mixing efficacy.
 
We measure the ability of a given stirring protocol to suppress concentration
fluctuations by the mixing efficacy.
 Similar ideas are often applied to the advection-diffusion equation~\cite{Thiffeault2004,Shaw2006,Thiffeault2006,Thiffeault2007}.
  We define the dimensionless mixing efficacy
\[
\eta_p\equiv\frac{m_p^{\mathrm{min}}\left(V_0=0\right)}{m_p\left(V_0\right)}.
\]
%
%
%
%
For a given flow, the number $\eta_p$ quantifies
the flow's ability to suppress concentration fluctuations.  In a well-mixed
flow, the local deviation of $c\left(\bm{x},t\right)$ away from the mean
will be small; a small $m_p$-value is a signature of a well-mixed flow.
We are therefore justified in calling $\eta_p$ the mixing efficacy.
We obtain some control over the mixing efficacy $\eta_4$ from the inequalities
of Secs.~\ref{sec:estimating_mixedness:measures} and~\ref{sec:estimating_mixedness:lower_bds}.
 Based on these inequalities, the mixing efficacy is bounded above and below,
\[
\eta_4^{\mathrm{min}}\equiv\frac{m_4^{\mathrm{min}}\left(V_0=0\right)}{m_4^{\mathrm{max}}\left(V_0\right)}
\leq\eta_4
\leq\frac{m_4^{\mathrm{min}}\left(V_0=0\right)}{m_4^{\mathrm{min}}\left(V_0\right)}\equiv\eta_4^{\mathrm{max}}.
\]
We have plotted the upper and lower bounds on the mixing efficacy for the
case of monochromatic sources in Fig.~\ref{fig:min_max_sineflow}~(b).  The
maximum efficacy
always exceeds unity in this case, which implies the possibility of finding
stirring protocols that homogenize the fluid.  On the other hand, the minimum
efficacy is less than unity, which indicates the possibility of finding
stirring protocols that actually enhance concentration fluctuations, and
this enhancement depends weakly on the maximum stretching $\Deform_\infty$.
 This
latter case is not surprising, given that a uniform shear flow causes the
domains of the Cahn--Hilliard fluid to align, rather than to break up.
The sine-flow efficacy is a nondecreasing function of the stirring parameter
$V_0$.  At small values of $V_0$, small increases in the vigour of stirring
lead to small increases in the mixing efficacy.  There is a window
of intermediate $V_0$-values for which the mixing efficacy increases sharply
with increasing $V_0$.  At higher values of $V_0$, the efficiency saturates,
so that further increases in the vigour of stirring have no effect on concentration
fluctuations.  In the saturated case, the concentration field mirrors the
structure of the source function, as in Fig.~\ref{fig:C_flow}~(c).
%
%

\subsection*{Constant flow}

\noindent We study the flow $\left(v_x,v_y\right)=\left(V_0,0\right)$,
where $V_0$ is a constant.  We choose a nondimensionalization that is set
by the diffusion time $T_D=L^2/D$, and obtain the following parametric version
of Eq.~\eqref{eq:ch},
\begin{equation}
\frac{\partial c}{\partial t'}+V_0'\frac{\partial c}{\partial x'}=\Delta'\left(c^3-c-\gamma'\Delta'
c\right)+\sqrt{2}S_0'\sin\left(k_{\mathrm{s}}'x'\right),
\label{eq:const_flow}
\end{equation}
where $V_0'=LV_0/D$, $\gamma'=\gamma/L^2$, and $S_0=S_0'L^2/D$. We immediately
drop the primes from Eq.~\eqref{eq:const_flow}. 
We fix $\gamma$ and $S_0$ and vary the flow strength $V_0$. 
As in the case of no flow, we choose $S_0$ such that the morpohology is qualitatively
different from the sourceless one; in the units used here, we set $S_0=50$.
This flow does not satisfy
 the time-correlation relations~\eqref{eq:statistical_hit},
 although the maximum stretching has the simple form $\Deform_\infty=0$.
  The
\begin{figure}[htb]
\centering
\subfigure[]{
 \includegraphics[width=.25\textwidth]{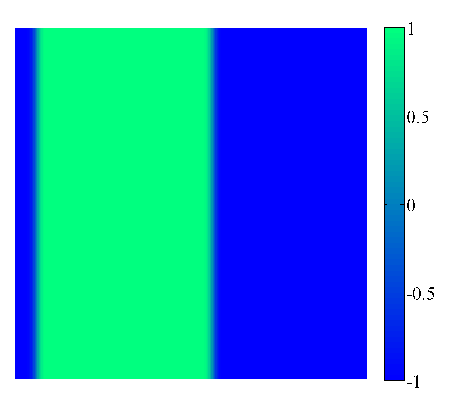}
}
\subfigure[]{
 \includegraphics[width=.25\textwidth]{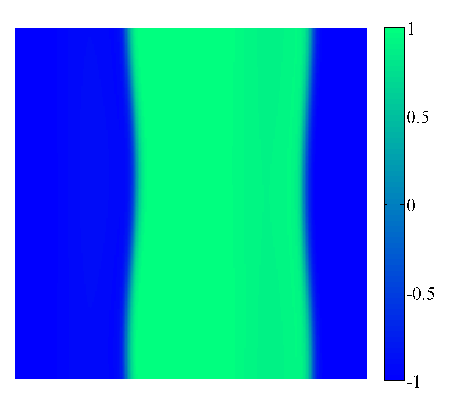}
}
\subfigure[]{
 \includegraphics[width=.25\textwidth]{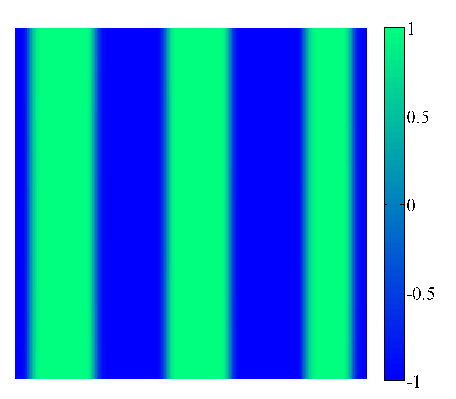}
}
\caption{A snapshot of the steady-state concentration field for (a)
$V_0=10$; (b) $V_0=100$; (c) $V_0=1000$.   The source parameters are $S_0=50$
and $k_{\mathrm{s}}=2\pi$.}
\label{fig:C_const_flow}
\end{figure}
upper bound obtained in Eq.~\eqref{eq:eqn_for_x} is therefore independent
of the flow strength.
  
We solve Eq.~\eqref{eq:const_flow} numerically for various values of $V_0$
and present the results in Fig.~\ref{fig:C_const_flow}.  For small $V_0$,
the morphology of the concentration field is similar to the flowless case
seen in Fig.~\ref{fig:no_flow}~(d),
except now the domains are uniformly advected in a direction perpendicular
to the source variation.  The small-$V_0$ case is shown in Fig.~\ref{fig:C_const_flow}~(a).
 As $V_0$ increases, the domain boundaries are
distorted due to the advection, while for large $V_0$, the advection is
sufficiently strong to precipitate the formation of narrower lamellae,
although the domain structure persists.


The $m_4$ measure of mixedness extracted from the numerical simulations is
almost constant across the range of stirring
parameters $V_0\in\left[0,1000\right]$, and resides in the range defined
by the bounds in Sec.~\ref{sec:estimating_mixedness:lower_bds}.
 For $V_0=1000$, $m_4$
 is slightly smaller than its value at $V_0=0$, due to the presence of
 more interfaces.  This difference is small however, and increasing $V_0$
 does little to mix the fluid.  This is not surprising, given that local
 shears are necessary to break up domain structures~\cite{Berti2005}, and
 that such shears are
 absent from constant flows.  What this example shows however, is the difference
 between a miscible mixture, and a phase-separating mixture.  For a diffusive
 mixture with the sinusoidal source we have studied, the constant flow discussed
 here is optimal for mixing~\cite{Thiffeault2007}; for a phase-separating
 mixture, the constant flow badly fails to homogenize the mixture.

\section{Summary}

\noindent We have introduced the advective Cahn--Hilliard equation with a
mean-zero driving term as a way of describing a stirred, phase-separating
fluid, in the presence of sources and sinks.  By specializing to symmetric
mixtures, we have studied a more tractable problem, although one with many
applications.

Our goal was to investigate
stirring protocols numerically and analytically, and to determine the
best way to break up the domains in the Cahn--Hilliard fluid and achieve
homogenization.
 To this end, we introduced the $m_p$ measure of concentration fluctuations.
  Since in a well-mixed fluid, the concentration exhibits small spatial fluctuations
  about the mean,
  with better mixing leading to smaller fluctuations, we used $m_p$ as a
  measure of mixedness or homogeneity.  We proved the existence of $m_p$
  for long times,
  for $p\in[1,4]$, and obtained \emph{a priori} upper and lower bounds on
  $m_4$, as an explicit function
  of the imposed flow $\bm{v}\left(\bm{x},t\right)$, and the source $s\left(\bm{x}\right)$.

  We compared the level homogeneity achieved by the random-phase sine flow
  and the constant flow with the lower bound, and found that the sine flow
  is effective at homogenizing the binary fluid,
  while the constant flow fails in this task.  This
  is not surprising, since it is known that differential shears are needed
  to break up
  binary fluid domains, although it is radically different from the advection-diffusion
  case, where the constant flow was the optimal mixer.  The question of whether
  or not a flow is a good mixer in this context was discussed using the mixing
  efficacy, defined in Sec.~\ref{sec:estimating_mixedness:numerics}.  Given
  such a definition,
  it is possible to compare stirring protocols and find the optimal protocol
  for mixing a binary fluid.  Our upper bound on the efficacy provides
  a meaningful notion of this optimality.  This result may be useful
  in applications where the homogenization of a binary fluid is desirable,
  since we have set a lower limit on precisely how much homogeneity can be
  achieved.

%% file: analysis_thin_films/analysis_thin_films.tex
\chapter{Nonlinear dynamics of phase separation in ultra-thin films: Analysis}
\label{ch:analysis_thin_films}

\section{Overview}
\label{sec:analysis_thin_films:overview}
In this chapter we recall the Navier--Stokes Cahn--Hilliard (NSCH) equations
derived
in Sec.~\ref{sec:background:NSCH}, shifting focus from the passive tracer,
to the active one.  We study these equations in the incompressible
limit.  We shall specialize to thin films, in which the binary liquid forms
a thin layer on a substrate.  Then, if the vertical gradients are small compared
to horizontal ones, a long-wavelength version of the NSCH equations can be
obtained, representing a much simplified problem.  We present the
derivation of these long-wavelength or thin-film Stokes Cahn--Hilliard equations.
 We analyze these equations and obtain results concerning the existence,
 regularity, and uniqueness of solutions.
\section{The thin-film Stokes Cahn--Hilliard equations}
\label{sec:model}
\noindent 
In this section we recall the incompressible Navier--Stokes Cahn--Hilliard
equations from Sec.~\ref{sec:background:NSCH} and discuss the passage from
these equations to the long-wavelength approximation in two dimensions.
We enumerate the scaling rules necessary to obtain the simplified
equations.  
Finally, we arrive at a set of equations that describe phase separation in
a thin film acted on by arbitrary body forces.

The incompressible NSCH equations derived in Sec.~\ref{sec:background:NSCH}
describe the coupled effects of phase separation
and flow in a density-matched binary fluid.  If $\bm{v}$ is the fluid velocity
and $c$ is the concentration of the mixture, where $c=\pm1$ indicates total
segregation, then these fields evolve in the following way,
\begin{subequations}
\begin{gather}
        \frac{\partial \bm{v}}{\partial t}+\bm{v}\cdot\nabla\bm{v}=\nabla\cdot\bm{T}-\frac{1}{\rho_0}\nabla\phi,\\
        \frac{\partial c}{\partial t}+\bm{v}\cdot\nabla c=D \Delta\left(c^3-c-\gamma\Delta{c}\right),\\
        \nabla\cdot\bm{v}=0,
\end{gather}%
\label{eq:NSCH}%
\end{subequations}%
where
\begin{equation}
        {T}_{ij} =-\frac{p}{\rho_0}\delta_{ij}+\nu\left(\frac{\partial
        v_i}{\partial
        x_j}+\frac{\partial
        v_j}{\partial x_i}\right)-\beta\gamma\frac{\partial c}{\partial x_i}\frac{\partial
        c}{\partial x_j}
\label{eq:NSCH_tensor}%
\end{equation}%
is the stress tensor, $p$ is the fluid pressure, $\phi$ is the body potential
and $\rho_0$ is the constant mixture density.  The constant $\nu$ is the kinematic
viscosity, $\nu=\eta/\rho_0$, where $\eta$ is the viscosity.  Additionally,
$\beta=\varepsilon_0/\rho_0$ is a constant with units of $[\mathrm{Energy}][\mathrm{Mass}]^{-1}$,
$\sqrt{\gamma}$ is a constant that gives the typical width of interdomain
transitions, and $D$ a diffusion coefficient with dimensions $[\mathrm{Length}]^2[\mathrm{Time}]^{-1}$.

If the system has a free surface in the vertical or $z$-direction and has
infinite or periodic boundary conditions (BCs) in the lateral or $x$-direction,
then the vertical BCs we impose are 
\begin{subequations}
\begin{equation}
u = w = c_z = c_{zzz}\text{ on }z=0,
\end{equation}
while on the free surface $z=h\left(x,t\right)$ they are
\begin{equation}
\hat{n}_i \hat{n}_j{T}_{ij} = -\Gamma\kappa,\qquad \hat{n}_i \hat{t}_j {T}_{ij} =
-\frac{\partial\Gamma}{\partial s},
\label{eq:BC_stress}
\end{equation}
\begin{equation}
w=\frac{\partial h}{\partial t}+u\frac{\partial h}{\partial x},
\label{eq:BC_height}
\end{equation}
\begin{equation}
\hat{n}_i \partial_i c = 0, \qquad \hat{n}_i \partial_i \Delta c = 0,
\label{eq:BC_c}
\end{equation}%
\label{eq:BCs}%
\end{subequations}%
where $\hat{\bm{n}}$ is the unit normal to the surface, $\hat{\bm{t}}$,
is the unit vector tangent to the surface, $s$ is the
surface coordinate, $\Gamma$ is the surface tension and $\kappa$ is the
mean curvature,
\begin{multline*}
\phantom{aaaaa}
\kappa=\nabla\cdot\hat{\bm{n}}
\\
=\left(\partial_x,\partial_z\right)\cdot\left(\frac{-\partial_x{h}}{\sqrt{1+\left(\partial_x{h}\right)^2}},\frac{1}{\sqrt{1+\left(\partial_x{h}\right)^2}}\right)=\frac{\partial_{xx}h}{\left[1+\left(\partial_x{h}\right)^2\right]^{\frac{3}{2}}}.
\phantom{aaaaa}
\end{multline*}
This choice of BCs guarantees the conservation of the total mass and volume,
respectively 
\begin{equation}
M =\int_{\text{Dom}\left(t\right)} c\left(\bm{x},t\right)d^2x,\qquad V=\int_{\text{Dom}\left(t\right)}d^2x.
\label{eq:mass_consv}
\end{equation}
Here 
$\text{Dom}\left(t\right)
$
represents the time-varying domain of integration, whose time dependence
is due to the variable nature of the free surface height.  Note that in view
of the concentration BC~\eqref{eq:BC_c}, the stress BC~\eqref{eq:BC_stress}
does not contain $c\left(\bm{x},t\right)$ or its derivatives.

\begin{figure}
\centering
    \includegraphics[width=.8\textwidth]{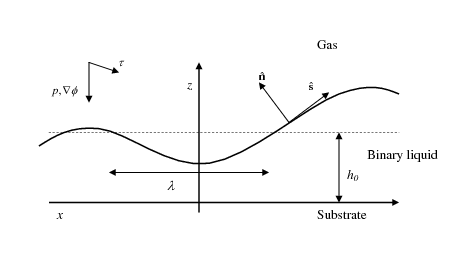}
\caption{Definition sketch for the long-wavelength approximation in two dimensions.
 The approximation relies on the smallness of the parameter $\Smalll=h_0/\lambda$.}
\label{fig:defn_sketch}
\end{figure}

These equations admit a great simplification if the fluid forms a thin layer
of mean thickness $h_0$, for then the scale of lateral variations
$\lambda$ is large compared with the scale of vertical variations $h_0$.
 This is shown in Fig.~\ref{fig:defn_sketch}.
 Specifically, the parameter $\Smalll = h_0/\lambda$ is small, and after nondimensionalization
 of Eq.~\eqref{eq:NSCH}, we expand its solution in terms of this parameter,
 keeping only the lowest-order terms.  For a review of this method and its
 applications, see~\cite{Oron1997}.  For simplicity, we work in two
 dimensions, but the generalization to three dimensions is easily effected.
In terms of the small parameter $\Smalll$, the equations nondimensionalize
as follows.   The diffusion timescale is $t_0 = \lambda^2 / D = h_0^2 /
\left(\Smalll^2 D\right)$ and we choose this to be the unit of time.  Then
the unit of horizontal velocity is $u_0 = \lambda/ t_0 = \Smalll D / h_0$
so that $u = \left(\Smalll D / h_0\right)U$, where variables in upper
 case denote dimensionless quantities.  Similarly the vertical velocity
 is $w = \left(\Smalll^2 D / h_0\right)W$.  For the equations of
 motion to be half-Poiseuille at $O\left(1\right)$ (in the absence of the
 backreaction) we choose $p = \left(\eta D / h_0^2\right)P$ and $\phi = \left(\eta
 D / h_0^2\right)\Phi$.  Using these scaling rules, the dimensionless momentum
 equations are
\begin{multline}
\Smalll Re \left(\frac{\partial U}{\partial T} + U\frac{\partial
U}{\partial X}+ W\frac{\partial U}{\partial Z}\right) 
= -\frac{\partial}{\partial X}\left(P+\Phi\right)+\Smalll^2\frac{\partial^2
U}{\partial X^2}+\frac{\partial^2 U}{\partial Z^2} \\
-\tfrac{1}{2}\frac{\beta\gamma}{\nu D}\frac{\partial}{\partial
X}\bigg[\Smalll^2\left(\frac{\partial c}{\partial X}\right)^2
+\left(\frac{\partial c}{\partial Z}\right)^2\bigg]-\frac{\beta\gamma}{\nu
D}\frac{\partial c}{\partial X}\bigg[\Smalll^2\frac{\partial^2 c}{\partial X^2}+\frac{\partial^2 c}{\partial Z^2}\bigg],
\label{eq:thin_film1}
\end{multline}
\begin{multline}
\Smalll^3 Re \left(\frac{\partial W}{\partial T} + U\frac{\partial
W}{\partial X} + W\frac{\partial W}{\partial Z}\right) 
= -\frac{\partial}{\partial Z}\left(P+\Phi\right)+\Smalll^4\frac{\partial^2
W}{\partial X^2}+\Smalll^2\frac{\partial^2 W}{\partial Z^2}  \\
-\tfrac{1}{2}\frac{\beta\gamma}{\nu D}\frac{\partial}{\partial
Z}\bigg[\Smalll^2\left(\frac{\partial c}{\partial X}\right)^2+\left(\frac{\partial
c}{\partial Z}\right)^2\bigg]
-\frac{\beta\gamma}{\nu D}\frac{\partial c}{\partial Z}\bigg[\Smalll^2\frac{\partial^2
c}{\partial X^2}+\frac{\partial^2 c}{\partial Z^2}\bigg],
\label{eq:thin_film2}
\end{multline}
where
\begin{equation}
Re = \frac{\Smalll D}{\nu} = \frac{\left(\Smalll D / h_0\right)h_0}{\nu} =
O\left(1\right).
\end{equation}
The choice of ordering for the Reynolds number $Re$ allows us to recover
half-Poiseuille flow at $O\left(1\right)$.  We delay choosing the ordering
of the dimensionless group $\beta\gamma/D\nu$ until we have examined the
concentration equation, which in nondimensional form is
\begin{multline}
\Smalll^2 \left(\frac{\partial c}{\partial T} + U\frac{\partial c}{\partial
X} + W\frac{\partial c}{\partial Z}\right)\\
= \Smalll^2 \frac{\partial^2}{\partial X^2}\left(c^3-c\right)+\frac{\partial^2}{\partial Z^2}\left(c^3-c\right)
-\Smalll^4 C_{\mathrm{n}}^2\frac{\partial^4c}{\partial X^4}-C_{\mathrm{n}}^2\frac{\partial^4
c}{\partial Z^4}
-2\Smalll^2 C_{\mathrm{n}}^2 \frac{\partial^2}{\partial X^2}\frac{\partial
c}{\partial Z^2},
\label{eq:conc0}
\end{multline}
where $C_{\mathrm{n}}^2=\gamma/h_0^2$.  By switching off the backreaction in
the momentum equations (corresponding
to $\beta\gamma/D\nu\rightarrow0$), we find the trivial solution to the
momentum equations, $U = W = \partial_X\left(P+\Phi\right) = \partial_Z\left(P+\Phi\right)
= 0$, $H = 1$.  The concentration boundary conditions are then $c_Z = c_{ZZZ}
= 0$ on $Z = 0,1$ which forces $c_Z\equiv0$ so that the Cahn--Hilliard equation
is simply
\[
\frac{\partial c}{\partial T} = \frac{\partial^2}{\partial X^2}\left(c^3-c\right)-\Smalll^2 C_{\mathrm{n}}^2\frac{\partial^4c}{\partial
X^4}.
\]
To make the lubrication approximation consistent, we take 
\begin{equation}
\Smalll C_{\mathrm{n}}= \tilde{C}_{\mathrm{n}}= \Smalll\sqrt{\gamma} /
h_0 = O\left(1\right).
\label{eq:gamma_h}
\end{equation}
We now carry out a long-wavelength approximation to Eq.~\eqref{eq:conc0},
writing
$U=U_0+O\left(\Smalll\right)$, $W=W_0+O\left(\Smalll\right)$, $c=c_0+\Smalll
c_1+\Smalll^2 c_2+...$.  We examine the boundary conditions on $c\left(\bm{x},t\right)$
first.  They are $\hat{\bm{n}}\cdot\nabla c = \hat{\bm{n}}\cdot\nabla\Delta
c=0$
on $Z=0,H$; on $Z=0$ these conditions are simply $\partial_Z c = \partial_{ZZZ}
c=0$, while on $Z=H$ the surface derivatives are determined by the relations
\[
\hat{\bm{n}}\cdot\nabla\propto-\Smalll^2 H_X\partial_X+\partial_Z,
\]
\[
\hat{\bm{n}}\cdot\nabla\Delta\propto-\Smalll^4
H_X\partial_{XXX}-\Smalll^2H_X\partial_X\partial_{ZZ}+\Smalll^2\partial_{XX}\partial_Z+\partial_{ZZZ}.
\]
Thus, the BCs on $c_0$ are simply $\partial_Z c_0 = \partial_{ZZZ} c_0=0$
on $Z=0,H$, which forces $c_0=c_0\left(X,T\right)$.  Similarly, we
find  $c_1=c_1\left(X,T\right)$, and
%
%
\[
\frac{\partial c_2}{\partial Z}=Z\frac{H_X}{H}\frac{\partial
c_0}{\partial X},\qquad
\frac{\partial^2 c_2}{\partial Z^2}=\frac{H_X}{H}\frac{\partial
c_0}{\partial X},\qquad \text{for any }Z\in \left[0,H\right].
\]
In the same manner, we derive the results $\partial_{ZZZZ}c_2=\partial_{ZZZZ}c_3=0$.
Using these facts, equation~\eqref{eq:conc0} becomes
%
%
%
%
%
%
%
%
%
%
%
%
%
%
%
%
%
%
%
%
%
\begin{multline*}
\frac{\partial c_0}{\partial T} + U_0\frac{\partial c_0}{\partial
X} =
\\
\frac{\partial^2}{\partial X^2}\left(c_0^3-c_0\right)
-\tilde{C}_{\mathrm{n}}^2\frac{\partial^4c}{\partial X^4}
+\left(3c_0^2-1\right)\frac{H_X}{H}\frac{\partial c_0}{\partial X}
-2\tilde{C}_{\mathrm{n}}^2\frac{\partial^2}{\partial X^2}\frac{H_X}{H}\frac{\partial
c_0}{\partial X}
-{\tilde{C}_{\mathrm{n}}^2}\frac{\partial^4 c_4}{\partial Z^4}.
\end{multline*}
We now integrate this equation from $Z=0$ to $Z=H$ and use the boundary conditions
\[
\frac{\partial^3 c_4}{\partial Z^3}=0\qquad\text{on }Z=0,
\]
\begin{eqnarray*}
\frac{\partial^3 c_4}{\partial Z^3}&=&
%
%
%
%
%
%
%
%
H_X\frac{\partial^3 c_0}{\partial X^3}+
H_X\frac{\partial}{\partial X}\left(\frac{H_X}{H}\frac{\partial c_0}{\partial
X}\right)
-H\frac{\partial^2}{\partial X^2}\left(\frac{H_X}{H}\frac{\partial c_0}{\partial
X}\right)\qquad\text{on }Z=H.
\end{eqnarray*}
After rearrangement, the concentration equation becomes
\begin{multline*}
H\frac{\partial c_0}{\partial T}+H\overline{U_0}\frac{\partial c_0}{\partial
X} = 
\\
H\frac{\partial^2}{\partial X^2}\left[c_0^3-c_0-\tilde{C}_{\mathrm{n}}^2\frac{\partial^2
c_0}{\partial X^2}-\tilde{C}_{\mathrm{n}}^2\frac{H_X}{H}\frac{\partial c_0}{\partial
X}\right]
+\frac{\partial H}{\partial X}\frac{\partial}{\partial X}\left[c_0^3-c_0-\tilde{C}_{\mathrm{n}}^2\frac{\partial^2
c_0}{\partial X^2}-\tilde{C}_{\mathrm{n}}^2\frac{H_X}{H}\frac{\partial c_0}{\partial
X}\right],
\end{multline*}
where
\[
\overline{U_0} = \frac{1}{H}\int_0^H U_0\left(X,Z,T\right)dZ,
\]
is the vertically-averaged velocity.  Introducing
\[
\mu 
%
%
=c_0^3-c_0-\frac{\tilde{C}_{\mathrm{n}}^2}{H}\frac{\partial}{\partial X}\left(H\frac{\partial
c_0}{\partial X}\right),
\]
the thin-film Cahn--Hilliard equation becomes
\begin{equation}
\frac{\partial c_0}{\partial T}+\overline{U_0}\frac{\partial c_0}{\partial
X} = \frac{1}{H}\frac{\partial}{\partial X}\left(H\frac{\partial\mu}{\partial
X}\right).
\end{equation}

We are now able to perform the long-wavelength approximation to Eqs.~\eqref{eq:thin_film1}
and~\eqref{eq:thin_film2}.  At lowest order, Eq.~\eqref{eq:thin_film2} is
\[
\frac{\partial}{\partial Z}\left(P+\Phi\right)+r\frac{\partial
c_0}{\partial Z}\left[\frac{\partial^2 c_0}{\partial X^2}+\frac{\partial^2
c_2}{\partial Z^2}\right]+\tfrac{1}{2}r\frac{\partial}{\partial
Z}\left(\frac{\partial c_0}{\partial X}\right)^2=0,\qquad
r=\frac{\Smalll^2\beta\gamma}{D\nu},
\]
which is simply $\partial_Z\left(P+\Phi\right)=0$, and hence
\[
P+\Phi = P_{\mathrm{surf}}+\Phi_{\mathrm{surf}}\equiv P\left(X,H\left(X,T\right),T\right)+\Phi\left(X,H\left(X,T\right),T\right).
\]
Here
\begin{equation}
r=\frac{\Smalll^2\beta\gamma}{D\nu}=O\left(1\right)
\end{equation}
is the dimensionless measure of the backreaction strength.  Thus, Eq.~\eqref{eq:thin_film1}
becomes
\[
\frac{\partial^2 U_0}{\partial Z^2}=\frac{\partial}{\partial X}\left(P_{\mathrm{surf}}+\Phi_{\mathrm{surf}}\right)+r\frac{\partial}{\partial
X}\left(\frac{\partial c_0}{\partial X}\right)^2+r\frac{\partial c_0}{\partial
X}\frac{\partial^2 c_2}{\partial Z^2},
\]
and using $\partial_{ZZ}c_2=\left(H_X/H\right)\left(\partial c_0/\partial
X\right)$ this is
\begin{equation}
\frac{\partial^2 U_0}{\partial Z^2}=\frac{\partial}{\partial X}\left(P_{\mathrm{surf}}+\Phi_{\mathrm{surf}}\right)+\frac{r}{H}\frac{\partial}{\partial
X}\left[H\left(\frac{\partial c_0}{\partial X}\right)^2\right].
\label{eq:d2U}
\end{equation}
At lowest order, the BC~\eqref{eq:BC_stress} becomes
\begin{equation}
\frac{\partial U_0}{\partial Z}=\frac{\partial\tilde{\Gamma}}{\partial{X}}\qquad\text{on
}Z=H,
\label{eq:gamma_bc}
\end{equation}
where $\tilde{\Gamma}$ is the dimensionless, spatially-varying surface tension.
 Combining Eqs.~\eqref{eq:d2U} and~\eqref{eq:gamma_bc} yields the relation
\[
\frac{\partial U_0}{\partial Z}=\frac{\partial\tilde{\Gamma}}{\partial X}+\left(Z-H\right)\bigg\{
\frac{\partial}{\partial X}\left(P_{\mathrm{surf}}+\Phi_{\mathrm{surf}}\right)+\frac{r}{H}\frac{\partial}{\partial
X}\left[H\left(\frac{\partial c_0}{\partial X}\right)^2\right]
\bigg\}.
\]
Making
use of the BC $U_0=0$ on $Z=0$ and integrating again, we obtain the result
\begin{equation}
U_0\left(X,Z,T\right)=Z\frac{\partial\tilde{\Gamma}}{\partial X}+\left(\tfrac{1}{2}Z^2-HZ\right)\bigg\{
\frac{\partial}{\partial X}\left(P_{\mathrm{surf}}+\Phi_{\mathrm{surf}}\right)+\frac{r}{H}\frac{\partial}{\partial
X}\left[H\left(\frac{\partial c_0}{\partial X}\right)^2\right]
\bigg\}.
\label{eq:U_0}
\end{equation}
The vertically-averaged velocity is therefore
\[
\overline{U_0} = \tfrac{1}{2}H\frac{\partial\tilde{\Gamma}}{\partial X}-\tfrac{1}{3}H^2\bigg\{
\frac{\partial}{\partial X}\left(P_{\mathrm{surf}}+\Phi_{\mathrm{surf}}\right)+\frac{r}{H}\frac{\partial}{\partial
X}\left[H\left(\frac{\partial c_0}{\partial X}\right)^2\right]
\bigg\}.
\]
Using the standard Laplace--Young free-surface boundary condition, this becomes
\[
\overline{U_0} = \tfrac{1}{2}H\frac{\partial\tilde{\Gamma}}{\partial X}-\tfrac{1}{3}H^2\bigg\{
\frac{\partial}{\partial X}\left(-\frac{1}{C}\frac{\partial^2 H}{\partial X^2}+\Phi_{\mathrm{surf}}\right)+\frac{r}{H}\frac{\partial}{\partial
X}\left[H\left(\frac{\partial c_0}{\partial X}\right)^2\right]
\bigg\},
\]
where
\begin{equation}
C = \frac{\nu\rho D}{h_0\sigma_0\Smalll^2}=O\left(1\right).
\label{eq:C}
\end{equation}
Finally, by integrating the continuity equation in the $Z$-direction, we
obtain, in a standard way, an equation for free-surface variations,
\[
\frac{\partial H}{\partial X}+\frac{\partial}{\partial X}\left(H\overline{U_0}\right)=0.
\]

Let us assemble our results, restoring the lower-case fonts and
omitting ornamentation over the nondimensional quantities,
\begin{subequations}
\begin{equation}
\frac{\partial h}{\partial t}+\frac{\partial J}{\partial x}=0,\\
\end{equation}
\begin{equation}
\frac{\partial}{\partial t}\left(c h\right)+\frac{\partial}{\partial x}\left(Jc\right)=\frac{\partial}{\partial{x}}\left(h\frac{\partial\mu}{\partial{x}}\right),
\end{equation}
where
\begin{equation}
J=\tfrac{1}{2}h^2\frac{\partial\Gamma}{\partial{x}}-\tfrac{1}{3}h^3\bigg\{\frac{\partial}{\partial{x}}\left(-\frac{1}{C}\frac{\partial^2{h}}{\partial{x}^2}
+\phi\right)+\frac{r}{h}\frac{\partial}{\partial{x}}\left[h\left(\frac{\partial{c}}{\partial{x}}\right)^2\right]\bigg\},
\end{equation}
\begin{equation}
\mu=c^3-c-\frac{C_{\mathrm{n}}^2}{h}\frac{\partial}{\partial{x}}\left(h\frac{\partial{c}}{\partial{x}}\right),
\end{equation}%
\label{eq:analysis_thin_films:model}%
\end{subequations}%
and where we have the following nondimensional constants,
\[
r=\frac{\Smalll^2\beta\gamma}{D\nu},\qquad C_{\mathrm{n}}=\frac{\Smalll\sqrt{\gamma}}{h_0},\qquad
C=\frac{\nu\rho D}{h_0\sigma_0\Smalll^2}.
\]
These are the thin-film Stokes Cahn--Hilliard (SCH) equations.  The integral
quantities $M$ and $V$ defined in Eq.~\eqref{eq:mass_consv} are manifestly
conserved, while the free surface and concentration are coupled.

We note that the relation $C_{\mathrm{n}}=\Smalll\sqrt{\gamma}/h_0=O\left(1\right)$
is the condition that the mean thickness of the film be much smaller than
the transition layer thickness.  In experiments involving the smallest film
thicknesses attainable ($10^{-8}$ m)~\cite{Sung1996}, this condition is automatically
satisfied.
The condition is also realized in ordinary thin films when external effects
such as the air-fluid and fluid-substrate interactions do not prefer one
binary fluid component or another.  In this case, the vertical extent of the domains
becomes comparable to the film thickness at late times, the thin film
behaves in a quasi two-dimensional way, and the model equations
are applicable.
%
%

The choice of potential $\phi$ determines the behaviour of solutions.
If interactions between the fluid and the substrate and air interfaces are
important, the potential should take account of the Van der Waals forces
present.  A simple model potential is thus
\[
\phi = Ah^{-p},
\]
where $A$ is the dimensionless Hamakar constant and typically $p=3$~\cite{Oron1997}.
 Here $A$ can be positive or negative, with positivity indicating a net attraction
 between the fluid and the substrate and negativity indicating a net repulsion.
  This choice of potential can also have a regularizing effect, preventing
  a singularity or rupture from occurring in Eq.~\eqref{eq:model}.  
  
For $\phi=-|A|/h^3$ (repulsive Van der Waals interaction), the system of
equations~\eqref{eq:model} has a Lyapunov functional $F=F_1+F_2$, where
\[
F_1=\int_{\Omega}{dx}\, \left[\frac{1}{2C}\left(\frac{\partial{h}}{\partial{x}}\right)^2+\tfrac{1}{2}\frac{|A|}{h^2}\right],\qquad
F_2 = \frac{r}{C_{\mathrm{n}}^2}\int_{\Omega}{dx}\,  h\left[\tfrac{1}{4}\left(c^2-1\right)^2+\tfrac{1}{2}{C_{\mathrm{n}}^2}\left(\frac{\partial{c}}{\partial{x}}\right)^2\right],
\]  
and where $\Omega=\left[0,L\right]$ defines the lateral extent of the system.
 In this chapter we take $\Omega$ to be a periodic domain, which fixes the
 boundary conditions, although other choices of boundary conditions are possible.
By differentiating these expressions with respect to time, we obtain the
relation
\begin{multline}
\dot{F}_1+\dot{F}_2
\\
=-\tfrac{1}{3}\int_\Omega{dx}\,  h^3\bigg\{\frac{\partial}{\partial x}\left(\frac{1}{C}\frac{\partial^2h}{\partial
x^2}+\frac{|A|}{h^3}\right)-\frac{r}{h}\frac{\partial}{\partial x}\left[h\left(\frac{\partial
c}{\partial x}\right)^2\right]  \bigg\}^2
-\int_\Omega{dx}\,  h\left(\frac{\partial\mu}{\partial x}\right)^2,
\label{eq:fe_decay}
\end{multline}
which is nonpositive for nonnegative $h$.  This fact holds the key to
the analytic results obtained in the next section.
\section{Existence and regularity of solutions}
\label{sec:analysis_thin_films:existence}
In this section we prove that solutions to the model equations do indeed
exist.  We focus on the one-dimensional equation set
\[
\frac{\partial h}{\partial t}=
-\frac{\partial}{\partial x}\left[f\left(h\right)\frac{\partial^{3}h}{\partial{x^3}}\right]+
\frac{\partial}{\partial x}\left[\frac{1}{g\left(h\right)}\frac{\partial
h}{\partial{x}}\right]+
\frac{\partial}{\partial x}\bigg\{\frac{f\left(h\right)}{g\left(h\right)}\frac{\partial}{\partial{x}}\left[g\left(h\right)\left(\frac{\partial{c}}{\partial{x}}\right)^2\right]\bigg\},\nonumber\\
\]
\vskip -0.2in
\begin{multline}
\frac{\partial}{\partial t}\left(c g\left(h\right)\right)=
-\frac{\partial}{\partial x}\left[cf\left(h\right)\frac{\partial^{3}h}{\partial{x^3}}\right]+
\frac{\partial}{\partial x}\left[\frac{c}{g\left(h\right)}\frac{\partial
h}{\partial{x}}\right]+
\frac{\partial}{\partial x}\bigg\{c\frac{f\left(h\right)}{g\left(h\right)}\frac{\partial}{\partial{x}}\left[g\left(h\right)\left(\frac{\partial{c}}{\partial{x}}\right)^2\right]\bigg\}
\\
+\frac{\partial}{\partial x}\bigg\{g\left(h\right)\frac{\partial}{\partial
x}\left[c^3-c-\frac{1}{g\left(h\right)}\frac{\partial}{\partial
x}\left(g\left(h\right)\frac{\partial c}{\partial x}\right)\right]\bigg\},
\label{eq:model_proof}
\end{multline}
where
\[
f\left(h\right)=h^3,\qquad g\left(h\right)=h.
\]
The application we have in mind involves a periodic domain $\Omega=[0,L]$.
 We shall prove our result for smooth initial conditions,
\begin{multline}
h\left(x,0\right)=h_0\left(x\right)>0,\qquad c\left(x,0\right)=c_0\left(x\right),\qquad
\left(h_0\left(x\right),c_0\left(x\right)\right)\in H^{2,2}\left(\Omega\right).
\label{eq:initial_data}
\end{multline}
We shall prove that the solutions to this equation pair exist in the strong
sense; however, we shall need the following definition of weak solutions:
\begin{quote}
A pair $\left(h,c\right)$ is a weak solution of Eq.~\eqref{eq:model_proof}
if the following integral relations hold,
\begin{multline*}
\int_0^{T_0}dt\int_\Omega {dx}\, \varphi_t h=
\\
\int_0^{T_0}dt\int_\Omega {dx}\, \varphi_x\bigg\{-f\left(h\right)\frac{\partial^{3}h}{\partial{x^3}}+
\frac{1}{g\left(h\right)}\frac{\partial h}{\partial{x}}+
\frac{f\left(h\right)}{g\left(h\right)}\frac{\partial}{\partial{x}}\left[g\left(h\right)\frac{\partial}{\partial{x}}\left(\frac{\partial{c}}{\partial{x}}\right)^2\right]\bigg\},
\end{multline*}
\begin{multline}
\int_0^{T_0}dt\int_\Omega {dx}\, \psi_t cg\left(h\right)=
\\
\int_0^{T_0}dt\int_\Omega {dx}\, \psi_x\bigg\{-cf\left(h\right)\frac{\partial^{3}h}{\partial{x^3}}+
\frac{c}{g\left(h\right)}\frac{\partial h}{\partial{x}}+
c\frac{f\left(h\right)}{g\left(h\right)}\frac{\partial}{\partial{x}}\left[g\left(h\right)\frac{\partial}{\partial{x}}\left(\frac{\partial{c}}{\partial{x}}\right)^2\right]\bigg\}
\\
+\int_0^{T_0}dt\int_\Omega {dx}\, \psi_x \bigg\{g\left(h\right)\frac{\partial}{\partial
x}\left[c^3-c-\frac{1}{g\left(h\right)}\frac{\partial}{\partial{x}}\left(g\left(h\right)\frac{\partial{c}}{\partial{x}}\right)\right]\bigg\},
\label{eq:weak_sln}
\end{multline}
where $T_0>0$ is any time, and $\varphi\left(x,t\right)$ and $\psi\left(x,t\right)$
are arbitrary differentiable test functions that are periodic on $\Omega$
and vanish at $t=0$ and $t=T_0$. 
\end{quote}
We address the following statement,
\begin{quote}
\textit{Given the initial data in Eq.~\eqref{eq:initial_data}, the equations~\eqref{eq:model_proof}
possess a unique strong solution endowed with the following regularity properties:
\[
\left(h,c\right)\in L^\infty\left(0,T_0,H^{2,2}\left(\Omega\right)\right)
\cap L^2\left(0,T_0;H^{4,2}\left(\Omega\right)\right)
\cap C^{\frac{3}{2},\frac{1}{8}}\left(\Omega\times\left[0,T_0\right]\right).
\]
}
\end{quote}
Recall that the function $f$ resides in the class $H^{q,p}\left(\Omega\right)$
if
\[
\Big(\|f\|_p^p+\|\partial_x f\|_p^p+...+\|\partial^q_x f\|_p^p\Big)^{\frac{1}{p}}<\infty.
\]
The proof is given in the following steps.


\subsection{Regularization of the problem}
\label{sec:regularization}

\begin{figure}
\centering
\includegraphics[width=0.35\textwidth]{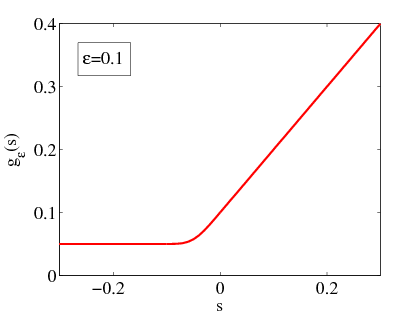}
\caption{A sketch of the regularized function $g_\varepsilon\left(s\right)$
obtained in Eq.~\eqref{eq:regularized_g}.}
\label{fig:regularized_g}
\end{figure}

\noindent We introduce regularized functions $f_\varepsilon\left(s\right)$
and $g_\varepsilon\left(s\right)$ such that 
\[
\lim_{\varepsilon\rightarrow0} f_\varepsilon\left(s\right)= f\left(s\right),\qquad
\lim_{\varepsilon\rightarrow0,s\geq0} g_\varepsilon\left(s\right)={g}\left(s\right).
\]
For now we do not specify $f_\varepsilon\left(s\right)$,
although we mention that a suitable choice of $f_\varepsilon\left(s\right)$
will cure the degeneracy of the fourth-order term in the height equation.
 On the other hand, we require that the function $g_\varepsilon\left(s\right)$
 have the following properties:
\begin{itemize}
\item $g_\varepsilon\left(s\right)=s+\varepsilon$, for $s\geq0$,
\item $g_\varepsilon\left(s\right)>0$, for $s<0$,
\item $\lim_{s\rightarrow-\infty}g_\varepsilon\left(s\right)=\tfrac{1}{2}\varepsilon$,
\item $g_\varepsilon\left(s\right)$ has as many derivatives as necessary.
\end{itemize}
One way of obtaining such a function is to define 
\begin{equation}
g_\varepsilon\left(s\right)=\Bigg\{\begin{array}{cc}
\frac{\varepsilon}{2}, &s\leq-\varepsilon, \\
s+\frac{s^2}{2\varepsilon}-\frac{\varepsilon}{4\pi^2}\cos\left(\frac{2\pi{s}}{\varepsilon}+\pi\right)+\varepsilon-\frac{\varepsilon}{4\pi^2},
& -\varepsilon\leq s\leq 0,\\
\varepsilon+s, & s\geq 0,
\end{array}
\label{eq:regularized_g}
\end{equation}
which is $C^{3}$ in the variable $s$, and has Lipschitz third derivative; inspection of Eq.~\eqref{eq:model_proof}
shows that this degree of smoothness is sufficient to regularize the equations.
 A sketch of this regularization is shown in Fig.~\ref{fig:regularized_g}.

The regularized PDEs we study are thus
\begin{eqnarray}
h_t&=&-J_{\varepsilon,x},\qquad 
J_\varepsilon=f_\varepsilon\left(h\right)h_{xxx}-\frac{1}{g_\varepsilon\left(h\right)}h_x-\frac{f_\varepsilon\left(h\right)}{g_\varepsilon\left(h\right)}\left(g_\varepsilon\left(h\right)c_{x}^2\right)_x,
\nonumber\\
\left(cg_\varepsilon\left(h\right)\right)_t&=&-\left(c J_\varepsilon\right)_x-\left(g_\varepsilon\left(h\right)\mu_{\varepsilon,x}\right)_x,
\label{eq:reg_pde}
\end{eqnarray}
where
\[
\mu_\varepsilon=c^3-c-\frac{1}{g_\varepsilon\left(h\right)}\left(g_\varepsilon\left(h\right)c_x\right)_x.
\]
The $c$-equation can also be written as
\[
c_t=-\frac{1}{g_\varepsilon\left(h\right)}J_\varepsilon c_x+\frac{1}{g_\varepsilon\left(h\right)}\left(g_\varepsilon\left(h\right)\mu_{\varepsilon,x}\right)_x-\frac{c}{g_\varepsilon\left(h\right)}\left[J_{\varepsilon,x}+g_\varepsilon'\left(h\right)h_t\right].
\]

\subsection{The Galerkin approximation}
\label{sec:galerkin}

\noindent We choose a complete orthonormal basis on the interval $\Omega$,
with periodic boundary conditions.  Let us denote the basis by $\{\phi_i\left(x\right)\}_{i\in\mathbb{N}_0}$.
 We consider the finite-dimensional vector space $\text{span}\{\phi_0,...\phi_\ord\}$.
  For convenience, let us take the $\phi_i\left(x\right)$'s to be
  the
  eigenfunctions of the Laplacian on $\left[0,L\right]$ with periodic boundary
  conditions, with corresponding eigenvalues $-\lambda_i^2$.  Moreover, let
  $\phi_0$ be the constant eigenfunction.  We construct approximate solutions
  to the PDEs~\eqref{eq:reg_pde} as finite sums,
\[
h_\ord\left(x,t\right)=\sum_{i=0}^\ord\eta_{\ord,i}\left(t\right)\phi_i\left(x\right),
\qquad c_\ord\left(x,t\right)=\sum_{i=0}^\ord\gamma_{\ord,i}\left(t\right)\phi_i\left(x\right).
\]
If the (smooth) initial data are given as
\[
h\left(x,0\right)=h_0\left(x\right)=\sum_{i=0}^\infty\eta_i^0\phi_i\left(x\right)>0,\qquad
c\left(x,0\right)=c_0\left(x\right)=\sum_{i=0}^\infty\gamma_i^0\phi_i\left(x\right),
\]
then the initial data for the Galerkin approximation are
\[
h_\ord\left(x,0\right)=h_\ord^0\left(x\right)=\sum_{i=0}^\ord\eta_{i}^0\phi_i\left(x\right),\qquad
c_\ord\left(x,0\right)=c_\ord^0\left(x\right)=\sum_{i=0}^\ord\gamma_{i}^0\phi_i\left(x\right),
\]
and the initial data of the Galerkin approximation converge strongly in the
$L^2\left(\Omega\right)$ norm to the initial data of the unapproximated problem.
 Thus, there is a $\ord_0\in\mathbb{N}$ such that $h_\ord^0\left(x\right)>0$,
 everywhere in $\Omega$, for all $\ord>\ord_0$.  Henceforth we work with
 Galerkin approximations with $\ord>\ord_0$.
 
Substitution of $h_\ord=\sum_{i=0}^\ord\eta_{\ord,i}\phi_i$  into a weak
form of the $h$-equation yields
\begin{equation}
\frac{d}{dt}\left(\left(h_\ord,\phi_j\right)\right)=\left(\left(J_{\varepsilon,\ord},\phi_{j,x}\right)\right),
\label{eq:weak_h}
\end{equation}
where
\[
J_\varepsilon\left(h_\ord,c_\ord\right)=f_\varepsilon\left(h_\ord\right)h_{\ord,xxx}-\frac{1}{g_\varepsilon\left(h_\ord\right)}h_{\ord,x}-\frac{f_\varepsilon\left(h_\ord\right)}{g_\varepsilon\left(h_\ord\right)}\left(g_\varepsilon\left(h_{\ord}\right)c_{\ord,x}^2\right)_x,
\]
is the flux for the regularized $h$-equation, and in this chapter, $\left(\left(\phi,\psi\right)\right)$
denotes the pairing $\int_\Omega\phi^{*}\psi{dx}$.
This is recast as
\[
\frac{d\eta_{\ord,j}}{dt}=\left(\left(J_{\varepsilon,\ord},\phi_{j,x}\right)\right)=\Phi_{\ord,j}\left(\eta_\ord,\gamma_\ord\right),
\]
where the function $\Phi_\ord\left(\eta_\ord,\gamma_\ord\right)$ depends
on
\[
\eta_\ord=\left(\eta_{\ord,0},...,\eta_{\ord,\ord}\right),\qquad
\gamma_\ord=\left(\gamma_{\ord,0},...,\gamma_{\ord,\ord}\right),
\]
and is locally Lipschitz in its variables.  This Lipschitz property arises
from the fact that the regularized flux, evaluated at the Galerkin approximation,
is a composition of Lipschitz continuous functions, and therefore, is itself
Lipschitz continuous.


Similarly, substitution of $c_\ord=\sum_{i=0}^\ord\gamma_{\ord,i}\phi_i$
into the weak form of the $c$-equation yields
\begin{equation}
\frac{d}{dt}\left(\left(g\left(h_\ord\right)c_\ord,\phi_j\right)\right)=\left(\left(K_{\varepsilon,\ord},\phi_{j,x}\right)\right),
\label{eq:weak_c}
\end{equation}
where
\begin{eqnarray*}
K_\varepsilon\left(h_\ord,c_\ord\right)&=&c_{\ord}f_\varepsilon\left(h_\ord\right)h_{\ord,xxx}-\frac{c_\ord}{g_\varepsilon\left(h_\ord\right)}h_{\ord,x}\\
&\phantom{a}&\phantom{aaaaaaaaaaaaaaaaaa}
-c_{\ord}\frac{f_\varepsilon\left(h_{\ord}\right)}{g_\varepsilon\left(h_\ord\right)}\left(g_\varepsilon\left(h_{\ord}\right)c_{\ord,x}^2\right)_x-g\left(h_\ord\right)\mu_{\varepsilon,\ord,x},\\
&=&c_{\ord}J_{\varepsilon,\ord}-g_\varepsilon\left(h_\ord\right)\mu_{\varepsilon,\ord,x},
\end{eqnarray*}
is the flux for the regularized $c$-equation.   Rearranging gives
\[
\left(\left(g\left(h_\ord\right)c_{\ord,t},\phi_j\right)\right)=\left(\left(K_{\varepsilon,\ord},\phi_{j,x}\right)\right)-\left(\left(g'\left(h_\ord\right)c_{\ord}h_{\ord,t},\phi_j\right)\right),
\]
and the left-hand side can be recast in matrix form as $\sum_{\alpha=0}^{\ord}M_{\alpha{j}}\dot{\gamma}_{\ord,\alpha}$,
where
\[
M_{\alpha j}=\int_\Omega{dx}\,  g_\varepsilon\left(\sum_i \eta_{\ord,i}\phi_i\right)\phi_\alpha\phi_j,
\] 
which is manifestly symmetric.  It is positive definite because given a nonzero vector
$\left(\xi_0,...,\xi_\ord\right)$, we have the relation
\begin{eqnarray*}
\sum_{\alpha,j}\xi_\alpha M_{\alpha j}\xi_j&=&
\int_\Omega {dx}\, g_\varepsilon\left(\sum_i\eta_{\ord,i}\phi_i\right)\sum_{\alpha{j}}\left(\phi_\alpha\xi_\alpha\right)\left(\phi_j\xi_j\right)\\
&=&\int_\Omega {dx}\, g_\varepsilon\left(\sum_i\eta_{\ord,i}\phi_i\right)\left(\sum_{j}\phi_j\xi_j\right)^2\\
&>&0\qquad\text{for }\left(\xi_0,...,\xi_\ord\right)\neq0,
\end{eqnarray*}
which follows from the positivity of the regularized function $g_\varepsilon\left(s\right)$.
 We therefore have the following equation for $\gamma_{\ord,i}\left(t\right)$,
\begin{equation}
\frac{d\gamma_{\ord,j}}{dt}=\sum_{\alpha=0}^{\ord} M_{\alpha j}^{-1}\Big[\left(\left(K_{\varepsilon,\ord},\phi_{\alpha,x}\right)\right)-\left(\left(g_\varepsilon'\left(h_\ord\right)c_{n}h_{\ord,t},\phi_\alpha\right)\right)
\Big].
\label{eq:gamma_dot}
\end{equation}
Inspecting the expression for $M_{\alpha j}$, $g\left(h_\ord\right)$, and
$K_{\varepsilon,\ord}$, we see that $\eta_\ord$ and $\gamma_\ord$ appear
in a (locally) Lipschitz-continuous way in the
expression $\sum_{\alpha=0}^n M_{\alpha j}^{-1}\left(\left(K_{\varepsilon,\ord},\phi_{\alpha,x}\right)\right)$,
while owing to the imposed smoothness of $g_\varepsilon\left(s\right)$, the
variables $\eta_\ord$, $\gamma_\ord$ and $\dot{\eta}_{\ord}=\left(\dot{\eta}_{\ord,0},...,\dot{\eta}_{\ord,\ord}\right)$
appear in a Lipschitz-continuous way in the quantity 
\[
\sum_{\alpha=0}^{\ord}M_{\alpha j}^{-1}\left(\left(g_\varepsilon'\left(h_\ord\right)c_{\ord}h_{\ord,t},\phi_\alpha\right)\right).
\]
The vector $\dot{\eta}_\ord$ can be replaced by the function $\Phi_\ord\left(\eta_\ord,\gamma_\ord\right)$
and thus we obtain a relation
\[
\frac{d\gamma_{\ord,j}}{dt}=\Psi_{\ord,j}\left(\eta_\ord,\gamma_\ord\right),
\]
 in place of Eq.~\eqref{eq:gamma_dot}, where $\Psi_{\ord,j}\left(\eta_\ord,\gamma_\ord\right)$
 is Lipschitz.  We therefore have a system of Lipschitz-continuous equations
\[
\frac{d\eta_{\ord,j}}{dt}=\Phi_{\ord,j}\left(\eta_\ord,\gamma_\ord\right),\qquad\frac{d\gamma_{\ord,j}}{dt}=\Psi_{\ord,j}\left(\eta_\ord,\gamma_\ord\right),
\]
and thus local existence theory guarantees a solution for the $\eta_{\ord,i}$'s
and $\gamma_{\ord,i}$'s for all times $t$ in a finite interval $0<t<\sigma$.
 This solution is, moreover, unique and continuous.

\subsection{\emph{A priori} bounds on the Galerkin approximation}
\label{sec:a_priori_bds}

\noindent We identify the free energy
\[
F=\int_{\Omega}{dx}\, \left[\tfrac{3}{2}h_x^2+G_\varepsilon\left(h\right)\right]
+\int_{\Omega}{dx}\, g_\varepsilon\left(h\right)\left[\tfrac{1}{4}\left(c^2-1\right)^2+\tfrac{1}{2}c_x^2\right],
\]
where $G_{\varepsilon}''\left(s\right)=1/\left[f_\varepsilon\left(s\right)g_\varepsilon\left(s\right)\right]$.
 Since the Galerkin approximation satisfies the weak form of the PDEs given
 in Eqs.~\eqref{eq:weak_h} and~\eqref{eq:weak_c}, it is possible to obtain
 the following free-energy decay law,
%
%
%
\begin{multline}
\frac{dF}{dt}\left(t\right)=
\\
-\int_{\Omega-\Omega_-}{dx}\,  f_\varepsilon\left(h_\ord\right)\left[-h_{\ord,xxx}+\frac{h_{\ord,x}}{f_\varepsilon\left(h_\ord\right)g_\varepsilon\left(h_\ord\right)}+\frac{1}{g_\varepsilon\left(h_\ord\right)}\left(g_\varepsilon\left(h_\ord\right)c_{\ord,x}^2\right)_x\right]^2
\\
-\int_\Omega{dx}\,  g_\varepsilon\left(h_\ord\right)\mu_{\varepsilon,\ord,x}^2
+\int_{\Omega_-}{dx}\, J_{\varepsilon,\ord}\left[-h_{\ord,xxx}+\frac{h_{\ord,x}}{f_\varepsilon\left(h_\ord\right)g_\varepsilon\left(h_\ord\right)}\right]
\\
+\int_{\Omega_-}{dx}\, \Big\{g_\varepsilon'\left(h_\ord\right)J_{\varepsilon,\ord}\left(c^3-c-c_{\ord,xx}\right)
%
%
%
-\mu_{\varepsilon,\ord}c_\ord
\left[g_\varepsilon'\left(h_\ord\right)+J_{\varepsilon,\ord,x}\right]-J_{\varepsilon,\ord}\mu_{\varepsilon,\ord}c_{\ord,x}\Big\},
\\
0\leq t<\sigma,
\label{eq:free_energy_decay}
\end{multline}
%
%
%
where $\Omega_-\left(t\right)=\{x\in\Omega|h_\ord\left(x,t\right)<0\}$. 
Now given the time continuity of $h_\ord\left(x,t\right)$ in $(0,\sigma)$,
and the initial condition $h_\ord^0\left(x\right)>0$ (since $\ord>\ord_0$),
there is a time $\sigma_1$, such that $0<\sigma_1\leq\sigma$, and such that
$h_\ord\left(x,t\right)>0$ for all $x\in\Omega$ and all $t\in\left(0,\sigma_1\right)$.
Therefore, $\Omega_-\left(t\right)=\emptyset$ for $t\in\left(0,\sigma_1\right)$,
and hence
\[
F\left[c_\ord\left(x,t\right),h_\ord\left(x,t\right)\right]\leq F\left[c_\ord\left(x,0\right),h_\ord\left(x,0\right)\right]\leq\sup_{\varepsilon,\ord\in\left[0,\infty\right)}F\left[c_\ord\left(x,0\right),h_\ord\left(x,0\right)\right]<\infty,
\]
for $0< t<\sigma_1$.  Consequently, we obtain the bound $\|h_{\ord,x}\|_2\leq
k_1$, where $0<t<\sigma_1$, and where $k_1$ depends only on the initial
conditions.  We have Poincar\'e's inequality for $h_{\ord,x}$,
\[
\|h_\ord\|_2^2\leq\left[\int_\Omega{dx}\,  h_\ord\left(x\right)\right]^2+\left(\frac{L}{2\pi}\right)^2\|h_{\ord,x}\|_2^2.
\]
Now $\int_\Omega{dx}\,h_\ord\left(x,t\right)=L\eta_{\ord,0}\left(t\right)$.
 Inspection of Eq.~\eqref{eq:weak_h} shows that $\eta_{\ord,0}\left(t\right)=\eta_{\ord,0}\left(0\right)=\eta_{0}^0$.
  Thus,
\[
\|h_\ord\|_2^2\leq L^2|\eta_{0}^0|^2+\left(\frac{L}{2\pi}\right)^2k_1\equiv
k_2<\infty.
\] 
Using the following inequality from Sec.~\ref{sec:background:inequalities},
\[
\|\varphi\|_\infty\leq \frac{1}{L}\|\varphi\|_1 + \|\varphi_x\|_1\leq\frac{1}{\sqrt{L}}\|\varphi\|_2+\sqrt{L}\|\varphi_x\|_2,
\]
we obtain the bound
\[
\|h_\ord\|_\infty\leq\frac{1}{\sqrt{L}}\|h_\ord\|_2+\sqrt{L}\|h_{\ord,x}\|_2,
\]
and hence $\|h_{\ord}\|_\infty\leq k_3$.
Additionally, the following properties hold:
\begin{itemize}
\item The function $h_{\ord}$ is H\"older continuous in space, with exponent
$\tfrac{1}{2}$,
\item $\int_\Omega{dx}\,  G_{\varepsilon}\left(h_\ord\right)\leq k_4$,
\end{itemize}
where these results hold in $0<t<\sigma_1$, and where the constants $k_1$,
$k_2$, $k_3$, and $k_4$ are independent of $\varepsilon$, $\ord$, $\sigma$,
and $\sigma_1$, and in fact depend only on the functions $h_0\left(x\right)$
and $c_0\left(x\right)$.

In order to continue the estimates to the whole interval $\left(0,\sigma\right)$,
we need to prove that $h_\ord\left(\cdot,\sigma_1\right)>0$ almost everywhere
(a.e.).  If this is
true, there is a new interval $\left[\sigma_1,\sigma_2\right)$, $\sigma_1<\sigma_2\leq\sigma$,
on which $h_\ord\left(\cdot,t\right)>0$ a.e., and we can then provide \emph{a
priori} bounds on $h_\ord\left(\cdot,t\right)$ and $c_\ord\left(\cdot,t\right)$
on
the interval $\left[\sigma_1,\sigma_2\right)$.  It is then possible to show
that $h_\ord\left(\cdot.,\sigma_2\right)>0$ a.e. and thus, by iteration,
we extend
the proof to the whole interval $\left(0,\sigma\right)$, and find that $h_\ord\left(.,t\right)>0$
a.e. on $\left(0,\sigma\right)$.

We have the bound
\begin{equation}
\int_\Omega{dx}\,  G_{\varepsilon}\left(h_\ord\left(\cdot,t\right)\right)\leq
k_4,
\label{eq:sigma_1_bound}
\end{equation}
where $k_4$ depends only on the initial conditions,
and where $0<t<\sigma_1$.  We now specify $G_\varepsilon\left(s\right)$,
in more detail.  This function satisfies the condition
\[
G_\varepsilon''\left(s\right)=\frac{1}{f_\varepsilon\left(s\right)g_\varepsilon\left(s\right)}.
\]
We take $g_\varepsilon\left(s\right)$ to be as defined previously, while
$f_\varepsilon\left(s\right)$ can be regularized as $g_\varepsilon\left(s\right)^3$,
which is Lipschitz continuous.
By defining
\[
\tilde{G}_\varepsilon\left(s\right)=-\int_s^\infty\frac{dr}{f_\varepsilon\left(r\right)g_\varepsilon\left(r\right)},\qquad
G_\varepsilon\left(s\right)=-\int_s^{\infty}{dr}\, \tilde{G}_\varepsilon\left(r\right),
\]
we obtain a function $G_\varepsilon\left(s\right)$ that is positive for all
$s\in\left(-\infty,\infty\right)$, and
%
%
%
%
%
%
\[
G_{\varepsilon}\left(s\right)=\tfrac{1}{6}\frac{1}{\left(s+\varepsilon\right)^2},\qquad
s\geq0.
\]
Using the boundedness of $G_{\varepsilon}\left(s\right)$,
and the time continuity of $h_n\left(\cdot,t\right)$, we employ the Dominated
Convergence Theorem,
\[
\lim_{t\rightarrow\sigma_1}\int_\Omega {dx}\, G_{\varepsilon}\left(h_\ord\left(\cdot,t\right)\right)=
\int_\Omega {dx}\, \lim_{t\rightarrow\sigma_1}G_{\varepsilon}\left(h_\ord\left(\cdot,t\right)\right)=
\int_{\Omega}{dx}\, G_{\varepsilon}\left(h_\ord\left(\cdot,\sigma_1\right)\right)
\leq k_4.
\] 
Similarly, since the constant $k_1$ in the inequality $\|h_{\ord,x}\|_2\leq
k_1$, $0\leq t<\sigma_1$ depends
only on the initial data, we extend this last inequality to $t=\sigma_1$,
and thus $h_\ord\left(x,\sigma_1\right)$ is H\"older continuous in space.
%
%


In the worst-case scenario, the time $\sigma_1$ is the first time at which
$h_\ord\left(x,t\right)$
touches down to zero, and and we therefore assume for contradiction that
$h_\ord\left(x_0,\sigma_1\right)=0$,
and that $h_\ord\left(x,\sigma_1\right)\geq0$ elsewhere.
 Then by the H\"older continuity, for any $x\in\Omega$ we have the bound
 $0\leq h_\ord\left(x,\sigma_1\right)\leq k_1\left|x-x_0\right|^{\frac{1}{2}}$,
 and thus
\begin{multline*}
\int_\Omega {dx}\, G_{\varepsilon}\left(h_\ord\left(\cdot,\sigma_1\right)\right)=\tfrac{1}{6}\int_\Omega\frac{dx}{\left[h_\ord\left(x,\sigma_1\right)+\varepsilon\right]^2}
\\
\geq\frac{k_1}{6}\int_0^L\frac{dx}{\left[|x-x_0|^{\frac{1}{2}}+\varepsilon\right]^2}\geq\frac{k_1}{6}\int_0^L\frac{dx}{|x-x_0|+\varepsilon\left(2\sqrt{L}+\varepsilon\right)}.
\end{multline*}
Hence,
\begin{multline*}
\frac{6}{k_1}\int_\Omega{dx}\, G_{\varepsilon}\left(h_\ord\left(\cdot,\sigma_1\right)\right)
\\
\geq-2\log\left[\varepsilon\left(2\sqrt{L}+\varepsilon\right)\right]+\log\bigg\{\left[L-x_0+\left(2\sqrt{L}+\varepsilon\right)\varepsilon\right]\left[x_0+\left(2\sqrt{L}+\varepsilon\right)\varepsilon\right]\bigg\}.
\end{multline*}
Thus, the integral $\int_\Omega G_\varepsilon\left(h_\ord\left(x,\sigma_1\right)\right)dx$
can be made arbitrarily large, which contradicts the $\varepsilon$-independent
bound for this quantity, obtained in Eq.~\eqref{eq:sigma_1_bound}.
We therefore have the strong condition that set on which $h_\ord\left(\cdot,\sigma_1\right)\leq0$
is empty.  Iterating the argument, we have the following important result:
\begin{quote}
The set on which $h_\ord\left(\cdot,t\right)\leq0$ is empty, for $0<t<\sigma$.
\end{quote}
Using the same argument, we have an estimate on the minimum value of $h_\ord\left(x,t\right)$,
\[
h_{\mathrm{min}}=\min_{x\in\Omega,t\in\left(0,\sigma\right]}h_\ord\left(x,t\right),
\]
namely,
\[
h_{\mathrm{min}}+\varepsilon\geq -k_1\sqrt{L}+\sqrt{k_1^2L+\frac{k_1^2L}{e^{k_4k_1^2}-1}},
\]
%
%
%
%
for all small positive $\varepsilon$.  Thus,
\begin{equation}
h_{\mathrm{min}}\geq B :=-k_1\sqrt{L}+\sqrt{k_1^2L+\frac{k_1^2L}{e^{k_4k_1^2}-1}},
\label{eq:min_h}
\end{equation}
a result that depends only on the initial data $c_0\left(x\right)$ and $h_0\left(x\right)$.
Now, using Eq.~\eqref{eq:min_h} and the boundedness result 
\[
\int_\Omega{dx}\, g_\varepsilon\left(h_\ord\right)\left[\tfrac{1}{4}\left(c_\ord^2-1\right)^2+\tfrac{1}{2}c_{\ord,x}^2\right]\leq
k_5
\]
where $k_5$ depends only on the initial data, we obtain \emph{a priori}
bounds on $\|c_{\ord,x}\|_2^2$,
\[
\int_\Omega{dx}\, c_{\ord,x}^2 \leq \frac{2k_5}{B},
\]
It is also possible to derive a bound on $\|c_\ord\|_2^2$.  We have the relation
\[
\int_\Omega{dx}\,\tfrac{1}{4}\left(c_\ord^2-1\right)^2\leq \frac{k_5}{B},
\]
which gives the inequality
$
\|c_\ord\|_4^4\leq 2\|c_\ord\|_2^2+\left({4k_5}/{B}\right).
$
Using the H\"older relation $\|c_\ord\|_2\leq |\Omega|^{\frac{1}{4}}\|c_\ord\|_4$,
we obtain a quadratic inequality in $\|c_\ord\|_2^2$,
\[
\|c_\ord\|_2^4\leq 2|\Omega|\|c_\ord\|_2^2+\frac{4|\Omega|k_5}{B},
\]
with solution
\[
\|c_\ord\|_2^2\leq |\Omega|+\frac{4|\Omega|k_5}{B},
\]
as required.  From the boundedness of $\|c_{\ord,x}\|_2$ and $\|c_{\ord}\|_2$
follows
the relation $\|c_\ord\|_\infty \leq k_6<\infty$, a result that depends only
on the initial conditions.
Let us recapitulate these results:
\begin{itemize}
\item $\|h_{\ord,x}\|_2$, is uniformly bounded;
\item $\|h_{\ord}\|_\infty$ is uniformly bounded;
\item The function $h_{\ord}$ is H\"older continuous in space, with exponent
$\tfrac{1}{2}$;
\item The function $h_{\ord}$ is nonzero everywhere and never decreases
below a certain value $B>0$, independent of $\ord$, $\varepsilon$, and $\sigma$.
\item $\|c_{\ord,x}\|_2$ is uniformly bounded;
\item $\|c_{\ord}\|_\infty$ is uniformly bounded;
\item The function $c_{\ord}$ is H\"older continuous in space, with exponent
$\tfrac{1}{2}$,
\end{itemize}
where these results are independent of $\ord$, $\varepsilon$ and $\sigma$,
and hold for $0<t<\sigma$.
\subsection{Equicontinuity of the Galerkin approximation; convergence
of Galerkin approximation}
\label{sec:equicontinuity}

\noindent  Using Eq.~\eqref{eq:free_energy_decay}, we obtain the bound
\begin{small}
\begin{multline*}
\int_0^t{dt'}\int_\Omega{dx}\,\bigg\{f_\varepsilon\left(h_\ord\right)\left[-h_{\ord,xxx}+\frac{h_{\ord,x}}{g_\varepsilon\left(h_\ord\right)f_\varepsilon\left(h_\ord\right)}+\frac{1}{g_\varepsilon\left(h_\ord\right)}\left(g_\varepsilon\left(h_\ord\right)c_{\ord,x}^2\right)_x\right]^2+g_\varepsilon\left(h_\ord\right)\mu_{\varepsilon,\ord,x}^2\bigg\}
\\
\leq F\left(0\right),\qquad 0<t<\sigma,
\end{multline*}
\end{small}
a bound that is independent of $\ord$, $\sigma$, and $\varepsilon$.  Since
the
quantity $\|f_\varepsilon\left(h_\ord\right)\|_\infty=\left(\|h_\ord\|_\infty+\varepsilon\right)^3$
is bounded above by a constant ${A}_1$ independent of $\ord$, $\varepsilon$,
and $\sigma$, we have the following string of inequalities,
\begin{small}
\begin{multline*}
\int_0^tdt'\int_\Omega {dx}\, J_{\varepsilon,\ord}^2\\
=\int_0^t{dt'}\int_\Omega{dx}\,\bigg\{f_\varepsilon\left(h_\ord\right)\left[-h_{\ord,xxx}+\frac{h_{\ord,x}}{g_\varepsilon\left(h_\ord\right)f_\varepsilon\left(h_n\right)}+\frac{1}{g_\varepsilon\left(h_n\right)}\left(g_\varepsilon\left(h_\ord\right)c_{\ord,x}^2\right)_x\right]\bigg\}^2\\
\leq\int_0^t{dt'}\|f_\varepsilon\left(h_\ord\right)\|_\infty\int_\Omega{dx}\,\bigg\{f_\varepsilon\left(h_\ord\right)\left[-h_{\ord,xxx}+\frac{h_{\ord,x}}{g_\varepsilon\left(h_\ord\right)f_\varepsilon\left(h_\ord\right)}+\frac{1}{g_\varepsilon\left(h_\ord\right)}\left(g_\varepsilon\left(h_\ord\right)c_{\ord,x}^2\right)_x\right]^2\bigg\}\\
\leq {A}_1\int_0^t{dt'}\int_\Omega{dx}\,
\bigg\{f_\varepsilon\left(h_\ord\right)\left[-h_{\ord,xxx}+\frac{h_{\ord,x}}{g_\varepsilon\left(h_\ord\right)f_\varepsilon\left(h_\ord\right)}+\frac{1}{g_\varepsilon\left(h_\ord\right)}\left(g_\varepsilon\left(h_\ord\right)c_{\ord,x}^2\right)_x\right]^2\bigg\}
\\
\leq {A}_1 F\left(0\right)\equiv A_2,
\end{multline*}
\end{small}
and thus
\[
\int_0^t{dt'}\,\|J_{\varepsilon,\ord}\|_2^2\leq A_2,\qquad 0< t<\sigma,
\]
where $A_2$ is independent of $\ord$, $\varepsilon$, and $\sigma$.  Similarly,
\[
\int_0^t{dt'}\,\|g_\varepsilon\left(h_\ord\right)\mu_{\varepsilon,\ord,x}\|_2^2\leq
A_3,\qquad 0<{t}<\sigma,
\]
and
\[
\int_0^t{dt'}\,\|c_\ord J_{\varepsilon,\ord}\|_2^2\leq A_2\|c_\ord\|_\infty^2\leq
A_2k_6^2,\qquad 0<t<\sigma,
\]
where $A_2$ and $A_3$ are independent of $\ord$, $\varepsilon$, and $\sigma$.
 By rewriting the evolution equations as
\[
h_{\ord,t}=-J_{\varepsilon,\ord,x},\qquad \left(g_\varepsilon\left(h_\ord\right)c\right)_t=-K_{\varepsilon,\ord,x},
\]
where $K_{\varepsilon,\ord}=c_\ord J_{\varepsilon,n}-g_\varepsilon\left(h_\ord\right)\mu_{\varepsilon,\ord,x}$,
we see that there are uniform bounds for $\int_0^t{dt'}\,\|J_{\varepsilon,\ord}\|_2^2$
and $\int_0^t{dt'}\,\|K_{\varepsilon,\ord}\|_2^2$, which depend only on the
initial data $c_0\left(x\right)$ and $h_0\left(x\right)$.
We mention the following result due Bernis and Friedman~\cite{Friedman1990}.
\begin{quote}
\textit{[Bernis and Friedman, 1990] Let $\varphi_i\left(x,t\right)$ be a sequence
of functions that weakly satisfy the equation
\[
\varphi_{i,t}=-J_{i,x},\qquad J_i=J\left(\varphi_i\right).
\]
If $\varphi_i\left(x,\cdot\right)$ is H\"older continuous (exponent $\tfrac{1}{2}$),
and if the fluxes $J_i$ satisfy
\[
\int_0^t dt'\,\|J_i\|_2^2\leq A_1,\qquad 0<t<\sigma,
\]
where $A_1$ is a number independent of the index $i$ and the time $\sigma$,
then there is a constant $A_2$, independent of $i$ and $\sigma$, such that
\[
\left|\varphi_i\left(\cdot,t_2\right)-\varphi_i\left(\cdot,t_1\right)\right|\leq
A_2\left|t_2-t_1\right|^{\frac{1}{8}},
\]
for all $t_1$ and $t_2$ in $\left(0,\sigma\right)$.  
}
\end{quote}
We now observe that the fluxes $J_{\varepsilon,\ord}$, and $K_{\varepsilon,\ord}$
satisfy the conditions of this theorem, and thus
\begin{quote}
The functions $h_{\ord}\left(\cdot,t\right)$ and $c_\ord\left(\cdot,t\right)$
are H\"older continuous (exponent $\tfrac{1}{8}$), for $0<t<\sigma$.
\end{quote}
We therefore have a uniformly bounded and equicontinuous family of functions
$\{\left(h_\ord,c_\ord\right)\}_{\ord=\ord_0+1}^\infty$.
We also have a recipe for constructing a uniformly bounded and equicontinuous
approximate solution $\left(h_\ord\left(x,t\right),c_\ord\left(x,t\right)\right)$,
in
a small interval $\left(0,\sigma\right)$.  The recipe can be iterated step-by-step,
and we obtain a uniformly bounded and equicontinuous family of approximate
solutions $\{\left(h_\ord,c_\ord\right)\}_{\ord=\ord_0+1}^\infty$, on $\left(0,T_0\right]\times\Omega$,
for an arbitrary time $T_0$.
Then, using the Arzel\`a--Ascoli theorem, we obtain the following convergence
result:
\begin{quote}
There is a subsequence of the family $\{\left(h_\ord,c_\ord\right)\}_{\ord=\ord_0+1}^\infty$
that converges uniformly to a limit $\left(h,c\right)$, in $\left(0,T_0\right]\times\Omega$.
\end{quote}
In contrast to Ch.~\ref{ch:background}, this convergence result is uniform
and true everywhere in $x\in \Omega$ (uniform strong convergence).  We prove
several facts about the pair $\left(h,c\right)$.
\begin{quote}
\textit{Let $\left(h,c\right)$ be the limit of the family of functions $\{\left(h_\ord,c_\ord\right)\}_{\ord=\ord_0+1}^\infty$
constructed in Steps~\eqref{sec:regularization}--\eqref{sec:equicontinuity}.
 Then the following properties hold for this limit:
\begin{enumerate}
\item The functions $h\left(x,t\right)$ and $c\left(x,t\right)$ are uniformly
H\"older continuous in space (exponent $\tfrac{1}{2}$), and uniformly H\"older
continuous in time (exponent $\tfrac{1}{8}$);
\item The initial condition $\left(h,c\right)\left(x,0\right)=\left(h_0,c_0\right)\left(x\right)$
holds;
\item $\left(h,c\right)$ satisfy the boundary conditions of the original
problem (periodic boundary conditions);
\item The derivatives $\left(h,c\right)_t$, $\left(h,c\right)_x$, $\left(h,c\right)_{xx}$,
$\left(h,c\right)_{xxx}$, and $\left(h,c\right)_{xxxx}$ are continuous in
the set $\left(0,T_0\right]\times\Omega$;
\item The function pair $\left(h,c\right)$ satisfy the following weak form
of the PDEs,
\begin{multline*}
\int\int_{Q_{T_0}}dt{dx}\, h\varphi_t+\int\int_{Q_{T_0}}dt{dx}\,J_\varepsilon\varphi_x=0,\\
J_\varepsilon=f_\varepsilon\left(h\right)h_{xxx}-\frac{1}{g_\varepsilon\left(h\right)}h_x-\frac{f_\varepsilon\left(h\right)}{g_\varepsilon\left(h\right)}\left(g_\varepsilon\left(h\right)c_{x}^2\right)_x,
\end{multline*}
\vskip -0.5in
\begin{multline*}
\int\int_{Q_{T_0}}dt{dx}\, g_\varepsilon\left(h\right)c\psi_t+\int\int_{Q_{T_0}}dt{dx}\,K_\varepsilon\psi_x=0,\\
K_\varepsilon= cJ_\varepsilon-g_\varepsilon\left(h\right)\left[c^3-c-\frac{1}{g_\varepsilon\left(h\right)}\left(g_\varepsilon\left(h\right)c_x\right)_x\right],
\end{multline*}
where $\varphi\left(x,t\right)$ and $\psi\left(x,t\right)$ are suitable test
functions.
%
%
\end{enumerate}
}
\end{quote}
The statements~(1),~(2), and~(3) are obvious.  Now, any pair $\left(h_\ord\left(x,t\right),c_\ord\left(x,t\right)\right)$
satisfies the equation set
\begin{multline*}
\int\int_{Q_{T_0}}dt{dx}\, h_\ord\varphi_t+\int\int_{Q_{T_0}}dt{dx}\, J_{\varepsilon,\ord}\varphi_x=0,
\\
J_{\varepsilon,\ord}=f_{\varepsilon}\left(h_\ord\right)h_{\ord,xxx}-\frac{1}{g_\varepsilon\left(h_\ord\right)}h_{\ord,x}-\frac{f_\varepsilon\left(h_\ord\right)}{g_\varepsilon\left(h_\ord\right)}\left(g_\varepsilon\left(h_{\ord}\right)c_{\ord,x}^2\right)_x,
\end{multline*}
\begin{multline*}
\int\int_{Q_{T_0}}dt{dx}\, g_\varepsilon\left(h_\ord\right)c_\ord\psi_t+\int\int_{Q_{T_0}}dt{dx}\,
K_{\varepsilon,\ord}\psi_x=0,\\
K_{\varepsilon,\ord}= c_{\ord}J_{\varepsilon,\ord}-g_\varepsilon\left(h_\ord\right)\left[c_\ord^3-c_\ord-\frac{1}{g_\varepsilon\left(h_\ord\right)}\left(g_\varepsilon\left(h_\ord\right)c_{\ord,x}\right)_x\right],
\end{multline*}
and from the boundedness of the fluxes $J_{\varepsilon,\ord}$ and $K_{\varepsilon,\ord}$
in $L^2\left(0,T_0,L^2\left(\Omega\right)\right)$, it follows that
\[
\left(J_{\varepsilon,\ord},K_{\varepsilon,\ord}\right)\rightharpoonup\left(J_\varepsilon,K_\varepsilon\right),\qquad\text{weakly
in }L^2\left(0,T_0,L^2\left(\Omega\right)\right),
\]
for a subsequence.  Using the regularity
theory for uniformly parabolic equations and the uniform H\"older continuity
of the $\left(h_{\ord},c_{\ord}\right)$'s, it follows that
\begin{quote}
The derivatives 
$\left(h_\ord,c_\ord\right)_t$, 
$\left(h_\ord,c_\ord\right)_x$, $\left(h_\ord,c_\ord\right)_{xx}$, $\left(h_\ord,c_\ord\right)_{xxx}$
and $\left(h_\ord,c_\ord\right)_{xxxx}$ 
are uniformly convergent in any compact
subset of $\left(0,T_0\right]\times\Omega$.
\end{quote}
Thus,
\[
J_\varepsilon=f_\varepsilon\left(h\right)h_{xxx}-\frac{1}{g_\varepsilon\left(h\right)}h_x-\frac{f_\varepsilon\left(h\right)}{g_\varepsilon\left(h\right)}\left(g_\varepsilon\left(h\right)c_{x}^2\right)_x,
\]
\[
K_\varepsilon=cJ_\varepsilon-g_\varepsilon\left(h\right)\left[c^3-c-\frac{1}{g_\varepsilon\left(h\right)}\left(g_\varepsilon\left(h\right)c_x\right)_x\right],
\]
on $\left(0,T_0\right]\times\Omega$, and therefore Claims~(4) and~(5)
follow.

\subsection{Convergence of regularized problem, as $\varepsilon\rightarrow0$}
\label{sec:convergence}

\noindent The result in Step~\eqref{sec:equicontinuity} produced a solution
$\left(h_\varepsilon,c_\varepsilon\right)$ to the regularized problem.  Due
to the result
\begin{equation}
h_{\varepsilon}\left(x,t\right)\geq h_{\mathrm{min}}\geq B=-k_1\sqrt{L}+\sqrt{k_1^2L+\frac{k_1^2L}{e^{k_4k_1^2}-1}}>0,\\
\label{eq:no_rupture}
\end{equation}
independent of $\varepsilon$, the argument of Step~\eqref{sec:equicontinuity}
can be recycled to produce a solution $\left(h,c\right)$ to the unregularized
problem.  
This solution is constructed as a limit $\left(h,c\right)=\lim_{\varepsilon\rightarrow0}\left(h_\varepsilon,c_\varepsilon\right)$,
and the results of the theorem in Step~\eqref{sec:equicontinuity} apply again
to $\left(h,c\right)$.
 The result~\eqref{eq:no_rupture} applies to $h$ constructed as $h=\lim_{\varepsilon\rightarrow0}h_\varepsilon$,
 and thus all the derivatives $\left(h,c\right)_t$, $\left(h,c\right)_x$,
 $\left(h,c\right)_{xx}$, $\left(h,c\right)_{xxx}$, and $\left(h,c\right)_{xxxx}$
 are continuous on the whole space $\left(0,T_0\right]\times\Omega$ and
 therefore, the weak solution $\left(h,c\right)$ is in fact a strong one.
 
\subsection{Regularity properties of the solution $\left(h,c\right)$}
\label{sec:analysis_thin_films:regularity}

Using a bootstrap argument, we show that the solution $\left(h,c\right)$
belongs to the regularity
classes $L^{\infty}\left(0,T_0;H^{2,2}\left(\Omega\right)\right)$ and $L^2\left(0,T_0;H^{4,2}\left(\Omega\right)\right)$.
 From Step~\eqref{sec:a_priori_bds} it follows immediately that
\[
\|h_x\|_2, \|c_x\|_2 < \infty,
\]
with time-independent bounds.  Thus, using Poincar\'e's inequality, it follows
that $h,c\in H^{1,2}\left(\Omega\right)$ and, moreover, 
\[
\sup_{\left[0,T_0\right]}\|h_x\|_2, \sup_{\left[0,T_0\right]}\|c_x\|_2 <
\infty.
\]
From Step~\eqref{sec:equicontinuity},
it follows that $J$ and $\mu_x$ belong to the regularity class $L^2\left(0,T_0;L^2\left(\Omega\right)\right)$,
and hence $J$, $\mu_x\in L^2\left(0,T_0;L^1\left(\Omega\right)\right)$. 
The functions $J$, $\mu$, and $\mu_x$ take the form
\[
J=h^3 h_{xxx}-h_xh^{-1}-h^2\left(h_x c_x^2+2h c_x c_{xx}\right),
\]
\[
\mu = c^3-c - h^{-1}\left(h_x c_x + hc_{xx}\right),
\]
and
\[
\mu_x= \left(3c^2-1\right)c_x + h^{-2}h_x^2 c_x - h^{-1}h_{xx}c_x - h^{-1}h_xc_{xx}
-c_{xxx},
\]
respectively.
The following results will help us in our demonstration,
\begin{itemize}
\item We shall use the boundedness of $h\left(x,t\right)$,
\[
0< h_{\mathrm{min}}\leq h\left(x,t\right)\leq h_{\mathrm{max}}<\infty,
\]
and the boundedness of $c\left(x,t\right)$, $\|c\|_\infty <\infty$. 
\item Since $\mu_x\in L^2\left(0,T_0;L^2\left(\Omega\right)\right)$, it follows
that $\mu\in L^2\left(0,T_0; L^2\left(\Omega\right)\right)$, by Poincar\'e's
inequality.
\item From this it follows that $c_x h_x + hc_{xx}$ is in the class $L^2\left(0,T_0;L^2\left(\Omega\right)\right)$.
\item  We have the inequality 
\[
\|h\mu c_x\|_2^2\leq h_{\mathrm{max}}\|\mu\|_\infty^2\|c_x\|_2^2\leq
h_{\mathrm{max}}\|c_x\|_2^2\left[\frac{1}{\sqrt{L}}\|\mu\|_2+\sqrt{L}\|\mu_x\|_2\right]^2,
\]
and thus $h_x c_x^2 + h c_x c_{xx}\in L^2\left(0,T_0;L^2\left(\Omega\right)\right)$.
\item Similarly, since $\int_0^{T_0}{dt}\,\|h\mu h_x\|_2^2<\infty$, we have the result
$h_x^2c_x +hh_x c_{xx}\in  L^2\left(0,T_0;L^2\left(\Omega\right)\right)$.
\end{itemize}

Now inspection of $\mu$ shows that $c_{xx}$ is in $L^2\left(0,T_0;L^1\left(\Omega\right)\right)$,
from which follows the result $h_{xxx},c_{xxx}\in L^1\left(0,T_0;L^1\left(\Omega\right)\right)$.
 By repeating the same argument, we find that $c_{xx}\in L^2\left(0,T_0;L^2\left(\Omega\right)\right)$.
  We also have the result that $\|h_x^2c_x\|_2$ is almost always bounded.
   To show that $h_{xx}\in  L^2\left(0,T_0;L^2\left(\Omega\right)\right)$,
   we take the evolution equation for $h\left(x,t\right)$, multiply it by
   $h$, and integrate, obtaining
\[
-\tfrac{1}{2}\frac{d}{dt}\int_\Omega{dx}\,h^2-3\int_\Omega{dx}\,h^2h_x^2h_{xx}-\int_\Omega{dx}\,h_xJ_0=\int_\Omega{dx}\,h^3h_{xx}^2,
\]
where $J_0=h_xh^{-1}+2h^3c_xc_{xx}+h^2h_xc_x^2$.   The time integral of the
first term on the
left-hand side of this equation is obviously bounded in time.  Let us examine
the time integral of the second term,
\begin{eqnarray*}
\int_0^{T_0}{dt}\int_\Omega{dx}\,h^2h_x^2h_{xx}&\leq& h_{\mathrm{max}}^2\left(\sup_{\left[0,T_0\right]}\|h_x\|_2^2\right)\int_0^{T_0}{dt}\,\|h_{xx}\|_\infty\\
&\leq& h_{\mathrm{max}}^2\left(\sup_{\left[0,T_0\right]}\|h_x\|_2^2\right)\int_0^{T_0}{dt}\,\left(L^{-1}\|h_{xx}\|_1+\|h_{xxx}\|_1\right)<\infty.
\end{eqnarray*}
The third term on the left-hand side is despatched with in a similar way,
so that $\int_0^{T_0}{dt}\|h_{xx}\|_2^2<\infty$.  We have now shown that
\[
h,c\in L^2\left(0,T_0;H^{2,2}\left(\Omega\right)\right).
\]
Using this result, together with the previous facts gathered together in
this section, it is readily shown that
\[
h,c\in L^2\left(0,T_0;H^{3,1}\left(\Omega\right)\right),
\]
And finally, using this result, it follows that
\[
h,c\in L^2\left(0,T_0;H^{3,2}\left(\Omega\right)\right).
\]
For example, 
\begin{eqnarray*}
\int_0^{T_0}{dt}\int_\Omega{dx}\,h_x^2c_{xx}^2&\leq& \left(\sup_{\left[0,T_0\right]}\|h_x\|_2^2\right)\int_0^{T_0}{dt}\,\|c_{xx}\|_\infty^2\\
&\leq&\left(\sup_{\left[0,T_0\right]}\|h_x\|_2^2\right)\int_0^{T_0}{dt}\,\left(L^{-1}\|c_{xx}\|_1+\|c_{xxx}\|_1\right)^2<\infty.
\end{eqnarray*}
This bound, together with the result $h\mu h_x\in L^2\left(0,T_0;L^2\left(\Omega\right)\right)$,
implies that 
\[
h_x^2 c_x\in L^2\left(0,T_0;L^2\left(\Omega\right)\right).
\]
Similarly,
\[
\int_0^{T_0}{dt}\int_\Omega{dx}\, h_x^2c_{xx}^2\leq\left(\sup_{\left[0,T_0\right]}\|h_x\|_2^2\right)\int_0^{T_0}{dt}\,\left(L^{-1}\|c_{xx}\|_1+\|c_{xxx}\|_1\right)^2,
\]
Hence $h_x c_{xx}\in   L^2\left(0,T_0;L^2\left(\Omega\right)\right)$, and
by
symmetry, $c_{x} h_{xx} \in  L^2\left(0,T_0;L^2\left(\Omega\right)\right)$.
Since
\[
\mu_x= \left(3c^2-1\right)c_x + h^{-2}h_x^2 c_x - h^{-1}h_{xx}c_x - h^{-1}h_xc_{xx}
-c_{xxx},
\]
is in $L^2\left(0,T_0;L^2\left(\Omega\right)\right)$, it follows that $c_{xxx}$
is in this class too.  A similar argument holds for $h_{xxx}$.  
Thus, the solution $\left(h,c\right)$ belongs to the following regularity
class:
\begin{equation}
\left(h,c\right)\in L^{\infty}\left(0,T_0;H^{1,2}\left(\Omega\right)\right)
\cap L^2\left(0,T_0;H^{3,2}\left(\Omega\right)\right)
\cap C^{\frac{1}{2},\frac{1}{8}}\left(\Omega\times\left[0,T_0\right]\right).
\label{eq:regularity0}
\end{equation}

Extra regularity is obtained by writing the equation pair as
\begin{eqnarray*}
\frac{\partial h}{\partial t}+h^3h_{xxxx}&=&-3h^2h_xh_{xxx}+\varphi_1+\varphi_2\equiv
\varphi\left(x,t\right),\\
\frac{\partial c}{\partial t}+c_{xxxx}&=&-\frac{2}{h}h_xc_{xxx}+\psi_1+\psi_2\equiv
\psi\left(x,t\right),
\end{eqnarray*}
where 
\begin{eqnarray*}
\int_0^{\tau}{dt}\,\|\varphi_1\|_2&\leq& \left(\sup_{\left[0,\tau\right]}\|c_{xx}\|_2\right)\int_0^{\tau}{dt}\,|\nu_1|,\qquad
\nu_1\in L^2\left(\left[0,\tau\right]\right),\\
\int_0^{\tau}{dt}\,\|\psi_1\|_2&\leq& \left(\sup_{\left[0,\tau\right]}\|c_{xx}\|_2\right)\int_0^{\tau}{dt}\,|\nu_2|,\qquad
\nu_2\in L^2\left(\left[0,\tau\right]\right),
\end{eqnarray*}
for any $\tau\in \left(0,T_0\right]$, and where $\varphi_2$ and $\psi_2$
belong to
the class $L^2\left(0,T_0;L^2\left(\Omega\right)\right)$.  By multiplying
the height and concentration equations by $h_{xxxx}$ and $c_{xxxx}$ respectively,
and by integrating over space and time, it is readily shown that
\[
\left(h,c\right)\in L^\infty\left(0,T_0,H^{2,2}\left(\Omega\right)\right),
\]
and hence
\[
\left(\varphi,\psi\right)\in L^2\left(0,T_0,L^{2}\left(\Omega\right)\right),
\]
from which follows the regularity result
\begin{equation}
\left(h,c\right)\in L^{\infty}\left(0,T_0;H^{2,2}\left(\Omega\right)\right)
\cap L^2\left(0,T_0;H^{4,2}\left(\Omega\right)\right)
\cap C^{\frac{3}{2},\frac{1}{8}}\left(\Omega\times\left[0,T_0\right]\right).
\label{eq:regularity}
\end{equation}
\subsection{Uniqueness of solutions}
\label{sec:analysis:thin_films:uniqueness}
Let us consider two solution pairs $\left(h,c\right)$ and $\left(h',c'\right)$
and form the difference $\left(\delta h,\delta c\right)=\left(h-h',c-c'\right)$.
 Given the initial conditions $\left(\delta c\left(x,0\right),\delta h\left(x,0\right)\right)=\left(0,0\right)$,
 we show that $\left(\delta h,\delta c \right)=\left(0,0\right)$ for all
 time, that is, that the solution we have constructed is unique.
We observe that that the equation for the difference $\delta c$ can be written
in the form
\begin{equation}
\frac{\partial}{\partial t}\delta{c}+\frac{\partial^4}{\partial{x}^4}\delta{c}=\delta
\varphi\left(x,t\right),
\label{eq:delta_c_unique}
\end{equation}
where 
%
%
%
%
%
%
%
%
%
%
%
$\delta \varphi\left(x,t\right) \in L^2\left(0,T_0,L^2\left(\Omega\right)\right)$,
and where $\delta \varphi\left(\delta c=0\right)=0$.
Thus, using the semigroup theory of Sec.~\ref{sec:background:Galerkin},
we find that Eq.~\eqref{eq:delta_c_unique} has a unique solution.
%
%
%
%
%
%
Since $\delta{c}=0$ satisfies Eq.~\eqref{eq:delta_c_unique}, and since $\delta{c}\left(x,0\right)=0$,
it follows that $\delta{c}=0$ for all times $t\in\left[0,T_0\right]$.

It is now possible to formulate an equation for the difference $\delta{h}$
by subtracting the evolution equations of $h$ and $h'$ from one another,
mindful that $\delta c=0$.  We multiply the resulting equation by $\delta{h}_{xx}$
and integrate over space, obtaining the result
\begin{multline*}
\tfrac{1}{2}\frac{d}{dt}\int_\Omega{dx}\,\delta h_x^2 +\int_\Omega{dx}\,h^3\delta
h_{xxx}^2=
\int_\Omega{dx}\,\delta h_{xxx}^2\delta h_x\left(h^{-1}+h'^2c_x^2\right)\\
%
%
%
%
-\int_\Omega{dx}\,\left(h^3-h'^3\right)\left(h'_{xxx}+4c_xc_{xx}\right)\delta
h_{xxx}\\
%
%
%
%
+
\int_\Omega{dx}\,\delta h_{xxx}\left[h'_x\left(h^{-1}-h'^{-1}\right)+h_xc_x^2\left(h^2-h'^2\right)\right].
\end{multline*}
Using the lower bound on $h\left(x,t\right)\geq h_{\mathrm{min}}>0$ and Young's
first inequality, this equation is transformed into an inequality,
\begin{multline*}
\tfrac{1}{2}\frac{d}{dt}\int_\Omega{dx}\,\delta h_x^2 +h_{\mathrm{min}}^3\int_\Omega{dx}\,\delta
h_{xxx}^2\leq
\kappa_1\int_\Omega{dx}\,\delta h_{xxx}^2+\frac{1}{4\kappa_1}\int_\Omega{dx}\,\delta{h}_x^2\left(h^{-1}+h'^2c_x^2\right)^2\\
+\kappa_2\int_\Omega{dx}\,\delta{h}_{xxx}^2+\frac{1}{4\kappa_2}\int_\Omega{dx}\,\left(h^3-h'^3\right)^2\left(h'_{xxx}+4c_xc_{xx}\right)^2\\
+
\kappa_3\int_\Omega{dx}\,\delta h_{xxx}^2+\frac{1}{4\kappa_3}\int_\Omega{dx}\,\left[h'_x\left(h^{-1}-h'^{-1}\right)+h_xc_x^2\left(h^2-h'^2\right)\right]^2,
\end{multline*}
where $\kappa_1$, $\kappa_2$, and $\kappa_3$ are arbitrary positive constants.
 By choosing $\kappa_1+\kappa_2+\kappa_3=h_{\mathrm{min}}^3$, the inequality
 simplifies,
\begin{multline*}
\tfrac{1}{2}\frac{d}{dt}\int_\Omega{dx}\,\delta h_x^2\leq
\frac{1}{4\kappa_1}\int_\Omega{dx}\,\delta{h}_x^2\left(h^{-1}+h'^2c_x^2\right)^2\\
+\frac{1}{4\kappa_2}\int_\Omega{dx}\,\left(h^3-h'^3\right)^2\left(h'_{xxx}+4c_xc_{xx}\right)^2\\
+
\frac{1}{4\kappa_3}\int_\Omega{dx}\,\left[h'_x\left(h^{-1}-h'^{-1}\right)+h_xc_x^2\left(h^2-h'^2\right)\right]^2,
\end{multline*}
which in turn reduces to
\begin{multline*}
2\kappa\frac{d}{dt}\int_\Omega{dx}\,\delta h_x^2\leq
\int_\Omega{dx}\,\delta{h}_x^2\left(h^{-1}+h'^2c_x^2\right)^2\\+
\int_\Omega{dx}\,\delta{h}^2\left[\left(h'_{xxx}+4c_xc_{xx}\right)^2+\left(h_x'+h_xc_x^2\right)^2\right],
%
%
%
\end{multline*}
where $\kappa$ is another positive constant.  We integrate over the time
interval $\left[0,T\right]$, and use the fact that $\|\delta{h}_x\|_2\left(0\right)=0$
to obtain the string of inequalities
%
%
%
%
%
%
%
%
%
%
%
\begin{multline*}
2\kappa\sup_{\tau\in\left[0,T\right]}\|\delta{h}_x\|_2^2\left(\tau\right)\\
\leq\int_0^T{dt}\int_\Omega{dx}\,\delta{h}_x^2\left(h^{-1}+h'^2c_x^2\right)^2\phantom{aaaaaaaaaaaaaaaaaaddddddddaaaaaaaaaaaa}\\
+\int_0^T{dt}\int_\Omega{dx}\,\delta{h}^2\left[\left(h'_{xxx}+4c_xc_{xx}\right)^2+\left(h_x'+h_xc_x^2\right)^2\right],
\end{multline*}
\vskip -0.7in
\begin{multline*}
\leq\int_0^T{dt}\|\delta{h}_x\|_2^2\|h^{-1}+h'^2c_x^2\|_\infty^2\\
+\int_0^T{dt}\|\delta{h}\|_\infty^2\int_\Omega{dx}\,\left[\left(h'_{xxx}+4c_xc_{xx}\right)^2+\left(h_x'+h_xc_x^2\right)^2\right],
\end{multline*}
\vskip -0.7in
\begin{multline*}
\leq\sup_{\tau\in\left[0,T\right]}\|\delta{h}_x\|_2^2\left(\tau\right)\int_0^T{dt}\|h^{-1}+h'^2c_x^2\|_\infty^2\\
+\sup_{\tau\in\left[0,T\right]}\|\delta{h}\|_\infty^2\left(\tau\right)\int_0^T{dt}\int_\Omega{dx}\,\left[\left(h'_{xxx}+4c_xc_{xx}\right)^2+\left(h_x'+h_xc_x^2\right)^2\right].
\end{multline*}
The Poincar\'e inequality can be combined with the one-dimensional differential
inequalities discussed in Ch.~\ref{ch:background} to yield the relation $\|f\|_\infty\leq
\kappa_P\|f_x\|_2$, where $f$ is some mean-zero function and $\kappa_P$ is
an $f$-independent constant.  We therefore arrive at the inequality
\begin{multline*}
2\kappa\sup_{\tau\in\left[0,T\right]}\|\delta{h}_x\|_2^2\left(\tau\right)
\leq\sup_{\tau\in\left[0,T\right]}\|\delta{h}_x\|_2^2\left(\tau\right)\int_0^T{dt}\|h^{-1}+h'^2c_x^2\|_\infty^2\\
+\kappa_P^2\sup_{\tau\in\left[0,T\right]}\|\delta{h}_x\|_2^2\left(\tau\right)\int_0^T{dt}\left(\|h'_{xxx}+4c_xc_{xx}\|_2^2+\|h_x'+h_xc_x^2\|_2^2\right).
\end{multline*}
Using the results of Sec.~\ref{sec:analysis_thin_films:regularity}, it is
readily shown that $h^{-1}+h'^2c_x^2\in L^2\left(0,T;L^{\infty}\left(\Omega\right)\right)$,
and that the functions $h'_{xxx}+4c_xc_{xx}$ and $h_x'+h_xc_x^2$ belong to
the class $L^2\left(0,T;L^2\left(\Omega\right)\right)$.  By choosing $T$
sufficiently small, it is possible to impose the inequality
\[
\frac{1}{2\kappa}\left[\int_0^T{dt}\|h^{-1}+h'^2c_x^2\|_\infty^2+\kappa_P^2\int_0^T{dt}\left(\|h'_{xxx}+4c_xc_{xx}\|_2^2+\|h_x'+h_xc_x^2\|_2^2\right)\right]<1,
\]
which in turn forces $\sup_{\tau\in\left[0,T\right]}\|\delta{h}_x\|_2^2=0$,
and hence the solution is unique.

\section{Summary}
\label{sec:analysis_thin_films:conclusionos}

In this chapter we have switched focus from the passive to the active tracer.
 The jump in complexity is significant, and involves the Navier--Stokes Cahn--Hilliard
 equations.  We focus on the case where the active system forms a thin layer
 on a substrate, and obtain simplified thin-film Stokes Cahn--Hilliard equations
 that are valid when the system's vertical gradients are small compared to
 the those in the lateral directions.  We analyze these equations in one
 lateral dimension, using the tools developed in Ch.~\ref{ch:background},
 and prove existence, regularity, and uniqueness results for the equations.
  While this
 analysis yields many precise
 results, it is unable to provide qualitative information about the solutions,
 in particular the effect of the backreaction on the concentration and free
 surface height, and we therefore turn to numerical simulations in the next
 chapter.

%% file: simulations_thin_films/simulations_thin_films.tex
\chapter{Nonlinear dynamics of phase separation in thin films: Qualitative
features}

\label{ch:simulations_thin_films}

\section{Overview}
\label{sec:simulations_thin_films:overview}

We take the thin-film Stokes Cahn--Hilliard (SCH) equations in two and
three dimensions as a starting point and investigate their qualitative features.
 By inspection of Eqs.~\eqref{eq:analysis_thin_films:model}, we can write
 down a three-dimensional
 set of SCH equations for the free surface
 $h\left(\bm{x},t\right)$, and for the phase-separating concentration field
 $c\left(\bm{x},t\right)$, and it is these we shall study in this chapter.
\begin{subequations}
\begin{equation}
\frac{\partial h}{\partial t}+\nabla_{\perp}\cdot\bm{J}=0,\\
\end{equation}
\begin{equation}
\frac{\partial}{\partial t}\left(c h\right)+\nabla_{\perp}\cdot\left(c\bm{J}\right)=\nabla_{\perp}\cdot\left(h\nabla_{\perp}\mu\right),
\end{equation}
where
\begin{equation}
\bm{J}=\tfrac{1}{2}h^2\nabla_{\perp}\Gamma-\tfrac{1}{3}h^3\bigg\{\nabla_{\perp}\left(-\frac{1}{C}\nabla_{\perp}^2h
+\phi\right)+\frac{r}{h}\nabla_{\perp}\left[h\left|\nabla_{\perp}c\right|^2\right]\bigg\},
\end{equation}
\begin{equation}
\mu=c^3-c-\frac{C_{\mathrm{n}}^2}{h}\nabla_{\perp}\cdot\left(h\nabla_{\perp}c\right).
\end{equation}%
\label{eq:simulations_thin_films:model}%
\end{subequations}%
We have used the derivative operator in the lateral directions, $\nabla_{\perp}=\left(\partial_x,\partial_y\right)$,
and have introduced the following nondimensional constants,
\[
r=\frac{\Small^2\beta\gamma}{D\nu},\qquad C_{\mathrm{n}}=\frac{\Small\sqrt{\gamma}}{h_0},\qquad
C=\frac{\nu\rho D}{h_0\sigma_0\Small^2},\qquad \varepsilon=\frac{h_0}{\lambda},
\]
where $\varepsilon$ is the ratio of the horizontal to the vertical lengthscales,
assumed to be small.  In Sec.~\ref{sec:simulations_thin_films:eqm_model} we shall study the equilibrium
version of these equations and, inspired by Sec.~\ref{sec:chaotic_advection:1D},
we shall reduce
this problem to a boundary-value problem.  In Sec.~\ref{sec:simulations_thin_films:2D}
we shall
carry out simulations
of the equations in three dimensions and study mechanisms for controlling
the phase separation.  Crucially, we find that these mechanisms depend on
the value of the backreaction strength, and we discuss this fact in the context
of microfabrication applications.
\section{Equilibrium solutions of the model equations}
\label{sec:simulations_thin_films:eqm_model}
 
\noindent In this section we discuss equilibrium solutions of the model equations~\eqref{eq:simulations_thin_films:model}.
 We use this knowledge to show that in the presence of a Van der Waals regularizing
 potential, the system exhibits the spinodal instability with an ultimate
 state of phase separation and a free-surface height slaved to the concentration
 field.  For simplicity, in this section we shall restrict to one lateral
 dimension. 
 
Let us set $\phi=A/h^3$ and $\Gamma=\text{constant}$.  We first of all investigate
the stability criteria
for the constant solution $\left(h_0,c_0\right)$ to Eq.~\eqref{eq:simulations_thin_films:model}
by identifying small deviations away from the base state, $\delta h \propto
e^{\sigma_h t}e^{i kx}$, $\delta c \propto e^{\sigma_c t}e^{i k x}$. 
Insertion of this ansatz into Eq.~\eqref{eq:simulations_thin_films:model}
gives
\begin{equation}
\begin{split}
\sigma_h &= -\frac{h_0^3 k^2}{3} \left(\frac{k^2}{C}-\frac{A}{h_0^{4}}\right),\\
\sigma_c &= \left(3 c_0^2-1\right)k^2 - C_{\mathrm{n}}^2 k^4.
\end{split}
\end{equation}
The system is stable to perturbations in the height field if $A<0$.  For
$A>0$, the height field will rupture in finite time~\cite{Oron1997,Davis_book}.
 For $\left|c_0\right|<1/\sqrt{3}$ however, the spinodal instability is accessible
 independently of the sign of $A$, which
 suggests that a critical mixed state will phase separate in a manner similar
 to the classical Cahn--Hilliard fluid, as described in Sec.~\ref{sec:background:ch}.

We are interested in the interplay between concentration gradients and the
height profile at late times and so we set $A=-|A|$ and study the system
in equilibrium, in one dimension.  By inspection of Eq.~\eqref{eq:simulations_thin_films:model},
the equilibrium conditions are $\mu=\mathrm{constant}$, $u=0$ which gives
equations
\begin{subequations}
\begin{equation}
\frac{1}{C}\frac{\partial^2 h}{\partial x^2}= C_{\mathrm{n}}^2\left|A\right|\left(1-\frac{1}{h^3}\right)
+r\left[\tfrac{1}{4}\left(c^2-1\right)^2+\tfrac{1}{2}\left(\frac{\partial{c}}{\partial{x}}\right)^2\right],
\end{equation}
\begin{equation}
\frac{\partial^2 c}{\partial x^2}=c^3-c-\frac{1}{h}\frac{\partial h}{\partial
x}\frac{\partial c}{\partial x},
\end{equation}%
\label{eq:eqm}%
\end{subequations}%
where we have enforced the boundary conditions $h\left(\pm\infty\right)=1$,
$\mu\left(\pm\infty\right)=0$ and have rescaled lengths by $C_{\mathrm{n}}$.
 
For the case $C=\infty$ and $\rho\equiv r/C_{\mathrm{n}}^2|A|\ll1$, Eq.~\eqref{eq:eqm}
has an asymptotic solution.  We find the $h$-equation
\begin{equation}
h=\biggl\{1+\rho\left[\tfrac{1}{4}\left(c^2-1\right)^2+\tfrac{1}{2}\left(\frac{\partial
c}{\partial x}\right)^2\right]\biggr\}^{-1/3},
\label{eq:r_dep}
\end{equation}
Hence, the $c$-equation is 
\begin{equation}
\frac{\partial^2 c}{\partial x^2}=\frac{1+\frac{1}{4}\rho\left(c^2-1\right)^2+\frac{5}{6}\rho\left(\frac{\partial
c}{\partial x}\right)^2}
{1+\frac{1}{4}\rho\left(c^2-1\right)^2+\frac{1}{6}\rho\left(\frac{\partial
c}{\partial x}\right)^2}\left(c^3-c\right).
\label{eq:c_eqn_no_st}
\end{equation}
For small $\rho$, the solution is $c=\tanh\left(x/\sqrt{2}\right)+O\left(\rho\right)$
and hence
\begin{equation}
h=1-\tfrac{1}{3}\rho\mathrm{sech}^4\left(\frac{x}{\sqrt{2}}\right)+O\left(\rho^2\right),\qquad\rho\ll1.
\label{eq:h_large_r}
\end{equation}
Thus, in this limiting case, the height profile is approximately constant
($h=1$) except in the transition region of the concentration field, where
the height profile possesses a valley.

\begin{figure}[htb]
\centering
\subfigure[]{
\includegraphics[width=0.3\textwidth]{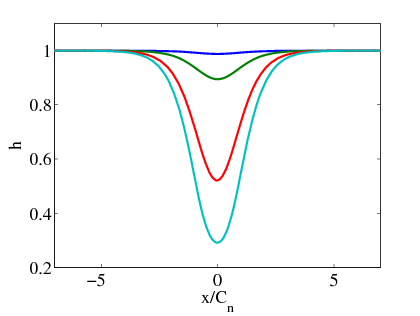}
}
\subfigure[]{
\includegraphics[width=0.3\textwidth]{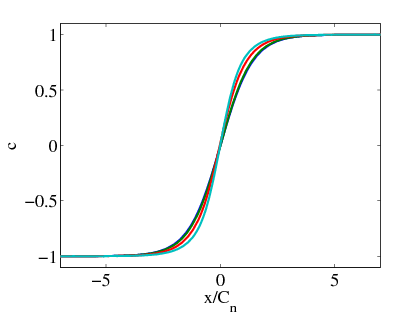}
}
\subfigure[]{
\includegraphics[width=0.3\textwidth]{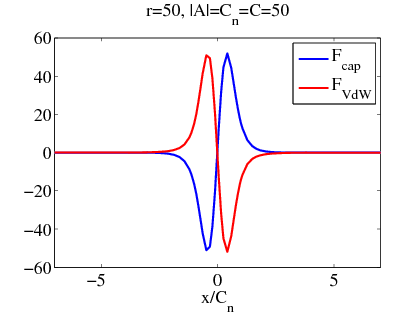}
}
\caption{Equilibrium solutions of the thin-film equations in one dimension,
obtained from solving a boundary-value problem with parameter values $C=C_{\mathrm{n}}^2|A|=1$
and $r=0.1,1,10,50$.  In (a) the valley deepens with increasing $r$ although
the film never ruptures, while in (b) the front steepens with increasing
$r$.  In (c) we plot the forces $F_{\mathrm{cap}}$ and $F_{\mathrm{vdW}}$
for $C=C_{\mathrm{n}}^2=|A|=1$ and $r=50$ to verify that they have opposite
sign.
}
\label{fig:hc}
\end{figure}

For nonzero surface tension, two parameters characterize the problem, for
now the equations are
\begin{subequations}
\begin{equation}
\frac{\partial^2 h}{\partial x^2}= CC_{\mathrm{n}}^2\left|A\right|\left(1-\frac{1}{h^3}\right)
+Cr\left[\tfrac{1}{4}\left(c^2-1\right)^2+\tfrac{1}{2}\left(\frac{\partial
c}{\partial x}\right)^2\right],
\end{equation}
\begin{equation}
\frac{\partial^2 c}{\partial x^2}=c^3-c-\frac{1}{h}\frac{\partial h}{\partial
x}\frac{\partial c}{\partial x},
\end{equation}%
\label{eq:eqm_parameters}%
\end{subequations}%
the two independent parameters being $C C_{\mathrm{n}}^2\left|A\right|$ and
$Cr$.  In Fig.~\ref{fig:hc} we present numerical
solutions
exhibiting the dependence of the solutions on these parameters.  As before,
the height field possesses peaks and valleys, where the valleys occur in
the transition region of concentration.  While the valley increases in depth
for
large $r$, rupture never takes place.  This result follows 
%
%
%
%
%
\begin{figure}[htb]
\centering
\subfigure[]{
\includegraphics[width=0.32\textwidth]{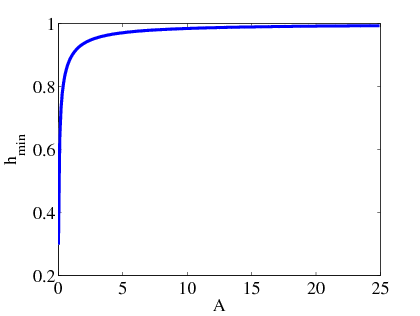}
}
\subfigure[]{
\includegraphics[width=0.32\textwidth]{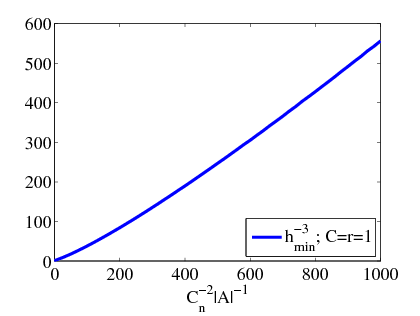}
}
\subfigure[]{
\includegraphics[width=0.32\textwidth]{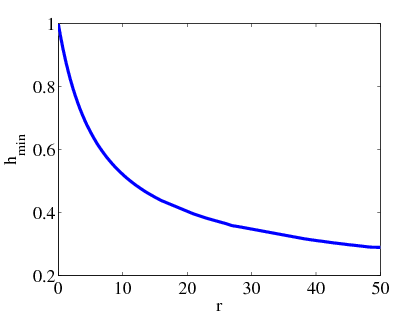}
}
\subfigure[]{
\includegraphics[width=0.32\textwidth]{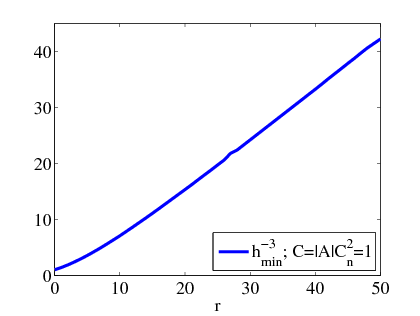}
}
\caption{The dependence of dip magnitude $h_{\mathrm{min}}$ on the problem
parameters.  Subfigures (a) and (b) show the dip magnitude's dependence on
the  parameter $C_{\mathrm{n}}^2|A|$; Subfigures (c) and (d) give the behaviour
of the dip magnitude as a function of $r$.
 In both cases, the approximate relation $h_{\mathrm{min}}^{-3}\sim r/{C_{\mathrm{n}}^2|A|}$
 is discernable.}
\label{fig:scaling_r_A}
\end{figure}
from the inequality
$h''\left(0\right)>0$, since $x=0$ is a local minimum.  Thus, from Eq.~\eqref{eq:eqm_parameters}
we obtain the bound
\begin{equation}
0<\bigg\{1+\frac{r}{C_{\mathrm{n}}^2|A|}\left[\tfrac{1}{4}\left(c\left(0\right)^2-1\right)^2+\tfrac{1}{2}c'\left(0\right)^2\right]\bigg\}^{-1}<\left[h\left(0\right)\right]^3.
\label{eq:no_rupture_bvp}
\end{equation}
The repulsive Van der Waals potential therefore has a regularizing effect on
the solutions.  We carried out a further numerical integration to find the
general dependence of the magnitude of the dip on the parameters of Eq.~\eqref{eq:eqm_parameters}.
 There is no clear scaling law for this dependence, although the approximate
 rule $h_{\mathrm{min}}^{-3}\sim r/{C_{\mathrm{n}}^2|A|}$ is evidenced by
 Fig.~\ref{fig:scaling_r_A},
 a relationship that is exact when $C=\infty$, as seen in Eq.~\eqref{eq:r_dep}.
  Thus, we have characterized the equilibrium solution as fully as possible.
   We now consider the physical meaning of these results.

The formation of the valley in the height field has the physical interpretation
of a balance between the Van der Waals and backreaction effects.  From Fig.~\ref{fig:hc}~(c)
we see that the backreaction or capillary force $F_{\mathrm{cap}}=-rh^{-1}\partial_x\left[h\left(\partial_xc\right)^2\right]$
and the Van der Waals force $F_{\mathrm{VdW}}=\left|A\right|\partial_x h^{-3}$
have opposite sign.  The repulsive Van der Waals force acts like a nonlinear
diffusion and inhibits rupture, and therefore $F_{\mathrm{cap}}$ promotes
rupture, a result
seen in experiments~\cite{WangH2000}.  The valley in the height field therefore
represents a balance between the smoothening and rupture-inducing effects.

\begin{figure}[htb]
\centering
\includegraphics[width=0.35\textwidth]{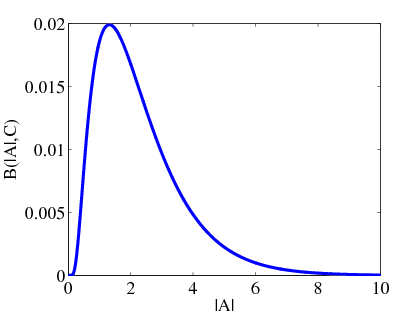}
\caption{A typical plot of $B\left(|A|,C\right)$ for $F_0=F_1=\tfrac{1}{2}$
and $C=1$.
 This theoretical lower bound has a different shape from those in Fig.~\ref{fig:scaling_r_A},
 which suggests that while $B\left(|A|,C\right)$ plays an important role
 in the analysis of the model equations, it does not capture the physics
 of film thinning.}
\label{fig:M_A}
\end{figure}

We compare the no-rupture condition in Eq.~\eqref{eq:no_rupture_bvp} with the
analytical results obtained in Ch.~\ref{ch:analysis_thin_films}.
In terms of the physical parameters of the system, the no-rupture condition
of Ch.~\ref{ch:analysis_thin_films} is
\[
h_{\mathrm{min}}\geq B\equiv\sqrt{2CL(F_0+F_1|A|)}\left(\sqrt{\frac{e^{4C|A|^{-1}\left(F_0+F_1|A|\right)^2}}{e^{4C|A|^{-1}\left(F_0+F_1|A|\right)^2}-1}}-1\right)>0,
\]
where $F_1=\tfrac{1}{2}\int_\Omega{dx}\left[h\left(x,0\right)\right]^{-2}\neq0$,
and $F_0=F\left(0\right)-F_1$.  The function $B\left(|A|,C\right)$ has no
explicit $r$-dependence: although $F_0$ depends on $r$, it is possible to
find initial data to remove this dependence.
We show a representative plot of $B\left(|A|,C\right)$ in Fig.~\ref{fig:M_A}.
Although a comparison between Fig.~\ref{fig:scaling_r_A} and Fig.~\ref{fig:M_A}
is not exact, since the boundary conditions and domains are different in
both cases, we see that the shape of the bound in Fig.~\ref{fig:scaling_r_A}
is different from that in Fig.~\ref{fig:M_A}.  Since the bound in Fig.~\ref{fig:scaling_r_A}
is obtained from numerical simulations, and is intuitively correct, we conclude
that it has the correct shape and that the bound of Fig.~\ref{fig:M_A}, while
mathematically indispensable, is not sharp enough to be useful in determining
the parametric dependence of the dip in free-surface height.

\begin{figure}[htb]
\centering
\subfigure[]{
\includegraphics[width=0.32\textwidth]{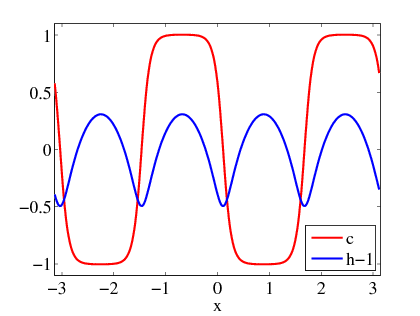}
 }
\subfigure[]{
\includegraphics[width=0.32\textwidth]{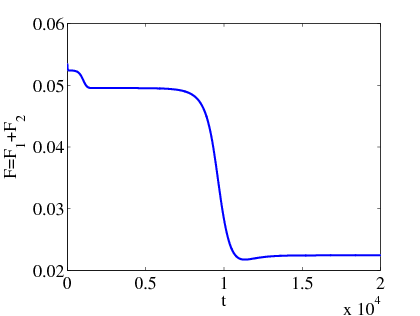}
 }
\caption{(a) A numerical solution of the SCH equations in one dimension,
 with $C=-A=1$ and $r=1$ (large backreaction); (b)
 The evolution of the free-energy functional.  The free energy decays monotonically,
 in agreement with the theory, except at late time, where there is a slight
 loss of
 monotonicity due to numerical error.}
\label{fig:hc_t}
\end{figure}

The procedure of identifying a phase-separating instability and studying
non-trivial equilibrium solutions is recurrent in Cahn--Hilliard dynamics,
and is discussed in Ch.~\ref{ch:background}.
 In that case, the nontrivial one-dimensional equilibrium solution hints
 at the late-time solution in several dimensions.  Indeed, the one-dimensional
 antikink solution suggests that the late-time solution in several dimensions
 consists of domains with transition regions of the same width as the antikink
 width.  Thus, we expect the equilibrium solution in Fig.~\ref{fig:hc} to
 reflect the late-time structure of a general phase-separating fluid in a
 thin film.  This is found to be the case in one-dimensional, time-dependent
 numerical simulations, of which Fig.~\ref{fig:hc_t} is an example; here
 the concentration forms domains $c\approx\pm1$, separated by smooth transition
 regions, while the free-surface height forms peaks and valleys, with valleys
 occurring in the transition regions.
The flatness of the peaks depends on the Van der Waals parameter $\left|A\right|$,
with large $\left|A\right|$ producing flat peaks, while the depth of the
valleys depends on the value of the backreaction strength $r$. 
The periodicity in Fig.~\ref{fig:hc_t} is due to the periodicity of the initial
conditions, in particular the period-four initial condition for the height
field.  
%
%
%
%
Purely random initial conditions give rise to a similar but aperiodic pattern
of domains, peaks, and valleys.

We have found by one-dimensional calculations, that the ultimate state of
the equation system Eq.~\eqref{eq:model} is a concentration field comprising
domains, with a height profile that dips at domain boundaries.  In order
to pursue this claim further, and to examine the dynamics of domain growth,
we turn to simulations in two dimensions.

\section{Solutions in two dimensions}
\label{sec:simulations_thin_films:2D}

\noindent In this section we outline a numerical method for the two-dimensional
thin-film Cahn--Hilliard equations and apply it in two
distinct cases: constant surface tension only, and spatially-varying surface
tension (or applied tangential stress).  In each case, the regularizing Van
der Waals potential is included to avoid film rupture.  The introduction
of appropriate structure functions enables us to measure the lengthscales
of the concentration morphology.

To measure the typical scale of the domains we introduce a structure
function $s\left(k,t\right)$ in what follows.  Because the concentration
field is dependent only on the lateral coordinates $x$ and $y$, we work
in two dimensions only.  Thus, as in Sec.~\ref{sec:chaotic_advection:model}
a measure of the correlation of concentration between two neighbouring points
in Fourier space is given by the expression
\[
S\left(\bm{k},t\right) = \frac{1}{L^2}\int_{\left[0,L\right]^2} d^2 x\int_{\left[0,L\right]^2}{d^2x'}e^{-i\bm{k}\cdot\bm{x}}\left[c\left(\bm{x}+\bm{x'}\right)c\left(\bm{x}\right)-\overline{c^{\phantom{2}}}^2\right],
\]
where $L$ is the box size, $\bm{x}=\left(x,y\right)$ and where $\overline{\left(\cdot\right)}$
denotes the spatial average.  We normalize this function and compute its
spherical average to obtain the structure function
\[
s\left(k,t\right)=\frac{1}{\left(2\pi\right)^2}\frac{\tilde{S}\left(k,t\right)}
{\overline{c^2}-\overline{c^{\phantom{2}}}^2},
\]
where the spherical average $\tilde{\phi}\left(k\right)$ any function $\phi\left(\bm{k}\right)$
is defined as
\[
        \tilde{\phi}\left(k\right) = \frac{1}{2\pi}\int_0^{2\pi} d\theta
        \phi\left(\bm{k}\right).
\]
Hence $\tilde{c}_k$ is the spherical average of the Fourier
coefficient $c_{\bm{k}}$.   Thus we identify the dominant lengthscale
$R\left(t\right)$ with the reciprocal of the most important
$k$-value, defined as the first moment of the distribution $s\left(k,t\right)$,
\begin{equation}
R\left(t\right)^{-1}=k_1=\frac{\int_0^\infty k s\left(k,t\right)dk}{\int_0^\infty
s\left(k,t\right)dk}.
\label{eq:R_b}
\end{equation}
In performing the spherical average, we have assumed that the system is isotropic.
 If the isotropy is broken, for example, by the presence of an external shear
 flow, then a vector-valued measure of the domain lengthscales is given
 by
\[
\left(k_x,k_y\right)=\frac{\int d^2k\bm{k}S\left(\bm{k},t\right)}{\int
d^2k S\left(\bm{k},t\right)}.
\]
Using the initial conditions and the scale analysis we have outlined, we
carry out the simulations for a variety of external conditions.
\subsection*{Numerical method}
To simulate the thin-film SCH equations~\eqref{eq:simulations_thin_films:model},
we use a semi-implicit finite-difference algorithm in two dimensions,
although here we sketch the method in one
dimension only.  The height field is updated first, using semi-implicit finite
differences~\cite{Grun2002}.  The height field of Eq.~\eqref{eq:model} is
discretized in time as follows,
\begin{equation}
\frac{h^{n+1}-h^n}{\Delta t} = -\frac{1}{3C}\left(h^{n+1}\right)^3\frac{\partial^4
h^{n+1}}{\partial x^4}+S_h\left(h^n,c^n\right),
\label{eq:euler_h}
\end{equation}
where 
\[
S_h\left(h,c\right)=h^2\frac{\partial h}{\partial x}\frac{\partial}{\partial x}\underbrace{\left(-\frac{1}{C}\frac{\partial^2
h}{\partial x^2}-\frac{|A|}{h^3}\right)}_{=p_0}-\frac{|A|}{3}h^3\frac{\partial^2}{\partial
x^2}\left(\frac{1}{h^3}\right)+\frac{r}{3}\frac{\partial}{\partial x}\bigg\{h^2\frac{\partial}{\partial
x}\left[h\left(\frac{\partial c}{\partial x}\right)^2\right]\bigg\}
\]  
contains third-order derivatives and lower.  Equation~\eqref{eq:euler_h}
is solved for $h^{n+1}$ using the Newton method for root-finding.  Using
the updated value of the height, we update the concentration, making use
of the following discretization in time,
\begin{equation}
\frac{c^{n+1}-c^n}{\Delta t}=-C_{\mathrm{n}}^2\frac{\partial^4 c^{n+1}}{\partial
x^4}+S_c\left(h^{n},c^n\right),
\label{eq:euler_c}
\end{equation}
where
\[
S_c\left(h,c\right)=-u\frac{\partial c}{\partial x}+\frac{1}{h}\frac{\partial
h}{\partial x}\frac{\partial\mu}{\partial x}+\frac{\partial^2}{\partial x^2}\left(c^3-c\right)-C_{\mathrm{n}}^2\frac{\partial^2}{\partial
x^2}\left(\frac{1}{h}\frac{\partial h}{\partial x}\frac{\partial c}{\partial
x}\right)
\]
contains third-order derivatives and lower.  Here $u$ is the velocity and
$\mu$ is the chemical potential, defined in Sec.~\ref{sec:model}.  Equation~\eqref{eq:euler_c}
is solved using a spectral method~\cite{Zhu_numerics}.

We solve the equations on a $128\times128$ square grid with a timestep of
$\Delta t = 5\times10^{-4}$, and the results do not change markedly upon
increasing
the resolution or decreasing the timestep.  Following standard practice~\cite{chaos_Berthier,Berti2005,ONaraigh2006},
the Cahn number $C_{\mathrm{n}}$ is taken to be $\sqrt{\Delta x^4/\Delta
t}$ so that the transition region is just barely resolved.  The boundary
conditions are periodic in the lateral directions.  The initial condition
for the height field is a sinusoidal perturbation while the concentration
field is given a random perturtion about zero.  In contrast to the usual
Cahn--Hilliard equation~\cite{ONaraigh2006} however, the  zero spatial mean
of the concentration field does not persist, since it is the quantity 
$\int
d^2 x c\left(\bm{x},t\right)h\left(\bm{x},t\right)$, 
not $\int d^2x c\left(\bm{x},t\right)$ that is conserved.
     
\subsection*{Constant surface tension}  
We let $\Gamma=\mathrm{constant}$, and study the velocity field
\begin{equation}
\bm{u}=\tfrac{1}{3}h^2\bigg[\nabla_{\perp}\left(\frac{1}{C}\nabla_{\perp}^2
h +\frac{|A|}{h^3}\right)-\frac{r}{h}\nabla_{\perp}\left(h\left|\nabla_{\perp}c\right|^2\right)\bigg].
\label{eq:u_cst_st}
\end{equation}
We carry out simulations according
to the scheme discussed above and find that the system evolves towards towards
its late-time configuration in the following way.  At early times, domains
of like concentration start to form,
\begin{figure}[htb]
\centering
\includegraphics[width=0.35\textwidth]{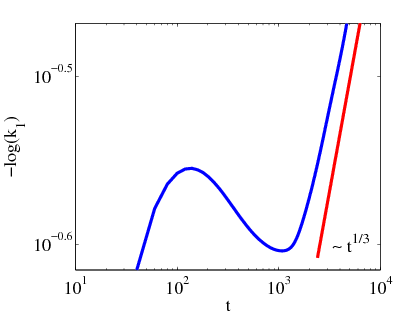}
\caption{Log-log plot indicating the time dependence of the typical domain
size $R\left(t\right)$.  This dependence is close to the Lifshitz--Slyozov
growth law, $R\sim
t^{1/3}$.}
\label{fig:L_t}
\end{figure}
while the initial configuration of the free surface is destroyed due to
nonlinear diffusive effects in the height equation.    At later times, the
domains continue to form and then to grow, while the free-surface height
becomes
slaved to the concentration field.  That is, the spatial and temporal variations
of the free surface depend on the concentration level.  The free-surface
height
possesses local maxima in regions where domains of concentration have formed
(peaks)  while at the transition between domains of different binary fluid
components, the free-surface height decreases markedly (valleys), as in the
one-dimensional
simulations of Sec.~\ref{sec:simulations_thin_films:eqm_model}.  The location
of the surface peaks
over the concentration domains has been seen in experiments involving polymeric
thin films~\cite{Jandt1996}.  As in the case of ordinary Cahn--Hilliard dynamics,
the late-time configuration selected is the one that tends to minimize the
free energy.  Now the free energy has the form
\[
F\left[c,h\right]=\int_{\left[0,L\right]^2}\left[\frac{1}{2C}\left|\nabla h\right|^2+\tfrac{1}{2}\frac{|A|}{h^2}\right]d^2x+
\frac{r}{C_{\mathrm{n}}^2}\int_{\left[0,L\right]^2} h\left[\tfrac{1}{4}\left(c^2-1\right)^2+\tfrac{1}{2}{C_{\mathrm{n}}^2}\left|\nabla
c\right|^2\right]d^2x,
\]
with $\dot{F}\leq0$, and the dynamics outlined do indeed decrease this free
energy.

We characterize the growth of the domains by examining
the function $R\left(t\right)$ in Fig.~\ref{fig:L_t}.  After transient effects
have passed, the lengthscale
\begin{figure}[htb]
\centering
\subfigure[]{
\includegraphics[width=0.24\textwidth]{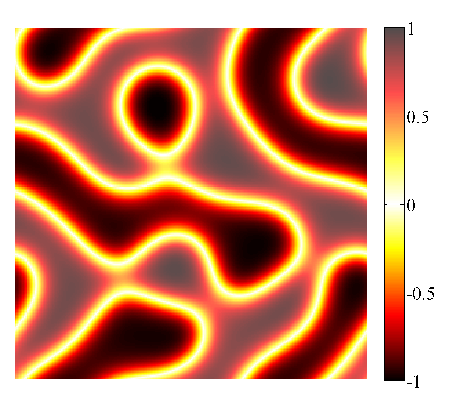}
}
\subfigure[]{
\includegraphics[width=0.25\textwidth]{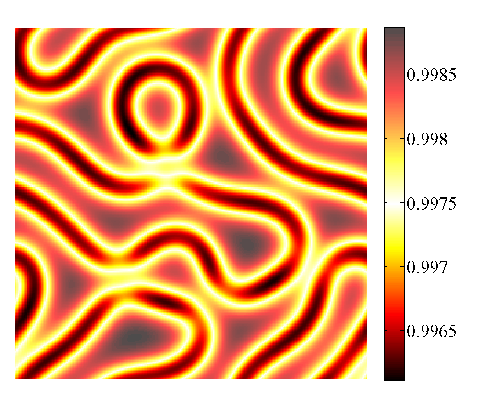}
}
\caption{(a) Concentration field at $t=5,000$, $r = 0.01$, $C=1$, and $A=-10$;
(b) The height field for the same parameter values.  Note that the height
field is slaved to the concentration field.}
\label{fig:patterns}
\end{figure}
$R\left(t\right)$ grows approximately as $t^{1/3}$, which is the Lifshitz--Slyozov
growth rate for phase separation in the absence of flow~\cite{LS}.  Several
numerical simulations with different values of $\left|A\right|$, $C$, and
$r$, indicate that the scaling law is independent of these parameters.  A
better estimate of the growth law could be obtained by increasing the resolution,
although this is computationally costly.  However, the qualitative features
of the system are well described by low-resolution simulations.  
The recovery of the growth law $R\left(t\right)\sim t^{1/3}$ demonstrates
that
the Lifshitz--Slyozov mechanism
of evaporation-condensation dominates over other effects,
such as viscosity-driven phase separation~\cite{Bray_advphys}.  This is not
surprising, since the timescale that determined the asymptotic ordering
in the derivation of Eq.~\eqref{eq:model} was the diffusion time $t_{\mathrm{diff}}\sim
L^2/D$.  In this sense, our  model equations demonstrate the importance of
variations in the free-surface height, rather than hydrodynamics.  At very
late times, the scaling rules are spoiled by finite-size effects. 

\subsection*{Variable surface tension}

\noindent We let $\Gamma=\Gamma_0\sin k x$, $\Gamma_0$ is a dimensionless
amplitude, $k=\left(2\pi/L\right)n = k_0 n$
is the spatial scale of the surface tension variation, and $n$ is an integer.
Then the velocity that drives the system becomes
\begin{equation}
\bm{u}=\tfrac{1}{2}h\left(k\Gamma_0\cos k x,0\right)+\tfrac{1}{3}h^2\bigg[\nabla_{\perp}\left(\frac{1}{C}\nabla_{\perp}^2
h +\frac{|A|}{h^3}\right)-\frac{r}{h}\nabla_{\perp}\left(h\left|\nabla c\right|^2\right)\bigg].
\label{eq:u_var_st}
\end{equation}
By inspection of Eq.~\eqref{eq:BC_stress},
this velocity field may also be obtained by imposing a shear stress $\bm{\tau}$
at the surface, provided $\bm{\tau}=\nabla_{\perp}\Gamma$.  The imposed shear
stress acts as a driving force.

\begin{figure}[htb]
\centering
\includegraphics[width=0.5\textwidth]{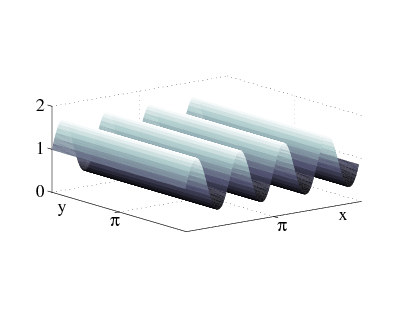}
\caption{The free-surface height for $r=0$ and $t=15000$ aligns with the applied
surface tension.  The height field at $t=15000$ for $r=\tfrac{1}{4}$ is similar.}
\label{fig:height_st}
\end{figure}

\begin{figure}[htb]
\centering
\subfigure[]{
\includegraphics[width=0.35\textwidth]{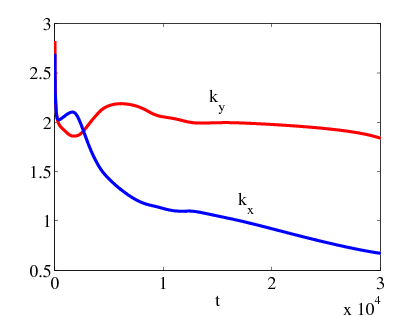}
}
\subfigure[]{
\includegraphics[width=0.35\textwidth]{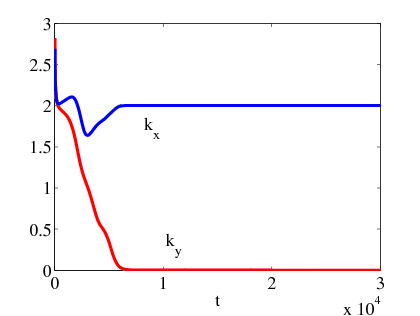}
}
\caption{Time dependence of the dominant spatial scales $k_x$ and $k_y$ for
(a) $r=0$, where the secular
behaviour of $k_x$ and $k_y$ has a small drift whose rate we do not report;
(b) $r=\tfrac{1}{4}$, where $k_x\rightarrow2$ and $k_y\rightarrow0$.}
\label{fig:scales_st}
\end{figure}

This choice of velocity field leads to control of phase separation in the
following manner.  For small values of the backreaction strength, with $r\rightarrow0$,
the height field quickly aligns with the surface tension profile as in
Fig.~\ref{fig:height_st},
\begin{figure}[htb]
\centering
\subfigure[]{
 \includegraphics[width=.2\textwidth]{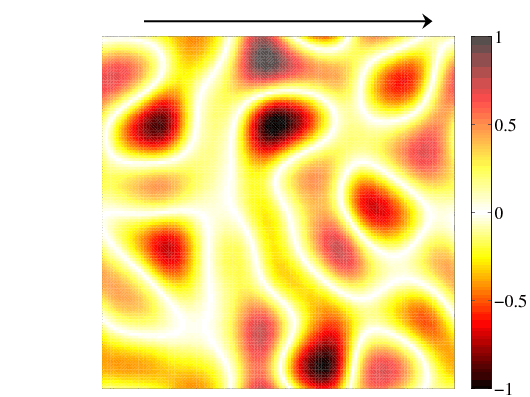}
}
\subfigure[]{
 \includegraphics[width=.2\textwidth]{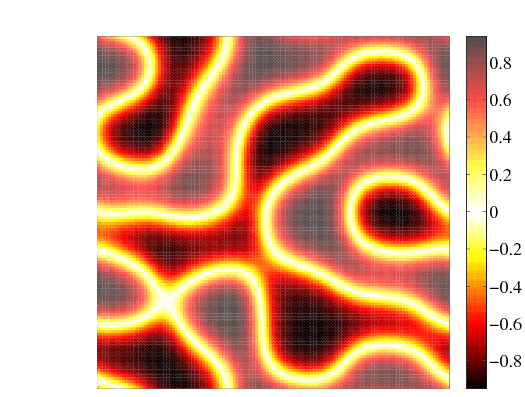}
}
\subfigure[]{
 \includegraphics[width=.2\textwidth]{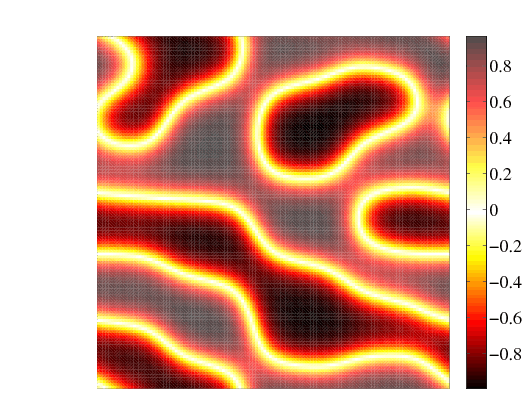}
}
\subfigure[]{
 \includegraphics[width=.2\textwidth]{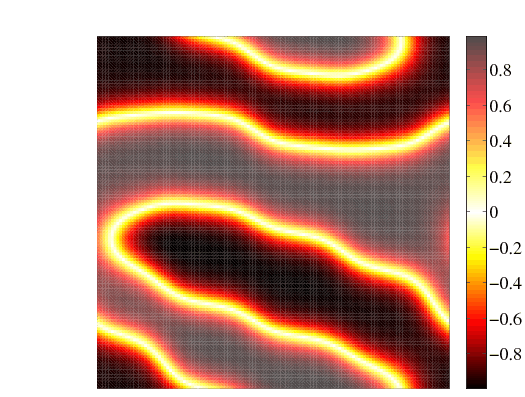}
}
\caption{Domain structure at $t=1000$, $t=3750$, $t=7500$, and $t=15000$
for $C=1$, $A=-10$, and $r=0$.  The surface tension gradient is parallel
to the arrow and $\Gamma=\Gamma_0\sin\left(k x\right)$, $\Gamma_0=20$ and
$k=4k_0$. For $r=0$, the domains align along the arrow.
}
\label{fig:domains_r0}
\end{figure}
since the strong effect of the Van der Waals diffusion destroys the unforced
part of $h\left(\bm{x},t\right)$. At the same time, the concentration field
begins to form domains.  At later times, when $k_x(t),k_y\left(t\right)\sim
k$,
the domains align with the gradient of the forcing term.   The growth of
the domains continues in this direction and is arrested (or slowed down considerably)
in the direction perpendicular to the forcing.  The domains are string-like,
with kinks occurring along lines where $\Gamma\left(x,y\right)$
is minimized, as evidenced by Fig.~\ref{fig:domains_r0}.  The decay
of $k_x$ and $k_y$ is shown in Fig.~\ref{fig:scales_st}.  It is not clear
whether the decay of $\left(k_x,k_y\right)$ is arrested or continues slowly,
and we do not report the decay rate here.

For moderate values of the backreaction strength with $r\sim O(1)$, the height
field again assumes a profile aligned with the surface tension, while
domains of concentration now
align in a direction perpendicular to the forcing gradient.  Domain growth
\begin{figure}[htb]
\centering
\subfigure[]{
 \includegraphics[width=.21\textwidth]{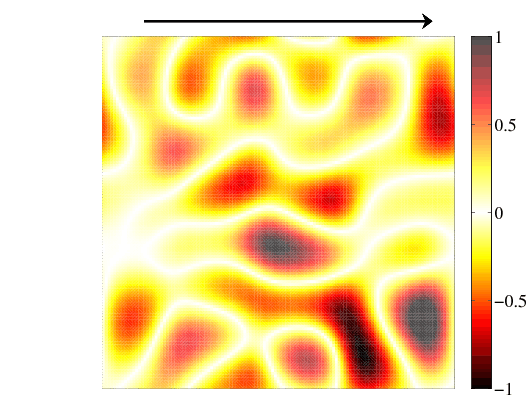}
}
\subfigure[]{
 \includegraphics[width=.15\textwidth]{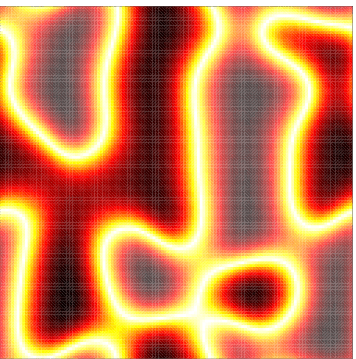}
}
\subfigure[]{
 \includegraphics[width=.15\textwidth]{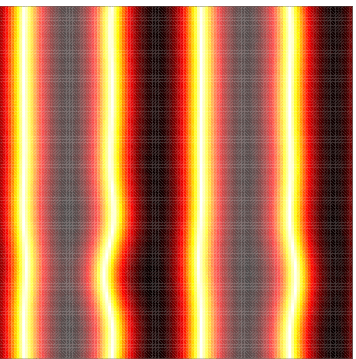}
}
\subfigure[]{
 \includegraphics[width=.15\textwidth]{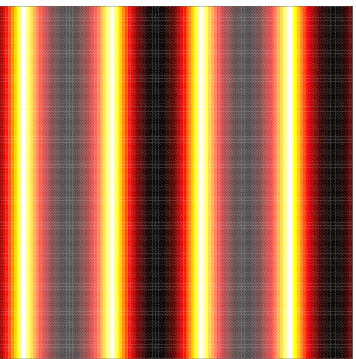}
}
\caption{Domain structure at $t=1000$, $t=3750$, $t=7500$, and $t=15000$
for $C=1$, $A=-10$, and $r=\tfrac{1}{4}$.  The surface tension gradient is
parallel to the arrow and $\Gamma=\Gamma_0\sin\left(k x\right)$, $\Gamma_0=20$
and $k=4k_0$.  The system reaches a steady state where the domains align
perpendicular to the arrow.
}
\label{fig:domains_r}
\end{figure}
continues in the perpendicular direction and is arrested in the direction
of the driving-force gradient.  A pattern
of string-like domains emerges, with domain boundaries forming along lines
where both $\sigma\left(x,y\right)$ and $h\left(x,y,t\right)$ are maximized.
Eventually, the domain boundaries align perfectly with the surface tension
maxima, as evidenced in Fig.~\ref{fig:domains_r}.  

The control of phase separation therefore depends crucially on the backreaction.
 This result is amplified by the existence of a no-rupture condition only
 for the $r=0$ case, where there is no backreaction.  This condition relies
 on the alignment of the height and surface tension profiles, which is exact
 only when the backreaction is zero.  Then at late times, the system evolves
 towards equilibrium and is described by the steady state $\nabla\cdot\left[\tfrac{1}{2}h^2\nabla_{\perp}\Gamma+\tfrac{1}{3}h^3\nabla_{\perp}\left(C^{-1}\nabla_{\perp}^2h+|A|h^{-3}\right)\right]=0$,
 which by the alignment property reduces to the one-dimensional equation
\begin{equation}
h^2\left[\tfrac{1}{2}\frac{\partial\Gamma}{\partial x}+\tfrac{1}{3}h\frac{\partial}{\partial
x}\left(\frac{1}{C}\frac{\partial^2h}{\partial x^2}+\frac{\left|A\right|}{h^3}\right)\right]=\mathrm{constant}.
\label{eq:eqm_st}
\end{equation}
By multiplying both sides of the expression by
$h$, differentiating and then evaluating the result at $x_0$, a coincident
minimum of height and surface tension, we obtain the condition
\[
\bigl[h\left(x_0\right)\bigr]^3\left[\frac{1}{3C}h\left(x_0\right)h^{\left(4\right)}\left(x_0\right)-\tfrac{1}{2}k^2\Gamma_0\right]=\left|A\right|h''\left(x_0\right).
\]
Since $x_0$ is a minimum of height, $h''\left(x_0\right)>0$, which prevents
$h\left(x_0\right)$ from being zero.  On the other hand, for $r$ and $\Gamma_0$
sufficiently large, the alignment of height and surface tension profiles
is not exact, the reduction of the steady-state condition
$\nabla\cdot\left[\tfrac{1}{2}h^2\nabla_{\perp}\Gamma+\tfrac{1}{3}h^3\nabla_{\perp}\left(C^{-1}\nabla_{\perp}^2h+|A|h^{-3}\right)\right]=0$
to Eq.~\eqref{eq:eqm_st} is no longer possible, and our calculation fails.
 In that case, simulations show that the film ruptures in finite time.
  
We have outlined, by numerical simulations and calculations, three
possible outcomes for the phase separation, depending on the backreaction
strength $r$.  For $r\ll1$, the concentration forms string-like domains,
aligned
with the applied force.  For $r=O\left(1\right)$ the concentration forms
domains that align perfectly in a direction perpendicular to the applied
force, while for $r\gg1$, the forcing causes the film to rupture.  Loosely
speaking, the different regimes occur because for $r=0$ the concentration
field couples to the gradient $\nabla_{\perp}\Gamma$, while for $r\sim O\left(1\right)$,
the concentration field couples to the patterned surface tension $\Gamma$.
 Thus, the interfacial tension or backreaction can therefore be tuned to
 achieve the desired outcome.
\section{Summary}
We have derived a thin-film model of phase separation based
on the NSCH equations, in which the reaction of
concentration gradients on the flow is important.  We have used this model
to give a qualitative picture of the features of phase separation in real
thin films, in particular the tendency of concentration gradients to promote
rupture in the film, and to produce peaks and valleys in the free surface
that mirror the underlying domain morphology.  We have found that in the
presence of a unidirectional sinusoidal variation in surface tension, the
strength of the backreaction determines the direction in which the domains
align.  This result could prove useful in microfabrication applications
where control of phase separation is required~\cite{Krausch1994}.

Because the lubrication model suppresses vertical variations in the concentration
field, we are limited to the case where the binary fluid components interact
identically with the boundaries at the substrate and free surface.
 However, the model quite generally gives an accurate description of surface
 roughening arising from Van der Waals forces.  More detailed models based
 on this approach, involving different boundary conditions that better
 reflect wetting behaviour and a concentration-dependent Hamakar coefficient,
 could capture a wider range of thin-film behaviour, although the tradeoff
 could be a loss of mathematically interesting features, such as the Lyapunov
 functional.

%% file: singular_solutions/gilbert_eqn.tex
\chapter{Singular solutions of aggregation equations}

\label{ch:singular_solutions}

\section{Overview}
\label{sec:singular_solutions:intro}

In this chapter we introduce a generalized gradient flow equation that serves
as a bridge from the Cahn--Hilliard theory we have discussed, to models
of aggregation that are currently used in nanoscale physics.  We then focus
on
a particular aggregation model that describes
a nonlocal magnetic system and admits singular solutions
that emerge from smooth initial data.
The magnetic system comprises a Gilbert-type equation for an orientation
vector, and a scalar density equation.  We study the uncoupled Gilbert-type
equation and 
find that the evolution
depends strongly on the lengthscales of the nonlocal effects.  We then
pass to
a coupled density-magnetization model and perform a linear stability analysis,
noting the effect of the lengthscales of nonlocality on the system's stability
properties.  We carry out numerical simulations of the coupled system and
find that singular solutions emerge from smooth initial data.  The singular
solutions represent a collection of interacting particles (clumpons).  By
restricting ourselves to the two-clumpon case, we are reduced to a two-dimensional
dynamical system that is readily analyzed, and thus we classify the different
clumpon interactions possible.
 
\section{Gradient flow, variable mobility, and nonlocality}

In this section we introduce a generalized aggregation equation that includes
the Cahn--Hilliard, Keller--Segel, and (scalar) Holm--Putkaradze equations
as special cases.  

Recall that the Cahn--Hilliard equation for an order parameter $\ordp$ is
derived from the gradient of a free-energy functional in $n$ dimensions,
\[
F_{\mathrm{CH}}\left[\ordp\right]=\int_\Omega d^n x\left[f_{\mathrm{tsm}}\left(\ordp\right)+\tfrac{1}{2}\gamma\left|\nabla{\ordp}\right|^2\right],
\]
where $f_{\mathrm{tsm}}\left(\ordp\right)$ is a potential with two stable
minima.
 In the
previous chapters, we chose the double-well potential $f_{\mathrm{tsm}}\left(\ordp\right)=\tfrac{1}{4}\left(\ordp^2-1\right)^2$,
and this choice simplified the analysis of the model.  However, any potential
with two stable minima and one unstable maximum will do.  Therefore, in this
chapter, we choose to work with the so-called double-obstacle potential
\[
f_{\mathrm{tsm}}\left(\ordp\right)= \xi W\left(\ordp\right)+\ordp\left(1-\ordp\right),\qquad
W\left(\ordp\right)=\ordp\log\ordp+\left(1-\ordp\right)\log\left(1-\ordp\right),
\] 
where $\xi$ is a positive parameter and the function $W$ is minus the
system's entropy density.  Recall that the quartic double well was invariant
under the component
relabelling transformation $\ordp\rightarrow-\ordp$.  The logarithmic double well
function has a similar symmetry: it is invariant under the relabelling transformation
$\ordp\rightarrow 1-\ordp$.  We can now write down the Cahn--Hilliard equation in
nonlocal form,
\[
F_{\mathrm{CH}}\left[\ordp\right]=\xi\int_\Omega d^n x W\left(\ordp\right)+\int_\Omega
d^n x{\ordp}\left(\bm{x},t\right)\int_\Omega d^n y K_{\mathrm{CH}}\left(\left|\bm{x}-\bm{y}\right|\right)\left(1-\ordp\left(\bm{y},t\right)\right),
\]
where $K_{\mathrm{CH}}\left(\left|x-y\right|\right)$ is the kernel whose
Fourier transform is
\[
\tilde{K}\left(\bm{k}\right)=1-\tfrac{1}{2}\gamma k^2.
\]
It is possible to prescribe a different kernel: other choices can lead to
anisotropic or nonlocal interactions between fluid particles.  Let us therefore
examine the possibility of nonlocal interactions, and leave the kernel unspecified.
 We shall, however, restrict to central interactions, $K\left(\bm{x}\right)=K\left(|\bm{x}|\right)$.
 The free energy functional is thus
\begin{equation}
F\left[\ordp\right]=\xi\int_\Omega d^n x W\left(\ordp\right)+\int_\Omega
d^n x{\ordp}\left(\bm{x},t\right)\int_\Omega
d^n y K\left(\left|\bm{x}-\bm{y}\right|\right)\left(1-\ordp\left(\bm{y},t\right)\right).
\label{eq:fe_nonlocal}
\end{equation}
As discussed in Ch.~\ref{ch:background}, the simplest rotationally-invariant,
mass-preserving evolution equation that decreases this free energy is the
gradient flow equation
\[
\frac{\partial \ordp}{\partial t}=\nabla\cdot\left( M\left(\ordp\right)\nabla\frac{\delta
F}{\delta\ordp}\right),
\]
where $M\left(\ordp\right)$ is a nonnegative function called the \emph{mobility}
and $\mu={\delta F}/{\delta\ordp}$ is the chemical potential. 
This evolution decreases the free energy:
\[
\frac{dF}{dt}=-\int_\Omega d^n x M\left(\ordp\right)\left|\nabla\mu\right|^2.
\]
The mobility function controls the strength of the particle diffusion and
depends on the local density of the fluid.  If the evolution is to respect
the component relabelling symmetry, then the mobility must be a function
of the combination $\ordp\left(1-\ordp\right)$.  The simplest possible choice
of mobility is thus
\[
M\left(\ordp\right)=\ordp\left(1-\ordp\right).
\]
The evolution of the system preserves the positivity of this function because
the energy cost of attaining the either of the states $\ordp=0,1$ is infinite,
as expressed in Eq.~\eqref{eq:fe_nonlocal}.

The nonlocal Cahn--Hilliard concentration for the concentration $\ordp\left(\bm{x},t\right)$
is thus
\begin{equation}
\frac{\partial \ordp}{\partial t}=\nabla\cdot\left[\ordp\left(1-\ordp\right)\nabla\left(\xi
W'\left(\ordp\right)-\int_\Omega{d^ny}K\left(|\bm{x}-\bm{y}|\right)\ordp\left(\bm{y},t\right)\right)\right].
\label{eq:ch_nonlocal}
\end{equation}
By specifying a particular kernel, or by modifying the mobility, we recover
a number of models of aggregation.  In particular,
\begin{itemize}
\item If the Fourier transform of the kernel $K$ is $1-\tfrac{1}{2}\gamma
k^2$, we recover the Cahn--Hilliard equation with the double-obstacle potential
and variable mobility;
\item If the Fourier transform of the kernel $K$ is $\left(\lambda^2 k^2+1\right)^{-1}$,
where $\lambda$ is the lengthscale of the potential energy $\int_\Omega
d^n y K\left(\left|\bm{x}-\bm{y}\right|\right)\left(1-\ordp\left(\bm{y},t\right)\right)$,
that is, if $K$ is the kernel of the Helmholtz operator $1-\lambda^2\Delta$,
we recover the regularized Keller--Segel equation\footnote{The regularized
Keller--Segel equation is discussed in Appendix~B};
\item If $K$ is any nonlocal kernel, $\xi=0$, and the mobility $\ordp\left(1-\ordp\right)$
is replaced with $\ordp\left(1-H*\ordp\right)$, where $H*\ordp$ is the smoothened
density
\[
\left(H*\ordp\right)\left(x,t\right)=\int_\Omega{d^ny}H\left(|\bm{x}-\bm{y}|\right)\ordp\left(\bm{y},t\right)
\]
we obtain the scalar Holm--Putkaradze model~\cite{Darryl_eqn6}.
\end{itemize}
The scalar Holm--Putkaradze model was introduced in~\cite{Darryl_eqn6,Darryl_eqn5}
and describes the evolution of a single scalar density.  In particular, given
the (negative) free energy
\[
F_{\mathrm{HP}}\left[\ordp\right]=-\tfrac{1}{2}\int_\Omega d^nx \ordp\left(\bm{x},t\right)\left(1-\alpha^2\Delta\right)^{-1}\ordp\left(\bm{x},t\right),
\]
the model admits singular solutions that emerge from smooth initial data.
 The singular solutions satisfy a set of ordinary differential equations.
 The model can therefore be used to describe self-assembly in nanoscale
 physics, in which
 atoms or molecules aggregate to from larger, mesoscale structures with
 particle-like properties.  These particles are governed by the ordinary
 differential equations to which the aggregation equation reduces.
The aggregation equation~\eqref{eq:ch_nonlocal} that we have formulated thus
provides a bridge from the Cahn--Hilliard theory we have discussed, to this
more recent application in nanoscale physics.
The authors Holm, Putkaradze, and Tronci have extended this model to a coupled
scalar-vector system in~\cite{Darryl_eqn1}, although they simply formulate
the equations, and do not examine the properties of the model in any detail.
 Therefore, using the numerical tools that we have developed, and our knowledge
 of aggregation
through energy minimization, we turn to this coupled scalar-vector system,
and provide qualitative and quantitative descriptions of its behaviour.

 We treat the initial state of the system as a continuum, a good approximation
 in nanophysics applications~\cite{Forest2007}.
 One realization of this problem is in nanoscale magnetic systems, in which
 particles
 with a definite magnetic moment collapse and form mesoscale structures,
 that in turn have a definite magnetic moment.  Thus, in this chapter we
 refer to the orientation vector in our continuum picture as the \emph{magnetization}.

\section{The nonlocal Gilbert equation}
\label{sec:gilbert}

In this section we study a magnetization equation that in form is similar
to the Gilbert equation, that is, the Landau--Lifshitz--Gilbert equation
in
the over-damped limit~\cite{GilbertIEEE,Weinan2000}.  The equation we focus
on incorporates
nonlocal effects, and was introduced in~\cite{Darryl_eqn1}.  We study the
evolution and energetics of this equation, and examine the importance of
the problem lengthscales in determining the evolution.

We study the following nonlocal Gilbert (NG) equation
\begin{equation}
\frac{\partial\Mag}{\partial t} = \Mag\times\left(\bm{\mu}_\Subm\times\frac{\delta{E}}{\delta\Mag}\right),
\label{eq:mag_eqn}
\end{equation}
where the magnetization vector $\Mag$, its smoothened version $\bm{\mu}_m$,
and the variational derivative $\delta E/\delta\Mag$ are functions of the
form
\[
\Mag,\bm{\mu}_m,\delta E/\delta\Mag:\left(\Omega\subset\mathbb{R}\right)\times\left(\mathbb{R}^+\cup\{0\}\right)\rightarrow\mathbb{R}^3.
\]
The smoothened magnetization $\bm{\mu}_\Subm$ is
defined as
\[
\bm{\mu}_\Subm = \left(1-\beta^2\partial_x^2\right)^{-1}\Mag,
\]
and $\delta E/\delta\Mag$ is prescribed as follows,
\[
\frac{\delta E}{\delta\Mag} = \left(1-\alpha^2\partial_x^2\right)^{-1}\Mag.
\]
The smoothened magnetization $\bm{\mu}_\Subm$ and the force $\delta{E}/\delta{\bm{m}}$
can be computed using the theory of Green's functions.  In particular,
\[
\bm{\mu}_\Subm\left(x,t\right)=\int_\Omega{dy} H_{\beta}\left(x-y\right)\Mag\left(y,t\right):=H_{\beta}*\Mag\left(x,t\right).
\]
Here $*$ denotes the convolution of functions, and the kernel $H_{\beta}\left(x\right)$
satisfies the equation
\begin{equation}
\left(1-\beta^2\frac{d^2}{dx^2}\right)H_\beta\left(x\right)=\delta\left(x\right).
\label{eq:kernel}
\end{equation}
The function $\delta\left(x\right)$ is the Dirac delta function.  Equation~\eqref{eq:kernel}
is solved subject to conditions imposed on the boundary of the domain $\Omega$.
 In this chapter we shall work with a periodic domain $\Omega=\left[-L/2,L/2\right]$
 or $\Omega=\left[0,L\right]$,
 although other boundary conditions are possible.  Since $e^{\pm ikx}$, $k\in\mathbb{R}$
 is
 a simultaneous eigenfunction of the operators $\left(1-\alpha^2\partial_x^2\right)^{-1}$
 and $\left(1-\beta^2\partial_x^2\right)^{-1}$, equation~\eqref{eq:mag_eqn}
 has a family of nontrivial equilibrium states given by
\[
\Mag_{\mathrm{eq}}\left(x\right)=\Mag_0\sin\left(kx+\phi_0\right),
\]
where $\Mag_0$ is a constant vector, $k$ is some wavenumber, and $\phi_0$
is a constant phase.  The derivation of this solution is subject to the boundary
conditions discussed in Sec.~\ref{sec:mag_dens}.

By setting $\beta=0$ and replacing $\left(1-\alpha^2\partial_x^2\right)^{-1}$
with $-\partial_x^2$, we recover the more familiar Landau--Lifshitz--Gilbert
equation, in the overdamped limit~\cite{GilbertIEEE},
\begin{equation}
\frac{\partial\Mag}{\partial t} = -\Mag\times\left(\Mag\times\frac{\partial^2\Mag}{\partial{x}^2}\right).
\label{eq:gilbert}
\end{equation}
%
%
%
%
%
%
%
%
Equation~\eqref{eq:mag_eqn} possesses several features that will be useful
in understanding the numerical simulations.  There is an energy
functional
\begin{equation}
E\left(t\right)=\tfrac{1}{2}\int_\Omega{dx}\Mag\cdot\left(1-\alpha^2\partial_x^2\right)^{-1}\Mag,
\end{equation}
which evolves in time according to the relation
\begin{eqnarray}
\frac{dE}{dt}&=&\int_\Omega{dx}\left[\bm{\mu}_\Subm\cdot\left(1-\alpha^2\partial_x^2\right)^{-1}\Mag\right]\left[\Mag\cdot\left(1-\alpha^2\partial_x^2\right)^{-1}\Mag\right]\nonumber\\
&\phantom{a}&\phantom{aaaaaaaaaaaaaaaaaaaaaadaaaa}
-\int_\Omega{dx}\left(\bm{\mu}_\Subm\cdot\Mag\right)\left[\left(1-\alpha^2\partial_x^2\right)^{-1}\Mag\right]^2,\nonumber\\
&=&-\int_\Omega{dx}\left[\Mag\times\left(1-\alpha^2\partial_x^2\right)^{-1}\Mag\right]\cdot\left[\bm{\mu}_\Subm\times\left(1-\alpha^2\partial_x^2\right)^{-1}\Mag\right].
\label{eq:dt_energy}
\end{eqnarray}
This is not necessarily a nonincreasing function of time, although setting $\beta=0$
gives
\begin{eqnarray}
\left(\frac{dE}{dt}\right)_{\beta=0}&=&\int_\Omega{dx}\left[\Mag\cdot\left(1-\alpha^2\partial_x^2\right)^{-1}\Mag\right]^2-
\int_\Omega{dx}\Mag^2\left[\left(1-\alpha^2\partial_x^2\right)^{-1}\Mag\right]^2,\nonumber\\
&=&\int_\Omega{dx}\Mag^2\left[\left(1-\alpha^2\partial_x^2\right)^{-1}\Mag\right]^2\left(\cos^2\varphi-1\right)\leq0,
\label{eq:dt_energy_beta0}
\end{eqnarray}
where $\varphi$ is the angle between $\Mag$ and $\left(1-\alpha^2\partial_x^2\right)^{-1}\Mag$.
 In the special case when $\beta\rightarrow0$, we therefore expect $E\left(t\right)$
  to be a nonincreasing function of time.  On the other hand, inspection
  of Eq.~\eqref{eq:dt_energy} shows that as $\alpha\rightarrow0$, the energy
  tends to a constant.
Additionally, the magnitude of the vector $\Mag$ is conserved.  This can
 be shown by multiplying Eq.~\eqref{eq:mag_eqn} by $\Mag$, and by exploiting
 the antisymmetry of the cross product.  Thus, we are interested only in
 the orientation of the vector $\Mag$; this can be parametrized by two
 angles on the sphere:
\begin{figure}
\centering
\includegraphics[width=0.35\textwidth]{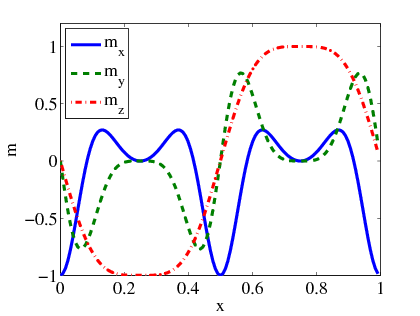}
\caption{The initial data for the magnetization equation~\eqref{eq:mag_eqn}.
 This initialization is obtained by allowing the orientation angles
 of the magnetization vector to vary sinusoidally in space, as in Eq.~\eqref{eq:initial}.
  Here the wavenumber of the variation is equal to the fundamental wavenumber
  $2\pi/L$.}
\label{fig:initial}
\end{figure}
the azimuthal angle $\theta\left(\bm{x},t\right)$, and the polar angle $\phi\left(\bm{x},t\right)$,
where
\begin{equation}
m_x=|\Mag|\cos\phi\sin\theta,\qquad
m_y=|\Mag|\sin\phi\sin\theta,\qquad
m_z=|\Mag|\cos\theta,
\label{eq:spherical_polars}
\end{equation}
and where $\phi\in\left[0,2\pi\right)$, and $\theta\in\left[0,\pi\right]$.

We carry out numerical simulations of Eqs.~\eqref{eq:mag_eqn} and~\eqref{eq:gilbert}
on a periodic domain $\left[0,L\right]$, and outline the findings in
what
follows.  Motivated by the change of coordinates~\eqref{eq:spherical_polars},
we choose the initial data
\begin{equation}
\phi_0\left(x\right)=\pi\left(1+\sin\left(2r\pi x/L\right)\right),\qquad
\theta_0\left(x\right)=\tfrac{1}{2}\pi\left(1+\sin\left(2\pi s x/L\right)\right),
\label{eq:initial}
\end{equation}
where $r$ and $s$ are integers.  These data are shown in Fig.~\ref{fig:initial}.

\begin{figure}[htb]
\centering
\subfigure[]{
\includegraphics[width=0.3\textwidth]{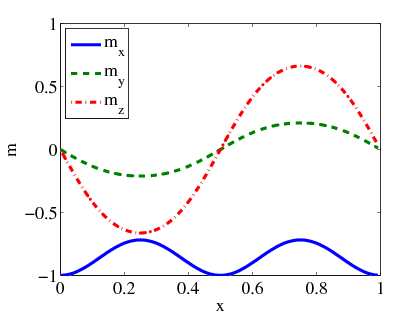}
}
\subfigure[]{
\includegraphics[width=0.3\textwidth]{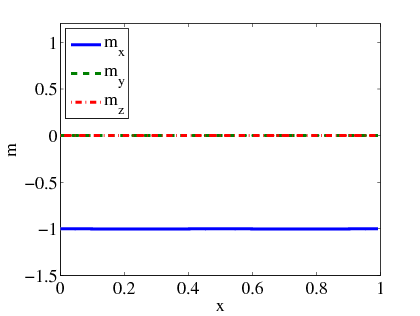}
}
\subfigure[]{
\includegraphics[width=0.3\textwidth]{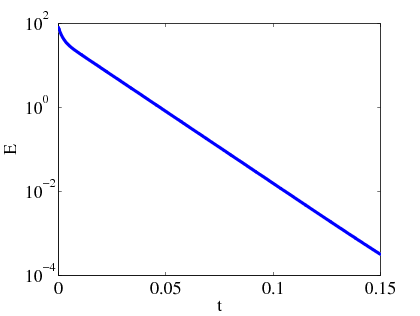}
}
\caption{Numerical simulations of Case~(1), the Landau--Lifshitz--Gilbert
equation in the over-damped limit.  In this case, the magnetization decays
to a constant state.
%
%
%
%
Subfigures (a) and (b) show the magnetization at times $t=0.03$ and $t=0.15$
respectively;
(c) is the energy functional, which exhibits exponential decay after some
transience.  The final orientation is $\left(\phi,\theta\right)=\left(\pi,\pi/2\right)$.}
\label{fig:alpha_LL} 
\end{figure}

\emph{Case 1: Numerical simulations of Eq.~\eqref{eq:gilbert}.}
Equation~\eqref{eq:gilbert} is usually solved by explicit or implicit finite
differences~\cite{Weinan2000}.  We solve the equation by these methods, and
by the explicit spectral method~\cite{Zhu_numerics}.  The accuracy and computational
cost
is roughly the same
in each case, and for simplicity, we therefore employ explicit finite differences;
 it is this method we use throughout the chapter.
 Given the initial conditions~\eqref{eq:initial}, each component of the magnetization
 $\Mag=\left(m_x,m_y,m_z\right)$ tends to a constant, the energy
\[
E = \tfrac{1}{2}\int_\Omega{dx}\left|\frac{\partial\Mag}{\partial{x}}\right|^2
\]
decays with time, and $\left|\Mag\right|^2$ retains its initial value $|\Mag|^2=1$.
 After some transience, the decay of the energy functional becomes exponential
 in time.  These results are shown in Fig.~\ref{fig:alpha_LL}

\begin{figure}[htb]
\centering
\subfigure[]{
\includegraphics[width=0.3\textwidth]{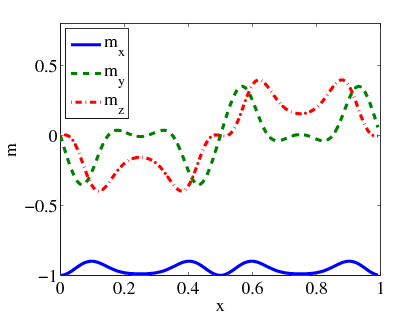}
}
\subfigure[]{
\includegraphics[width=0.3\textwidth]{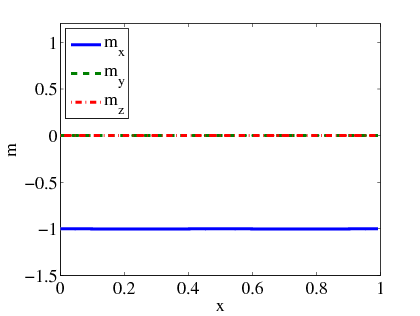}
}
\subfigure[]{
\includegraphics[width=0.3\textwidth]{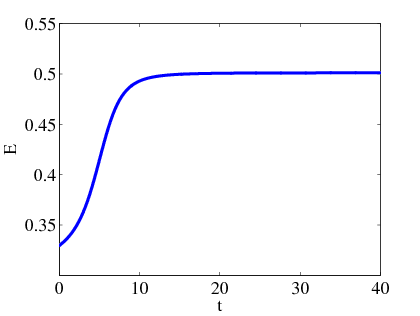}
}
\caption{Numerical simulations of Case~(2), the nonlocal
Gilbert equation
with with $\alpha<\beta$.  In this case, the energy increases to a constant
value, and the magnetization becomes constant.
%
%
%
%
Subfigures (a) and (b) show the magnetization
at times $t=8$ and $t=40$; (c) is the energy functional.  The final orientation
is $\left(\phi,\theta\right)=\left(\pi,\pi/2\right)$.}
\label{fig:alpha_small} \end{figure}

\emph{Case 2: Numerical simulations of Eq.~\eqref{eq:mag_eqn} with
$\alpha<\beta$}.  Given the smooth initial data~\eqref{eq:initial},
in time each component of the magnetization $\Mag=\left(m_x,m_y,m_z\right)$
decays to zero, while the energy
\[
E=\tfrac{1}{2}\int_\Omega{dx}\Mag\cdot\left(1-\alpha^2\partial_x^2\right)^{-1}\Mag
\]
tends to a constant value.  Given our choice of initial conditions, the energy
in fact \emph{increases} to attain this constant value.  Again the quantity
$\left|\Mag\right|^2$ stays constant.  These results are shown in Fig.~\ref{fig:alpha_small}.
We find similar results when we set $\alpha=0$.

\begin{figure}[htb]
\centering
\subfigure[]{
\includegraphics[width=0.3\textwidth]{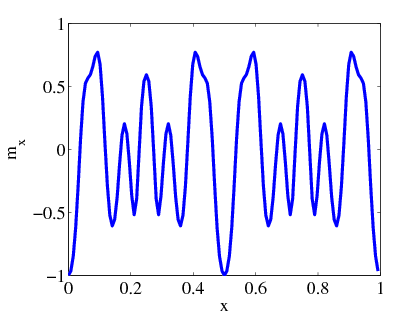}
}
\subfigure[]{
\includegraphics[width=0.3\textwidth]{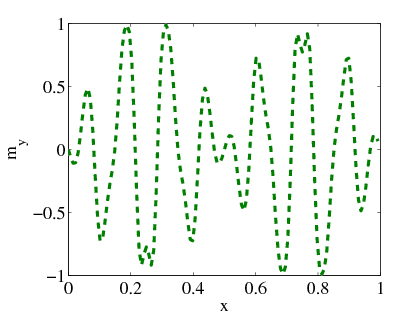}
}
\subfigure[]{
\includegraphics[width=0.3\textwidth]{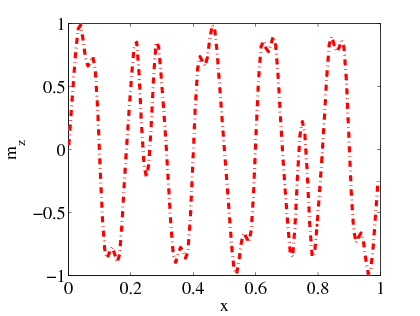}
}
\subfigure[]{
\includegraphics[width=0.3\textwidth]{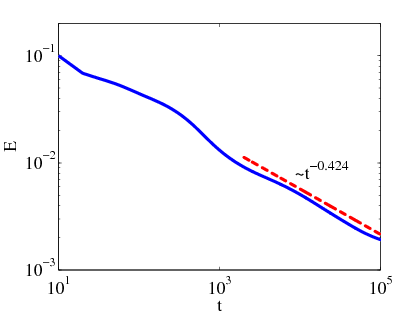}
}
\subfigure[]{
\includegraphics[width=0.3\textwidth]{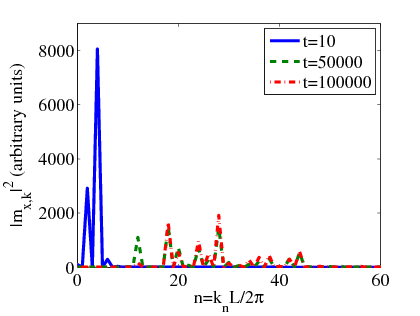}
}
\caption{Numerical simulations of Case~(3), the nonlocal
Gilbert equation
with $\alpha>\beta$.  In this case, the energy decreases indefinitely, and
the magnetization vector develops finer and finer scales.
%
%
%
%
Subfigures (a), (b), and (c) show the magnetization at time $t=10000$; (d)
is the energy functional, which decreases in time as a power law at late
times.  Subfigure (e) shows the power spectrum of $m_x$;
the integer index $n$ labels the spatial scales: if $k_n$ is a wavenumber,
then the corresponding integer label is $n=k_nL/2\pi$.
}
\label{fig:alpha_big} 
\end{figure}

\emph{Case 3: Numerical simulations of Eq.~\eqref{eq:mag_eqn} with
$\alpha>\beta$}.  Given the smooth initial data~\eqref{eq:initial}, in time
each component of the magnetization $\Mag=\left(m_x,m_y,m_z\right)$ develops
finer and finer scales.  
The development of small scales is driven by the decreasing nature of the
energy functional, which decreases as power law at late times, and is reflected
in snapshots of the power spectrum of the magnetization vector, shown in
Fig.~\ref{fig:alpha_big}.  As the system evolves, there
\begin{table}[h!b!p!]
\centering
\begin{tabular}{|c|c|c|c|c|}
\hline
Case&Lengthscales&Energy&Outcome as $t\rightarrow\infty$&Linear Stability\\
\hline
\hline
 &$\beta=0$, & & & \\[-1ex]
\raisebox{1.5ex}{(1)}&$\delta{E}/\delta{\Mag}=-\partial_x^2\Mag$&\raisebox{1.5ex}{Decreasing}&\raisebox{1.5ex}{Constant
state}&\raisebox{1.5ex}{Stable}\\
\hline
 & & & & \\[-1ex]
\raisebox{1.5ex}{(2)}&\raisebox{1.5ex}{$\alpha<\beta$}&\raisebox{1.5ex}{Increasing}&\raisebox{1.5ex}{Constant
state}&\raisebox{1.5ex}{Stable}\\
\hline
 & & & Development of finer & \\[-1ex]
\raisebox{1.5ex}{(3)}&\raisebox{1.5ex}{$\alpha>\beta$}&\raisebox{1.5ex}{Decreasing}&and
finer scales&\raisebox{1.5ex}{Unstable}\\
\hline
\end{tabular}
\caption{Summary of the forms of Eq.~\eqref{eq:mag_eqn} studied.}
\label{tab:table_summary}
\end{table}
is a transfer
of large amplitudes to higher wavenumbers.  This transfer slows down at
late
times, suggesting that the rate at which the solution roughens tends to zero,
as $t\rightarrow\infty$.
The evolution preserves the symmetry of the magnetization
vector $\Mag\left(x,t\right)$ under parity transformations.  This is
seen by comparing Figs.~\ref{fig:initial} and~\ref{fig:alpha_big}.
The energy is a decaying function of time, while the quantity
$\left|\Mag\right|^2$ stays constant.  We find similar results for the case
when $\beta=0$.

These results can be explained qualitatively as follows.  In Case~(1), the
energy functional exacts a penalty for the formation of gradients.  The energy
decreases with time and the system evolves into a state in
which no magnetization gradients are present, that is, a constant state.
On the other hand, we have demonstrated that in Case~(2), when $\alpha<\beta$,
the energy increases to a constant value.  Since in the nonlocal model, the
energy functional represents the cost of forming smooth spatial structures,
an increase in energy produces a smoother magnetization field, a process
that continues until the  magnetization reaches a constant value.  
Finally, in Case~(3), when $\alpha>\beta$, the energy functional
decreases, and this decrease corresponds to a roughening of the magnetization
field, as seen in Fig.~\ref{fig:alpha_big}. 
In Sec.~\ref{sec:mag_dens} we
shall show that Case~(2) is stable to small perturbations around a constant
state, while Case~(3) is unstable.  
%
%
%
These results are summarized in Table~\ref{tab:table_summary}.

The solutions of Eqs.~\eqref{eq:mag_eqn} and~\eqref{eq:gilbert} do not become
singular.  This is not surprising: the manifest
conservation of $\left|\Mag\right|^2$ in Eqs.~\eqref{eq:mag_eqn}
and~\eqref{eq:gilbert} provides a pointwise bound on the magnitude of the
solution, preventing blow-up.  Any addition
to Eq.~\eqref{eq:mag_eqn} that breaks this conservation law gives rise to
the possibility of singular solutions, and it is to this possility that we
now turn.

\section{Coupled density-magnetization equations}
\label{sec:mag_dens}
In this section we study a coupled density-magnetization equation
pair that admit singular solutions.
 We investigate the linear stability
of the equations and examine the conditions for instability.  We find that
the stability or otherwise of a constant state is controlled by that state's
magnetization and density, and by the relative magnitude of the problem
lengthscales.  Using numerical and analytical techniques, we investigate
the emergence and self-interaction of singular solutions.

The equations we study are as follows,
\begin{subequations}
\begin{equation}
\frac{\partial\ordp}{\partial t}=\frac{\partial}{\partial{x}}\left[\ordp\left(\mu_\ordp\frac{\partial}{\partial{x}}\frac{\delta{E}}{\delta\ordp}+\bm{\mu}_{\Subm}\cdot\frac{\partial}{\partial{x}}\frac{\delta{E}}{\delta\bm{m}}\right)\right],
\end{equation}
\begin{equation}
\frac{\partial\Mag}{\partial t} = \frac{\partial}{\partial x}\left[\Mag\left(\mu_\ordp\frac{\partial}{\partial{x}}\frac{\delta{E}}{\delta\ordp}+\bm{\mu}_{\Subm}\cdot\frac{\partial}{\partial{x}}\frac{\delta{E}}{\delta\Mag}\right)\right]+\Mag\times\left(\bm{\mu}_{\Subm}\times\frac{\delta{E}}{\delta\Mag}\right),
\end{equation}%
\label{eq:mag_dens}%
\end{subequations}%
where we set
\[
\mu_\ordp=1,\qquad\frac{\delta{E}}{\delta\ordp}=-\left(1-\alpha_\rho^2\partial_x^2\right)^{-1}\ordp,
\]
and, as before,
\[
\bm{\mu}_{\Subm} = \left(1-\beta_\Subm^2\partial_x^2\right)^{-1}\Mag,\qquad
\frac{\delta E}{\delta\Mag} = \left(1-\alpha_\Subm^2\partial_x^2\right)^{-1}\Mag.
\]
The order parameter $\ordp\left(\bm{x},t\right)$ has the interpretation of
a particle density, while the orientation vector $\bm{m}\left(\bm{x},t\right)$
has the interpretation of a local magnetization field.
These equations have been introduced by Holm, Putkaradze and Tronci in~\cite{Darryl_eqn1},
using a kinetic-theory description.
%
%
The density and the magnetization vector are driven by the velocity
\begin{equation}
V=\mu_\ordp\frac{\partial}{\partial{x}}\frac{\delta{E}}{\delta\ordp}+\bm{\mu}_{\Subm}\cdot\frac{\partial}{\partial{x}}\frac{\delta{E}}{\delta\bm{m}}.
\label{eq:velocity}
\end{equation}
The velocity advects the ratio $|\Mag|/\ordp$ by
\[
\left(\frac{\partial}{\partial{t}}-V\frac{\partial}{\partial{x}}\right)\frac{|\Mag|}{\ordp}=0.
\]
We have the system energy
\begin{equation}
E = \tfrac{1}{2}\int_\Omega{dx}\Mag\cdot\left(1-\alpha_\Subm^2\partial_x^2\right)^{-1}\Mag
- \tfrac{1}{2}\int_\Omega{dx}\ordp\left(1-\alpha_\ordp^2\partial_x^2\right)^{-1}\ordp,
\label{eq:energy}
\end{equation}
and, given a nonnegative density, the second term is always nonpositive.
 This represents an energy of attraction, and we therefore expect singularities
 in the magnetization vector to arise from a collapse of the particle density
 due to the ever-decreasing energy of attraction.  There are three lengthscales
 in the problem that control the time evolution: the
 ranges $\alpha_\Subm$ and $\alpha_\ordp$ of the potentials in Eq.~\eqref{eq:energy},
 and the smoothening length $\beta_\Subm$.
\subsection*{Linear stability analysis}
We study the linear stability of the constant state $\left(\Mag,\ordp\right)=\left(\Mag_0,\ordp_0\right)$.
We evaluate the smoothened values of this constant solution as follows,
\begin{eqnarray*}
\left(1-\alpha_\ordp^2\partial_x^2\right)^{-1}\ordp_0&=&f\left(x\right),\\
\ordp_0&=&f\left(x\right)-\alpha_\ordp^2\frac{d^2 f}{dx^2},\\
f\left(x\right)&=&\ordp_0+A\sinh\left(x/\alpha_\ordp\right)+B\cosh\left(x/\alpha_\ordp\right).
\end{eqnarray*}
For periodic or infinite boundary conditions, the constants $A$ and
$B$ are in fact zero and thus 
\begin{equation}
\left(1-\alpha_\ordp\partial_x^2\right)^{-1}\ordp_0=\ordp_0,
\label{eq:smooth_const}
\end{equation}
and similarly $\bm{\mu}_0=\left(\delta E/\delta\Mag\right)_{\Mag_0}=\Mag_0$.
 The result~\eqref{eq:smooth_const} guarantees that the constant state $\left(\Mag_0,\ordp_0\right)$
 is indeed a solution of Eq.~\eqref{eq:mag_dens}.
 
We study a solution $\left(\Mag,\ordp\right)=\left(\Mag_0+\delta\Mag,\ordp_0+\delta\ordp\right)$,
which represents a perturbation away from the constant state.  By assuming
that $\delta\Mag$ and $\delta\ordp$ are initially small in magnitude, we obtain
the following linearized equations for the perturbation density and magnetization,
\begin{subequations}
\begin{equation*}
\frac{\partial}{\partial t}\delta\ordp=-\ordp_0\frac{\partial^2}{\partial{x}^2}\left(1-\alpha_\ordp^2\partial_x^2\right)^{-1}\delta\ordp+\ordp_0\frac{\partial^2}{\partial{x}^2}\left(1-\alpha_\Subm^2\partial_x^2\right)^{-1}\Mag_0\cdot\delta\Mag,
\end{equation*}
\begin{multline*}
\frac{\partial}{\partial t}\delta\Mag =\Mag_0\left[-\frac{\partial^2}{\partial{x}^2}\left(1-\alpha_\ordp^2\partial_x^2\right)^{-1}\delta\ordp+\frac{\partial^2}{\partial{x}^2}\left(1-\alpha_\Subm^2\partial_x^2\right)^{-1}\Mag_0\cdot\delta\Mag\right]\\
+\Mag_0\times\Big\{\Mag_0\times\left[\left(1-\alpha_m^2\partial_x^2\right)^{-1}\delta\Mag-\left(1-\beta_m^2\partial_x^2\right)^{-1}\delta\Mag\right]\Big\}.
\end{multline*}%
\label{eq:mag_dens_linear}%
\end{subequations}%
 For $\Mag_0\neq0$ we may choose two unit vectors $\hat{\bm{n}}_1$ and $\hat{\bm{n}}_2$
 such that $\Mag_0/|\Mag_0|$, $\hat{\bm{n}}_1$ and $\hat{\bm{n}}_2$ form
 an orthonormal triad (that is, we have effected a change of basis).  We
 then study the quantities $\delta\ordp$, $\delta\chi$, $\delta\xi_1$ and
 $\delta\xi_2$, where
\[
\delta\chi=\Mag_0\cdot\delta\Mag,\qquad\delta\xi_1=\hat{\bm{n}}_1\cdot\delta\Mag,\qquad\delta\xi_2=\hat{\bm{n}}_2\cdot\delta\Mag.
\]
We obtain the linear equations
\begin{equation*}
\frac{\partial}{\partial t}\delta\ordp=-\ordp_0\frac{\partial^2}{\partial{x}^2}\left(1-\alpha_\ordp^2\partial_x^2\right)^{-1}\delta\ordp+\ordp_0\frac{\partial^2}{\partial{x}^2}\left(1-\alpha_\Subm^2\partial_x^2\right)^{-1}\delta\chi,
\end{equation*}
\begin{equation*}
\frac{\partial}{\partial t}\delta\chi =-|\Mag_0|^2\frac{\partial^2}{\partial{x}^2}\left(1-\alpha_\ordp^2\partial_x^2\right)^{-1}\delta\ordp+|\Mag_0|^2\frac{\partial^2}{\partial{x}^2}\left(1-\alpha_\Subm^2\partial_x^2\right)^{-1}\delta\chi,
\end{equation*}
\begin{equation*}
\frac{\partial}{\partial t}\delta\xi_i=\left[\left(1-\beta_\Subm^2\partial_x^2\right)^{-1}-\left(1-\alpha_\Subm^2\partial_x^2\right)^{-1}\right]\delta\xi_i,\qquad
i=1,2.
\end{equation*}%
By focussing on a single-mode disturbance with wavenumber $k$ we obtain the
following system of equations
\begin{equation*}
\frac{d}{dt}\left(\begin{array}{c}\delta\ordp \\ \delta\chi \\ \delta\xi_1
\\ \delta\xi_2\end{array}\right)=
\left(\begin{array}{cccc}
\frac{\ordp_0k^2}{1+\alpha_\ordp^2k^2}&-\frac{\ordp_0k^2}{1+\alpha_\Subm^2k^2}&0&0\\
\frac{|\Mag_0|^2k^2}{1+\alpha_\ordp^2k^2}&-\frac{|\Mag_0|^2k^2}{1+\alpha_\Subm^2k^2}&0&0\\
0&0&\frac{1}{1+\beta_\Subm^2k^2}-\frac{1}{1+\alpha_\Subm^2k^2}&0\\
0&0&0&\frac{1}{1+\beta_\Subm^2k^2}-\frac{1}{1+\alpha_\Subm^2k^2}
\end{array}\right)
\left(\begin{array}{c}\delta\ordp \\ \delta\chi \\ \delta\xi_1
\\ \delta\xi_2\end{array}\right),
\end{equation*}
%
%
%
%
%
%
with eigenvalues
\begin{equation}
\sigma_0=0,\qquad 
\sigma_1=\frac{\ordp_0k^2}{1+\alpha_\ordp^2k^2}-\frac{|\Mag_0|^2k^2}{1+\alpha_\Subm^2k^2},\qquad
\sigma_2=\frac{1}{1+\beta_\Subm^2k^2}-\frac{1}{1+\alpha_\Subm^2k^2}.
\end{equation}
The eigenvalues are the growth rate of the disturbance $\left(\delta\ordp,\delta\chi,\delta\xi_1,\delta\xi_2\right)$~\cite{ChandraFluids}.
 There are two routes to instability, when $\sigma_1>0$, or when $\sigma_2>0$.
  The first route leads to an instability when
\[
\sigma_1>0,\qquad \frac{\ordp_0}{|\Mag_0|^2}>\frac{1+\alpha_\ordp^2k^2}{1+\alpha_\Subm^2k^2},
\]
while the second route leads to instability when
\[
\sigma_2>0,\qquad \alpha_\Subm>\beta_\Subm.
\] 
%
%
%
%
%
%
%
 
\begin{figure}[htb]
\centering
\subfigure[]{
\includegraphics[width=0.35\textwidth]{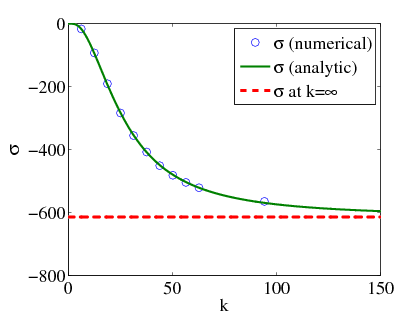}
}
\subfigure[]{
\includegraphics[width=0.35\textwidth]{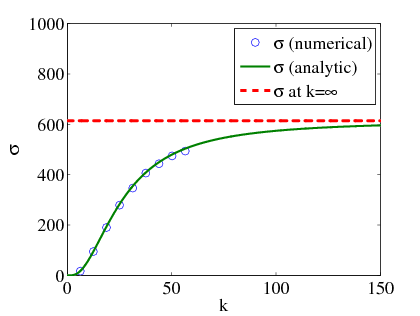}
}
\caption{The first route to instability.   Subfigure (a) shows
the growth
rate $\sigma_1$ for $\alpha_\Subm<\beta_\Subm<\alpha_\ordp$, with negativity
indicating a stable equilibrium; (b) gives the growth rate $\sigma_1$
for $\alpha_\ordp<\alpha_\Subm<\beta_\Subm$, with positivity indicating an
unstable equilibrium.  We have set $|\Mag_0|=\ordp_0=1$.
}
\label{fig:growth_rates} 
\end{figure}

\begin{figure}[htb]
\centering
\subfigure[]{
\includegraphics[width=0.35\textwidth]{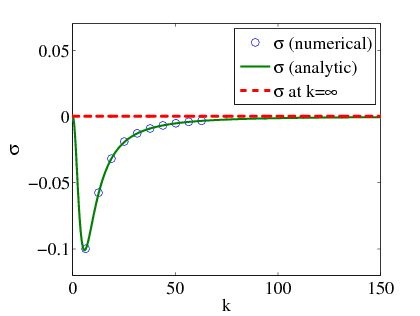}
}
\subfigure[]{
\includegraphics[width=0.35\textwidth]{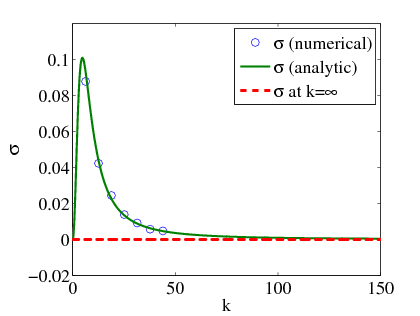}
}
\caption{The second route to instability.   Subfigure (a)
shows the growth
rate $\sigma_2$ for $\alpha_\Subm<\beta_\Subm$, with negativity
indicating a stable equilibrium; (b) gives the growth rate $\sigma_2$
for $\alpha_\Subm>\beta_\Subm$, with positivity indicating an unstable equilibrium.
 We have set $|\Mag_0|=\ordp_0=1$.   
}
\label{fig:growth_rates2} 
\end{figure}

We have plotted the growth rates for the case when $\ordp_0=|\Mag_0|^2=1$,
and compared the theory with numerical simulations.  There is excellent agreement
at low wavenumbers, although the numerical simulations become less accurate
at high wavenumbers.  This can be remedied by increasing the resolution
of the simulations.  These plots are shown in Figs.~\ref{fig:growth_rates}
and~\ref{fig:growth_rates2}.
The growth rates $\sigma_{1,2}$ are parabolic in $k$ at small $k$; $\sigma_1$
saturates at large $k$, while $\sigma_2$ attains a maximum and decays at
large $k$.  The growth rates can be positive or negative, depending on the
initial configuration,
and on the relationship between the problem lengthscales.
 In contrast to some standard instabilities of pattern formation (e.g. Cahn--Hilliard~(Sec.~\ref{sec:background:ch})
 or Swift--Hohenberg~\cite{OjalvoBook}), the $\sigma_1$-unstable state
 becomes more unstable at higher wavenumbers (smaller scales), thus preventing
 the `freezing-out' of the instability by a reduction of the box size~\cite{Argentina2005}.
  The growth at small scales is limited,
however, by the saturation in $\sigma$ as $k\rightarrow\infty$.  Heuristically,
 this can be explained as follows: at higher wavenumber, the disturbance
  $\left(\delta\ordp,\delta\chi,\delta\xi_1,\delta\xi_2\right)$ gives rise
  to more and more peaks per unit length.  This makes merging events increasingly
  likely, so that peaks combine to form larger peaks, enhancing the growth
  of the disturbance.

Recall in Sec.~\ref{sec:gilbert} that the different behaviours of the magnetization
equation~\eqref{eq:mag_eqn} are the result of a competition between the lengthscales
$\alpha_\Subm$ and $\beta_\Subm$.  For $\alpha_\Subm<\beta_\Subm$ the initial
(large-amplitude) disturbance tends to a constant, while for $\alpha_\Subm>\beta_\Subm$
the initial disturbance develops finer and finer scales.  In this section,
we have shown that the coupled density-magnetization equations are linearly
stable when $\alpha_\Subm<\beta_\Subm$, while the reverse case is unstable.
 In contrast to the first route to instability, the growth rate $\sigma_2$,
 if positive, admits a maximum.  This is obtained by setting $\sigma_2'\left(k\right)=0$.
  Then the maximum growth rate occurs at a scale
\begin{equation*}
\lambda_{\mathrm{max}}:=2\pi k_{\mathrm{max}}^{-1}=2\pi\sqrt{\alpha_\Subm\beta_\Subm}.
\end{equation*}
Thus, the scale at which the disturbance is most unstable is determined
by the geometric mean of $\alpha_\Subm$ and $\beta_\Subm$.  Given a disturbance
$\left(\delta\ordp,\delta\chi,\delta\xi_1,\delta\xi_2\right)$ with a range
of modes initially present, the instability selects the disturbance on the
scale $\lambda_{\mathrm{max}}$.  This disturbance develops a large amplitude
and a singular solution subsequently emerges.  It is to this aspect of the
problem that we now turn.

\subsection*{Singular solutions}
%
%
%
%
 In this section we show that a finite weighted sum of delta functions satisfies
 the partial differential equations~\eqref{eq:mag_dens}.  Each delta function
 has the
 interpretation of a particle or \emph{clumpon}, whose weights and positions satisfy
 a finite set of ordinary differential equations.  We investigate the two-clumpon
 case analytically and show that the clumpons tend to a state in which they
 merge, diverge, or are separated by a fixed distance.  In each case, we
 determine the final state of the two clumpons' relative orientation.
 

To verify that singular solutions are possible, let us substitute the ansatz
\begin{equation}
\ordp\left(x,t\right)=\sum_{i=1}^M a_i\left(t\right)\delta\left(x-x_i\left(t\right)\right),\qquad
\Mag\left(x,t\right)=\sum_{i=1}^M \bm{b}_i\left(t\right)\delta\left(x-x_i\left(t\right)\right),
\label{eq:clumpon_sln}
\end{equation}
into the weak form of equations~\eqref{eq:mag_dens}.  Here we sum over the
different components of the singular solution (which we call clumpons).
In this section we work on the infinite domain $x\in\left(-\infty,\infty\right)$.
 The weak form of the equations is obtained by testing Eqs.~\eqref{eq:mag_dens}
with once-differentiable functions $f\left(x\right)$ and $\bm{g}\left(x\right)$,
\begin{subequations}
\begin{equation}
\frac{d}{dt}\Int{dx}\ordp\left(x,t\right)f\left(x\right)=-\Int{dx}f'\left(x,t\right)\left(\mu_\ordp\frac{\partial}{\partial{x}}\frac{\delta{E}}{\delta\ordp}+\bm{\mu}_{\bm{m}}\cdot\frac{\partial}{\partial{x}}\frac{\delta{E}}{\delta\bm{m}}\right),
\end{equation}
\begin{multline}
\frac{d}{dt}\Int{dx}\Mag\left(x,t\right)\cdot\bm{g}\left(x\right) 
= -\Int{dx}\bm{g}'\left(x\right)\cdot\Mag\left(x,t\right)\left(\mu_\ordp\frac{\partial}{\partial{x}}\frac{\delta{E}}{\delta\ordp}
+\bm{\mu}_{\Subm}\cdot\frac{\partial}{\partial{x}}\frac{\delta{E}}{\delta\Mag}\right)
\\
+\Int{dx}\bm{g}\left(x\right)\cdot\left[\Mag\times\left(\bm{\mu}_{\Subm}\times\frac{\delta{E}}{\delta\Mag}\right)\right],
\end{multline}%
\label{eq:mag_dens_weak}%
\end{subequations}%
Substitution of the ansatz~\eqref{eq:clumpon_sln} into the weak equations~\eqref{eq:mag_dens_weak}
yields the relations
\begin{equation}
\frac{da_i}{dt}=0,\qquad \frac{dx_i}{dt}=-V\left(x_i\right),\qquad\frac{d\bm{b}_i}{dt}=\bm{b}_i\times\left(\bm{\mu}_\Subm\times\frac{\delta{E}}{\delta\Mag}\right)\left(x_i\right),\qquad
i\in\{1,...,M\},
\label{eq:clumpon_evolution}
\end{equation}
where $V$ and $\left(\bm{\mu}_\Subm\times\left({\delta{E}}/{\delta\Mag}\right)\right)$
are obtained from the ansatz~\eqref{eq:clumpon_sln} and are evaluated at
$x_i$.  Note that the density weights $a_i$ and the magnitude of the
weights $\bm{b}_i$ remain constant in time.

We develop further understanding of the clumpon dynamics  by studying the
two-clumpon version of Eqs.~\eqref{eq:clumpon_evolution}.  Since the weights
$a_1$, $a_2$, $|\bm{b}_1|$, and $|\bm{b}_2|$ are constant, two variables
%
%
%
%
suffice to describe the interaction: the relative separation $x=x_1-x_2$
of the clumpons,
and the cosine of the angle between the clumpon magnetizations, $\cos\varphi=\bm{b}_1\cdot\bm{b}_2/|\bm{b}_1||\bm{b}_2|$.
Using the properties
of the kernel $H\left(0\right)=1$, $H'\left(0\right)=0$, we derive the equations
%
%
%
%
%
\begin{subequations}
\begin{equation}
\frac{dx}{dt}=M H_{\alpha_\ordp}'\left(x\right)- B_1 H'_{\alpha_m}\left(x\right)H_{\beta_m}\left(x\right)
- B_2 H_{\alpha_m}'\left(x\right)y,\qquad y=\cos\varphi
\end{equation}
\begin{equation}
\frac{dy}{dt}=B_2\left(1-y^2\right)\left[H_{\beta_m}\left(x\right)-H_{\alpha_m}\left(x\right)\right],
\label{eq:xtheta_b}
\end{equation}%
\label{eq:xtheta}%
\end{subequations}%
where $M=a_1+a_2$, $B_1=|\bm{b}_1|^2+|\bm{b}_2|^2$, and $B_2=2|\bm{b}_1||\bm{b}_2|$
are constants.  
In this framework, $H_\alpha\left(x\right)$ is the Green's function of the
Helmholtz operator $\left(1-\alpha^2\Delta\right)^{-1}$, $H_\alpha\left(x\right)\propto
e^{-\alpha|x|}$.  Equations~\eqref{eq:xtheta} form a dynamical system whose
properties we now investigate using phase-plane analysis~\cite{StrogatzBook}.
 We note first
\begin{figure}[htb]
\centering
\subfigure[]{
\includegraphics[width=0.35\textwidth]{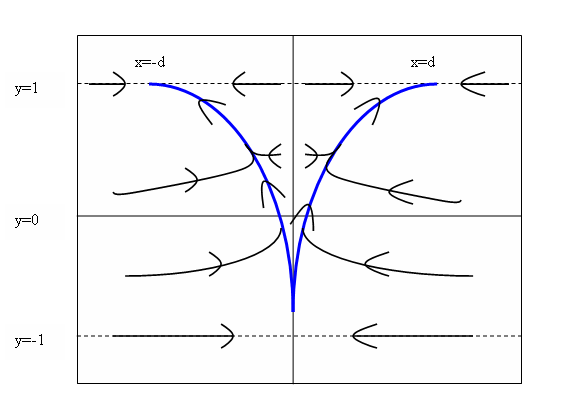}
}
\subfigure[]{
\includegraphics[width=0.35\textwidth]{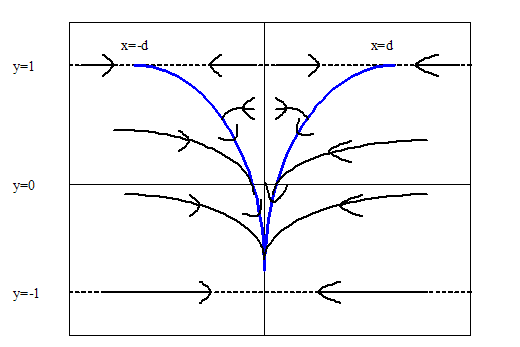}
}
\caption{ The nullcline $dx/dt=0$ of the two-clumpon dynamical
system with $\alpha_m<\alpha_\rho$.  The
region contained inside the dotted lines $y=\pm1$ gives the allowed values
of the dynamical variables $\left(x,y\right)$.
 Subfigure~(a)
shows the case when $\beta_m<\alpha_m$.  The stable equilibria of the
system are $\left(x,y\right)=\left(\pm{d},1\right)$ and the line $x=0$.
 All initial conditions flow into one of these equilibrium states; subfigure
 (b) shows the case when $\alpha_m<\beta_m$.  Initial conditions confined
 to the
 line $y=1$ flow into the fixed point $\left(\pm{d},1\right)$,
  while all other initial conditions flow into the line $x=0$.}
\label{fig:nullclines} 
\end{figure} 
of all that the $|y|>1$ region of the phase plane is forbidden, since the
$y$-component of the vector field $\left(dx/dt,dy/dt\right)$ vanishes at
$|y|=1$.  The vertical lines $x=0$ and $x=\pm\infty$ are equilibria, although
\begin{figure}[htb]
\centering
\subfigure[]{
\includegraphics[width=0.35\textwidth]{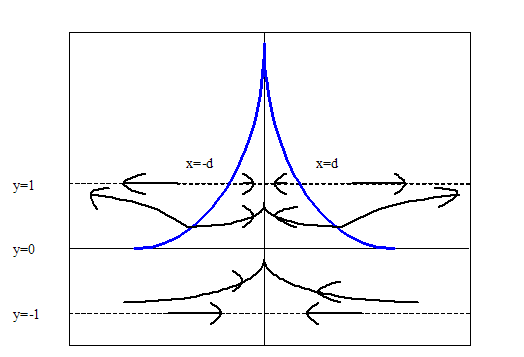}
}
\subfigure[]{
\includegraphics[width=0.35\textwidth]{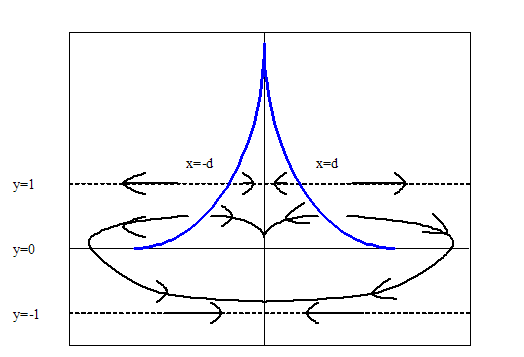}
}
\caption{The nullcline $dx/dt=0$ of the two-clumpon dynamical
system with $\alpha_\rho<\alpha_m$.  The
region contained inside the dotted lines $y=\pm1$ gives the allowed values
of the dynamical variables $\left(x,y\right)$.
 Subfigure~(a)
shows the case when $\beta_m<\alpha_m$.  The lines $x=0$ and $x=\pm\infty$
form the stable equilibria of the system.  All initial conditions flow
 into one of these states; subfigure (b)
 shows the case when $\alpha_m<\beta_m$.  
Initial conditions confined to the  line $y=1$ flow into the fixed points
$\left(0,1\right)$ and $\left(\pm\infty,1\right)$, while all other initial
conditions flow into the line $x=0$. 
}
\label{fig:nullclines2} 
\end{figure} 
their stability will depend on the value of the parameters $\left(\alpha_m,\alpha_\ordp,\beta_m,B_1,B_2,M\right)$.
The curve across which $dx/dt$ changes sign is called the nullcline.  This
is given by
\[
y=\frac{MH_{\alpha_\ordp}'\left(x\right)-B_1H'_{\alpha_m}\left(x\right)H_{\beta_m}\left(x\right)}{B_2H'_{\alpha_m}\left(x\right)},
\]
which on the domain $x\in\left(-\infty,\infty\right)$ takes the form
\[
y=\frac{\alpha_m}{B_2}\left[\frac{M}{\alpha_\ordp}e^{-|x|\left(\frac{1}{\alpha_\ordp}-\frac{1}{\alpha_m}\right)}-\frac{B_1}{\alpha_m}e^{-\frac{1}{\beta_m}|x|}\right].
\]
Several qualitatively different behaviours are possible, depending on the
\begin{table}[h!b!p!]
\centering
\begin{tabular}{|c|c|c|c|c|}
\hline
Case&$\alpha_m$ vs. $\alpha_\ordp$&$\alpha_m$ vs. $\beta_m$&Equilibria&Flow\\
\hline
\hline
%
%
 & & &$\left(x,y\right)=\left(\pm{d},1\right)$; & Flow into $x=0$ and\\[-1ex]
\raisebox{1.5ex}{(1)}&\raisebox{1.5ex}{$\alpha_m<\alpha_\ordp$}&\raisebox{1.5ex}{$\beta_m<\alpha_m$}&
$x=0$; $x=\pm\infty$&$\left(x,y\right)=\left(\pm{d},1\right)$\\
\hline
%
 & & &$\left(x,y\right)=\left(\pm{d},1\right)$ & Flow into $x=0$ and\\[-1ex]
\raisebox{1.5ex}{(2)}&\raisebox{1.5ex}{$\alpha_m<\alpha_\ordp$}&\raisebox{1.5ex}{$\alpha_m<\beta_m$}&
$x=0$; $x=\pm\infty$&$\left(x,y\right)=\left(\pm{d},1\right)$\\
\hline
%
%
%
 & & &$\left(x,y\right)=\left(\pm{d},1\right)$; & Flow into $x=0$\\[-1ex]
\raisebox{1.5ex}{(3)}&\raisebox{1.5ex}{$\alpha_\ordp<\alpha_m$}&\raisebox{1.5ex}{$\beta_m<\alpha_m$}&
$x=0$; $x=\pm\infty$&and
$x=\pm\infty$\\
\hline
%
%
%
 & & &$\left(x,y\right)=\left(\pm{d},1\right)$; & Flow into $x=0$\\[-1ex]
\raisebox{1.5ex}{(4)}&\raisebox{1.5ex}{$\alpha_\ordp<\alpha_m$}&\raisebox{1.5ex}{$\alpha_m<\beta_m$}&
$x=0$; $x=\pm\infty$&and
$x=\pm\infty$\\
\hline
\end{tabular}
\caption{Summary of the distinct phase portraits of Eq.~\eqref{eq:xtheta}
studied.}
\label{tab:table2_summary}
\end{table}
magnitude of the values taken by the parameters $\left(\alpha_m,\alpha_\ordp,\beta_m,B_1,B_2,M\right)$.
 Here we outline four of these behaviour types.
\begin{itemize} 
\item\emph{Case 1:}  The lengthscales are in the relation $\alpha_m<\alpha_\ordp$,
and $\beta_m<\alpha_m$.  The nullcline is shown in Fig.~\ref{fig:nullclines}.
There is flow into the fixed points $\left(x,y\right)=\left(\pm{d},1\right)$,
and into the line $x=0$,
while $y$ is a nondecreasing function of time, which follows
from Eq.~\eqref{eq:xtheta_b}.  The ultimate state of the system is thus $x=\pm{d}$,
$\varphi=0$ (alignment), or $x=0$ (merging). In the latter case the final
orientation is given by the integral of Eq.~\eqref{eq:xtheta_b},
\begin{multline}
\tan\left(\frac{\varphi}{2}\right)=\tan\left(\frac{\varphi_0}{2}\right)\exp\left[-B_2\int_0^\infty{dt}\left[H_{\beta_m}\left(x\left(t\right)\right)-H_{\alpha_m}\left(x\left(t\right)\right)\right]\right],\\
\varphi_0=\varphi\left(t=0\right).
\label{eq:theta_final}
\end{multline}
\item\emph{Case 2:}  The lengthscales are in the relation $\alpha_m<\alpha_\ordp$,
$\alpha_m<\beta_m$, and the nullcline is the same as Case~(2).  All flow
along the line $y=1$ is directed towards the fixed points $\left(\pm{d},1\right)$.
 All other initial conditions flow
 into the line $x=0$, since $y$ is now a
nonincreasing function of time.  The ultimate state of the
system is thus $x=\pm{d}$, $\varphi=0$ (alignment), or $x=0$ (merging). In
the latter case the final orientation is given by the formula~\eqref{eq:theta_final}.
\item\emph{Case 3:}  The lengthscales are in the relation $\alpha_\ordp<\alpha_m$
and $\beta_m<\alpha_m$.  The nullcline is shown in Fig.~\ref{fig:nullclines2}.
 Inside the region bounded by the line $y=0$ and the nullcline, the flow
 is into the line $x=0$ (merging), and the fixed points
 $\left(\pm{d},1\right)$ are unstable.  The flow below the line $y=0$ is
 towards the line
 $x=0$.  Outside of these regions, however, the flow is into the lines
 $x=\pm\infty$, which shows that for a suitable choice of parameters
 and initial conditions, the clumpons can be made to diverge.
\item\emph{Case 4:}   The lengthscales are in the relation $\alpha_\ordp<\alpha_m$
and $\alpha_m<\beta_m$, and the nullcline is the same as Case~(3).   The
quantity $y$ is a nonincreasing function of time. All flow along
the line $y=1$ is directed away from the fixed points $\left(\pm{d},1\right)$
and is into the fixed points $\left(0,1\right)$, or $\left(\pm\infty,1\right)$.
 All other initial conditions flow into $x=0$, although initial conditions
 that start above the curve formed by the nullcline flow in an arc and eventually
 reach a fixed point $\left(x=0,y<0\right)$.
\end{itemize}
We summarize the cases we have discussed in Table~\ref{tab:table2_summary}.
Using numerical simulations of Eqs.~\eqref{eq:xtheta}, we have verified that
Cases~(1)--(4) do indeed occur.  The list of cases we have considered is
not exhaustive: depending
on the parameters $B_1$, $B_2$, and $M$, other phase portraits may arise.
 Indeed, it is clear from Fig.~\ref{fig:nullclines} that through saddle-node
 bifurcations, the fixed points
 $\left(x,y\right)=\left(\pm{d},1\right)$ may disappear, or additional
 fixed points $\left(x,y\right)=\left(\pm{d'},-1\right)$ may appear.  Our
 analysis shows, however, that it is possible to choose
 a set of parameters $\left(\alpha_\ordp,\alpha_m,\beta_m,B_1,B_2,M\right)$
 such that two clumpons either merge, diverge, or are separated by a fixed
 distance.

\subsection*{Numerical Simulations}
To examine the emergence and subsequent interaction of the clumpons, we carry
out numerical simulations of Eq.~\eqref{eq:mag_dens} for a variety
of initial conditions.  We use
an explicit finite-difference algorithm with a small amount of artifical
diffusion.
 We solve the following weak form of Eq.~\eqref{eq:mag_dens}, obtained by
 testing Eq.~\eqref{eq:mag_dens_weak} with $H_{\beta_\Subm}$,
\[
\frac{\partial\overline{\ordp}}{\partial{t}}=D_{\mathrm{artif}}\frac{\partial^2\overline{\ordp}}{\partial{x^2}}+\int_\Omega{dy}H_{\beta_\Subm}'\left(x-y\right)\ordp\left(y,t\right)V\left(y,t\right),
\]
\begin{multline*}
\frac{\partial\mu_i}{\partial{t}}=D_{\mathrm{artif}}\frac{\partial^2\mu_i}{\partial{x^2}}+\int_\Omega{dy}H_{\beta_\Subm}'\left(x-y\right)\Mag_i\left(y,t\right)V\left(y,t\right)\\
+\int_\Omega{dy}H_{\beta_\Subm}\left(x-y\right)\bm{e}_i\cdot\left[\Mag\times\left(\bm{\mu}_\Subm\times\frac{\delta{E}}{\delta\Mag}\right)\right],
\end{multline*}
where $\overline{\ordp}=H_{\beta_m}*\ordp$ and $\bm{e}_i$ is the unit vector
in the $i^{\mathrm{th}}$ direction. 
We work on a periodic domain $\Omega=\left[-L/2,L/2\right]$, at a resolution of $250$
gridpoints; going to higher resolution does not noticeably increase
the accuracy of the results.

\begin{figure}[htb]
\centering
\subfigure[]{
\includegraphics[width=0.35\textwidth]{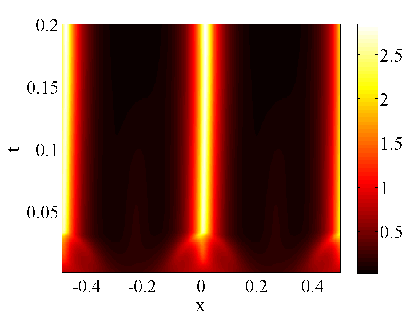}
}
\subfigure[]{
\includegraphics[width=0.35\textwidth]{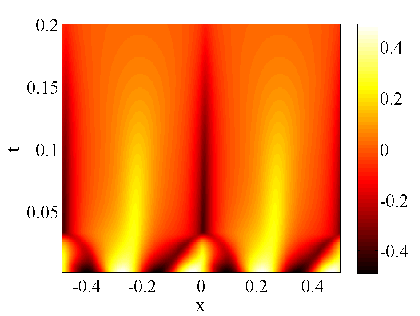}
}
\caption{Evolution of sinusoidally-varying initial conditions
for the density
and magnetization, as in Eq.~\eqref{eq:initial_conditions1}.
 Subfigure~(a) shows the evolution of $H_\ordp*\ordp$ for $t\in\left[0,0.15\right]$,
 by which time the initial data have formed two clumpons; (b) shows
 the evolution of $\mu_x$.  The profiles of $\mu_y$ and $\mu_z$ are similar.
  Note that
 the peaks in the density profile correspond to the troughs in the magnetization
 profile.  This agrees with the linear stability analysis, wherein disturbances
 in the density give rise to disturbances in the magnetization.
}
\label{fig:evolution_ic2} 
\end{figure}

%
%
The first set of initial conditions we study is the following,
\begin{eqnarray}
\Mag\left(x,0\right)&=&\left(\sin\left(4k_0x+\phi_x\right),\sin\left(4k_0x+\phi_y\right),\sin\left(4k_0x+\phi_z\right)\right),\nonumber\\
\ordp\left(x,0\right)&=&0.5+0.35\cos\left(2k_0x\right),
\label{eq:initial_conditions1}
\end{eqnarray}
where $\phi_x$, $\phi_y$, and $\phi_z$ are random phases in the interval
$\left[0,2\pi\right]$, and $k_0=2\pi/L$ is the fundamental wavenumber.  The
initial conditions for the magnetization vector are chosen to represent the
lack of a preferred direction in the problem.
The time evolution of
equations~\eqref{eq:mag_dens} for this set of initial conditions is shown
in Fig.~\ref{fig:evolution_ic2}.
After a short time, the initial data become singular, and subsequently,
the solution $\left(\ordp,\Mag\right)$ can be represented as a sum of clumpons,
\[
\ordp\left(x,t\right)=\sum_{i=1}^M a_i\delta\left(x-x_i\left(t\right)\right),\qquad
\Mag\left(x,t\right)=\sum_{i=1}^M \bm{b}_i\left(t\right)\delta\left(x-x_i\left(t\right)\right),\qquad
M=2.
\]
Here $M=2$ is the number of clumpons present at the singularity time.  
This
number corresponds to the number of maxima in the initial density profile.
 The forces exerted by each clumpon on the other balance
 because of the effect of the periodic boundary conditions.  Indeed,
 any number of equally-spaced, identical, interacting particles arranged
 on a ring are in equilibrium, although this equilibrium is unstable for
 an attractive force.  Thus, at late
 times, the clumpons are stationary, while the magnetization vector $\bm{\mu}_\Subm$
 shows alignment of clumpon magnetizations.
\begin{figure}[htb]
\centering
\subfigure[]{
\includegraphics[width=0.35\textwidth]{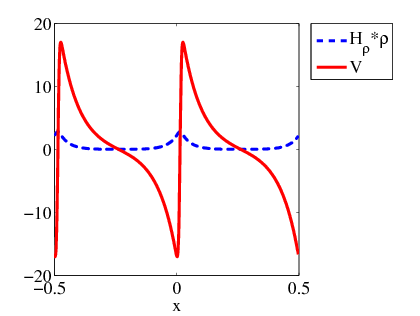}
}
\subfigure[]{
\includegraphics[width=0.35\textwidth]{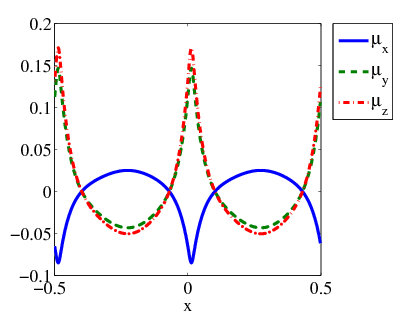}
}
\caption{Evolution of sinusoidally-varying initial conditions
for the density
and magnetization, as in Eq.~\eqref{eq:initial_conditions1}.
 Subfigure~(a) shows the system velocity $V$ given in Eq.~\eqref{eq:velocity},
just before the singularity time; (b) shows the magnetization $\bm{\mu}_\Subm$
at the same time.  The density maxima emerge at the locations where the convergence
of $-V$ (flow into $x=0$ and $x=\pm L/2$) occurs, and the magnetization develops
extrema there.}
\label{fig:snapshot_ic2} 
\end{figure} 

We gain further understanding of the formation of singular solutions by studying
the system velocity $V$ just before the onset of the singularity.  This is
done in Fig.~\ref{fig:snapshot_ic2}.
Figure~\ref{fig:snapshot_ic2}~(a) shows the development of the two clumpons
from the initial data.  Across each density maximum, the velocity has the
profile $V\approx\lambda\left(t\right)x$, where $\lambda\left(t\right)>0$
is an increasing function of time.  This calls to mind the advection problem
for the scalar $\theta\left(x,t\right)$, studied by Batchelor in the context
of passive-scalar mixing~\cite{Batchelor1959}
\[
\frac{\partial\theta}{\partial t}=\lambda_0x\frac{\partial\theta}{\partial{x}},\qquad
\lambda_0>0.
\]
Given initial data $\theta\left(x,0\right)=\theta_0 e^{-x^2/\ell_0^2}$, the
solution evolves in time as
\[
\theta\left(x,t\right)=\theta_0 e^{-x^2/\left(\ell_0^2e^{-2\lambda_0{t}}\right)},
\]
so that gradients are amplified exponentially in time,
\[
\frac{\partial\theta}{\partial{x}}=-\frac{2\theta_0}{\ell_0^2}xe^{\lambda_0{t}}
e^{-x^2/\left(\ell_0^2e^{-2\lambda_0{t}}\right)},
\]
in a similar manner to the problem studied.  

The evolution of the set of
initial conditions~\eqref{eq:initial_conditions1} has therefore demonstrated
the following:
the local velocity $V$ is such that before the onset of the singularity,
matter
is compressed into regions where $\ordp\left(x,0\right)$ is large, to such
an extent that the matter eventually accumulates at isolated points, and
the singular solution emerges.  Moreover, the density maxima, rather than
the magnetization
extrema, drive the formation of singularities.  This is not surprising, given
that the attractive part of the system's energy comes from density variations.
\begin{figure}[htb]
\centering
\subfigure[]{
\includegraphics[width=0.35\textwidth]{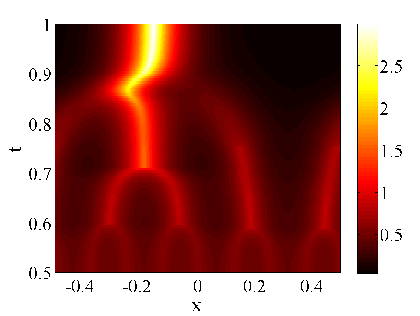}
}
\subfigure[]{
\includegraphics[width=0.35\textwidth]{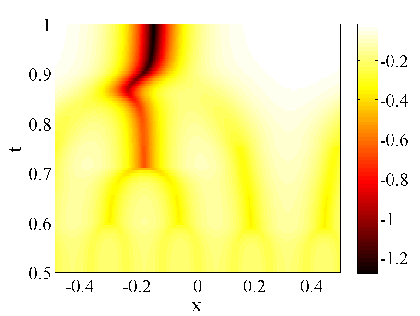}
}
\caption{Evolution of a flat magnetization field and a sinusoidally-varying
density, as in Eq.~\eqref{eq:initial_conditions2}.
 Subfigure~(a) shows the evolution of $H_\ordp*\ordp$ for $t\in\left[0.5,1\right]$;
 (b) shows the evolution of $\mu_x$.  The profiles of $\mu_y$ and $\mu_z$
 are similar.  At $t=0.5$, the initial data have formed eight equally spaced,
 identical clumpons, corresponding to the eight density maxima in the initial
 configuration.  By impulsively shifting the clumpon
 at $x=0$ by a small amount, the equilibrium is disrupted and the clumpons
 merge repeatedly until only one clumpon remains.
}
\label{fig:evolution_ic1} 
\end{figure}
%
%
%
%
%
%
%
%

\begin{figure}
\centering
\includegraphics[width=0.35\textwidth]{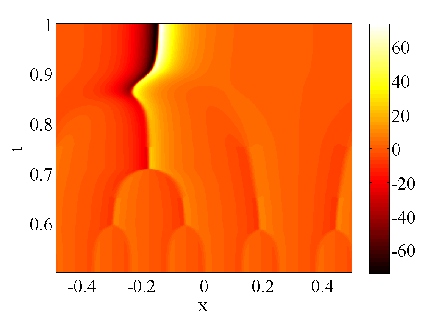}
\caption{Evolution of a flat magnetization field and a sinusoidally-varying
density, as in Eq.~\eqref{eq:initial_conditions2}.
 Shown is the velocity profile for $t\in\left[0.5,1\right]$; the system velocity
 is given by Eq.~\eqref{eq:velocity}.  The velocity $-V$ flows into each
 density
 maximum, concentrating matter at isolated points and precipitating the
 formation of eight equally-spaced identical clumpons.  On a periodic domain,
 such an arrangement is an equilibrium state, although it is unstable.  Thus,
 by impulsively shifting the clumpon at $x=0$ by a small amount, we force
 the clumpons to collapse into larger clumpons, until only a single clumpon
 remains.
 }
\label{fig:V_time_ic1}
\end{figure}

To highlight the interaction between clumpons, we examine the following set
of initial conditions,
\begin{equation}
\Mag\left(x,0\right)=\Mag_0=\text{const.},\qquad
\ordp\left(x,0\right)=0.5+0.35\cos\left(8k_0x\right),
\label{eq:initial_conditions2}
\end{equation}
where $k_0=2\pi/L$ is the fundamental wavenumber.   Since this set of initial
conditions contains a large number of density maxima, we expect a large number
of closely-spaced clumpons to emerge, and this will illuminate the
clumpon interactions.
The time evolution of equations~\eqref{eq:mag_dens} for this set of initial
conditions is shown in Fig.~\ref{fig:evolution_ic1}.
As before, the solution becomes singular after a short time, and is subsequently
represented  by a sum of clumpons,
\[
\ordp\left(x,t\right)=\sum_{i=1}^M a_i\delta\left(x-x_i\left(t\right)\right),\qquad
\Mag\left(x,t\right)=\sum_{i=1}^M \bm{b}_i\left(t\right)\delta\left(x-x_i\left(t\right)\right),\qquad
M=8.
\]
Here $M=8$ is the number of clumpons at the singularity time.  This number
corresponds to the number of maxima in the initial density profile.
As before, this configuration of equally spaced, identical clumpons is an
equilibrium state, due to periodic boundary conditions.  Therefore, once
the particle-like state has formed, we impulsively shift the clumpon at $x=0$
by a small amount, and precipitate the merging of clumpons.
The eight clumpons then merge repeatedly until only a single clumpon remains.
%
%
%
%
%
The tendency for the clumpons to merge is explained by the velocity $V$,
which changes sign across a clumpon.  Thus, if a clumpon is within the range
of the force exerted by its neighbours, the local velocity, if unbalanced,
 will advect a given clumpon
\begin{figure}[htb]
\centering
\subfigure[]{
\includegraphics[width=0.35\textwidth]{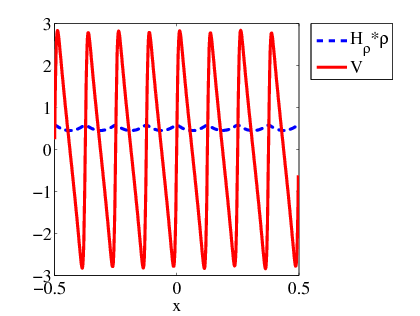}
}
\subfigure[]{
\includegraphics[width=0.35\textwidth]{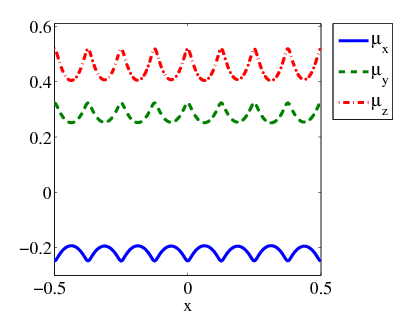}
}
\caption{Evolution of a flat magnetization field and a sinusoidally-varying
density, as in Eq.~\eqref{eq:initial_conditions2}.
 Subfigure~(a) shows the system velocity $V$ given in Eq.~\eqref{eq:velocity}
just before the singularity time; (b) gives the magnetization $ \bm{\mu}_\Subm$
at the same time.  The density maxima emerge at the locations where the convergence
of $-V$ occurs, and the magnetization develops extrema there.}
\label{fig:snapshot_ic1} 
\end{figure} 
in the direction of one of its neighbours, and the clumpons merge.  This
process is shown in Fig.~\ref{fig:V_time_ic1}.  

\section{Summary}
\label{sec:conclusions}
We have investigated the nonlocal Gilbert (NG) equation introduced by Holm,
Putkaradze, and Tronci in~\cite{Darryl_eqn1} using a combination of numerical
simulations and simple analytical arguments.  The NG equation contains two
competing lengthscales of nonlocality: there is a lengthscale $\alpha$
associated with the range of the interaction potential, and a lengthscale
$\beta$ that governs the smoothened magnetization vector that appears in
the equation.  When $\alpha<\beta$ all initial configurations of the magnetization
tend to a constant value, while for $\beta<\alpha$ the initial configuration
of the magnetization field develops finer and finer scales.  These two effects
are in balance when $\alpha=\beta$, and the system does not evolve away
from its initial state.  Furthermore, the NG equation conserves the norm
of the magnetization vector $\Mag$, thus providing a pointwise bound on the
solution and preventing the formation of singular solutions.

To study the formation of singular solutions, we couple the NG equation to
a
scalar density equation.  Associated with the scalar density is a negative
energy of attraction that drives the formation of singular solutions and
breaks the pointwise bound on the $\Mag$.  
Three lengthscales of nonlocality
now enter into the problem: the range of the force associated with the scalar
density, the range of the force due to the magnetization,
and the smoothening length.  As before, the competition of lengthscales is crucial to the evolution of the system; this is seen in the
linear stability analysis of the coupled equations, in which the relative
magnitude of the lengthscales determines the stability or otherwise of a
constant state.

Using numerical simulations, we
have demonstrated the emergence of singular solutions from smooth initial
data, and have explained this behaviour by the negative energy of attraction
produced by the scalar density.  The singular solution consists of a weighted
sum of delta functions, given in Eq.~\eqref{eq:clumpon_sln},
which we interpret 
as interacting particles or clumpons.  The clumpons evolve under simple finite-dimensional
dynamics.  We have shown that a system of two clumpons is governed by a two-dimensional
dynamical system that has a multiplicity of steady states.  Depending on
the lengthscales of nonlocality and the clumpon weights, the two clumpons
can merge, diverge, or align and remain separated by a fixed distance.

%% file: conclusion/conclusion.tex
\chapter{Conclusions}
\label{ch:conclusions}

The advective Cahn--Hilliard equation, describing the twin effects of advection
and phase separation, has been the subject of this report.  We have considered
cases with and without feedback from concentration gradients into the advecting
flow field.

In Ch.~\ref{ch:background} we introduced the theory necessary for the report,
and one new result: an existence theory for the passively-advected Cahn--Hilliard
equation.  Although an important exercise, such existence theories generally
tell us no more than the regularity class to which the solution belongs.
 Therefore, we resorted to numerical simulations
in Ch.~\ref{ch:chaotic_advection}, considering phase separation in the presence
of a passive, chaotic flow.  We characterized the model random-phase sine
flow by its Lyapunov exponent and correlation length.  We quantified the
phase separation using the morphology, lengthscales, and probability distribution
function of the concentration field.  We showed how a chaotic flow will not
only arrest the growth of the phase-separated domain structure, but will
lead to remixing of the binary fluid.  We highlighted the importance of this
result in applications where the coarsening tendency of the binary fluid
is undesirable.

Continuing in this vein, in Ch.~\ref{ch:estimating_mixedness} we defined
what it means for a symmetric binary fluid to be well mixed and, given
a smooth stirring velocity, and sources and sinks, we obtained very precise
bounds on the level of mixedness achievable by the flow.  The question then
arises, given a certain source term, is there a best flow for mixing a binary
fluid?  Unlike in the theory
of miscible liquids, we could not give a complete answer, but noted that
the random-phase sine flow of Ch.~\ref{ch:chaotic_advection} is efficacious,
but not necessarily optimal, at mixing.

In Ch.~\ref{ch:analysis_thin_films}, we switched focus to the active Cahn--Hilliard
equation, which is coupled to the Navier--Stokes equations.  By specializing
to the case where the binary fluid forms a thin layer on a substrate, it
was possible to obtain new, coupled equations, describing the coevolution
of the film's free-surface height and the fluid concentration, the so-called
Stokes
Cahn--Hilliard equations.  Since we were interested in the late-time behaviour
of the system, we focussed on the cases where film rupture was impossible,
by introducing a repulsive Van der Waals force.  In that case, we obtained
existence, regularity, and uniqueness results for solutions to the Stokes
Cahn--Hilliard
equations in two dimensions, and then passed over to numerical simulations
in two and three dimensions in Ch.~\ref{ch:simulations_thin_films}.  The
most salient result of this chapter was obtained by applying a spatially-varying
surface tension across the film.  Then, for moderate backreaction strengths,
the binary fluid domains aligned with the direction of this variation, while
for zero backreaction, the domains aligned in the perpendicular direction.
 This surprising result highlights the importance of coupled models, and
 gives a means of controlling phase separation.
 
In Ch.~\ref{ch:singular_solutions} we formulated a general theory of aggregation
equations that includes as special cases the Cahn--Hilliard, Keller--Segel,
and (scalar) Holm--Putkaradze equations.  Motivated by this link between
the theory we have developed and the nanoscale physics described by the Holm--Putkaradze
equation, we studied a coupled vector-scalar system of aggregation equations
that describes the formation of magnetic nanoparticles.  Although this system
was introduced by Holm, Putkaradze, and Tronci in~\cite{Darryl_eqn1}, its
properties had yet to be studied.  Therefore, using numerical and analytical
techniques, we highlighted the competition of
lengthscales in a purely magnetic system, and demonstrated that singular
solutions are never possible in such a model.  By passing to the coupled
vector-scalar model, we demonstrated the emergence of singular solutions
from smooth initial data.  We obtained a set of ordinary differential equations
satisified by the discrete elements of the singular solution (clumpons).
 By focussing on the two-clumpon case, we showed rigorously that the two
 clumpons can merge, diverge, or align and remain at a fixed distance.  We
 thus gave a full characterization of the Holm--Putkaradze--Tronci model
 of nanomagnetic particles.

There are many avenues for future work.
 The advective Cahn--Hilliard equation could be used to give a full picture
 the breakup and coalescence of droplets with high surface tension, an area
 in
 which there are many experiments but little theory, while more
 work could be done in finding an optimal flow for mixing the binary fluid.
  Computational topology~\cite{Wanner2004} could be used to give an extra
  dimension to the characterization of concentration morphology in the presence
  of flow.
  On the other hand, more experiments on and modelling of the thin-film problem
  would shed further light on this aspect of NSCH dynamics.  The relevance
  of Cahn--Hilliard and thin-film type equations to the theory of self-assembly
  could be probed further.  It is the author's hope to work on at least some
  of these problems in the future.

%% file: bibliography/bibliography.tex


%% file: appendix_a/appendix_a.tex
\chapter{The integral $\int_0^{\pi/2}{du}\log\cos\left(u\right)$}
\label{appendix_a}

\renewcommand{\theequation}{A-\arabic{equation}}
\renewcommand{\thefigure}{A-\arabic{figure}}
\setcounter{equation}{0}  
\setcounter{figure}{0}

We compute the integral
\begin{equation}
I=\int_0^{\pi/2}{du}\log\cos\left(u\right).
\end{equation}
Making the substitution $t=\cos^2\left(u\right)$, we have,
\begin{eqnarray*}
I&=&\tfrac{1}{4}\int_0^1{dt}\frac{\log\left(t\right)}{\sqrt{t}\sqrt{1-t}},\\
&=&\tfrac{1}{4}\left[\frac{d}{d\alpha}\int_0^1{dt} \left(1-t\right)^{-1/2}t^{\alpha-1/2}\right]_{\alpha=0},\\
&=&\tfrac{1}{4}\left[\frac{d}{d\alpha}B\left(\alpha+\tfrac{1}{2},\tfrac{1}{2}\right)\right]_{\alpha=0},\\
&=&\tfrac{1}{4}\left[\frac{d}{d\alpha}\frac{\Gamma\left(\tfrac{1}{2}\right)\Gamma\left(\tfrac{1}{2}+\alpha\right)}{\Gamma\left(1+\alpha\right)}\right]_{\alpha=0},\\
\end{eqnarray*}%
where $B\left(x,y\right)$ and $\Gamma\left(x\right)$ are the beta and gamma
functions respectively.  Using the standard identities
\[
\Gamma\left(1\right)=1,\qquad\Gamma\left(\tfrac{1}{2}\right)=\sqrt{\pi},\qquad
\Gamma'\left(\tfrac{1}{2}\right)=-\sqrt{\pi}\left(\gamma+2\log 2\right),\qquad \Gamma'\left(1\right)=-\gamma,
\]
where $\gamma$ is the Euler--Mascheroni constant, the integral $I$ becomes
\begin{equation}
I=-\tfrac{1}{2}\pi\log 2.
\end{equation}
%
%
%

%% file: appendix_b/appendix_b.tex
\chapter{The regularized Keller--Segel equation}

\renewcommand{\theequation}{B-\arabic{equation}}
\renewcommand{\thefigure}{B-\arabic{figure}}
\setcounter{equation}{0}  
\setcounter{figure}{0}

The authors Keller and Segel introduced their eponymous equation in a celebrated
paper that describes the aggregation of the slime mould
\emph{Dictyostelium discoideum}~\cite{KellerSegel1970}.  Here we outline
the derivation of a regularized Keller--Segel equation.  This particular
model has been considered by previous authors~\cite{Chavanis2006}, and general
discussions of the Keller--Segel model are given in~\cite{Segel1985,Horstmann}.

The Keller--Segel equation is an equation for the density $\rho\left(\bm{x},t\right)$
of a biological population that evolves in time through self-advection and
diffusion,
\[
\frac{\partial \rho}{\partial t}+\nabla\cdot\left(\bm{v}\rho\right)=\nabla\cdot\left(D\nabla\rho\right),
\]
where $\bm{v}=\chi\left(\rho\right)\nabla c$ is the motion of individual
elements due to gradients in the concentration field $c\left(\bm{x},t\right)$.
 This velocity equation can be thought of as a Darcy's Law for a potential
 $-c$, although in a biological context, it represents the tendency of the
 population to move towards regions rich in the chemical whose concentration
 is $c\left(\bm{x},t\right)$.  In the situation described by Keller and Segel,
 the chemical $c$ is in fact produced by individual elements in the population.
  That is, an element of the population produces a chemical that attracts
  its neighbours, which are then advected towards the producer.  This is
  encoded in the concentration equation
\[
\frac{\partial c}{\partial t}=D_c\Delta c + g\left(\rho,c\right),
\]
where the function $g\left(\rho,c\right)$ is a growth law, here taken to
be $g\left(\rho,c\right)=-bc+a\rho$.  Thus, the concentration
degrades linearly (in time) of its own accord, but increases in the presence
of a finite population $\rho$.  In many applications, this concentration
field relaxes quickly to a value for which $\partial_t c=0$ and hence
\[
c=\frac{a}{D_c}\left(-\Delta^2+\frac{b}{D_c}\right)^{-1}\rho,
\]
or
\[
c\left(\bm{x},t\right)=\int\frac{d^n k}{\left(2\pi\right)^n}\frac{a}{D_c}\frac{1}{k^2+\frac{b}{D_c}}e^{i\bm{k}\cdot\bm{x}}\tilde{\rho}_{\bm{k}}=K_{\mathrm{KS}}*\rho.
\]

We now examine a Keller--Segel model with a velocity field $\bm{v}=\chi_0\left(1-\rho/\sigma_0\right)\nabla
c$ so that the elements of the population do not move once a maximum density
$\sigma_0$ is attained; this takes into account the finite size of the individual
elements of the population.  This approach has
been used in several studies (see, for example, reference~\cite{Chavanis2006}),
and is known to prevent the
formation of delta-function singularities that would otherwise occur in
the density field $\rho\left(\bm{x},t\right)$.  To find out what the appropriate
free-energy functional should be, let us work backwards from the evolution
equation,
\begin{eqnarray*}
\frac{\partial\rho}{\partial t} &=&\nabla\cdot\left(D\nabla\rho-\chi_0\rho\left(1-\frac{\rho}{\sigma_0}\right)\nabla
c\right),\\
&=&\nabla\cdot\left[D\rho\left(1-\frac{\rho}{\sigma_0}\right)\left(\frac{\nabla\rho}{\rho\left(1-\rho/\sigma_0\right)}-\frac{\chi_0}{D}\nabla
c\right)\right],\\
&=&\nabla\cdot\left[D\rho\left(1-\frac{\rho}{\sigma_0}\right)\left(\nabla\log\left(\frac{\rho}{\sigma_0-\rho}\right)-\frac{\chi_0}{D}\nabla
c\right)\right].
\end{eqnarray*}
Let us take $\chi_0$ and $D$ to be constants.  Then,
\begin{eqnarray}
\frac{\partial\rho}{\partial t}&=&\nabla\cdot\left[D\rho\left(1-\frac{\rho}{\sigma_0}\right)\nabla\left(\log\left(\frac{\rho}{\sigma_0-\rho}\right)-\frac{\chi_0}{D}c\right)\right],\nonumber\\
&=&\nabla\cdot\left(\chi_0\rho\left(1-\frac{\rho}{\sigma_0}\right)\nabla\frac{\delta
F_{\mathrm{KS}}}{\delta \rho}\right),\nonumber\\
&=&\nabla\cdot\left(M\left(\rho\right)\nabla\mu_{\mathrm{KS}}\right),
\label{eq:grad_flow_KS}
\end{eqnarray}
Where we have identified the chemical potential 
\[
\mu=\delta F_{\mathrm{KS}}/\delta\rho=\log\left[\frac{\rho/\sigma_0}{1-\left(\rho/\sigma_0\right)}\right]-\frac{\chi_0}{D}c=W'\left(\rho/\sigma_0\right)-\frac{\chi_0}{D}c,
\]
and the mobility $M\left(\rho\right)=D\rho\left(1-\rho/\sigma\right)$.  The
free energy is thus
\[
F_{\mathrm{KS}}=\sigma_0\int d^n x W\left(\frac{\rho}{\sigma_0}\right)-\frac{\chi_0}{2D}\int
d^n x\rho\left(\bm{x},t\right)\int d^n y K_{\mathrm{KS}}\left(\left|\bm{x}-\bm{y}\right|\right)\rho\left(\bm{y},t\right),
\]
with
\[
\mu_{\mathrm{KS}}=W'\left(\rho/\sigma_0\right)-\frac{\chi_0}{D}K_{\mathrm{KS}}*\rho=W'\left(\rho/\sigma_0\right)-\frac{\chi_0}{D}c.
\]
Let us take $\psi=\rho/\sigma_0$ and $\tau=\chi_0 t$, $\xi = D/\chi_0$,
$\lambda^2 = D_c/\sigma_0 a$ and $B=b/\sigma_0 a$.  Then the model is
\begin{equation}
\frac{\partial\psi}{\partial\tau}=\nabla\cdot\left[\psi\left(1-\psi\right)\nabla\left(\xi
W'\left(\psi\right)-\int\frac{d^n k}{\left(2\pi\right)^n}\frac{\tilde{\psi}_{\bm{k}}e^{i\bm{k}\cdot\bm{x}}}{\lambda^2
k^2+B}\right)\right],
\end{equation}
which in form is similar to the Cahn--Hilliard equation with double-obstacle
potential and variable mobility
\[
\frac{\partial\psi}{\partial t}=\nabla\cdot\left[\psi\left(1-\psi\right)\nabla\left(\xi
W'\left(\psi\right)-\int\frac{d^n k}{\left(2\pi\right)^n}\tilde{\psi}_{\bm{k}}e^{i\bm{k}\cdot\bm{x}}\left(2-\gamma
k^2\right)\right)\right].
\]